\documentclass[11pt, english, singlespacing]{sty/thesis}

\let\svthefootnote\thefootnote
\usepackage{sty/mythesis}
\usepackage{textcomp}
\usepackage{gensymb}
\usepackage{pifont}
\usepackage{url}
\usepackage{xspace}
\usepackage{footmisc}
\usepackage[normalem]{ulem}
\usepackage{float}
\usepackage{graphicx}
\usepackage{booktabs}
\usepackage{multirow}
\usepackage{color}
\usepackage{xcolor}
\usepackage{pifont}
\usepackage{caption}
\usepackage{subcaption}
\usepackage{amsmath}
\usepackage[numbers, sort]{natbib}
\usepackage{setspace}

\let\oldcelsius\celsius
\renewcommand{\celsius}{~\oldcelsius\xspace}

\newcommand{\ra}{$\rightarrow$\xspace}

\newcommand{\numone}  {{\large\ding{202}}\xspace}
\newcommand{\numtwo}  {{\large\ding{203}}\xspace}
\newcommand{\numthree}{{\large\ding{204}}\xspace}
\newcommand{\numfour} {{\large\ding{205}}\xspace}
\newcommand{\numfive} {{\large\ding{206}}\xspace}

\newcommand{\squishlist} {
	\begin{list}{$\bullet$} {
		\setlength{\itemsep}{0pt}
		\setlength{\parsep}{3pt}
		\setlength{\topsep}{3pt}
		\setlength{\partopsep}{0pt}
		\setlength{\leftmargin}{1.0em}
		\setlength{\labelwidth}{1em}
		\setlength{\labelsep}{0.5em}
	}
}
\newcommand{\squishend} {
	\end{list}
}

\newcommand{\cmdact}{$\mathtt{ACTIVATION}$\xspace}
\newcommand{\cmdpre}{$\mathtt{PRECHARGE}$\xspace}
\newcommand{\cmdrd} {$\mathtt{READ}$\xspace}
\newcommand{\cmdwr} {$\mathtt{WRITE}$\xspace}

\newcommand{\trcd}    {$t_{\mathit{RCD}}$\xspace}
\newcommand{\tras}    {$t_{\mathit{RAS}}$\xspace}
\newcommand{\trp}     {$t_{\mathit{RP}}$\xspace}
\newcommand{\twr}     {$t_{\mathit{WR}}$\xspace}
\newcommand{\trc}     {$t_{\mathit{RC}}$\xspace}
\newcommand{\tcl}     {$t_{\mathit{CL}}$\xspace}
\newcommand{\tcwl}    {$t_{\mathit{CWL}}$\xspace}
\newcommand{\tbl}     {$t_{\mathit{BL}}$\xspace}
\newcommand{\trcdfar} {$t_{\mathit{RCDfar}}$\xspace}
\newcommand{\trcfar}  {$t_{\mathit{RCfar}}$\xspace}
\newcommand{\trcdnear}{$t_{\mathit{RCDnear}}$\xspace}
\newcommand{\trcnear} {$t_{\mathit{RCnear}}$\xspace}
\newcommand{\trasnear}{$t_{\mathit{RASnear}}$\xspace}
\newcommand{\trasfar} {$t_{\mathit{RASfar}}$\xspace}
\newcommand{\trpnear} {$t_{\mathit{RPnear}}$\xspace}
\newcommand{\trpfar}  {$t_{\mathit{RPfar}}$\xspace}

\newcommand{\vdd}  {$V_{\mathit{DD}}$\xspace}

\newcommand{\qvdd} {$\frac{1}{4}V_{\mathit{DD}}$\xspace}
\newcommand{\hvdd} {$\frac{1}{2}V_{\mathit{DD}}$\xspace}
\newcommand{\tfvdd}{$\frac{3}{4}V_{\mathit{DD}}$\xspace}

\newcommand{\msc}{SC\xspace}
\newcommand{\mwmc}{WMC\xspace}
\newcommand{\mbbc}{BBC\xspace}

\newcommand{\ALD}[0]{AL-DRAM\xspace}
\newcommand{\DIMMs}[0]{115\xspace}

\newcommand{\ReadHot}[0]{21.1}
\newcommand{\WriteHot}[0]{34.4}
\newcommand{\trcdHot}[0]{15.6}
\newcommand{\trasHot}[0]{20.4}
\newcommand{\twrHot}[0]{20.6}
\newcommand{\trpHot}[0]{28.5}
\newcommand{\ReadCold}[0]{32.7}
\newcommand{\WriteCold}[0]{55.1}
\newcommand{\trcdCold}[0]{17.3}
\newcommand{\trasCold}[0]{37.7}
\newcommand{\twrCold}[0]{54.8}
\newcommand{\trpCold}[0]{35.2}

\newcommand{\dimms}[1]{96\xspace}

\newcommand{\navdimms}[0]{24\xspace}

\newcommand{\mycolor}[1]{\textcolor{black}{#1}}

\renewcommand\footnotemark{}

\thesistitle{\bf {\LARGE Reducing DRAM Latency at Low Cost \\by Exploiting
Heterogeneity}}

\begin{document}

	\maketitle
	\pagenumbering{roman}

	\newpage

	\vspace*{35pt}

	\pagestyle{fancy}
	\renewcommand{\headrulewidth}{0pt}
	\lhead{}\rhead{}\cfoot{\thepage}

	\begin{center}
	\vspace*{20pt}
	\textbf{\LARGE Abstract} \\
	\vspace*{10pt}
	\end{center}

	\emph{
		In modern systems, DRAM-based main memory is significantly slower than the
processor. Consequently, processors spend a long time waiting to access data
from main memory, making the long main memory access latency one of the most
critical bottlenecks to achieving high system performance. Unfortunately, the
latency of DRAM has remained almost constant in the past decade. This is mainly
because DRAM has been optimized for cost-per-bit, rather than access latency.
As a result, DRAM latency is not reducing with technology scaling, and
continues to be an important performance bottleneck in modern and future
systems.

This dissertation seeks to achieve low latency DRAM-based memory systems at low
cost in three major directions. \mycolor{The key idea of these three major
directions is to enable and exploit latency heterogeneity in DRAM
architecture.} First, based on the observation that long bitlines in DRAM are
one of the dominant sources of DRAM latency, we propose a new DRAM
architecture, Tiered-Latency DRAM (TL-DRAM), which divides the long bitline
into two shorter segments using an isolation transistor, allowing one segment
to be accessed with reduced latency. Second, we propose a fine-grained DRAM
latency \mycolor{reduction} mechanism, Adaptive-Latency DRAM, which optimizes
DRAM latency for the common operating conditions \mycolor{for individual DRAM
module}. We observe that DRAM manufacturers incorporate a very large timing
margin as a provision against the worst-case operating \mycolor{conditions},
which is accessing the \mycolor{slowest} cell \mycolor{across all DRAM
products} with the worst latency at the highest temperature, even though
\mycolor{such a slowest cell and} such an operating condition \mycolor{are
rare}. Our mechanism dynamically optimizes DRAM latency to the current
operating condition \mycolor{of the accessed DRAM module, thereby} reliably
improving system performance. Third, we observe that cells closer to the
peripheral logic can be much faster than cells farther from the peripheral
logic (\mycolor{a phenomenon we call} architectural variation).  Based on this
observation, we propose a new technique, Architectural-Variation-Aware DRAM
(AVA-DRAM), which reduces DRAM latency at low cost, by profiling and
identifying only the inherently slower regions in DRAM to dynamically determine
the lowest latency DRAM can operate at without causing failures.

This dissertation provides a detailed analysis of DRAM latency by using both
circuit-level simulation with a detailed DRAM model and FPGA-based profiling of
real DRAM modules. Our latency analysis shows that our low latency DRAM
mechanisms enable significant latency reductions, leading to \mycolor{large
improvement in} both system performance and energy efficiency \mycolor{across}
a variety of workloads in our evaluated systems, while ensuring reliable DRAM
operation.

	}

	\vspace*{\fill}
	\newpage

	\newpage

	\newpage

	\vspace*{35pt}

	\pagestyle{fancy}
	\renewcommand{\headrulewidth}{0pt}
	\lhead{}\rhead{}\cfoot{\thepage}

	\begin{center}
	\vspace*{20pt}
	\textbf{\LARGE Acknowledgement} \\
	\vspace*{10pt}
	\end{center}

	The last five years at Carnegie Mellon University have been most exciting time
of my life, thanks to all of the fantastic people that I have met and worked
together with. First and foremost, I am grateful to my advisor, Prof. Onur
Mutlu. He provided a great opportunity for me to join his research group as a
Ph.D. student and provided great guidance to do not only all the works in this
dissertation but also many other works. He taught me to think differently and
thoroughly to determine real-world problems and to find better ways to solve
these problems, leading to making an impact on the real world. Prof. Mutlu
always supported me and encouraged me to improve all my works and my abilities.
I am also very thankful to Prof. Mutlu for providing me with a great research
environment. Under his support, providing all of the required resources, I
could focus on my research and collaborate greatly with many fantastic people.

I would like to thank my thesis committee members, Prof. Todd Mowry, Prof.
Kayvon Fatahalian, Prof. Shih-Lien Lu, and Prof. Mattan Erez for their time,
efforts, and comments in bringing this dissertation to completion. Special
thanks to Prof. Erez for his precise feedback on the entire dissertation.
Thanks to Prof. Mowry for his guidance on my research, which started from my
qualifying examination. Thanks to Prof. Lu for his interest in my research and
valuable feedback. Thanks to Prof. Kayvon Fatahalian for his valuable comments
on my research. I would also like to thank Prof. Rajeev Balasubramonian, Dr.
Michael Kozuch and Konrad Lai for their interests and great feedback on my
research.

The SAFARI group has been like my home and family. Without support from SAFARI
members, this dissertation could not have been completed. Yoongu Kim has always
been my good friend and mentor. I am really thankful to all his support over
every step of my graduate school career. I mostly followed his mentoring, which
made it possible to finish this dissertation. Thanks to Vivek Seshadri for his
incisive insight and plentiful helps, which improved all works in this
dissertation. His optimistic attitude on research and life impressed me.
Thanks to Lavanya Subramanian for all her great support, and for many valuable
discussions on research and life. Whenever I faced problems, she was always
there and provided priceless suggestions. Thanks to Samira Khan for her
kindness to listen to all of my problems and provide valuable suggestions. With
her dedicated support, I was able to keep working and finish this dissertation.
Thanks to Gennady Pekhimenko for his valuable comments on my ideas and
suggestions on research directions. His enthusiasm on research and endless
efforts to achieve each of his goals impressed me, which made me to put more
efforts into my research. Thanks to Hongyi Xin for being a great office mate
over four and half years. He taught me the basics of programming and provided
valuable guidance on being a programmer. He made my graduate life more exciting
and fruitful. Thanks to Saugata Ghose for his critiques. He provided great
feedback and helped to improve my work. Thanks to Rachata Ausavarungnirun for
his kindness to discuss anything on research and life. He always supported and
encouraged me to be a better researcher. Thanks to Chris Fallin for his critical
feedback on my research. He provided great help when I prepared for my qualifying
examination and conference presentations. His endless efforts and enthusiasm on
research impressed me. Thanks to Kevin Chang for his friendly nature and great
help in building the DRAM infrastructure. Thanks to Justin Meza, Hanbin Yoon, Ben
Jayiyen, Jamie Liu, Nandita Vijayakumar, Yang Li, Kevin Hsieh, Amirali
Boroumand, Jeremie Kim, Damla Senol, and Minesh Patel for many discussions and
feedback on my research.

Beside the members of the SAFARI group, many people have supported me as I have
finished all these works. Thanks to Prof. Jongmoo Choi for his valuable feedback
on my research from a system-level perspective. Thanks to Prof. Can Alkan,
Farhad Hormozdiari, and Faraz Hach for all their guidance on my DNA sequencing
research~\cite{lee-methods2014, xin-genomics2013}. Thanks to Ahmad Khairi,
Cheng-Yuan Wen, Jaewon Choi, Sandipan Kundu, Shadi Saberi, and Prof. Jeyanandh
Paramesh for their help on my first-year research. Thanks to Elaine Lawrence,
Samantha Goldstein, Karen Lindenfelser, Nathan Snizaski, Marilyn Patete, Olivia
Vadnis, and Jennifer Gabig for their administrative support. Thanks to John and
Claire Bertucci for providing me with their fellowship. Thanks to Prof. Andrzej
Strojwas and Prof. Brandon Lucia for providing TA opportunities. Thanks to
Jinkyu Kim, Soonho Kong, Yongjun Kim, Abhishek Sharma, and Hyoseung Kim for
their friendship, which has helped me complete this dissertation.

My family has always provided me with endless support. Thanks to my parents,
Seungjoon lee and Hanki Yang, for their encouragement, support, and love.
Thanks to my parents-in-law, Byunghwan Choi and Soonja Lee, for their
understanding and great support. My brothers and their families, Dongshin Lee,
Grace Kim, Hyunseok Choi, Jungyoon Heo, and Yewon Choi, also deserve many
thanks for all of their support.

Finally, I would like to thank to my wife, Woonjung Choi, for her devotion
supports. She has always been by my side and provided endless love and support
during my Ph.D. She always provided a comfortable environment for me to focus on
my research, understood me, and encouraged me to make progress on my work. I
could not have been completed any of the work in this dissertation without her
support. \\

\noindent Donghyuk Lee \\
\noindent April 2016, Pittsburgh, PA

	\vspace*{\fill}
	\newpage

	\newpage

	\tableofcontents
	\newpage

	\doublespacing
	\setcounter{page}{1}
	\pagenumbering{arabic}
	\pagestyle{thesis}

	\chapter{{Introduction}} \label{ch:intro}

\section{Problem} \label{ch:intro_problem}

Primarily due to its low cost-per-bit, DRAM has long been the choice substrate
for architecting main memory systems. In fact, DRAM's cost-per-bit has been
decreasing at a rapid rate as DRAM process technology scales to integrate ever
more DRAM cells into the same die area. As a result, each successive generation
of DRAM has enabled increasingly large-capacity main memory subsystems at low
cost.

\begin{figure}[h]
	\centering
	\includegraphics[width=0.6\linewidth]{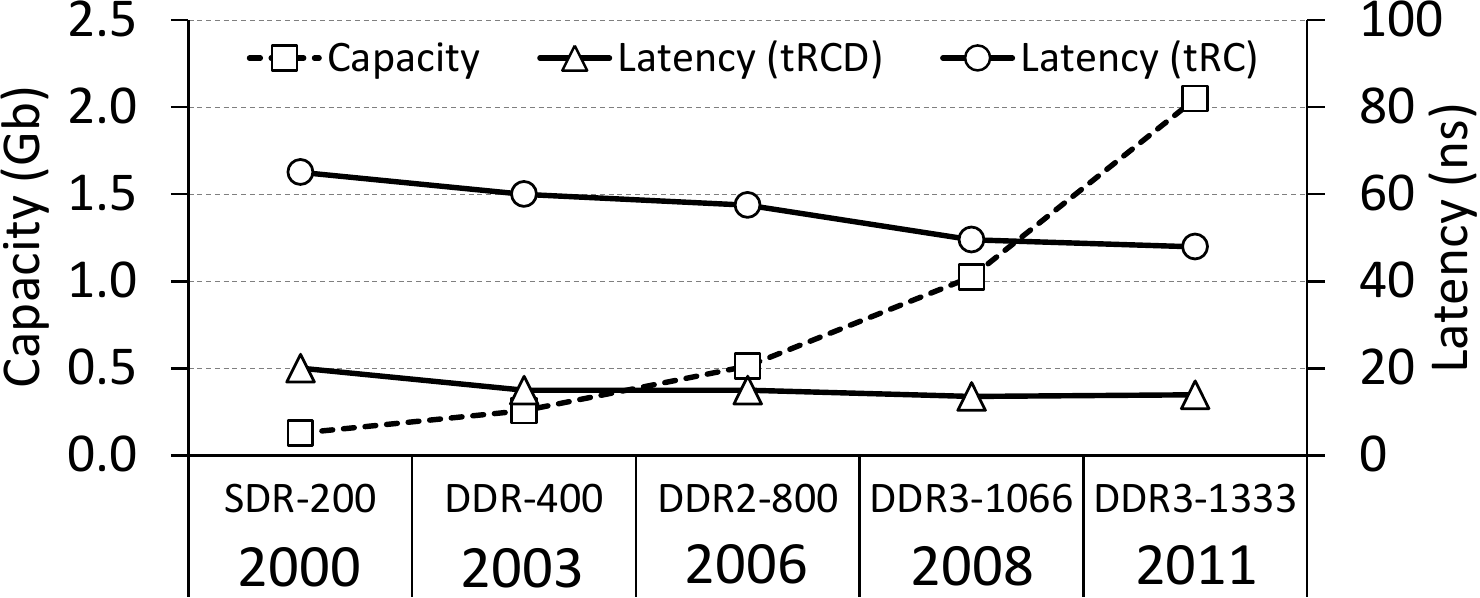} \\
	\footnotesize{ $\dagger$ We refer to the dominant DRAM chips during the
	period of time~\cite{borkar-cacm2011, samsung-roadmap}.}
	\caption{DRAM Capacity \& Latency Over Time~\cite{borkar-cacm2011,
	samsung-roadmap, latencylag, samsung-spec}}
	\label{fig:intro_trend}
\end{figure}

In stark contrast to the continued scaling of cost-per-bit, the {\em latency}
DRAM has remained almost constant, During the same 11-year interval in which
DRAM's cost-per-bit decreased by a factor of 16, DRAM latency (as measured by
the \trcd and \trc timing constraints) decreased by only 30.5\% and 26.3\%, as
shown in Figure~\ref{fig:intro_trend}. From the perspective of the processor,
an access to DRAM takes hundreds of cycles -- time during which a modern
processor is likely stalled, waiting for DRAM~\cite{mutlu-hpca2003,
mutlu-ieeemicro2003, ailamaki-vldb1999, mutlu-isca2005, mutlu-isca2008,
mutlu-ieeemicro2006, glew-asplos1998}. Such wasted time, which is more than
50-60\% of the execution time for many memory-intensive
workloads~\cite{mutlu-hpca2003, mutlu-isca2005, ailamaki-vldb1999,
mutlu-ieeemicro2006}, leads to large performance degradations commonly referred
to as the ``memory wall''~\cite{wulf-sigarch1995} or the ``memory
gap''~\cite{wilkes-sigarch2001}.

However, the high latency of commodity DRAM chips is in fact a {\em deliberate}
trade-off made by DRAM manufacturers. While process technology scaling has
enabled DRAM designs with both lower cost-per-bit {\em and} lower
latency~\cite{itrs}, DRAM manufacturers have usually sacrificed the latency
benefits of scaling in order to achieve even lower cost-per-bit. Hence, while
low-latency DRAM chips exist~\cite{kimuta-isscc1999, rldram, sato-vlsi1998},
their higher cost-per-bit relegates them to specialized applications such as
high-end networking equipment that \mycolor{demand} very low latency even
\mycolor{at the expense of} a very high cost~\cite{ciaran-ahs2007}.

{\em Our main research objective is to enable a low latency DRAM-based memory
system, in order to achieve high system performance,\mycolor{via} simple and
low cost DRAM architectures and memory control techniques.}

\section{Our Approach} \label{ch:intro_solution}

Towards achieving our goal of \mycolor{a} low latency DRAM-based memory system,
our major approach is to enable or exploit {\em latency heterogeneity} in DRAM.
Due to the large number of cells in DRAM, there already exists such latency
heterogeneity (latency variation) \mycolor{due to} multiple aspects. For
example, {\em i)} DRAM cells in different locations \mycolor{in the same chip
or across different chips} have different latencies, and {\em ii)} DRAM has
different latencies at different operating conditions (e.g., temperature).
However, to provide a simplified interface to access DRAM, DRAM has been
designed to improve only the {\em worst case latency}. One simple example is
that DRAM uses {\em identical} timing parameters for accessing {\em all} DRAM
cells \mycolor{from {\em any} DRAM chip} at {\em all} operating conditions,
even though the latency can be very different based on the location of each
cell, \mycolor{the characteristics of different chips, or the characteristics}
of each cell at different operating conditions. Therefore, DRAM has been
focused on either minimizing these latency variations in its architecture or
hiding them. Unfortunately, the latency variation in DRAM is expected to
increase with further DRAM cell scaling due to worsening process variation at
smaller process technology nodes. Therefore, enabling low DRAM latency is
expected to be more difficult in the future.

Our approach for lowering DRAM latency is enabling or exploiting \mycolor{such}
latency heterogeneity by {\em i)} rearchitecting DRAM \mycolor{at} low cost,
{\em ii)} enabling fine-grained DRAM control to optimize DRAM latency for the
common operating conditions, and {\em iii)} exploiting the latency variation
\mycolor{inherent in} the internal DRAM architecture.

\subsection{Lowering DRAM Latency by Rearchitecting DRAM Bitline Architecture
(Tiered-Latency DRAM)} \label{ch:intro_tldram}

DRAM has been developed to make its capacitor-based cell smaller for higher
capacity. The small size of this capacitor necessitates the use of an auxiliary
structure, called a {\em (local) sense amplifier}, to detect the small amount
of charge held by the cell and amplify it to a full digital logic value. But, a
local sense amplifier is approximately one hundred times larger than a
cell~\cite{rambus-power}. To amortize its large size, each local sense
amplifier is connected to many DRAM cells through a wire called a {\em
bitline}. Every bitline has an associated parasitic capacitance whose value is
proportional to the length of the bitline. Unfortunately, such parasitic
capacitance slows down DRAM operation for two reasons. First, it increases the
latency of the local sense amplifiers. When the parasitic capacitance is large,
a cell cannot quickly create a voltage perturbation on the bitline that could
be easily detected by the local sense amplifier. Second, it increases the
latency of charging and precharging the bitlines. Although the cell and the
bitline must be restored to their quiescent voltages during and after an access
to a cell, such a procedure takes much longer when the parasitic capacitance is
large. Due to the above reasons and a detailed analysis of the latency
break-down~\cite{lee-hpca2013}, we conclude that long bitlines are the dominant
source of DRAM latency.

The bitline length is a key design parameter that exposes the important
trade-off between latency and die-size (cost). Short bitlines (few cells per
bitline) constitute a small electrical load (parasitic capacitance), which
leads to low latency. However, they require more local sense amplifiers for a
given DRAM capacity, which leads to a large die-size. In contrast, long
bitlines have high latency and a small die-size. As a result, neither of these
two approaches can optimize for both latency and cost-per-bit. The goal in this
work is to design a new DRAM architecture to approximate the best of both
worlds (i.e., low latency and low cost), based on the key observation that long
bitlines are the dominant cause of DRAM latency.

To achieve the latency advantage of short bitlines and the cost advantage of
long bitlines, we propose the {\em Tiered-Latency DRAM} (TL-DRAM) architecture,
which divides the long bitline into two shorter segments using an {\em
isolation transistor}: the {\em near segment} (connected directly to the local
sense amplifier) and the {\em far segment} (connected \mycolor{to the local
sense amplifier only when the isolation transistor is turned on}). As a result,
the near segment has much lower latency than the far segment. To maximize the
latency benefits from this heterogeneous bitline architecture, we propose two
mechanisms -- {\em i)} using the near segment as \mycolor{a} hardware-managed
cache to the far segment, and {\em ii)} exposing the near segment to
\mycolor{the} operating system, \mycolor{which} places latency-critical data to
the near segment. We propose two new policies to manage the near segment cache
that specifically exploit the asymmetric latency characteristics of TL-DRAM.
Our most sophisticated cache management algorithm, Benefit-Based Caching
(\mbbc) improves system performance by an average of 12.8\% and reduces energy
consumption by an average of 23.6\% over a wide variety of
\mycolor{data-intensive} workloads.

In summary, TL-DRAM enables latency heterogeneity in DRAM by changing the
internal DRAM architecture with low area cost, and provides mechanisms to
maximize the latency benefits \mycolor{by enabling intelligent data placement}.

\subsection{Optimizing DRAM Latency to the Common Operating Conditions
(Adaptive-Latency DRAM)} \label{ch:intro_aldram}

When a DRAM chip is accessed, it requires a certain amount of time before
enough charge can move into the cell (or the bitline) for the data to be
reliably stored (or retrieved). To guarantee this behavior, DRAM manufacturers
impose a set of minimum latency restrictions on DRAM accesses, referred to as
{\em timing parameters}~\cite{jedec-ddr3}. Ideally, timing parameters should
provide just enough time for a DRAM chip to operate correctly. In practice,
however, DRAM manufacturers {\em pessimistically incorporate a very large
margin} into their timing parameters to ensure correct operation under {\em the
worst case \mycolor{conditions}} due to two major concerns. First, due to {\bf
{\em process variation}}~\cite{friedberg-isqed2005, lee-iedm1996,
smruti-tsm2008}, some outlier cells suffer from a larger delay than other
cells, and require more time to be charged. Although every cell is designed to
have a large capacitance (to hold more charge) and a small resistance (to
facilitate the flow of charge), some deviant cells may not be implemented in
such a manner. Second, due to {\bf {\em temperature dependence}}, all cells
suffer from a weaker charge-drive at high temperatures, and require more time
to charge the bitline. Consequently, to accommodate the combined effect of
process variation {\em and} temperature dependence (the worst case condition),
existing timing parameters prescribed by the DRAM manufacturers are set to a
very large value.

Our approach is to reduce the DRAM latency by \mycolor{reducing} the additional
{\em latency slack} in DRAM due to the pessimism on timing parameters. To study
the potential for reducing timing parameters for each DRAM module, we
characterize 115 DRAM modules from three manufacturers to expose the excessive
margin that is built into their timing parameters. We make two observations.
First, even at the highest temperature of 85\celsius, \mycolor{there is a high
potential for reducing the latency of DRAM modules} (21.1\% on average for read
and 34.4\% for write operations). Second, we observe that at lower temperatures
(e.g., 55\celsius) the potential for latency reduction is even greater (32.7\%
on average for read and 55.1\% on average for write operations). As a result,
we conclude that \mycolor{exploiting} process variation and lower temperatures
enable a significant potential to reduce DRAM latencies.

Based on our characterization, we propose Adaptive-Latency DRAM (AL-DRAM), a
mechanism that dynamically optimizes the timing parameters for different
modules at different temperatures. AL-DRAM exploits only the {\em additional
charge slack} in the common-case compared to the worst-case, thereby
maintaining the reliability \mycolor{of DRAM modules}. We evaluate AL-DRAM on a
real system~\cite{amd-4386, amd-bkdg} and show show that AL-DRAM improves the
performance of a wide variety of memory-intensive workloads by 14.0\% (on
average) without introducing \mycolor{any} errors.

In summary, AL-DRAM enables lower DRAM latency while maintaining memory
correctness and without requiring changes to the internal DRAM architecture or
the DRAM interface (low cost).

\subsection{Lowering DRAM Latency by Exploiting the Awareness of Internal
DRAM Architecture (Architectural-Variation-Aware DRAM)}
\label{ch:intro_avadram}

Modern DRAM consists of 2D cell arrays, each of which has \mycolor{its} own
accessing structure (e.g., wordline driver) and data sensing structure (e.g.,
local sense amplifier). We observe that there is variability \mycolor{across}
DRAM cells based on their locations in a DRAM cell array ({\em mat}). Some DRAM
cells can be accessed faster than others \mycolor{due to} their physical
location. We refer to this variability in cells' access times, caused by the
physical organization of DRAM, as {\em architectural variation}. Architectural
variation arises from the difference in the distance between the cells and the
peripheral logic that is used to access these cells. The wires connecting the
cells to peripheral logic exhibit large resistance and large
capacitance~\cite{lee-hpca2013, lee-hpca2015}. Consequently, cells experience
different RC delays based on their distance from the peripheral logic (e.g,
accessing and sensing structures). Cells closer to \mycolor{the} peripheral
logic experience smaller delay and can be accessed faster than the cells
located farther from the peripheral logic.

Architectural variation in latency is present in both vertical and horizontal
directions in a mat: {\em i)} Each vertical {\em column of cells} \mycolor{in a
mat} is connected to a {\em local sense amplifier} and {\em ii)} each
horizontal {\em row of cells} \mycolor{in} a mat is connected to a {\em
wordline driver}. Variations in the vertical and horizontal dimensions,
together, divide the cell array into heterogeneous latency regions, where cells
in some regions require larger latencies for reliable operation. This variation
in latency has direct impact on the reliability of the cells. Reducing the
latency {\em uniformly across all regions} in DRAM would improve performance,
but can introduce failures in the {\em inherently slower} regions that have to
be accessed longer for correct DRAM operation. We refer to these inherently
slower regions of DRAM as {\em architecturally vulnerable regions}. We first
experimentally demonstrate the existence of architectural variation in modern
DRAM chips and identify the architecturally vulnerable regions. We then propose
new mechanisms that leverage this variation to reduce DRAM latency while
providing reliability at low cost.

Based on our experimental study \mycolor{that characterizes} \dimms~DRAM
modules by using our FPGA-based DRAM testing infrastructure, we show that {\em
i)} modern DRAM chips exhibit architectural latency variation in both row and
column directions, and {\em ii)} architectural vulnerability gradually
increases in the row direction within a mat and repeats the variability pattern
in every mat. Then, we develop two new mechanisms that exploit the
architecturally vulnerable regions to enable low DRAM latency with high
reliability and at low cost ({\em Architectural-Variation-Aware DRAM} ({\em
AVA-DRAM})). The first mechanism, {\em AVA Profiling}, identifies the lowest
possible latency that ensures reliable operation \mycolor{at} low cost by
periodically profiling {\em only} the architecturally vulnerable regions. To
further reduce DRAM latency, the second mechanism, {\em AVA Shuffling},
distributes data from architecturally vulnerable regions to multiple ECC
codewords to make it correctable by \mycolor{using} ECC. AVA Profiling can
dynamically reduce the latencies of read/write operations by 35.1\%/57.8\% at
55\celsius~while ensuring reliable operation at low cost. AVA Shuffling on
average corrects 26\% of total errors which are not correctable by conventional
ECC, leading to further latency reduction for 24 DRAM modules out of 96 DRAM
modules. We show that the combination of our techniques, AVA-DRAM, leads to a
raw DRAM latency reduction of 40.0\%/60.5\% (read/write) and an overall system
performance improvement of 14.7\%/13.7\%/13.8\% (2-/4-/8-core) over a variety
of workloads in our evaluated systems, while ensuring reliable operation.

\mycolor{AVA-DRAM is the first work that exposes and experimentally
demonstrates the existence of architectural variation. We then propose two
mechanisms that leverage architectural variation towards achieving high
performance, energy efficiency, and reliability through dynamic profiling (AVA
Profiling) and data shuffling (AVA Shuffling).}

\section{Thesis Statement}

{\em DRAM latency can be reduced by enabling and exploiting latency
heterogeneity in DRAM architecture.}

\newpage
\section{Contributions}

This dissertation makes the following major contributions.

\squishlist

	\item This dissertation makes the observation that long internal wires
	(bitlines) are the dominant source of DRAM latency, and exposes the important
	trade-off between DRAM latency and area. Based on this, this dissertation
	proposes a new DRAM architecture, Tiered-Latency DRAM, which divides long
	bitline into fast and slow segments, enabling latency heterogeneity in DRAM.
	This dissertation quantitatively evaluates the latency, area, and power
	characteristics of Tiered-Latency DRAM through circuit simulations based on a
	publicly available 55nm DRAM process technology~\cite{rambus-power}.
	\mycolor{We show that the near segment latency (\trc) for a 32-row near
	segment can be 49\% lower than the modern DRAM standard latency.}

	\item This dissertation describes two major ways of leveraging TL-DRAM: {\em
	i)} \mycolor{by using the near segment as a hardware-managed cache without
	exposing it to software}, and {\em ii)} by exposing the near segment capacity
	to the OS and using hardware/software to map frequently accessed pages to the
	near segment. We propose two new policies to manage the near segment that
	specifically exploit the asymmetric latency characteristics of TL-DRAM. Our
	most sophisticated cache management algorithm, Benefit-Based Caching (\mbbc)
	improves system performance by an average of 12.8\% and reduces energy
	consumption by an average of 23.6\% over a wide variety of workloads.

	\item This dissertation provides a detailed analysis of why we can reduce
	DRAM timing parameters without sacrificing reliability in the common case. We
	show that the latency of a DRAM access depends on how quickly charge moves
	into or out of a cell. Compared to the worst-case cell operating at the
	worst-case temperature (85\celsius), a typical cell at a typical temperature
	allows much faster movement of charge, leading to shorter latency. This
	enables the opportunity to reduce timing parameters without introducing
	errors.

	\item This dissertation provides detailed DRAM profiling results (for \DIMMs
	DRAM modules, comprised of 920 DRAM chips, from three manufacturers) by using
	an FPGA-based DRAM testing infrastructure, and exposes the large margin built
	into their timing parameters. In particular, we identify four timing
	parameters that are the most critical during a DRAM access: \trcd, \tras,
	\twr, and \trp. At 55\celsius, we demonstrate that the parameters can be
	reduced by an average of \trcdCold\%, \trasCold\%, \twrCold\%, and \trpCold\%
	while still maintaining correctness.

	\item This dissertation proposes a practical mechanism, {\em Adaptive-Latency
	DRAM (\ALD)}, to take advantage of the \mycolor{extra margin built into DRAM
	latency.} The key idea is to dynamically adjust the DRAM timing parameters
	for each module based on its latency characteristics and temperature so that
	the timing parameters are dynamically optimized for the current operating
	condition \mycolor{and the current DRAM module}. We show that the hardware
	cost of \ALD is very modest, with no changes to DRAM. We evaluate \ALD on a
	real system~\cite{amd-4386, amd-bkdg} running real workloads by dynamically
	reconfiguring the timing parameters. \ALD improves system performance by an
	average of 14.0\% and a maximum of 20.5\% over a wide variety of
	memory-intensive workloads, without incurring \mycolor{any} errors.

	\item This dissertation exposes and experimentally demonstrates the
	phenomenon of {\em architectural \mycolor{latency} variation} in DRAM cell
	arrays, \mycolor{i.e., that the access latency of a cell depends on its
	location in the DRAM array.} This phenomenon \mycolor{causes} certain regions
	of DRAM to be \mycolor{inherently} more vulnerable to latency reduction than
	others based on their relative distance from the peripheral logic.

	\item This dissertation identifies the regions in DRAM that are most
	vulnerable to \mycolor{latency reduction} based on the internal hierarchical
	organization of DRAM bitlines and wordline drivers. \mycolor{We call such
	regions as {\em architecturally vulnerable regions}.} We experimentally
	demonstrate the existence of architecturally vulnerable regions in DRAM by
	testing and characterizing \dimms~real DRAM modules (768 DRAM chips).

	\item This dissertation develops two new mechanisms, called AVA Profiling and
	AVA Shuffling, that exploit architectural variation to improve performance
	and reliability of DRAM at low cost. \mycolor{AVA Profiling dynamically finds
	the lowest latency at which a DRAM chip can operate reliably.} AVA Profiling
	can dynamically reduce the latencies of read/write operations by
	35.1\%/57.8\% at 55\celsius~while ensuring reliable operation at low cost.
	\mycolor{AVA Shuffling distributes data from architecturally vulnerable
	regions to multiple ECC codewords to make it correctable by using ECC.} AVA
	Shuffling on average corrects 26\% of total errors which are not correctable
	by conventional ECC. We show that the combination of our techniques,
	AVA-DRAM, leads to a raw DRAM latency reduction of 40.0\%/60.5\% (read/write)
	and an overall system performance improvement of 14.7\%/13.7\%/13.8\%
	(2-/4-/8-core) over a wide variety of workloads in our evaluated systems,
	while ensuring reliable operation.

\squishend

\section{Dissertation Outline} \label{ch:outline}

This dissertation is organized into seven chapters. Chapter~\ref{ch:bak}
presents background on memory system organization and DRAM organization.
Chapter~\ref{ch:prev} discusses related prior work on techniques for reducing
\mycolor{and tolerating} DRAM latency. Chapter~\ref{ch:tldram} presents the
design of Tiered-Latency DRAM and mechanisms to leverage the Tiered-Latency
DRAM substrate for reducing overall DRAM latency. Chapter~\ref{ch:aldram} first
presents the latency slack in DRAM standard timing parameters that are dictated
by the worst case \mycolor{conditions} (accessing the smallest cell in DRAM
products at the worst \mycolor{case} operating temperature). It then proposes
and evaluates Adaptive-Latency DRAM that optimizes DRAM timing parameters for
the common case (accessing the common cells in each DRAM module at current
operating temperature). Chapter~\ref{ch:avadram} \mycolor{first} presents the
observation of architectural variation in a DRAM cell array (mat). It then
proposes and evaluates AVA-DRAM that leverages architectural variation for
reducing DRAM latency \mycolor{at} low cost, by periodically profiling only the
worst latency regions at low cost and distributing data \mycolor{in the worst
latency regions to multiple ECC codewords. Chapter~\ref{ch:system_guide}
introduces system design guidelines for future memory systems that have
heterogeneous latency.} Finally, Chapter~\ref{ch:conclusion} presents our
conclusions and future research directions that are enabled by this
dissertation.

	\chapter{{Background}} \label{ch:bak}

To understand the dominant sources of DRAM latency, we first provide the
necessary background on DRAM organization and operation.

\section{DRAM Organization} \label{sec:background}

\begin{figure}[h]
	\centering
	\subcaptionbox{DIMM (8 chips)\label{fig:dram_module}}[0.45\linewidth] {
		\includegraphics[height=1.80in]{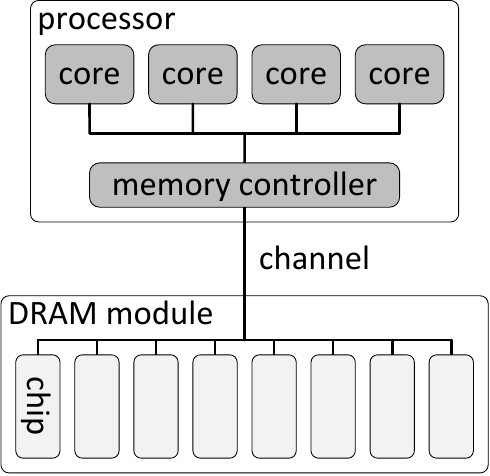}
	}
	\subcaptionbox{Chip (8 banks)\label{fig:dram_chip}}[0.45\linewidth] {
		\includegraphics[height=1.80in]{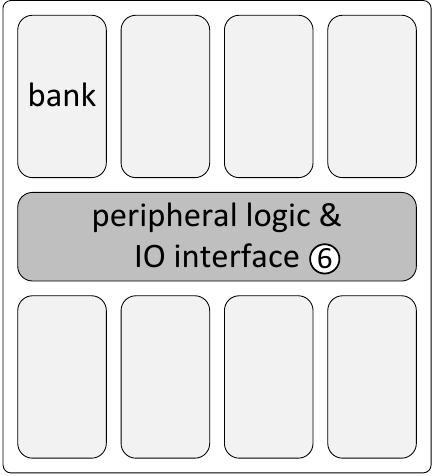}
	}

	\subcaptionbox{Bank\label{fig:dram_bank}}[0.45\linewidth] {
		\includegraphics[height=1.80in]{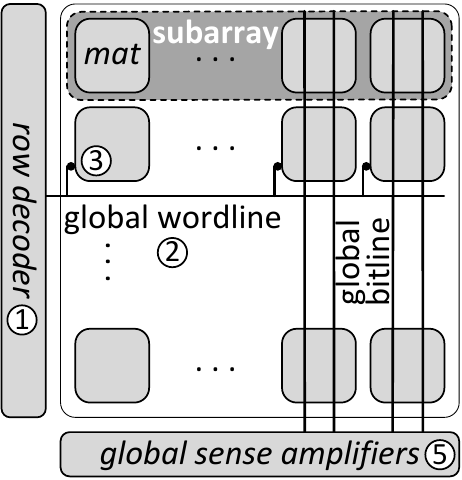}
	}
	\subcaptionbox{Mat (512$\times$512 cells)\label{fig:dram_mat}}[0.45\linewidth] {
		\includegraphics[height=1.80in]{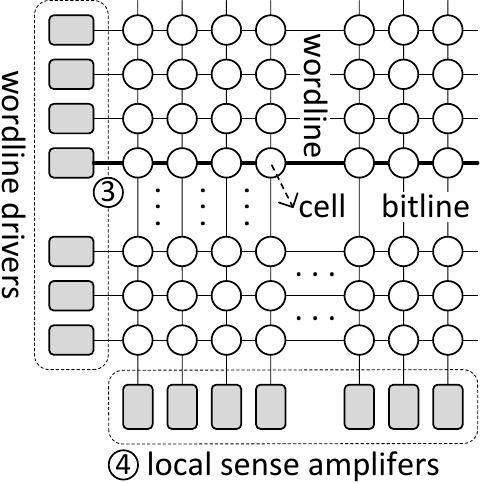}
	}
	\caption{Hierarchical Organization of a DRAM System}
	\label{fig:bak_dram_org}
\end{figure}

DRAM is organized in a hierarchical manner where each DRAM module consists of
multiple chips, banks, mats, and subarrays. Figure~\ref{fig:bak_dram_org} shows
the hierarchical organization of a typical DRAM-based memory system, where the
hierarchy consists of five levels
(Figure~\ref{fig:dram_module}--\ref{fig:dram_mat}).

\noindent {\bf Module.} At the highest level of the hierarchy (shown in
Figure~\ref{fig:dram_module}), a memory controller in a processor is connected
to a DRAM {\em module} over a memory channel. The memory channel has a 64-bit
data bus that is divided into eight 8-bit buses connected to eight chips in the
DRAM module. These eight chips operate in lock step while accessing the DRAM
module. For example, when there is a read request from the memory controller,
each DRAM chip transfers 8 bytes of data in 8 data bursts over the dedicated
8-bit data bus connected to it, transferring a total of 64 bytes (cache line
size) across all chips.

\noindent {\bf Chip.} A DRAM {\em chip} (shown in Figure~\ref{fig:dram_chip})
consists of {\em i)} multiple banks and {\em ii)} peripheral logic that is used
to transfer data to the memory channel through the IO interface.

\noindent {\bf Bank.} Each {\em bank} (shown in Figure~\ref{fig:dram_bank}), is
subdivided into multiple {\em mats}. In a bank, there are two global components
that are used to access the mats: {\em i)} a {\em row decoder} that selects a
row of cells {\em across} a subarray that consists of multiple mats and {\em
ii)} {\em global sense amplifiers} that transfer a fraction of data from the
row through the global bitlines, based on the column address.

\noindent {\bf Mat.} Figure~\ref{fig:dram_mat} shows the organization of a {\em
mat} that consists of three components: {\em i)} a 2-D cell array in which the
cells in each row are connected horizontally by a shared wire called the {\em
wordline}, and the cells in each column are connected vertically by a wire
called the {\em bitline}, {\em ii)} a column of wordline drivers that drive
each wordline to appropriate voltage levels in order to activate a row during
an access and {\em iii)} a row of {\em local sense amplifiers} that sense and
latch data from the activated row.

\noindent {\bf Subarray.} A row of mats in a bank forms a subarray, where cells
in a {\em subarray} are accessed simultaneously, managed by wordlines connected
to a global wordline, as shown in Figure~\ref{fig:dram_bank}.

\section{DRAM Cell Organization} \label{subsec:subarray_organization}

\begin{figure}[h]
	\centering
	\subcaptionbox{Cell\label{fig:background_cell}}[0.30\linewidth] {
		\centering
		\includegraphics[height=1.4in]{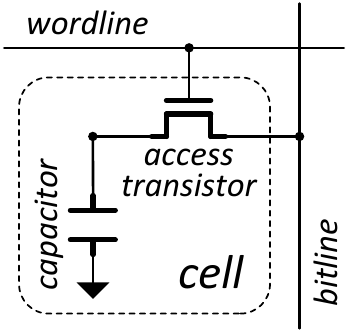}
	}
	\subcaptionbox{Bitline \& Local Sense Amplifier\label{fig:background_bitline}}[0.35\linewidth] {
		\centering
		\includegraphics[height=1.4in]{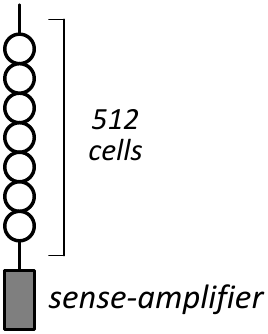}
	}
	\subcaptionbox{Simplified View\label{fig:model}}[0.30\linewidth] {
		\includegraphics[height=1.4in]{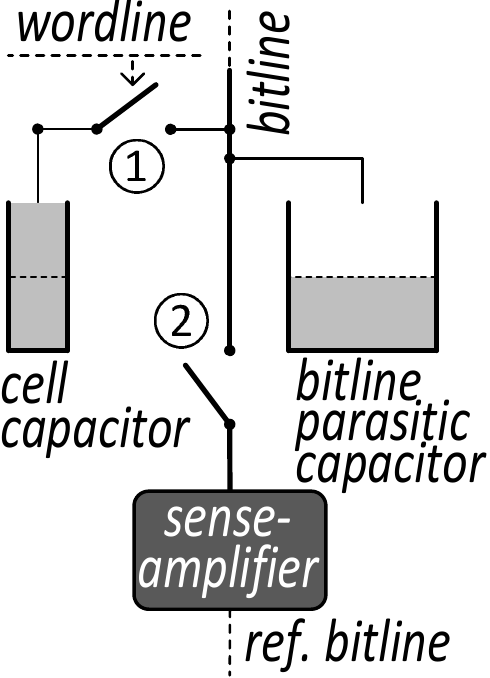}
	}
	\caption{DRAM Elementary Components}
	\label{fig:background_bitline_cell}
\end{figure}

As explained in Section~\ref{sec:background}, a DRAM subarray is a group of
mats in the horizontal direction and each mat is a 2-D array of elementary
units called {\em cells}. As shown in Figure~\ref{fig:background_cell}, a cell
consists of two components: {\em i)} a capacitor that represents binary data in
the form of stored electrical charge and {\em ii)} an access transistor that is
switched on/off to connect/disconnect the capacitor to a {\em bitline}. As
shown in Figure~\ref{fig:background_bitline}, there are approximately 512 cells
in the vertical direction (a ``column'' of cells), all of which share the same
bitline. For each bitline, there is a {\em local sense amplifier} whose main
purpose is to read from a cell by reliably detecting the very small amount of
electrical charge stored in the cell. When writing to a cell, on the other
hand, the local sense amplifier acts as an electrical driver and programs the
cell by filling or depleting its stored charge.

Numerous bitlines (and their associated local sense amplifiers) are laid
side-by-side in parallel to compose a subarray (Figures~\ref{fig:dram_bank}
and~\ref{fig:dram_mat}). All cells in the horizontal direction (a ``row'' of
cells) have their access transistors controlled by a {\em wordline}. When the
wordline voltage is raised to \vdd, all cells of a row are connected to their
respective bitlines and sensed in lockstep by the local sense amplifiers. This
is why the set of all local sense amplifiers in a subarray is also called a
{\em row buffer}. At any given time, at most one wordline in a subarray is ever
raised (i.e., at most one cell per column is connected to the bitline) --
otherwise, cells in the same column would corrupt each other's data.

Figure~\ref{fig:model} depicts a simplified view of a cell as well as its
bitline and local sense amplifier, in which electrical charge is represented in
gray. Switch \ding{192} represents the access transistor controlled by the
wordline, and switch \ding{193} represents the on/off state of the sense
amplifier.

\section{DRAM Access} \label{subsec:subarray_access}

As the timelines in Figure~\ref{fig:background_three_phases} show, a DRAM chip
access can be broken down into three distinct {\em phases}: {\em i)}
activation, {\em ii)} IO, and {\em iii)} precharging. Activation and
precharging occur entirely within the subarray, whereas IO occurs in the
peripheral logic and IO circuitry. All these operations consist of two levels
of accesses through: {\em i)} global structures across all subarrays within a
bank (global sense amplifiers, global wordlines and global bitlines) and {\em
ii)} local structures within a mat (local sense amplifiers, local wordlines,
and local bitlines). A DRAM access goes through multiple steps in the
global-local hierarchy:

\begin{figure}[h]
    \centering
    \includegraphics[width=0.7\linewidth]{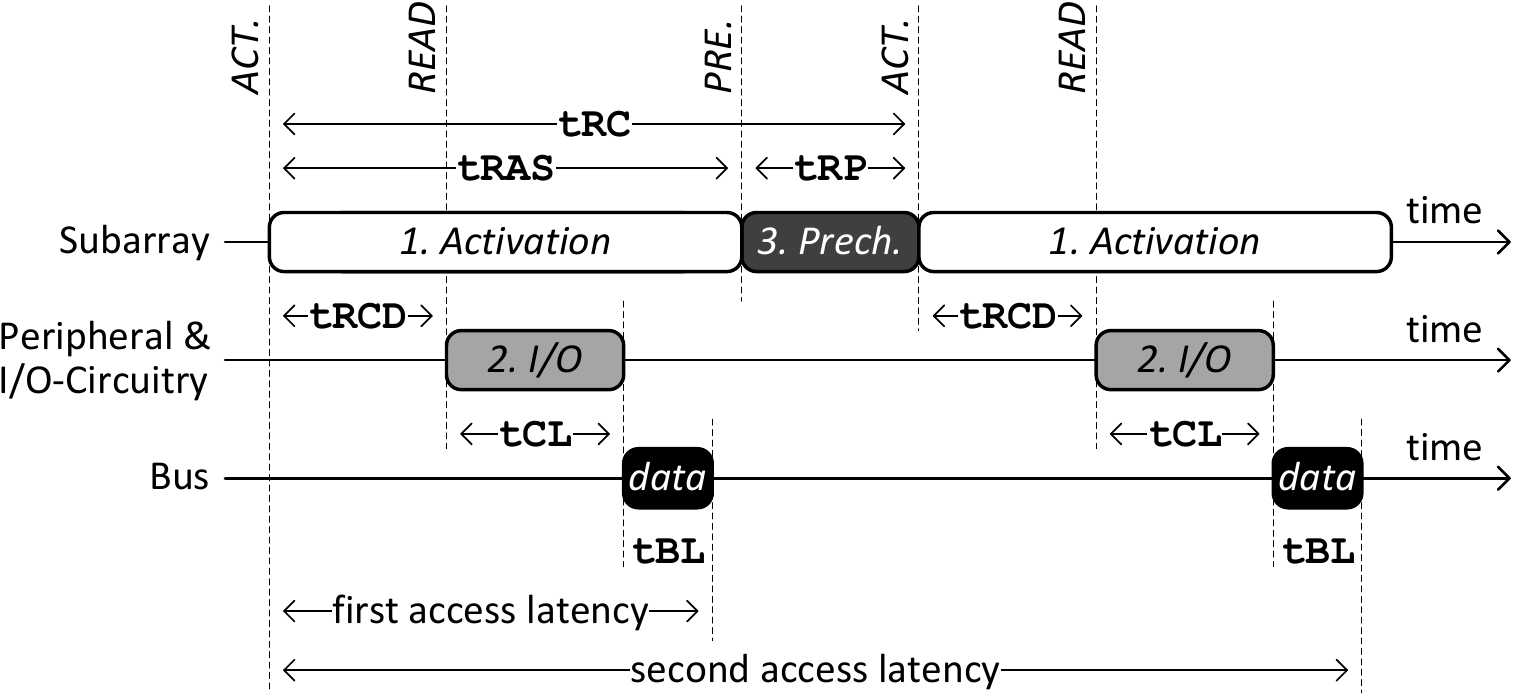}
    \captionof{figure}{Three Phases of DRAM Access}\label{fig:background_three_phases}
\end{figure}

\squishlist

	\item {\bf Activation.} During the {\em activation} phase, the row decoder
	in a bank receives a row address to activate (\ding{172} in
	Figure~\ref{fig:dram_bank}), then, it first activates the corresponding
	global wordline in the bank (\ding{173} in Figure~\ref{fig:dram_bank}). The
	global wordline, in turn, activates the corresponding wordline driver in
	each mat of the subarray. The wordline driver in each mat activates the
	corresponding local wordline connecting the row of cells to the bitlines
	(\ding{174} in Figures~\ref{fig:dram_bank} and~\ref{fig:dram_mat}). Soon
	thereafter, the data in the row of cells is copied (detected) to the local
	sense amplifiers of that subarray (\ding{175} in Figure~\ref{fig:dram_mat}).

	\item {\bf IO.} During the {\em IO} phase, the local sense amplifiers
	transfer the data to the global sense amplifiers, through the global bitlines
	(\ding{176} in Figure~\ref{fig:dram_bank}). Data from the global sense
	amplifiers is then sent to the memory channel through the IO interface of the
	chip (\ding{177} in Figure~\ref{fig:dram_chip}). From there, the data leaves
	the DRAM chip and is sent to the processor over the memory channel. As
	Figure~\ref{fig:background_three_phases} shows, the IO phase's latency is
	overlapped with the latency of the activation phase.

	\item {\bf Precharging.} During the {\em precharging} phase, the raised
	wordline in the subarray is lowered, disconnecting the row of cells from the
	bitlines. Also, the subarray's local sense amplifiers and bitlines are
	initialized (i.e., cleared of their data) to prepare for the next access to a
	new row of cells.

\squishend

{\bf Three DRAM Commands.} The DRAM controller (typically residing on the
processor die) issues {\em commands} to the DRAM chip to initiate the three
phases listed above. As shown in Figure~\ref{fig:background_three_phases},
there are three commands, one for each phase. In their respective order, they
are: \cmdact (ACT), \cmdrd/\cmdwr, and \cmdpre (PRE). Among the commands,
\cmdact and \cmdpre are subarray-related commands since they directly operate
on the subarray, whereas \cmdrd and \cmdwr are IO-related commands.

{\bf Timing Constraints.} After the DRAM controller issues a command to
initiate a phase, it must wait for a sufficient amount of time before issuing
the next command. Such restrictions imposed between the issuing of commands are
called {\em timing constraints}. DRAM timing constraints are visualized in
Figure~\ref{fig:background_three_phases} and summarized in
Table~\ref{tab:background_timing}. Two of the most important timing constraints
are \trcd ({\em row-to-column delay}) and \trc ({\em row-cycle time}). Every
time a new row of cells is accessed, the subarray incurs \trcd (15ns; \cmdact
\ra \cmdrd/\cmdwr) to copy the row into the local sense amplifiers. On the
other hand, when there are multiple accesses to different rows in the same
subarray, an earlier access delays all later accesses by \trc (52.5ns; \cmdact
\ra \cmdact). This is because the subarray needs time to complete the
activation phase (\tras) and the precharging phase (\trp) for the earlier
access, whose sum is defined as \trc ($=$ \tras $+$ \trp), as shown in
Figure~\ref{fig:background_three_phases}.

\begin{table}[h]
	\centering
	\small{
	\begin{tabular}{clcc}
	\toprule
	Phase & Commands & Name & Value \\
	\midrule
	\multirow{3}{*}{1} & \cmdact \ra\ \cmdrd & \multirow{2}{*}{{\normalsize \trcd}} & \multirow{2}{*}{15ns} \\
	& \cmdact \ra\ \cmdwr &	 &	 \\
	\cmidrule{2-4}
	& \cmdact \ra\ \cmdpre & {\normalsize \tras}	& 37.5ns \\
	\cmidrule{1-4}
	\multirow{3}{*}{2} & \cmdrd \ra\ {\em data} & {\normalsize \tcl} & 15ns \\
	& \cmdwr \ra\ {\em data} & {\normalsize \tcwl} & 11.25ns \\
	\cmidrule{2-4}
	& {\em data burst} & {\normalsize \tbl} & 7.5ns \\
	\cmidrule{1-4}
	3 & \cmdpre \ra\ \cmdact & {\normalsize \trp} & 15ns \\
	\cmidrule{1-4}
	\multirow{2}{*}{1 \& 3} & \multirow{2}{*}{\cmdact \ra\ \cmdact} & {\normalsize \trc} & \multirow{2}{*}{52.5ns} \\
	& & {(\tras $+$ \trp)} & \\
	\bottomrule
	\end{tabular}
	}
	\captionof{table}{Timing Constraints (DDR3-1066)~\cite{samsung-spec}}
	\label{tab:background_timing}
\end{table}

{\bf Access Latency.} Figure~\ref{fig:background_three_phases} illustrates how
the DRAM access latency can be decomposed into individual DRAM timing
constraints. Specifically, the figure shows the latencies of two read accesses
(to different rows in the same subarray) that are served one after the other.
From the perspective of the first access, DRAM is ``unloaded'' (i.e., no prior
timing constraints are in effect), so the DRAM controller immediately issues an
\cmdact on its behalf. After waiting for \trcd, the controller issues a \cmdrd,
at which point the data leaves the subarray and incurs additional latencies of
\tcl (peripherals and IO circuitry) and \tbl (bus) before it reaches the
processor. Therefore, the latency of the first access is 37.5ns (\trcd $+$ \tcl
$+$ \tbl). On the other hand, the second access is delayed by the timing
constraint that is in effect due to the first access (\trc) and experiences a
large ``loaded'' latency of 90ns (\trc $+$ \trcd $+$ \tcl $+$ \tbl).

\section{DRAM Cell Operation: A Detailed Look}
\label{subsec:subarray_operation}

\begin{figure}[h]
	\center
	\includegraphics[width=0.99\linewidth]{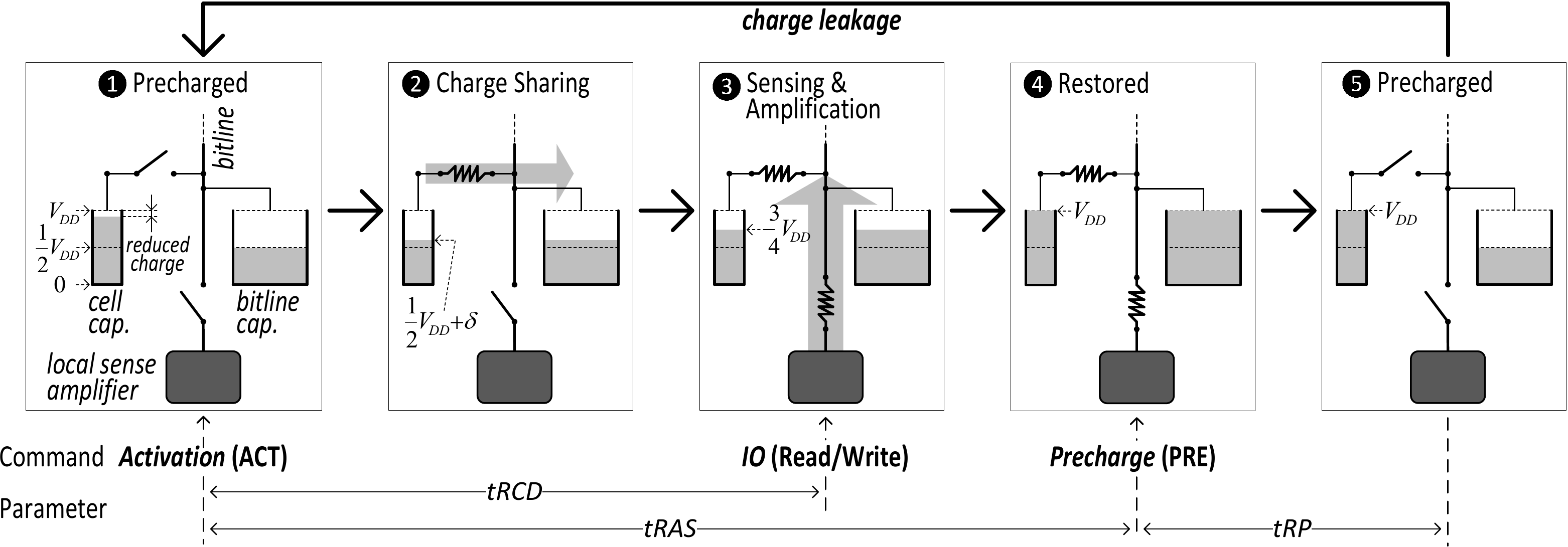}
	\caption{DRAM Operations, Commands and Parameters}
	\label{fig:operation}
\end{figure}

Subarray-related timing constraints (\trcd and \trc) constitute a significant
portion of the unloaded and loaded DRAM access latencies: 40\% of 37.5ns and
75\% of 90ns, respectively. Since \trcd and \trc exist only to safeguard the
timely operation of the underlying subarray, in order to understand why their
values are so large, we must first understand how the subarray operates during
the activation and the precharging phases. (As previously explained, the IO
phase does not occur within the subarray and its latency is overlapped with the
activation phase.) Specifically, we show how the bitline plays a crucial role
in both activation and precharging, such that it heavily influences both \trcd
and \trc. As shown in Figure~\ref{fig:operation}, a cell transitions through
five different states during each access.

\squishlist

	\item {\bf Precharged (Initial) State.} In the first state (State~\numone),
	which is called the {\em precharged} state, before it is accessed, the cell
	is initially ``fully'' charged, while the bitline is only halfway charged
	(i.e., the bitline voltage is maintained at \hvdd). In practice, the cell is
	usually not completely charged because of a phenomenon called {\em leakage},
	wherein the cell capacitor loses charge over time.

	\item {\bf Activation (Charge Sharing).} In order to access data from a cell,
	the DRAM controller issues a command called \cmdact. Upon receiving this
	command, DRAM increases the wordline voltage, thereby turning on access
	transistors connected to the wordline, leading to connecting their capacitors
	to the bitlines. Since the cell is at a higher voltage than the bitline, the
	charge in the cell capacitor (or the lack thereof) is slowly shared with the
	{\em bitline parasitic capacitor}, thereby perturbing the bitline voltage
	away from its quiescent value (\hvdd) in the positive (or negative) direction
	until their voltages are equalized at \hvdd$+$~$\delta$ (or
	\hvdd$-$~$\delta$). This is depicted in state \numtwo, which is called {\em
	charge sharing}. During charge sharing, note that the cell's charge is
	modified (i.e., data is lost) because it is shared with the bitline. But this
	is only temporary since the cell's charge is restored as part of the next
	step, as described below.

	\item {\bf Activation (Sensing \& Amplification).} After allowing sufficient
	time for the charge sharing to occur, the sense amplifier is turned on.
	Immediately, the sense amplifier ``senses'' (i.e., observes) the polarity of
	the perturbation on the bitline voltage. Then the sense amplifier
	``amplifies'' the perturbation by injecting (or withdrawing) charge into (or
	from) both the cell capacitor and the bitline parasitic capacitor. After a
	latency of \trcd, midway through amplification, enough charge has been
	injected (or withdrawn) such that the bitline voltage reaches a {\em
	threshold} state of \tfvdd (or \qvdd). At this point, data is considered to
	have been ``copied'' from the cell to the sense amplifier (State~\numthree).
	In other words, the bitline voltage is now close enough to \vdd (or {\em 0})
	for the sense amplifier to detect a binary data value of `1' (or `0') and
	transfer the data to the IO circuitry, allowing \cmdrd and \cmdwr commands to
	be issued. After completing the sense-amplification, the voltage of the
	bitline and the cell are fully amplified to \vdd or {\em 0} (State~\numfour).
	Only at this point is the charge in the cell fully restored to its original
	value. The latency to reach this {\em restored} state (State~\numfour) is
	\tras (which is one component of \trc). If there is a write operation, some
	additional time is required for the bitline and the cell to reach this state,
	which is expressed as a timing parameter called \twr.

	\item {\bf Precharging.} Before we can access data from a different cell
	connected to the same bitline, the sense amplifier must be taken back to the
	precharged state. This is done by issuing a \cmdpre command. Upon receiving
	this command, DRAM first decreases the wordline voltage, thereby turning off
	access transistors and disconnecting the cell from the bitline. By doing so,
	the cell becomes decoupled from the bitline and is not affected by any future
	changes in the bitline voltage. Next, DRAM disables the sense amplifier and
	withdraws (or injects) charge from the bitline parasitic capacitor such that
	the bitline voltage reaches the quiescent value of \hvdd (State~\numfive).
	Precharging is required to ensure that the next accessed cell can perturb the
	bitline voltage in either direction (towards \vdd or towards {\em 0}). This
	would not be possible if the bitline is left unprecharged at \vdd or {\em 0}.
	The time taken for the precharge operation is expressed as a timing parameter
	called \trp (which is the other component of \trc).

	\item {\bf Idle State (between Accesses).} At state \numfive, note that the
	cell is completely filled with charge. Subsequently, however, the cell
	slowly loses some of its charge until the next access (cycling back to state
	\numone). The length of time for which the cell can reliably hold its charge
	is called the cell's {\em retention time}. If the cell is not accessed for a
	long time, it may lose enough charge to invert its stored data, resulting in
	an error. To avoid data corruption, DRAM refreshes the charge in all of its
	cells at a regular interval, called the {\em refresh interval}. To avoid
	data corruption, modern DRAMs periodically access the cells to restore the
	lost amount of charge in all of its cells, called the {\em refresh}.  This
	refresh operation happens at a regular interval, called the {\em refresh
	interval}. DRAM refresh operations waste energy and also degrade performance
	by delaying memory requests. This negative impacts are expected to increase
	in the future high capacity DRAMs. \mycolor{To mitigate the energy and
	performance impact of DRAM refresh, many works performed experimental
	studies to analyze DRAM refresh characteristics~\cite{liu-isca2013,
	khan-sigmetrics2014, qureshi-dsn2015} and proposed hardware or software
	techniques~\cite{liu-isca2012, chang-hpca2014, qureshi-dsn2015}.}

\squishend

\noindent{\bf Summary.} Through our discussion, we have established a
relationship between the timing constraints (\trcd, \tras, \trp, and \trc) and
the internal DRAM architecture and cell operations. To summarize, these timing
constraints are determined by how quickly the bitline voltage can be driven --
for \trcd, from \hvdd to \tfvdd ({\em threshold}); for \tras, from \hvdd to
\vdd ({\em restored}) and back again to \hvdd; for \trp, from \vdd to \hvdd
{\em precharging}. In Chapter~\ref{ch:tldram}, we first show that the
drivability of the bitline is determined by the bitline parasitic capacitance,
whose value is a function of the bitline length, then proposes a new DRAM
architecture that enables latency heterogeneity by changing the DRAM bitline
architecture. In Chapter~\ref{ch:aldram}, we exploit the latency slack in the
internal DRAM architecture to optimize the timing constraints for the common
operation conditions. In Chapter~\ref{ch:avadram}, we exploit the latency
variation inherited from the internal DRAM architecture to reduce DRAM latency.

	\chapter{Related Prior Work} \label{ch:prev}

In this chapter, we describe prior proposals to reduce overall memory access
latency in different areas, {\em i)} new DRAM architectures for lowering DRAM
latency, {\em ii)} heterogeneous memory control techniques for reducing DRAM
latency, {\em iii)} new DRAM architectures for enabling more parallelism, {\em
iv)} memory scheduling to mitigate high DRAM latency, and {\em v)} other
related prior works for mitigating high DRAM latency.

\section{Low Latency DRAM Architecture} \label{sec:lld}

Prior works aim to reduce DRAM latency in three major directions. The first
approach is changing/optimizing DRAM architecture to reduce DRAM latency. The
second approach is integrating low latency SRAM cells in DRAM, which can be
used for caching recently accessed data. The third approach enables latency
variation in DRAM. We present more detail of these three approaches. The key
distinguishing factor of the proposals in this dissertation is the fact that
they are low cost, while most prior works have high cost.

{\bf DRAMs Optimized for Low Access Latency.} Some specialized DRAMs
\mycolor{provide} shorter latency than commodity DRAM by reducing the number of
cells-per-bitline in their cell array. Micron's RL-DRAM~\cite{rldram} and
Fujitsu's FCRAM~\cite{sato-vlsi1998} have much shorter bitlines than commodity
DRAM, enabling lower latency than conventional DRAM. Embedded
DRAM~\cite{meterelliyoz-vlsi2014, narasimha-iedm2012, keitel-dtc2001,
chao-dac2009} is recently introduced to use DRAM as the last level cache in the
processor. To this end, embedded DRAM uses very short \mycolor{bitlines,}
enabling both competitive latency and much larger capacity compared to
conventional SRAM-based on-chip \mycolor{caches}. Unfortunately, reducing
cells-per-bitline \mycolor{requires} more sense amplifiers to integrate
\mycolor{the same amount of storage} in a DRAM chip, \mycolor{leading to high
area cost.} We estimate the effective area of using shorter \mycolor{bitlines}
by using CACTI-D~\cite{shyamkumar-isca2008}. Our estimation shows that using
128 cells-per-bitline increases area by 30\% compared to using 512
cells-per-bitline (\mycolor{commonly used in modern DRAM chip})
\mycolor{because doing so requires four times the sense amplifiers of
conventional DRAM.} Therefore, this approach results in significantly higher
cost-per-bit than conventional DRAM. Compared to these approaches, all our
proposals reduce DRAM latency {\em without} significant area overhead and thus
achieve both high system performance and low implementation cost.

{\bf Cached DRAM -- Integrating SRAM Cache in DRAM.} Cached DRAM~\cite{esdram,
charles-compcon1994, hidaka-ieeemicro1990, hsu-isca1993, vc-sdram,
sartore-patent1999, zhang-ieeemicro2001} integrates an SRAM cache in a DRAM
chip. Due to the locality in memory accesses, many requests could hit recently
accessed data that is temporarily stored in \mycolor{the} on-DRAM cache.
Cached DRAM serves such requests from the low latency SRAM cells, reducing
overall memory access latency. However, \mycolor{such} cached DRAM approaches
have two major limitations. First, an SRAM cache incurs significant area
overhead. Based on DRAM area analysis using CACTI-D~\cite{shyamkumar-isca2008},
an SRAM-cached DRAM requires 145.3\% additional area \mycolor{(leading to
requiring 245.3\% area of the baseline conventional DRAM)} to integrate SRAM
cells (6\% of total DRAM capacity) in a DRAM chip. Second, transferring data
between the DRAM array and the SRAM cache requires \mycolor{the use of}
relatively narrow global I/O \mycolor{buses} within the DRAM chip, leading to
\mycolor{high latency to move the data from the DRAM array into the SRAM cache
(and vice versa).}

Compared to this approach, our proposals reduce DRAM latency {\em without}
significant area overhead and thus achieve both high system performance and low
implementation cost.

{\bf Enabling Latency Heterogeneity in DRAM.} Son et al.~\cite{son-isca2013}
proposed a low latency DRAM architecture in two major directions. First, it
integrates both short bitline subarrays, which \mycolor{have} low access
latency, and long bitline subarrays, which have high access latency, within a
DRAM chip. Second, it leverages the latency difference from the physical
locations of banks. For example, banks near an IO interface can be accessed
with lower latency than banks far from an IO interface. Therefore, this
approach enables latency heterogeneity in DRAM by enabling different bitline
lengths in different locations. For example, accessing a subarray, which is
near the IO interface and \mycolor{which} consists of short \mycolor{bitlines},
has the lowest access latency. This static \mycolor{partitioning} of different
latency regions limits the effectiveness of the approach. \mycolor{One
shortcoming of the design is the high latency required to move data between the
slow and fast regions.} This approach provides largest benefit \mycolor{if
latency critical data is statically allocated} to the low latency regions (the
low latency subarrays). \mycolor{Static identification of latency critical or
hot data could be difficult.}

Lu et al.~\cite{lu-micro2015} improves the heterogeneous subarray architecture
(having both long bitline subarrays and short bitline subarrays) by introducing
\mycolor{low-latency migration capability between slow and fast subarrays.} In
the open-bitline scheme~\cite{inoue-jssc1988}, even bitlines are connected to
the upper sense amplifiers and odd bitlines are connected to the lower sense
amplifiers. This work introduces new DRAM cells \mycolor{to connect} these two
bitlines (one connected to the upper sense amplifier and the other to the lower
sense amplifier), enabling data migration between \mycolor{slow and fast}
subarrays (e.g., long bitline subarrays and short bitline subarrays).
\mycolor{However, this approach requires specialized migration cells, leading
to higher manufacturing cost.}

Our first approach (TL-DRAM) reduces DRAM latency by introducing latency
heterogeneity {\em within} a subarray, which is different from the works of Son
et al. and Lu et al., which enable heterogeneity {\em across} subarrays.
\mycolor{As such, TL-DRAM can achieve very fast migration between the slow and
fast regions of the same subarray.}

Our two other approaches (AL-DRAM and AVA-DRAM) \mycolor{reduce} DRAM latency
by exploiting the existing latency slack in DRAM, which none of these prior
works has leveraged. Therefore, our proposals can be combined with these prior
works in a synergistic manner to achieve \mycolor{even lower DRAM latency than
all individual proposals}.

\section{Reducing DRAM Latency by Enabling Heterogeneous Memory Control}
\label{sec:hetero}

Prior works optimize DRAM latency for the operating conditions by exploiting
process variation~\cite{chandrasekar-date2014} or memory access
patterns~\cite{shin-hpca2014, hassan-hpca2016}. The key distinguishing factor
of our proposals is the fact that they provide mechanisms to maintain {\em
reliability} while reducing DRAM latency.

{\bf Heterogeneous Memory Control Based on Process and Voltage Variation.}
Chandrasekar et al.~\cite{chandrasekar-date2014} evaluate the potential of
relaxing some DRAM timing parameters to reduce DRAM latency. This work observes
latency variations across DIMMs as well as for a DIMM at different operating
temperatures. However, there is no explanation as to why this phenomenon exists
and no clear mechanism \mycolor{to exploit it in this prior work.}

{\bf Heterogeneous Memory Control Based on \mycolor{Memory Access Patterns.}}
Shin et al.~\cite{shin-hpca2014} show that DRAM leakage affects two DRAM timing
parameters (\trcd/\tras). \mycolor{As a result,} recently-refreshed rows have
more charge, \mycolor{and} can be accessed with lower latency than DRAM
standard. Based on this observation, they propose a mechanism to access
recently refreshed rows with reduced latency. However, this approach has two
limitations. First, this work focuses only on the latency variation between
refreshes, resulting in relatively small performance gains. Second, it is not
clear how the proposed mechanism affects DRAM reliability.

Compared to these prior works, our proposals identify and \mycolor{analyze} the
root cause of latency variation in detail, and provide mechanisms to maintain
DRAM reliability \mycolor{while} reducing DRAM latency.

{\bf Multi-Row Activation.} Choi et al.~\cite{choi-isca2015} propose
simultaneously activating multiple rows that have the same set of data.
\mycolor{This leads} to effectively higher amount of charge \mycolor{for the
data, thereby reducing access latency to the rows.} However, the drawback of
this approach is that it potentially sacrifices DRAM capacity because multiple
rows needs to have the same data to lower the latency to access the data.

Compared to this prior work, our proposals, AL-DRAM and AVA-DRAM, do not
\mycolor{sacrifice} capacity while reducing DRAM latency. This is because our
proposals leverage the already-existing latency variation in DRAM.

\section{Enabling More Parallelism in DRAM to Hide DRAM Latency}
\label{sec:parallelism}

Prior works have focused on mitigating DRAM latency by enabling more
parallelism in DRAM, with the goal of mitigating high DRAM latency. One major
example is \mycolor{the innovation of having multiple DRAM ``banks''.} Each
bank has its own DRAM \mycolor{cells} and access structures (e.g., row decoder
and sense amplifiers). In a multi-bank DRAM architecture, data in different
banks can be accessed simultaneously, reducing overall DRAM access latency.
Unfortunately, increasing the number of banks requires additional area for
additional access structures. There are two major approaches to enable more
parallelism in DRAM at low cost, with the goal of hiding DRAM latency, {\em i)}
enabling more accesses \mycolor{to occur} in parallel, and {\em ii)} enabling
\mycolor{accesses and refreshes to occur} in parallel. The key distinguishing
factor of the proposals in this dissertation is that they {\em directly} reduce
DRAM latency \mycolor{instead of trying to overlap latency by exploiting
parallelism.}

{\bf Enabling Accesses to Different Subarrays in Parallel.} Kim et
al.~\cite{kim-isca2012} propose a new DRAM architecture (SALP) that enables
\mycolor{the almost simultaneous access of many subarrays} in a DRAM chip at
low cost. The key observation behind this work is that each DRAM subarray (in a
DRAM bank) has its own cell access structures. By decoupling global structures
over subarrays in a bank, SALP \mycolor{enables} each subarray to operate
\mycolor{mostly} in parallel, enabling more requests to be served in parallel
\mycolor{(to be precise, in a pipelined manner)}.
\mycolor{Half-DRAM~\cite{zhang-isca2014} proposes a low-cost implementation of
SALP.}

{\bf Enabling More Ranks in a DRAM channel.} Several prior
works~\cite{ahn-cal2009, ware-iccd2006, zheng-micro2008} propose partitioning a
DRAM rank into multiple independent rank subsets that can be accessed in
parallel~\cite{ahn-taco2012}. All of these proposals reduce the frequency of
row buffer conflicts and bank conflicts, \mycolor{thereby potentially improving
DRAM access latency. However, these works require more cycles to transfer data
for a single access (due to the narrower bus allocated to each subset), which
increases the latency of a single DRAM access.}

{\bf Enabling Access and Refresh in Parallel.} DRAM needs to
\mycolor{periodically} perform refresh operations to restore \mycolor{the lost
charge} in DRAM cells. During this refresh operation, DRAM \mycolor{cannot} be
accessed, \mycolor{causing delays in the servicing of requests.} Chang et
al.~\cite{chang-hpca2014} propose a new mechanism that allows \mycolor{the
servicing of data access requests} and refresh operations in parallel by
leveraging partial array refresh (which already exists in DRAM) and SALP. As a
result, this work mitigates the negative impact of refresh to DRAM latency.

These prior works do {\em not} fundamentally reduce DRAM latency
\mycolor{(instead, they exploit parallelism to tolerate the latency)} and do
{\em not} \mycolor{provide latency overlapping benefits when there are rank,
bank, or subarray conflicts.} \mycolor{In contrast, our proposals directly
reduce DRAM latency even when there are rank, bank, or subarray conflicts.} We
believe that our proposals can be combined with these prior works in a
synergistic manner to further \mycolor{hide} memory access latency.

\section{Memory Scheduling for Mitigating High DRAM Latency}
\label{sec:scheduling}

To tolerate high DRAM latency, many memory scheduling techniques have been
proposed in two major directions. The first approach is increasing more
parallelism by exploiting DRAM interface. The second approach is prioritizing
latency critical requests with the awareness of applications' characteristics.
The key difference of our proposals from these prior approaches is that our
proposals directly reduce \mycolor{raw} DRAM latency, \mycolor{whereas memory
scheduling can reduce queuing or DRAM produced overhead latencies}.

{\bf Memory Scheduling for More Parallelism.} Several prior
works~\cite{rixner-isca2000, mutlu-isca2008, moscibroda-podc2008,
lee-micro2009, lee-techreport2010, frfcfs-patent} propose memory scheduling
techniques that enable more parallelism. FRFCFS (First-Ready First-Come
First-Serve)~\cite{rixner-isca2000, frfcfs-patent} prioritizes accesses to
already activated row. Serving more requests from the already activated row
amortizes the row activation and precharge latency over multiple requests.
Parallelism-Aware Batch Scheduling (PARBS)~\cite{mutlu-isca2008,
moscibroda-podc2008} forms batches of requests and serves requests for a row
together, maximizing parallelism in memory accesses. Lee et
al.~\cite{lee-micro2009} proposes two memory scheduling mechanisms that
maximize the bank-level parallelism (BLP). The first mechanism, BAPI (BLP-Aware
Prefetch Issues), manages Miss Status Holding Registers (MSHRs) to issue
prefetch requests to different banks for serving them in parallel. The second
mechanism, BPMRI (BLP-Preserving Multi-core Request Issue), manages the memory
controller's request buffers to issue requests from the same core together,
reducing interference from other cores in multi-core systems. DRAM-aware
writeback~\cite{lee-techreport2010} maximizes the row-buffer locality by
scheduling the write requests corresponding to the activated rows.
\mycolor{The dirty-block indexing~\cite{seshadri-isca2014}} reorganizes the
dirty bits in on-chip cache to efficiently gather per-row dirty bit
information, which helps to efficiently implement DRAM-aware scheduling
mechanisms (e.g., DRAM-aware writeback).

{\bf Application-Aware Memory Scheduling.} Several prior
works~\cite{nesbit-micro2006, moscibroda-usenix2007, mutlu-isca2008,
moscibroda-podc2008, kim-hpca2010, kim-micro2010, muralihara-micro2011,
ausavarungnirun-isca2012, subramanian-hpca2013, subramanian-iccd2014,
subramanian-tpds2016, subramanian-micro2015, ebrahimi-micro2011,
stfm-micro2007, zhao-micro2014, usui-taco2016, ausavarungnirun-pact2015,
das-hpca2013, jog-sigmetrics2016} propose application-aware memory scheduling
techniques that take into account the memory access characteristics of
applications with the goal of tolerating high DRAM latency. The key idea is
\mycolor{to prioritize latency-critical} requests over other requests.
PARBS~\cite{mutlu-isca2008, moscibroda-podc2008} batches the oldest requests
from applications and ranks individual applications based on the number of
outstanding requests from the application. Using this total rank order, PARBS
prioritizes requests of applications that have low-memory-intensity.
ATLAS~\cite{kim-hpca2010} ranks individual applications based on the amount of
long-term memory service each application receives, and prioritizes
applications that receive low memory service. TCM (Thread cluster memory
scheduling)~\cite{kim-micro2010} ranks individual applications by memory
intensity such that low-memory-intensity applications are prioritized over
high-memory-intensity applications. Prior works have exploited DRAM access
scheduling in the controller (e.g.~\cite{ausavarungnirun-isca2012,
ebrahimi-micro2011, stfm-micro2007, subramanian-hpca2013,
subramanian-iccd2014, subramanian-tpds2016, subramanian-micro2015,
zhao-micro2014, usui-taco2016, ipek-isca2008, hashemi-isca2016}), to mitigate
inter-application interference, queueing latencies, \mycolor{and DRAM produced
overheads} in multi-core systems.

Compared to these prior works, our proposals directly reduce \mycolor{raw} DRAM
latency. Thus, our proposals are fundamentally different from these memory
scheduling mechanisms. Our proposals can be also combined with these prior
works to further reduce DRAM latency.

\section{In-Memory Communication and Computation}
\label{sec:in-memory_pim}

Transferring \mycolor{significant amounts of} data over the memory channel
\mycolor{takes significant amount of time.} \mycolor{Therefore, it can
significantly delay} other data transfers, leading to high DRAM latency. To
mitigate the negative impacts of the memory channel contention, prior works
have focused on reducing the memory channel contention by {\em i)} offloading
bulk data movement to DRAM~\cite{seshadri-micro2013, seshadri-micro2015,
chang-hpca2016} or \mycolor{other parts of} the memory
system~\cite{lee-pact2015, seshadri-isca2015}, and {\em ii)} enabling in-memory
computation~\cite{stone-tc1970, kogge-icpp1994, patterson-ieeemicro1997,
seshadri-cal2015, ahn-isca2015a, ahn-isca2015b, gao-pact2015,
farmahini-hpca2015, guo-wondp2014, zhang-hpdc2014}. The key difference of our
proposals from these prior works is that our proposals directly reduce
\mycolor{raw} DRAM latency. \mycolor{Our proposals can be also combined with
these prior works.}

\noindent{\bf Bulk In-Memory or Across-Memory Communication.}
Rowclone~\cite{seshadri-micro2013} proposes in-memory bulk data copy and
initialization by enabling \mycolor{an} extremely high bandwidth interface
\mycolor{within} a DRAM subarray. In a conventional DRAM-based memory system,
to migrate a page data from the original row (source row) to the other row
(destination row), the memory controller first reads the data from the source
row, and writes the data back to the destination row, which takes very long
time (high latency). Using direct wire connections between cells in a subarray
(bitlines), Rowclone migrates data from a row to the other row directly with
low latency. While Rowclone is promising to enable lower latency for
page-granularity data transfer \mycolor{and initialization}, Rowclone
\mycolor{reduces latency only for page copy and initialization operations.}
Compared to this prior work, our proposals reduce DRAM latency in general, and
thus \mycolor{can} improve performance \mycolor{of} most memory intensive
applications.

LISA~\cite{chang-hpca2016} enables \mycolor{fast} in-memory bulk data movement
across the subarrays in a DRAM bank by connecting adjacent sense amplifiers
through bitlines. \mycolor{By placing a} fast subarray (having short bitlines)
and \mycolor{a} slow subarray (having long bitlines) side by side and
\mycolor{by} enabling data migrations between these different latency
subarrays, \mycolor{LISA} enables latency heterogeneity in a DRAM bank.
Compared to this work \mycolor{that combines different-latency} subarrays, our
TL-DRAM \mycolor{substrate} enables \mycolor{different-latency} segments {\em
within} a subarray, leading to \mycolor{a simpler and more} efficient subarray
architecture. Furthermore, AL-DRAM and AVA-DRAM do not change DRAM
architecture, having more productivity compared to the prior work.

Prior works~\cite{seshadri-micro2015, lee-pact2015} have focused on efficiently
managing IO interfaces by either transferring data from different
\mycolor{locations} in a DRAM row \mycolor{in a gather-scatter
manner}~\cite{seshadri-micro2015} or enabling direct transfer through Dual-Port
DRAM~\cite{lee-pact2015}. In this dissertation, we exploit latency
heterogeneity in DRAM, reducing overall DRAM access latency. Thus, our
proposals are fundamentally different from these works and can be combined with
them to further reduce DRAM latency.

\noindent{\bf Bulk In-Memory Computation.} Prior works have focused on
offloading simple \mycolor{computations} to the memory system, reducing the
load on the memory channel (e.g., \cite{stone-tc1970, kogge-icpp1994,
patterson-ieeemicro1997, seshadri-cal2015, ahn-isca2015a, ahn-isca2015b,
gao-pact2015, farmahini-hpca2015, guo-wondp2014, zhang-hpdc2014,
hsieh-isca2016}). The key distinguishing factor of the proposals in this
dissertation is the fact that they reduce \mycolor{raw} DRAM latency while the
most prior works \mycolor{in in-memory computation mainly} reduce DRAM
bandwidth requirement. Thus, our proposals can be combined with \mycolor{these
prior works} to further improve the memory system performance.

\section{Enabling Heterogeneity in the System to Optimize System Design}
\label{sec:other_discussion}

Heterogeneity is a fundamental approach to approximate the benefits of multiple
different components or technologies to enable the optimization of multiple
metrics~\cite{mutlu-superfri2015, mutlu-acar2010}. Many recent works exploited
heterogeneous designs in the system, including the memory system, for various
purposes. These include works that combine multiple different memory
technologies to achieve the benefits of multiple
technologies~\cite{lee-isca2009, lee-ieeemicro2010, qureshi-isca2009,
yoon-iccd2012, meza-cal2012, ramos-ics2011, phadke-date2011, meza-weed2013,
li-arxiv2015, ren-micro2015, nil-micro2012, luo-dsn2014}, works that combine
multiple different interconnects to achieve both high-performance and
energy-efficient interconnect designs~\cite{mishra-dac2013,
balasubramonian-hpca2005, muralimanohar-isca2007, grot-isca2011,
grot-ieeemicro2012, ausavarungnirun-sbacpad2014, ausavarungnirun-pc2016,
grot-hpca2009, seshadri-micro2015}, works that combine multiple different types
of cores to achieve both high serial performance and high parallel throughput
and high energy efficiency~\cite{annavaram-isca2005, kumar-micro2003,
suleman-asplos2009, suleman-isca2010, joao-isca2013, joao-asplos2012}, works
that combine multiple different types of execution paradigms within a single
core to improve single-thread performance at high
efficiency~\cite{lukefahr-micro2012, fallin-iccd2014}, works that incorporate
heterogeneity into thread/memory/interconnect scheduling, throttling and
partitioning algorithms to optimize for multiple metrics~\cite{kim-micro2010,
usui-taco2016, nychis-hotnets2010, nychis-sigcomm2012, das-micro2009,
das-isca2010, das-hpca2013, ausavarungnirun-isca2012, ausavarungnirun-pact2015,
luo-msst2015, kayiran-micro2014, chang-sbacpad2012, ebrahimi-asplos2010,
ebrahimi-hpca2009, muralihara-micro2011, thottethodi-hpca2001, zhao-micro2014,
grot-ieeemicro2012, grot-isca2011, grot-micro2009, jog-sigmetrics2016,
lee-micro2008, luo-dsn2014}, and works that exploit heterogeneous behavior due
to process variation in the system to improve energy efficiency, performance or
lifetime~\cite{liu-isca2012, khan-sigmetrics2014, cai-date2013, cai-iccd2012,
cai-sigmetrics2014, cai-iccd2013, cai-dsn2015, cai-hpca2015}. In this
dissertation, we exploit the notion of heterogeneity to reduce DRAM latency at
low cost, an approach that is new.

\section{Other Related Prior Works for Mitigating High DRAM Latency}
\label{sec:other_prior}

There are many other methods \mycolor{proposed for} reducing/hiding memory
latency.

\noindent{\bf Caching and Paging Techniques.} First, many prior works
(e.g.,~\cite{collins-micro1999, jaleel-isca2010, johnson-tc1999,
qureshi-isca2006, piquet-acsac2007, qureshi-isca2007, qureshi-micro2006,
seshadri-pact2012, khan-hpca2014, seshadri-taco2015}) have proposed
sophisticated cache management policies in the context of processor SRAM
caches. These techniques are potentially applicable to our TL-DRAM to manage
which rows get cached in the near segment. The TL-DRAM substrate also allows
the hardware or the operating system to exploit the asymmetric latencies of the
near and far segments in either hardware or software through intelligent page
placement and migration techniques~\cite{verghese-asplos1996,
chandra-asplos1994, muralihara-micro2011, sudan-asplos2010, das-hpca2013}.

\noindent{\bf Aggressive Prefetching.} Second, systems employ prefetching
techniques to preload data from memory before it is
needed~\cite{srinath-hpca2007, patterson-sosp1995, nesbit-pact2004,
ebrahimi-hpca2009, ebrahimi-micro2009, ebrahimi-isca2011, dahlgren-tpds1995,
alameldeen-hpca2007, cao-sigmetrics1995, lee-micro2008, mutlu-hpca2003,
mutlu-ieeemicro2003, mutlu-isca2005, mutlu-micro2005, dundas-ics1997,
cooksey-asplos2002, mutlu-ieeemicro2006}, which may hide some memory access
latencies. However, prefetching is not efficient for applications that have
irregular access patterns, consumes memory system bandwidth, and increases
interference in the memory system~\cite{ebrahimi-isca2011, ebrahimi-hpca2009,
ebrahimi-micro2009, lee-micro2008}.

\noindent{\bf Multithreading.} Third, systems employ
multithreading~\cite{thornton-fjcc1964, smith-icpp1978}. Multithreading enables
instruction-level parallelism, but, increases contention in the
memory system~\cite{moscibroda-usenix2007, ebrahimi-isca2011, mutlu-isca2008,
das-hpca2013} and does not aid to increase single-core
performance~\cite{joao-asplos2012, suleman-asplos2009}.

\noindent{\bf Value Prediction.} Fourth, systems \mycolor{can} employ value
prediction~\cite{lipasti-asplos1996, sazeides-micro1997, zhou-ics2003,
thwaites-pact2014, yazdanbakhsh-taco2016, mutlu-micro2005, yazdanbakhsh-dt2016,
mutlu-ieeetc2006} to predict results of instructions before loading required
data from memory, which might hide some memory access latencies. However,
incorrect value predictions require roll back, degrading performance, value
prediction requires additional complexity, and many instructions are difficult
to value-predict accurately.

\noindent{\bf Dynamic Instruction Reuse.} Fifth, systems \mycolor{can} employ
dynamic instruction reuse techniques~\cite{sodani-isca1997, connors-micro1999,
alvarez-ics2001, alvarez-tc2012, huang-hpca1999, citron-asplos1998,
arnau-isca2014}, which mostly reduces repeated computations and partially
reduces the memory bandwidth consumption. \mycolor{However, these techniques}
do not reduce memory access latency.

\noindent{\bf Memory/Cache Compression.} Sixth, systems employ
memory/cache compression~\cite{pekhimenko-micro2013, pekhimenko-pact2012,
shafiee-hpca2014, zhang-asplos2000, wilson-atec1999, dusser-ics2009,
douglis-usenix1993, decastro-sbacpad2003, alameldeen-tech2004,
alameldeen-isca2004, abali-ibm2001, pekhimenko-hpca2015, pekhimenko-hpca2016},
which reduces the memory bandwidth consumption in the memory system and
mitigates the memory channel contention. However, this approach does not
reduce \mycolor{raw} DRAM access latency.

Our proposals are largely orthogonal to all of these prior works and can be
combined with these prior works to further improve system performance by
reducing \mycolor{raw} DRAM latency.

	\chapter{Tiered-Latency DRAM:\\
Lowering Latency by Modifying the Bitline Architecture}
\label{ch:tldram}

\let\thefootnote\relax\footnotetext{This work has been published in HPCA
2013~\cite{lee-hpca2013}. This dissertation includes more discussions and
evaluation results in Sections~\ref{sec:power},~\ref{sec:result_near_length},
and~\ref{sec:result_dual} beyond the HPCA 2013 paper.}
\let\thefootnote\svthefootnote

The capacity and cost-per-bit of DRAM have historically scaled to satisfy the
needs of increasingly large and complex computer systems, while DRAM latency
has remained almost constant, making memory latency the performance bottleneck
in today's systems, as we demonstrated and discussed in Chapter~\ref{ch:intro}.
However, the high latency of commodity DRAM chips is in fact a {\em deliberate}
trade-off made by DRAM manufacturers. While process technology scaling has
enabled DRAM designs with both lower cost-per-bit {\em and} lower
latency~\cite{itrs}, DRAM manufacturers have usually sacrificed the latency
benefits of scaling in order to achieve even lower cost-per-bit, as we explain
below. Hence, while low-latency DRAM chips exist~\cite{kimuta-isscc1999,
rldram, sato-vlsi1998}, their higher cost-per-bit relegates them to specialized
applications such as high-end networking equipment that require very low
latency even at a very high cost~\cite{ciaran-ahs2007}.

\begin{figure}[h]
	\centering
	\subcaptionbox{Latency Optimized\label{fig:intro_specialized_dram}}[0.37\linewidth] {
		\includegraphics[width=1in]{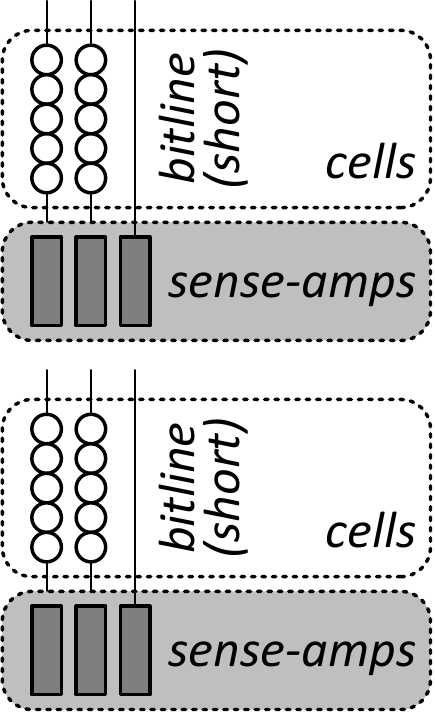}
	}
	\subcaptionbox{Cost Optimized\label{fig:intro_commodity_dram}}[0.20\linewidth] {
		\includegraphics[width=1in]{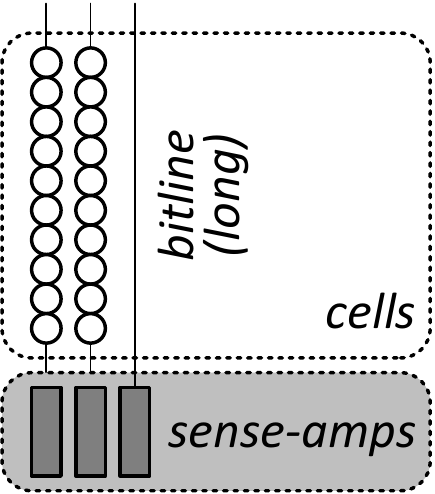}
	}
	\subcaptionbox{Our Proposal\label{fig:intro_tldram}}[0.37\linewidth] {
		\includegraphics[width=1in]{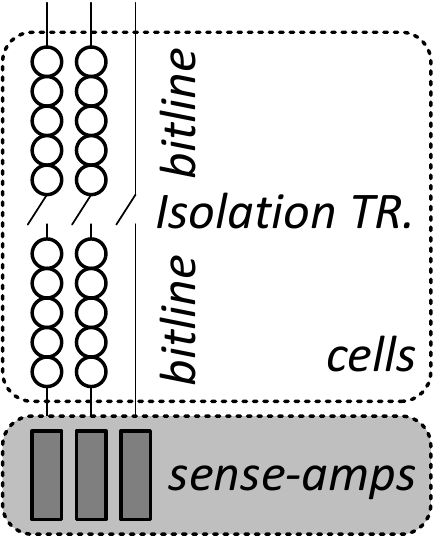}
	}
	\caption{DRAM: Latency vs. Cost Optimized, Our Proposal}
	\label{fig:intro_commodity_specialized}
\end{figure}

DRAM manufacturers trade latency for cost-per-bit by adjusting the length of
these bitlines. Shorter bitlines (fewer cells connected to the bitline)
constitute a smaller electrical load on the bitline, resulting in decreased
latency, but require a larger number of sense amplifiers for a given DRAM
capacity (Figure~\ref{fig:intro_specialized_dram}), resulting in higher
cost-per-bit. Longer bitlines (more cells connected to the bitline) require
fewer sense amplifiers for a given DRAM capacity
(Figure~\ref{fig:intro_commodity_dram}), reducing cost-per-bit, but impose a
higher electrical load on the bitline, increasing latency. As a result, neither
of these two approaches can optimize for both cost-per-bit and latency.

{\bf Our goal} is to design a new DRAM architecture that provides low latency
for the common case while still retaining a low cost-per-bit overall. Our
proposal, which we call {\em Tiered-Latency DRAM}, is based on the key
observation that {\em long bitlines are the dominant source of DRAM
latency}~\cite{sharroush-elek2012}.

{\bf Our key idea} is to adopt long bitlines to achieve low cost-per-bit, while
allowing their lengths to appear shorter in order to achieve low latency. In
our mechanism ({\em Tiered-Latency DRAM}), each long bitline is split into two
shorter segments using an {\em isolation transistor}, as shown in
Figure~\ref{fig:intro_tldram}: the segment that is connected directly to the
sense amplifier is called the {\em near segment}, whereas the other is called
the {\em far segment}. To access a cell in the near segment, the isolation
transistor is turned off, so that the cell and the sense amplifier see only the
portion of the bitline corresponding to the near segment (i.e., reduced
electrical load). Therefore, the near segment can be accessed quickly. On the
other hand, to access a cell in the far segment, the isolation transistor is
turned on to connect the entire length of the bitline to the sense amplifier.
In this case, however, the cell and the sense amplifier see the full electrical
load of the bitline in addition to the extra load of the isolation transistor.
Therefore, the far segment can be accessed slowly.

	\section{Motivation: Short vs. Long Bitlines} \label{sec:motivation}

The key parameter in the design of a DRAM subarray is the number of DRAM cells
connected to each bitline (cells-per-bitline) -- i.e., the number of DRAM rows
in a subarray. This number directly affects the length of the bitline, which in
turn affects both the access latency and the area of the subarray. As we
describe in this section, the choice of the number of cells-per-bitline
presents a crucial trade-off between the DRAM access latency and the DRAM
die-size.

\subsection{Latency Impact of Cells-per-Bitline}
\label{subsec:cells-per-bitline-latency}

Every bitline has an associated parasitic capacitance whose value is
proportional to the length of the bitline. This parasitic capacitance increases
the subarray operation latencies: {\em i)} charge sharing, {\em ii)} sensing \&
amplification, and {\em iii)} precharging, which we discussed in
Figure~\ref{fig:operation} in Chapter~\ref{ch:bak}.

First, the bitline capacitance determines the bitline voltage after charge
sharing. The larger the bitline capacitance, the closer its voltage will be to
\hvdd after charge sharing. Although this does not significantly impact the
latency of charge sharing, this causes the sense amplifier to take longer to
amplify the voltage to the final restored value (\vdd or {\em 0}).

Second, in order to amplify the voltage perturbation on the bitline, the sense
amplifier injects (or withdraws) charge into (or from) both the cell and the
bitline. Since the sense amplifier can do so only at a fixed rate, the
aggregate capacitance of the cell and the bitline determine how fast the
bitline voltage reaches the {\em threshold} and the {\em restored} states
(States \numthree and \numfour in Figure~\ref{fig:operation} in
Chapter~\ref{ch:bak}). A long bitline, which has a large parasitic capacitance,
slows down the bitline voltage from reaching these states, thereby lengthening
both \trcd and \tras, respectively.

Third, to precharge the bitline, the sense amplifier drives the bitline voltage
to the quiescent value of \hvdd. Again, a long bitline with a large
capacitance is driven more slowly and hence has a large \trp.

\subsection{Die-Size Impact of Cells-per-Bitline}
\label{subsec:cells-per-bitline-area}

Since each cell on a bitline belongs to a row of cells (spanning horizontally
across multiple bitlines), the number of cells-per-bitline in a subarray is
equal to the number of rows-per-subarray. Therefore, for a DRAM chip with a
given capacity (i.e., fixed total number of rows), one can either have many
subarrays with short bitlines (Figure~\ref{fig:intro_specialized_dram}) or few
subarrays with long bitlines (Figure~\ref{fig:intro_commodity_dram}). However,
since each subarray requires its own set of sense amplifiers, the size of the
DRAM chip increases along with the number of subarrays. As a result, for a
given DRAM capacity, its die-size is inversely proportional to the number of
cells-per-bitline (as a first-order approximation).

\subsection{Trade-Off: Latency vs. Die-Size}
\label{subsec:trade-off-latency-area}

\begin{figure}[h]
	\centering
  \includegraphics[width=0.6\linewidth]{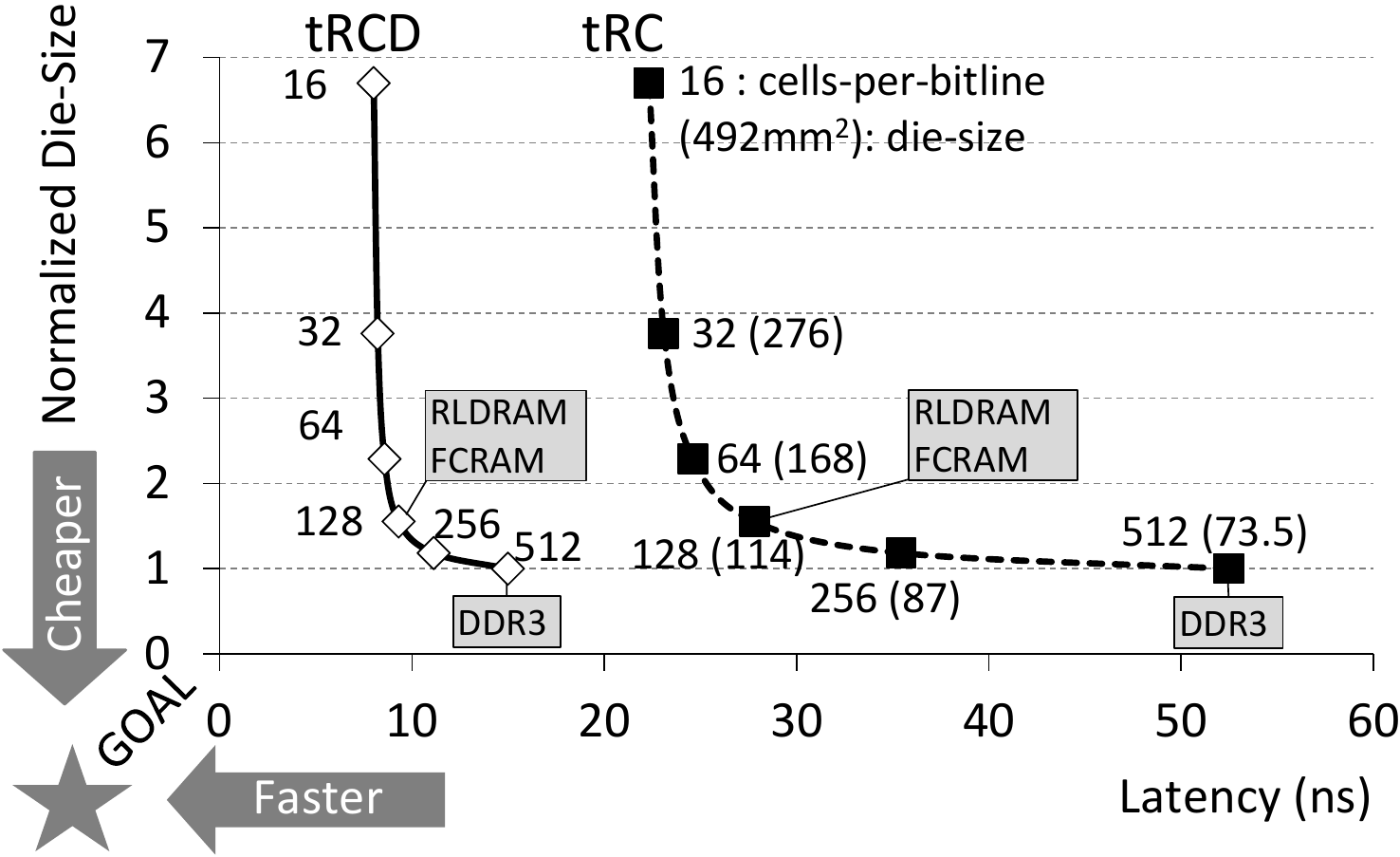} \\
  \footnotesize{$\dagger$ RLDRAM bitline length is estimated from its latency
  and die-size~\cite{dram-circuit-design,rldram}.} \\
  \footnotesize{$\dagger$ The reference DRAM is 55nm 2GB DDR3~\cite{rambus-power}.}	\\
  \caption{Bitline Length: Latency vs. Die-Size}
  \label{fig:cell-per-bitline-trade-off}
\end{figure}

From the above discussion, it is clear that a short bitline (fewer
cells-per-bitline) has the benefit of lower subarray latency, but incurs a
large die-size overhead due to additional sense amplifiers. On the other hand,
a long bitline (more cells-per-bitline), as a result of the large bitline
capacitance, incurs high subarray latency, but has the benefit of reduced
die-size overhead. To study this trade-off quantitatively, we ran
transistor-level circuit simulations based on a publicly available 55nm DRAM
process technology~\cite{rambus-power}.
Figure~\ref{fig:cell-per-bitline-trade-off} shows the results of these
simulations. Specifically, the figure shows the latency (\trcd and \trc) and
the die-size for different values of cells-per-bitline. The figure clearly
shows the above described trade-off between DRAM access latency and DRAM
die-size.

As Figure~\ref{fig:cell-per-bitline-trade-off} shows, existing DRAM
architectures are either optimized for die-size (commodity
DDR3~\cite{samsung-spec, moon-isscc2009}) and are thus low cost but high
latency, or optimized for latency (RLDRAM~\cite{rldram},
FCRAM~\cite{sato-vlsi1998}) and are thus low latency but high cost. {\bf Our
goal} is to design a DRAM architecture that achieves the best of both worlds --
i.e., low access latency and low cost.

	\section{Tiered-Latency DRAM (TL-DRAM)}
\label{sec:segmented}

\begin{figure}[h]
	\centering
	\includegraphics[width=0.5\linewidth]{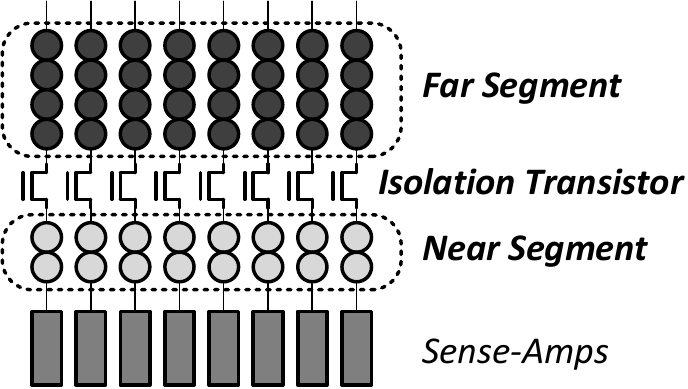}
	\caption{TL-DRAM: Near vs. Far Segments}
	\label{fig:substrate_tld}
\end{figure}

To obtain both the latency advantages of short bitlines and the cost advantages
of long bitlines, we propose the {\em Tiered-Latency DRAM} (TL-DRAM)
architecture, as shown in Figure~\ref{fig:substrate_tld}. The key idea of TL-DRAM
is to introduce an {\em isolation transistor} that divides a long bitline into
two segments: the {\em near segment}, connected directly to the
sense-amplifier, and the {\em far segment}, connected through the isolation
transistor. Unless otherwise stated, throughout the following discussion, we
assume, without loss of generality, that the isolation transistor divides the
bitline such that length of the near and far segments is 128 cells and 384
cells (=512-128), respectively. (Section~\ref{sec:latency_analysis_qualitative}
discusses the latency sensitivity to the segment lengths.)

\subsection{Latency Analysis (Overview)}
\label{sec:latency_analysis_qualitative}

The primary role of the isolation transistor is to electrically decouple the
two segments from each other. As a result, the effective bitline length (and
also the effective bitline capacitance) as seen by the cell and sense-amplifier
is changed. Correspondingly, the latency to access a cell is also changed --
albeit differently depending on whether the cell is in the near or the far
segment (Table~\ref{tab:segment_latency}), as will be explained next.

\begin{table}[h]
	\centering
	\small{
	\setlength{\tabcolsep}{12pt}
	\begin{tabular}{ccc}
	\toprule
	& Near Segment & Far Segment \\
	 & (128 cells) & (384 cells) \\ \cmidrule{1-3}
	\multirow{2}{*}{\normalsize \trcd} & \bf Reduced & \bf Reduced \\
	& (15ns \ra\ 9.3ns) & (15ns \ra\ 13.2ns) \\ \cmidrule{1-3}
	\multirow{2}{*}{\normalsize \trc} & \bf Reduced & Increased \\
	& (52.5ns \ra\ 27.8ns) & (52.5ns \ra\ 64.1ns) \\ \bottomrule
	\end{tabular}
	}
	\caption{Segmented Bitline: Effect on Latency}
	\label{tab:segment_latency}
\end{table}

\begin{figure}[h]
  \centering
  \subcaptionbox{Near Segment Access\label{fig:substrate_tldram_off}}[0.4\linewidth] {
		\centering
    \includegraphics[height=1.8in]{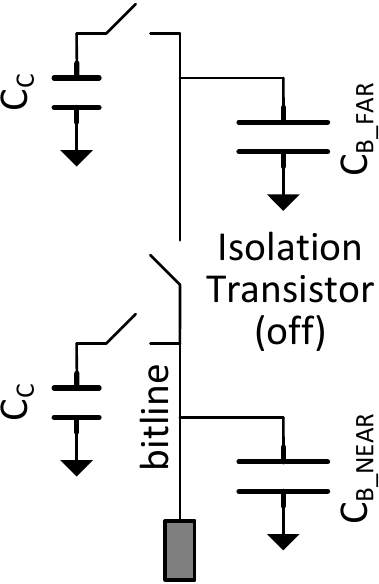}
	}
  \subcaptionbox{Far Segment Access\label{fig:substrate_tldram_on}}[0.4\linewidth] {
		\centering
    \includegraphics[height=1.8in]{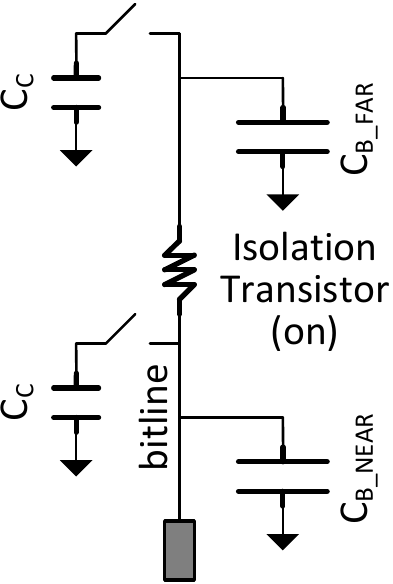}
	}
  \caption{Circuit Model of Segmented Bitline}
  \label{fig:substrate_tldram_model}
\end{figure}

{\bf Near Segment.} When accessing a cell in the near segment, the isolation
transistor is turned off, disconnecting the far segment
(Figure~\ref{fig:substrate_tldram_off}). Since the cell and the sense-amplifier
see only the reduced bitline capacitance of the shortened near segment, they
can drive the bitline voltage more easily. In other words, for the same amount
of charge that the cell or the sense-amplifier injects into the bitline, the
bitline voltage is higher for the shortened near segment compared to a long
bitline. As a result, the bitline voltage reaches the {\em threshold} and {\em
restored} states (Figure~\ref{fig:operation} in Chapter~\ref{ch:bak}) more
quickly, such that \trcd and \tras for the near segment is significantly
reduced. Similarly, the bitline can be precharged to \hvdd more quickly,
leading to a reduced \trp. Since \trc is defined as the sum of \tras and \trp
(Section~\ref{subsec:subarray_access} in Chapter~\ref{ch:bak}), \trc is reduced
as well.

{\bf Far Segment.} On the other hand, when accessing a cell in the far segment,
the isolation transistor is turned on to connect the entire length of the
bitline to the sense-amplifier. In this case, however, the isolation transistor
acts like a resistor inserted between the two segments
(Figure~\ref{fig:substrate_tldram_on}).

When the sense-amplifier is turned on during activation (as explained in
Section~\ref{subsec:subarray_operation} in Chapter~\ref{ch:bak}), it begins to
drive charge onto the bitline. In commodity DRAM, this charge is spread across
the large capacitance of the entire long bitline. However, in TL-DRAM, the
resistance of the isolation transistor limits how quickly charge flows to the
far segment, such that the reduced capacitance of the shortened near segment is
charged more quickly than that of the far segment. This has two key
consequences. First, because the near segment capacitance is charged quickly,
the near segment voltage rises more quickly than the bitline voltage in
commodity DRAM. As a result, the near segment voltage more quickly reaches
\tfvdd ({\em threshold} state, Section~\ref{subsec:subarray_operation} in
Chapter~\ref{ch:bak}) and, correspondingly, the sense-amplifier more quickly
detects the binary data value of `1' that was stored in the far segment cell.
That is why \trcd is lower in TL-DRAM than in commodity DRAM even for the far
segment. Second, because the far segment capacitance is charged more slowly, it
takes {\em longer} for the far segment voltage --- and hence the cell voltage
--- to be {\em restored} to \vdd or {\em 0}. Since \tras is the latency to
reach the {\em restored} state (Section~\ref{subsec:subarray_operation} in
Chapter~\ref{ch:bak}), \tras is increased for cells in the far segment.
Similarly, during precharging, the far segment voltage reaches \hvdd more
slowly, for an increased \trp. Since \tras and \trp both increase, their sum
\trc also increases.

\begin{figure}[h]
	\centering
	\subcaptionbox{Cell in Near Segment\label{fig:substrate_latency_near}}[0.45\linewidth] {
		\includegraphics[width=0.4\linewidth]{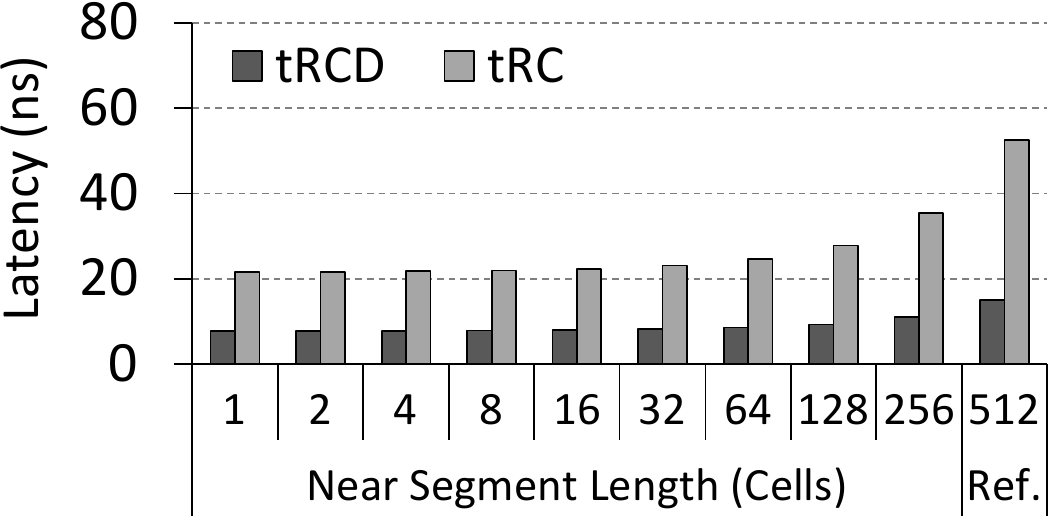}
	}
	\subcaptionbox{Cell in Far Segment\label{fig:substrate_latency_far}}[0.45\linewidth] {
		\includegraphics[width=0.4\linewidth]{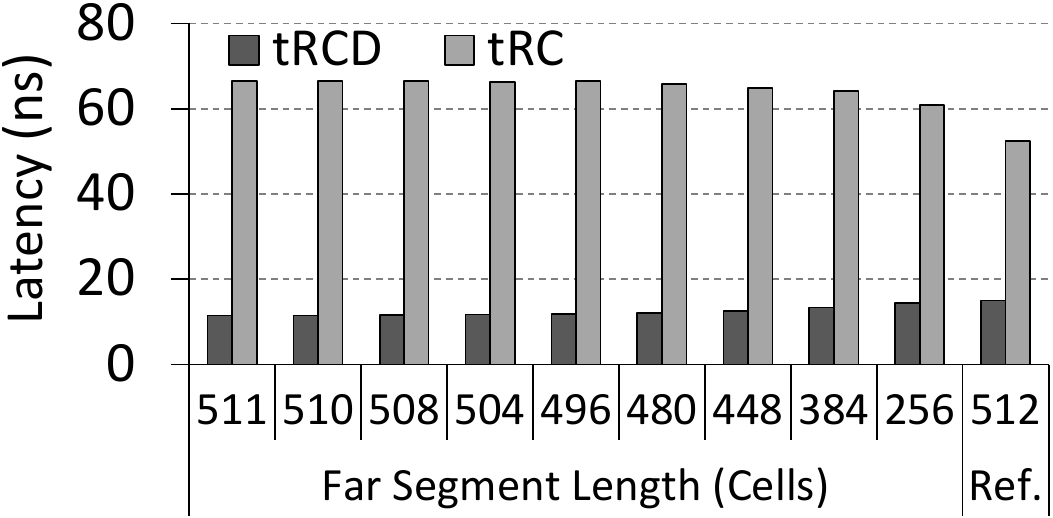}
	}
	\caption{Latency Analysis} \label{fig:substrate_latency}
\end{figure}

{\bf Sensitivity to Segment Length.} The lengths of the two segments are
determined by where the isolation transistor is placed on the bitline. Assuming
that the number of cells per bitline is fixed at 512 cells, the near segment
length can range from as short as a single cell to as long as 511 cells. Based
on our circuit simulations, Figure~\ref{fig:substrate_latency_near} and
Figure~\ref{fig:substrate_latency_far} plot the latencies of the near and far
segments as a function of their length, respectively. For reference, the
rightmost bars in each figure are the latencies of an unsegmented long bitline
whose length is 512 cells. From these figures, we draw three conclusions.
First, the shorter the near segment, the lower its latencies (\trcd and \trc).
This is expected since a shorter near segment has a lower effective bitline
capacitance, allowing it to be driven to target voltages more quickly. Second,
the longer the far segment, the lower the far segment's \trcd. Recall from our
previous discussion that the far segment's \trcd depends on how quickly the
near segment (not the far segment) can be driven. A longer far segment implies
a shorter near segment (lower capacitance), and that is why \trcd of the far
segment decreases. Third, the shorter the far segment, the smaller its \trc.
The far segment's \trc is determined by how quickly it reaches the full voltage
(\vdd or {\em 0}). Regardless of the length of the far segment, the current
that trickles into it through the isolation transistor does not change
significantly. Therefore, a shorter far segment (lower capacitance) reaches the
full voltage more quickly.

\subsection{Latency Analysis (Circuit Evaluation)}
\label{sec:latency_analysis_quantitative}

We model TL-DRAM in detail using SPICE simulations. Simulation parameters are
mostly derived from a publicly available 55nm DDR3 2Gb process technology
file~\cite{rambus-power} which includes information such as cell and bitline
capacitance and resistance, physical floorplanning, and transistor
dimensions. Transistor device characteristics were derived
from~\cite{narasimha-iedm2006} and scaled to agree with~\cite{rambus-power}.

\begin{figure}[h]
	\begin{minipage}[b] {0.58\linewidth}
		\centering
		\subcaptionbox{Cell in Near Segment (128 cells)\label{fig:substrate_sim_active_off}}[0.9\linewidth] {
			\centering
			{\footnotesize{\hspace{-10mm}\trcdnear\hspace{2mm}\trasnear\hspace{20mm}}}
			\includegraphics[height=1in]{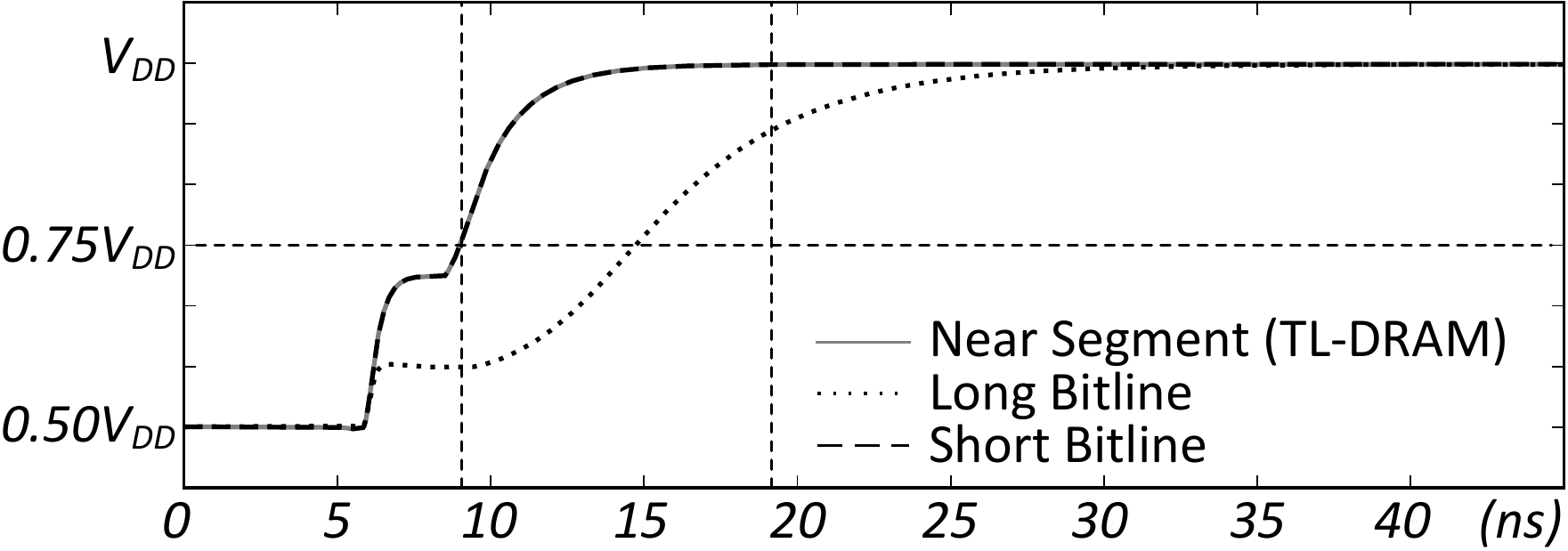}
		}

		\subcaptionbox{Cell in Far Segment (384 cells)\label{fig:substrate_sim_active_on}}[0.9\linewidth] {
			\centering
			\footnotesize{\hspace{23mm}\trcdfar\hspace{30mm}\trasfar\hspace{-15mm}}
			\includegraphics[height=1in]{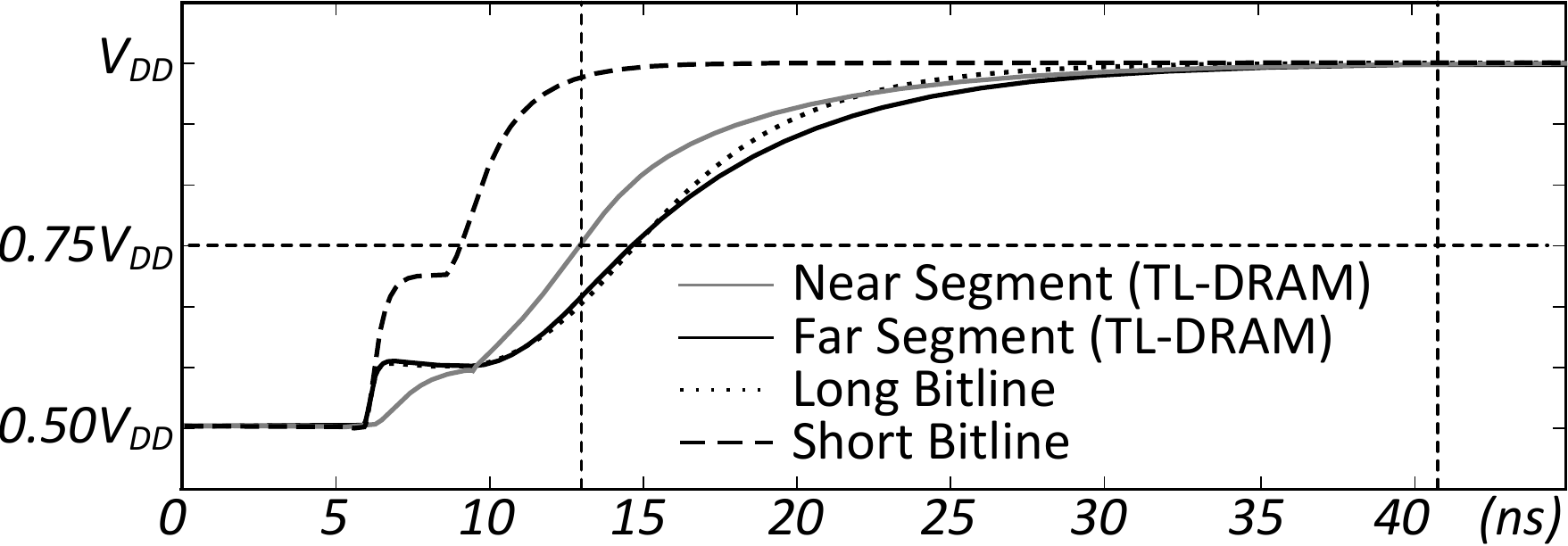}
		}
		\caption{Activation: Bitline Voltage} \label{fig:substrate_sim_active}
	\end{minipage}
	\begin{minipage}[b] {0.40\linewidth}
		\centering
		\subcaptionbox{Cell in Near Segment\label{fig:substrate_sim_prech_off}}[0.90\linewidth] {
			\centering
			{\footnotesize{\hspace{6mm}\trpnear}}
			\includegraphics[height=1in]{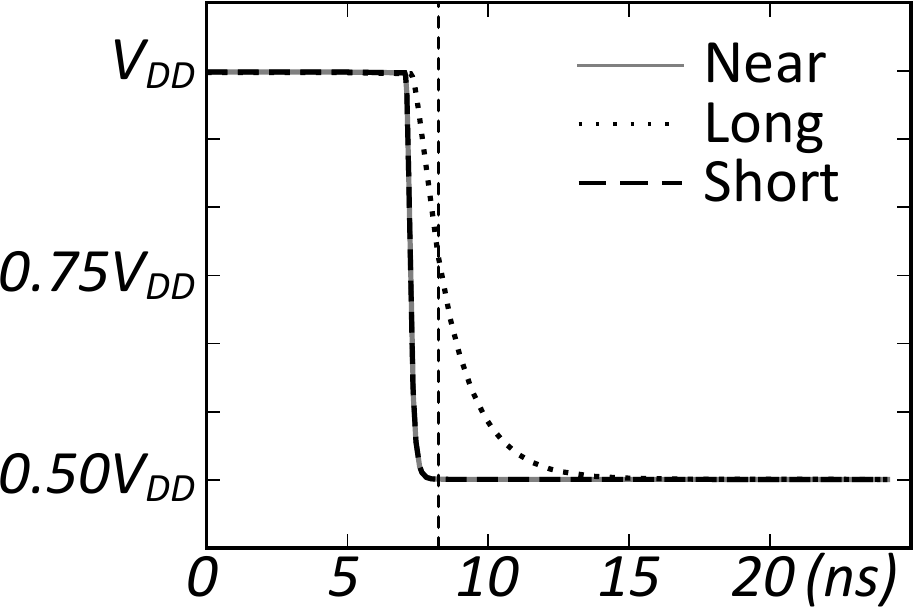}
		}

		\subcaptionbox{Cell in Far Segment\label{fig:substrate_sim_prech_on}}[0.90\linewidth] {
			\centering
			{\footnotesize{\hspace{28mm}\trpfar}}
			\includegraphics[height=1in]{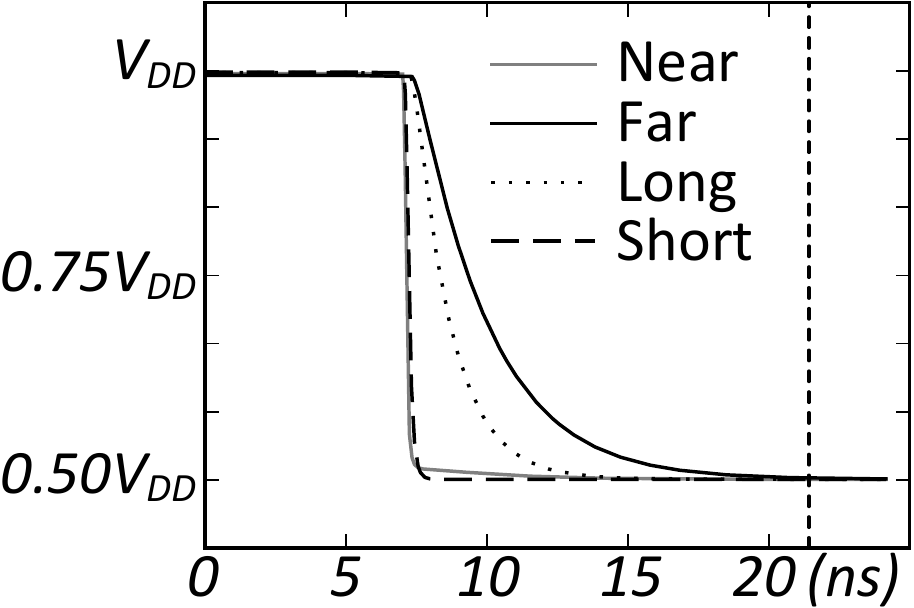}
		}
		\caption{Precharging} \label{fig:substrate_sim_prech}
		\end{minipage}
\end{figure}

Figure~\ref{fig:substrate_sim_active} and Figure~\ref{fig:substrate_sim_prech} show
the bitline voltages during activation and precharging respectively. The
$x$-axis origin (time 0) in the two figures correspond to when the subarray
receives the \cmdact or the \cmdpre command, respectively. In addition to the
voltages of the segmented bitline (near and far segments), the figures also
show the voltages of two unsegmented bitlines (short and long) for reference.

{\bf Activation} (Figure~\ref{fig:substrate_sim_active}). First, during an
access to a cell in the near segment
(Figure~\ref{fig:substrate_sim_active_off}), the far segment is disconnected
and is floating (hence its voltage is not shown). Due to the reduced bitline
capacitance of the near segment, its voltage increases almost as quickly as the
voltage of a short bitline (the two curves are overlapped) during both charge
sharing and sensing \& amplification. Since the near segment voltage reaches
\tfvdd and \vdd (the {\em threshold} and {\em restored} states) quickly, its
\trcd and \tras, respectively, are significantly reduced compared to a long
bitline. Second, during an access to a cell in the far segment
(Figure~\ref{fig:substrate_sim_active_on}), we can indeed verify that the
voltages of the near and the far segments increase at different rates due to
the resistance of the isolation transistor, as previously explained. Compared
to a long bitline, while the near segment voltage reaches \tfvdd more quickly,
the far segment voltage reaches \vdd more slowly. As a result, \trcd of the far
segment is reduced while its \tras is increased.

{\bf Precharging} (Figure~\ref{fig:substrate_sim_prech}). While precharging the
bitline after accessing a cell in the near segment
(Figure~\ref{fig:substrate_sim_prech_off}), the near segment reaches \hvdd
quickly due to the smaller capacitance, almost as quickly as the short bitline
(the two curves are overlapped). On the other hand, precharging the bitline
after accessing a cell in the far segment (Figure~\ref{fig:substrate_sim_prech_on})
takes longer compared to the long bitline baseline. As a result, \trp is reduced
for the near segment and increased for the far segment.

\subsection{Die-Size Analysis}
\label{sec:segment_diesize}

Adding an isolation transistor to the bitline increases only the height of the
subarray and not the width~\cite{lim-vlsic2001, saito-vlsic1996, keeth00,
dram-circuit-design}. Without the isolation transistor, the height of a
baseline subarray is equal to the sum of height of the cells and the
sense-amplifier. In the following analysis, we use values from the Rambus power
model~\cite{rambus-power}.\footnote{We expect the values to be of similar
orders of magnitude for other designs.} The sense amplifier and the isolation
transistor are respectively 115.2x and 11.5x taller than an individual cell.
For a subarray with 512 cells, the overhead of adding a single isolation
transistor is $\frac{11.5}{115.2 + 512}$ = 1.83\%.

\begin{figure}[h]
	\centering
	\subcaptionbox{Isolation Transistor in Open Bitline Scheme\label{fig:openbitline_org}}[0.45\linewidth] {
		\centering
		\includegraphics[height=2in]{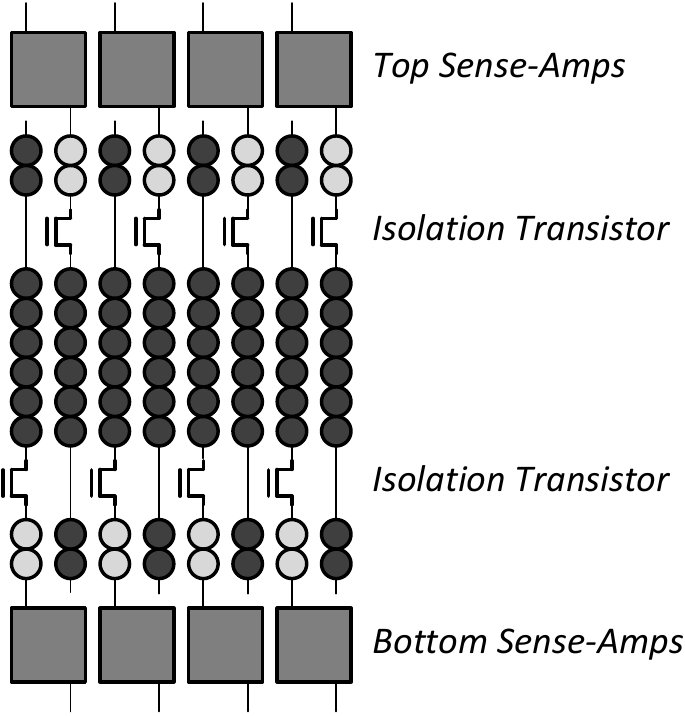}
	}
	\subcaptionbox{Far \& Near Segment in Open Bitline Scheme\label{fig:openbitline_segment}}[0.45\linewidth] {
		\centering
		\includegraphics[height=2in]{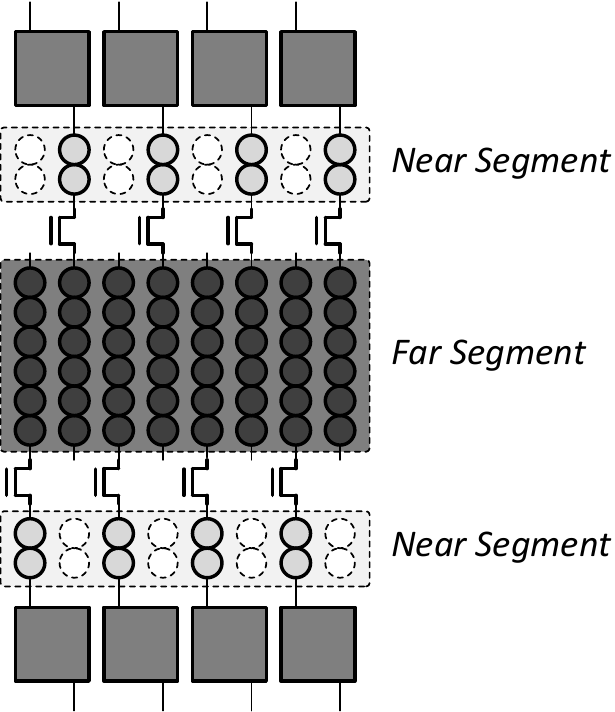}
	}\vspace{-0.2in}
	\caption{Tiered-Latency DRAM Integration in Open Bitline Scheme}
	\label{fig:openbitline_tldram}
\end{figure}

Until now we have assumed that all cells of a DRAM row are connected to the
same row of sense-amplifiers. However, \mycolor{in the open bitline scheme,} the
sense-amplifiers are twice as wide as an individual cell. Therefore, in
practice, only every other bitline (cell) is connected to a bottom row of
sense-amplifiers. The remaining bitlines are connected to the another row of
sense-amplifiers of the vertically adjacent (top) subarray. This allows for
tighter packing of DRAM cells within a subarray. As a result, each subarray
requires two sets of isolation transistors as shown in
Figure~\ref{fig:openbitline_org}. Therefore, the increase in subarray area due
to the two isolation transistors is 3.66\%. Once we include the area of the
peripheral and I/O circuitry which does not change due to the addition of the
isolation transistors, the resulting DRAM die-size area overhead is 3.15\%.

We plot the near segment and the far segment in
Figure~\ref{fig:openbitline_segment}. The portion directly connected to either
the top sense-amplifiers and the bottom sense-amplifiers form the near segment
and the remained portion is the far segment. To access the far segment, the
isolation transistors are turned on and activated an wordline of the far
segment, which are similar to the conventional DRAM's row access. To access the
near segment, isolation transistors are turned off, leading to connecting only
the two near segments are connected to sense-amplifiers (light grey portion in
Figure~\ref{fig:openbitline_segment}). In this case, half of \mycolor{the}
cells in the near segments can not be accessed since the dotted cells in
Figure~\ref{fig:openbitline_segment} are not connected to sense-amplifiers. As
a result, in \mycolor{the open bitline scheme}, \mycolor{half of the capacity}
in the near segment can not be available to use, leading to \mycolor{higher}
area overhead. For example, TL-DRAM \mycolor{with} a near segment size of 32
rows loses 3.125\% capacity, leading to 6.275\% area overhead in total.

\subsection{Enabling Inter-Segment Data Transfer} \label{sec:segment_migration}

One way of exploiting TL-DRAM's asymmetric latencies is to use the near segment
as a cache to the far segment. Compared to the far segment, the near segment is
smaller and faster. Therefore, frequently-used or latency-critical data can
benefit significantly from being placed in the near segment as opposed to the
far segment. The problem lies in enabling an efficient way of transferring
(copying or migrating) data between the two segments. Unfortunately, in
existing DRAM chips, even to transfer data from one row to another row within
the same subarray, the data must be read out externally to the DRAM controller
and then written back to the chip. This wastes significant amounts of power and
bandwidth on the memory bus (that connects the DRAM chip to the DRAM
controller) and incurs large latency.

In TL-DRAM, data transfer between the two segments occurs entirely within DRAM
without involving the external memory bus. TL-DRAM leverages the fact that the
bitline itself is essentially a bus that connects to all cells in both the near
and far segments. As an example, let us assume that we are accessing a cell in
the far segment ({transfer source}). When the bitline has reached the {\em
restored} state (Sec~\ref{subsec:subarray_operation}), the data in the cell is
fully copied onto the bitline. Normally, at this point, the DRAM controller
would issue a \cmdpre to clear the bitline voltage to \hvdd. Instead, TL-DRAM
allows the DRAM controller to issue another \cmdact, this time to a cell in the
near segment ({transfer destination}). Since the bitline is at a full voltage,
the bitline drives the near segment cell so that data is copied into it.
According to our simulations, writing into the destination cell takes about
4ns. In fact, the destination cell can be connected to the bitline even before
the source cell has reached the {\em restored} state, thereby overlapping the
copying latency with the \tras latency. More generally, the source and the
destination can be any two cells connected to the same bitline, regardless of
which segment they lie on.

	\section{Leveraging the TL-DRAM Substrate} \label{sec:mechanism}

One simple way of leveraging the TL-DRAM substrate is to use the near segment as
a hardware-managed cache for the far segment. In this approach, the memory
controller does not expose the near segment capacity to the operating system
(OS). While this approach reduces the overall available memory capacity, it
keeps both the hardware and the software design simple. Another alternative
approach is to expose the near segment capacity to the OS. As we will describe
in Section~\ref{sec:alternative}, effectively exploiting the TL-DRAM substrate
using this alternate approach may slightly increase the complexity of the
hardware or the software. We now describe our different mechanisms to leverage
the TL-DRAM substrate.

\subsection{Near Segment as an OS-Transparent Hardware-Managed Cache}
\label{sec:inclusive}

We describe three different mechanisms that use the near segment as a
hardware-managed cache to the far segment. In all three mechanisms, the memory
controller tracks the rows in the far segment that are cached in the near
segment (for each subarray). The three mechanisms differ in 1)~when they cache
a far-segment row in the near segment, and 2)~when they evict a row from the
near segment.

{\bf Mechanism 1: Simple Caching (\msc)}. Our first mechanism, {\em Simple
Caching} (SC), utilizes the near segment as an LRU cache to the far segment.
Under this mechanism, the DRAM controller categorizes a DRAM access into one of
three cases: {\em i)}~sense-amplifier hit: the corresponding row is already
activated; {\em ii)}~near segment hit: the row is already cached in the near
segment; and {\em iii)}~near segment miss: the row is not cached in the near
segment. In the first case, sense-amplifier hit (alternatively, row-buffer
hit), the access is served directly from the row-buffer. Meanwhile, the
LRU-ordering of the rows cached in the near segment remains unaffected. In the
second case, near segment hit, the DRAM controller quickly activates the row in
the near segment, while also updating it as the MRU row. In the third case,
near segment miss, the DRAM controller checks whether the LRU row (eviction
candidate) in the near segment is dirty. If so, the LRU row must first be
copied (or written back) to the far segment using the transfer mechanism
described in Section~\ref{sec:segment_migration}. Otherwise, the DRAM
controller directly activates the far segment row (that needs to be accessed)
and copies (or caches) it into the near segment and updates it as the MRU row.

{\bf Mechanism 2: Wait-Minimized Caching (\mwmc)}. When two accesses to two
different rows of a subarray arrive almost simultaneously, the first access
delays the second access by a large amount, \trc. Since the first access causes
the second access to {\em wait} for a long time, we refer to the first access
as a {\em wait-inducing access}. Assuming both rows are in the far segment, the
latency at the subarray experienced by the second access is
$t_{\mathit{RCfar}}+t_{\mathit{RCDfar}}$ (77.3ns). Such a large latency is
mostly due to the wait caused by the first access, \trcfar (64.1ns). Hence, it
is important for the second access that the wait is minimized, which can be
achieved by caching the {\em first} accessed data in the near segment. By doing
so, the wait is significantly reduced from \trcfar (64.1ns) to \trcnear
(27.8ns). In contrast, caching the {\em second} row is not as useful, since it
yields only a small latency reduction from \trcdfar (13.2ns) to \trcdnear
(9.3ns).

Our second mechanism, {\em Wait-Minimized Caching} (\mwmc), caches only {\em
wait-inducing rows}. These are rows that, while they are accessed, cause a
large wait (\trcfar) for the next access to a different row. More specifically,
a row in the far segment is classified as wait-inducing if the next access to a
different row arrives while the row is still being activated. \mwmc operates
similarly to our SC mechanism except for the following differences. First,
\mwmc copies a row from the far segment to the near segment {\em only if the
row is wait-inducing}. Second, instead of evicting the LRU row from the near
segment, \mwmc evicts the {\em least-recently wait-inducing} row. Third, when
a row is accessed from the near segment, it is updated as the
{\em most-recently wait-inducing} row only if the access would have caused the
next access to wait had the row been in the far segment. The memory controller
is augmented with appropriate structures to keep track of the necessary
information to identify wait-inducing rows (details omitted due to space
constraints).

{\bf Mechanism 3: Benefit-Based Caching (\mbbc)}. Accessing a row in the near
segment provides two benefits compared to accessing a row in the far segment:
1)~reduced \trcd (faster access) and 2)~reduced \trc (lower wait time for the
subsequent access). Simple Caching (\msc) and Wait-Minimized Caching (\mwmc)
take into account only one of the two benefits. Our third mechanism,
{\em Benefit-Based Caching} (\mbbc) explicitly takes into account both
benefits of caching a row in the near segment. More specifically, the memory
controller keeps track of a {\em benefit} value for each row in the near
segment. When a near segment row is accessed, its benefit is incremented by the
number of DRAM cycles saved due to reduced access latency and reduced wait time
for the subsequent access. When a far-segment row is accessed, it is
immediately promoted to the near segment, replacing the near-segment row with
the least benefit. To prevent benefit values from becoming stale, on every
eviction, the benefit for every row is halved. (Implementation details are
omitted due to space constraints.)

\subsection{Exposing Near Segment Capacity to the OS} \label{sec:alternative}

Our second approach to leverage the TL-DRAM substrate is to expose the near
segment capacity to the operating system. Note that simply replacing the
conventional DRAM with our proposed TL-DRAM can potentially improve system
performance due to the reduced \trcdnear, \trcdfar, and \trcnear, while not
reducing the available memory capacity. Although this mechanism incurs no
additional complexity at the memory controller or the operating system, we find
that the overall performance improvement due to this mechanism is low. To
better exploit the benefits of the low access latency of the near segment,
frequently accessed pages should be mapped to the near segment. This can be
done by the hardware or by the OS. To this end, we describe two different
mechanisms.

{\bf Exclusive Cache}. In this mechanism, we use the near segment as an
{\em exclusive} cache to the rows in the far segment. The memory controller
uses one of the three mechanisms proposed in Section~\ref{sec:inclusive} to
determine caching and eviction candidates. To cache a particular row, the data
of that row is {\em swapped} with the data of the row to be evicted from the
near segment. For this purpose, each subarray requires a {\em dummy-row}
(D-row). To swap the data of the to-be-cached row (C-row) and to-be-evicted row
(E-row), the memory controller simply performs the following three migrations:

\setlength{\abovedisplayskip}{0pt}
\setlength{\belowdisplayskip}{0pt}
\setlength{\abovedisplayshortskip}{0pt}
\setlength{\belowdisplayshortskip}{0pt}
\begin{align*}
 \textrm{C-row} &\rightarrow \textrm{D-row} &
 \textrm{E-row} &\rightarrow \textrm{C-row} &
 \textrm{D-row} &\rightarrow \textrm{E-row}
\end{align*}

\normalsize
The exclusive cache mechanism provides almost full main memory capacity (except
one dummy row per subarray, $<$~0.2\% loss in capacity) at the expense of two
overheads. First, since row swapping changes the mappings of rows in both the
near and the far segment, the memory controller must maintain the mapping of
rows in both segments. Second, each swapping requires three migrations,
which increases the latency of caching.

{\bf Profile-Based Page Mapping}. In this mechanism, the OS controls the
virtual-to-physical mapping to map frequently accessed pages to the near
segment. The OS needs to be informed of the bits in the physical address that
control the near-segment/far-segment mapping. Information about frequently
accessed pages can either be obtained statically using compiler-based
profiling, or dynamically using hardware-based profiling. In our evaluations,
we show the potential performance improvement due to this mechanism using
hardware-based profiling. Compared to the exclusive caching mechanism, this
approach requires much lower hardware storage overhead.

	\section{Implementation Details \& Further Analysis}

\subsection{Near Segment Row Decoder Wiring}
\label{subsec:near_segment_row_decoder}

To avoid the need for a large, monolithic row address decoder at each subarray,
DRAM makes use of {\em predecoding}. Outside the subarray, the row address is
divided into $M$ sets of bits. Each set, $N$ bits, is decoded into $2^N$ wires,
referred to as $N:2^N$ predecoding.\footnote{$N$ may differ between sets.} The
input to each row's wordline driver is then simply the logical AND of the $M$
wires that correspond to the row's address. This allows the per-subarray row
decoding logic to be simple and compact, at the cost of increasing the wiring
overhead associated with row decoding at each subarray. As a result, wiring
overhead dominates the cost of row decoding.

The inter-segment data transfer mechanism described in
Section~\ref{sec:segment_migration} requires up to two rows to be activated in the
same subarray at once, necessitating a second row decoder. However, since one
of the two activated rows is always in the near segment, this second row
decoder only needs to address rows in the near segment. For a near segment with
32 rows, a scheme that splits the 5 ($\log_{2}32$) near segment address bits
into 3-bit and 2-bit sets requires 12 additional wires to be routed to each
subarray (3:8 predecoding + 2:4 predecoding). The corresponding die-size
penalty is 0.33\%, calculated based on the total die-size and wire-pitch
derived from Vogelsang~\cite{rambus-power, vogelsang-micro2010}.

\subsection{Additional Storage in DRAM Controller} \label{subsec:controller}

The memory controller requires additional storage to keep track of the rows
that are cached in the near segment. In the inclusive caching mechanisms, each
subarray contains a near segment of length $N$, serving as an $N$-way
(fully-associative) cache of the far segment of length $F$. Therefore, each
near-segment row requires a $\lceil\log_2 F\rceil$-bit tag. Hence, each
subarray requires $N\lceil\log_2 F\rceil$ bits for tag storage, and a system
with $S$ subarrays requires $SN\lceil\log_2 F\rceil$ bits for tags. In the
exclusive caching mechanism, row swapping can lead to any physical page within
a subarray getting mapped to any row within the subarray. Hence, each row
within the subarray requires the tag of the physical page whose data is stored
in that row. Thus, each row in a subarray requires a $\lceil\log_2
(N+F)\rceil$-bit tag, and a system with $S$ subarrays requires
$S(N+F)\lceil\log_2 (N+F)\rceil$ bits for tags. In the system configuration we
evaluate (described in Section~\ref{sec:evaluation}), $(N, F, S) = (32, 480, 256)$,
the tag storage overhead is 9~KB for inclusive caching and 144~KB for exclusive
caching.

\msc and \mwmc additionally require $\log_2 N$ bits per near segment row for
replacement information ($SN\log_2 N$ bits total), while our implementation of
\mbbc uses an 8-bit benefit field per near segment row ($8SN$ bits total). For
our evaluated system, these are overheads of 5~KB and 8~KB respectively.

\subsection{Energy Consumption Analysis} \label{sec:power}

The non-I/O power consumption of the DRAM device can be broken into three
dominant components: {\em i)} raising and lowering the wordline during \cmdact
and \cmdpre, {\em ii)} driving the bitline during \cmdact, and {\em iii)}
transferring data from the sense amplifiers to the peripherals. The first two
of these components differ between a conventional DRAM and TL-DRAM, for two
reasons:

{\bf Reduced Power Due to Reduced Bitline Capacitance in Near Segment.} The
energy required to restore a bitline is proportional to the bitline's
capacitance. In TL-DRAM, the near segment has a lower capacitance than that of
a conventional DRAM's bitline, resulting in decreased power consumption.

{\bf Additional Power Due to Isolation Transistors.} The additional power
required to control the isolation transistors when accessing the far segment is
approximately twice that of raising the wordline, since raising the wordline
requires driving one access transistor per bitline, while accessing the far
segment requires driving two isolation transistors per bitline
(Sec~\ref{sec:segment_diesize}).

\begin{figure}[h]
	\centering
	\subcaptionbox{Power Consumption for ACTIVATE\label{fig:power_active}}[0.9\linewidth] {
		\centering
		\includegraphics[width=0.6\linewidth]{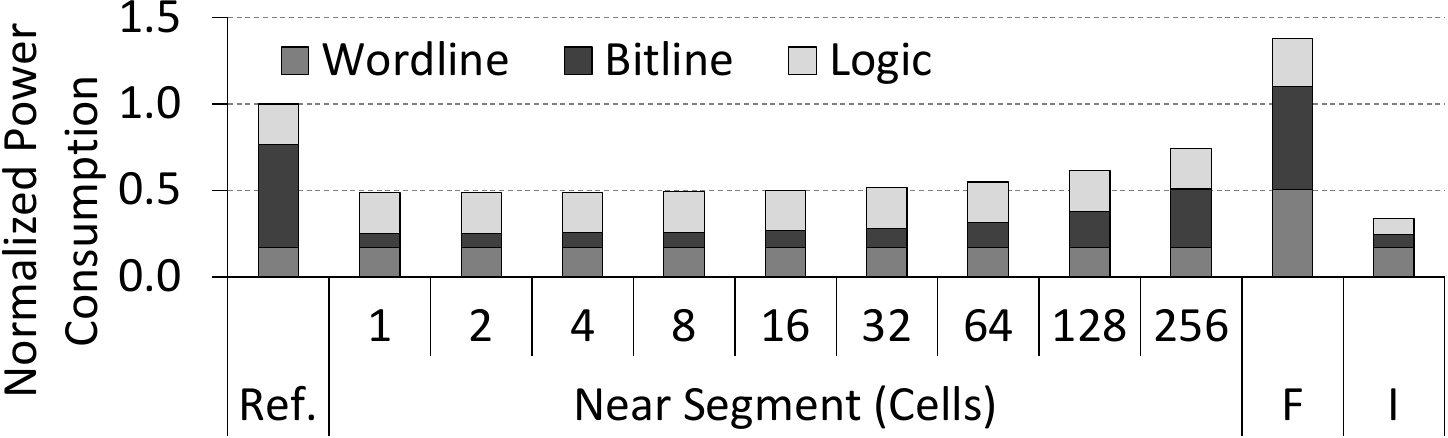}
	}

	\subcaptionbox{Power Consumption for PRECHARGE\label{fig:power_prech}}[0.9\linewidth] {
		\centering
		\includegraphics[width=0.6\linewidth]{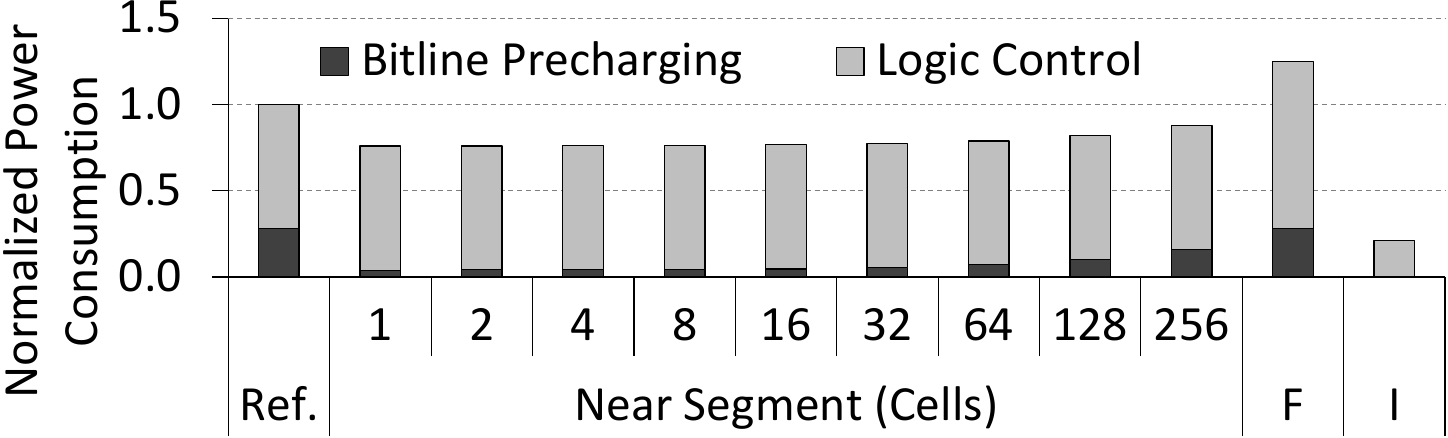}
	} \\
	\footnotesize {$\dagger$ Ref.: Long Bitline, F: Far Segment, I: Inter-Segment Data Transfer}
	\caption{Power Consumption vs. Bitline Length}
	\label{fig:power_active_precharge}
\end{figure}

Using DRAM power models from Rambus and Micron~\cite{micron-power,
rambus-power, vogelsang-micro2010}, we estimate power consumption of TL-DRAM
and conventional DRAM in Figure~\ref{fig:power_active_precharge}. Note that,
while the near segment's power consumption increases with the near segment
length, the far segment's power does not change as long as the total bitline
length is constant. Our evaluations in Section~\ref{sec:results} take these
differences in power consumption into account.

\subsection{More Tiers} \label{subsec:more_tiers}

Although we have described TL-DRAM with two segments per subarray, one can
imagine adding more isolation transistors to divide a subarray into more tiers,
each tier with its own latency and energy consumption characteristics. As a case
study, we evaluated the latency characteristics of TL-DRAM with three tiers per
subarray (near/middle/far with 32/224/256 cells). The normalized \trcd and \trc
of the near/middle/far segments are 54.8\%/70.7\%/104.1\% and
44.0\%/77.8\%/156.9\%. While adding more tiers can enable more fine-grained caching
and partitioning mechanisms, they come with the additional area cost of
3.15\% per additional tier, additional energy consumption to
control the multiple isolation transistors, and logic complexity in DRAM and the
DRAM controller.

	\section{Evaluation Methodology} \label{sec:evaluation}

{\bf Simulators.} We use a modified version of Ramulator~\cite{kim-cal2015}, a
fast cycle-accurate DRAM simulator that is available publicly~\cite{ramulator,
safari-tools}, and Ramulator releases an open-source implementation of TL-DRAM.
We use Ramulator combined with a cycle-level x86 multi-core simulator. We use
Ramulator as part of a cycle-level in-house x86 multi-core simulator, whose
front-end is based on Pin~\cite{luk-pldi2005}.

{\bf System Configuration.} The evaluated system is configured as shown in
Table~\ref{tab:methodology_syscfg}. Unless otherwise stated, the evaluated
TL-DRAM has a near segment size of 32 rows.

\begin{table}[h]
  \centering
	\small {
  \begin{tabular}{ll} \toprule
    Processor & 5.3~GHz, 3-wide issue, 8 MSHRs/core, \\
    & 128-entry instruction window \\ \midrule
    Last-Level  & 64B cache line, 16-way associative, \\
	Cache 		& 512kB private cache slice/core \\ \midrule
    Memory 		& 64/64-entry read/write request queue, \\
    Controller 	& row-interleaved mapping, closed-page policy, \\
    & FR-FCFS scheduling~\cite{rixner-isca2000} \\ \midrule
    DRAM & 2GB DDR3-1066,\\
    & 1/2/4 channel (@1-core/2-core/4-core), 1 rank/channel,\\
    & 8 banks/rank, 32 subarrays/bank, 512 rows/bitline\\
    & \trcd (unsegmented): 15.0ns, \trc (unsegmented): 52.5ns \\ \midrule
    TL-DRAM & 32 rows/near segment, 480 rows/far segment \\
    & \trcd (near/far): 8.2/12.1ns,
	 \trc (near/far): 23.1/65.8ns\\ \bottomrule
  \end{tabular}
	}
  \caption{Evaluated System Configuration} \label{tab:methodology_syscfg}
\end{table}

{\bf Parameters.} DRAM latency is as calculated in
Section~\ref{sec:latency_analysis_quantitative}. DRAM dynamic energy
consumption is evaluated by associating an energy cost with each DRAM command,
derived using the tools~\cite{micron-power, rambus-power, vogelsang-micro2010}
and the methodology given in Section~\ref{sec:power}.

{\bf Benchmarks.} We use 32 benchmarks from SPEC CPU2006, TPC~\cite{tpc},
STREAM~\cite{stream} and a {\em random} microbenchmark similar in behavior to
GUPS~\cite{gups}. We classify benchmarks whose performance is significantly
affected by near segment size as {\em sensitive}, and all other benchmarks as
{\em non-sensitive}. For single-core sensitivity studies, we report results
that are averaged across all 32 benchmarks. We also present multi-programmed
multi-core evaluations in Section~\ref{sec:result_dual}. For each multi-core
workload group, we report results averaged across 16 workloads, generated by
randomly selecting from specific benchmark groups ({\tt sensitive}, {\tt high},
{\tt low}, and {\tt random}).

{\bf Simulation and Evaluation.} We simulate each benchmark for 100 million
instructions. For multi-core evaluations, we ensure that even the slowest core
executes 100 million instructions, while other cores continue to exert pressure
on the memory subsystem. To measure performance, we use instruction throughput
(IPC) for single-core systems and {\it weighted
speedup}~\cite{snavely-asplos2000} for multi-core systems.

	\section{Results} \label{sec:results}

\subsection{Single-Core Results: Inclusive Cache} \label{sec:single_inclusive}

\begin{figure}[h]
	\centering
	\captionsetup[subfigure]{width=1\linewidth}
	\includegraphics[width=\linewidth]{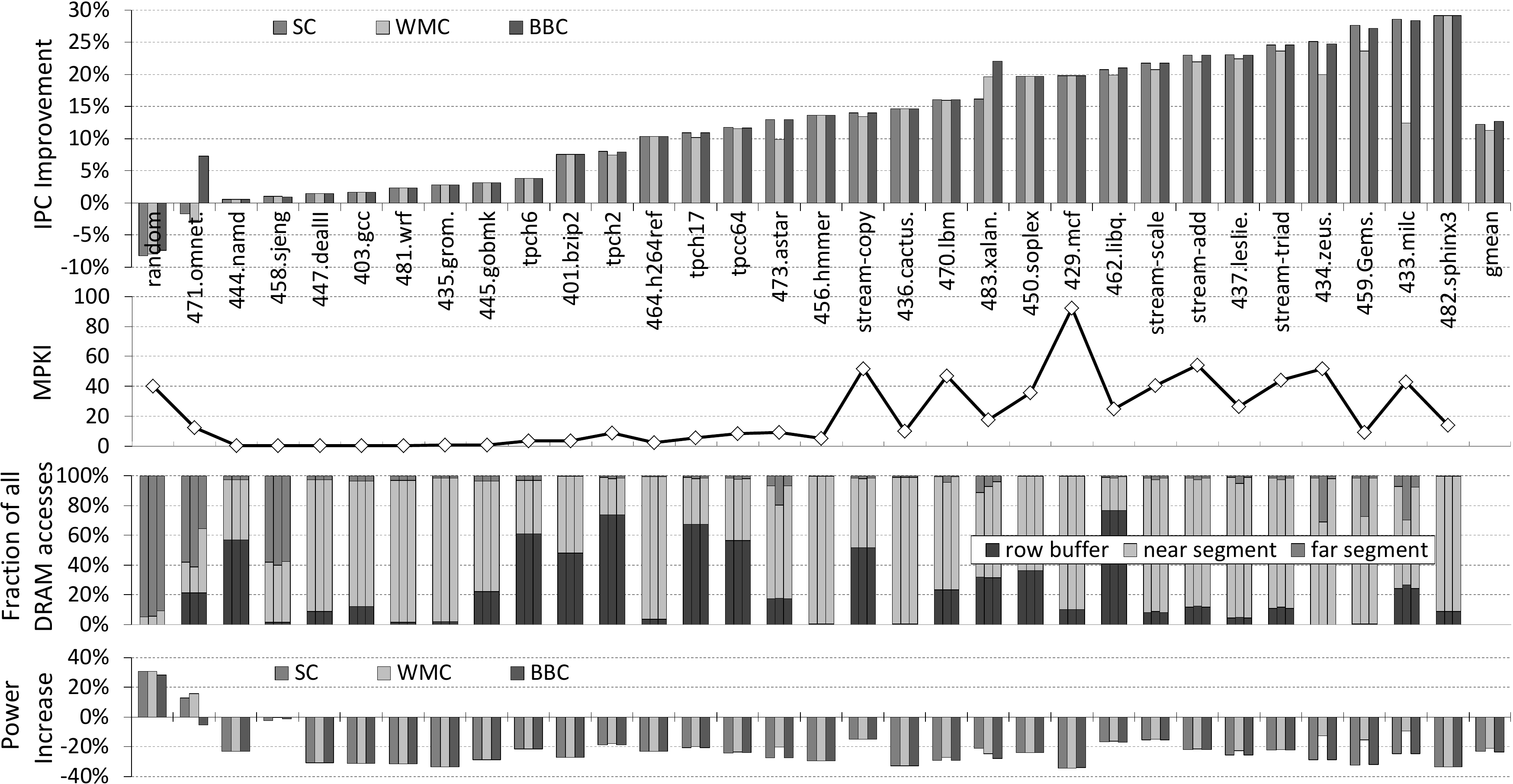}
	\caption{Single-core: IPC improvement, LLC MPKI, Fraction of accesses
	serviced at row buffer/near segment/far segment, Power consumption}
	\label{fig:result_single}
\end{figure}

Figure~\ref{fig:result_single} compares our proposed TL-DRAM based mechanisms
(\msc, \mwmc, and \mbbc) to the baseline with conventional DRAM. For each
benchmark, the figure plots four metrics: {\em i)}~performance improvement of
the TL-DRAM based mechanisms compared to the baseline, {\em ii)}~the number of
misses per instruction in the last-level cache, {\em iii)}~the fraction of
accesses that are served at the row buffer, near segment and the far segment
using each of the three proposed mechanisms, and {\em iv)}~the power
consumption of the TL-DRAM based mechanisms relative to the baseline. We draw
three conclusions from the figure.

First, for most benchmarks, all of our proposed mechanisms improve performance
significantly compared to the baseline. On average, \msc, \mwmc, and \mbbc
improve performance by 12.3\%, 11.3\% and 12.8\%, respectively, compared to the
baseline. As expected, performance benefits are highest for benchmarks with
high memory intensity (MPKI: Last-Level Cache Misses-Per-Kilo-Instruction) and
high near-segment hit rate.

Second, all three of our proposed mechanisms perform similarly for most
benchmarks. However, there are a few benchmarks where \mwmc significantly
degrades performance compared to \msc. This is because \mwmc only caches
wait-inducing rows, ignoring rows with high reuse that cause few conflicts.
\mbbc, which takes both reuse and wait into account, outperforms both \msc and
\mwmc. For example, \mbbc significantly improves performance compared to \msc
($>8$\% for {\em omnetpp}, $>5$\% for {\em xalancbmk}) by reducing wait and
compared to \mwmc ($>9$\% for {\em omnetpp}, $>2$\% for {\em xalancbmk}) by
providing more reuse.

Third, \mbbc degrades performance only for the microbenchmark {\em random}.
This benchmark has high memory intensity (MPKI = 40) and very low reuse
(near-segment hit rate $<10$\%). These two factors together result in frequent
bank conflicts in the far segment. As a result, most requests experience the
full penalty of the far segment's longer \trc. We analyze this microbenchmark
in detail in Section~\ref{sec:result_random_anal}.

{\bf Power Analysis.} As described in Section~\ref{sec:power}, power
consumption is significantly lower for near-segment accesses, but higher for
far-segment accesses, compared to accesses in a conventional DRAM. As a result,
TL-DRAM achieves significant power savings when most accesses are to the near
segment. The bottom-most plot of Figure~\ref{fig:result_single} compares the
power consumption of our TL-DRAM-based mechanisms to that of the baseline. Our
mechanisms produce significant power savings (23.6\% for \mbbc vs. baseline) on
average, since over 90\% of accesses hit in the row buffer or near segment for
most benchmarks.

\subsection{Effect of Near Segment Length: Inclusive Cache}
\label{sec:result_near_length}

\begin{figure}[h]
  \centering
  \includegraphics[width=0.6\linewidth]{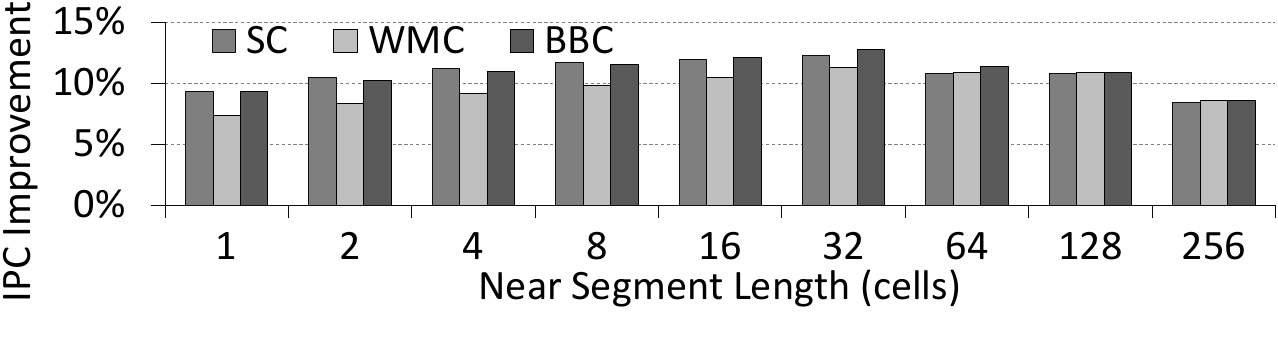}
  \caption{Varying Near Segment Capacity (Inclusive Cache)}
  \label{fig:result_single_sensitive}
\end{figure}

The number of rows in each near segment presents a trade-off, as increasing the
near segment's size increases its capacity but also increases its access
latency. Figure~\ref{fig:result_single_sensitive} shows the average
performance improvement of our proposed mechanisms over the baseline as we vary
the near segment size. As expected, performance initially improves as the
number of rows in the near segment is increased due to increased near segment
capacity. However, increasing the number of rows per near segment beyond 32
reduces the performance benefits due to the increased near segment access
latency.

\begin{figure}[h]
	\centering
	\includegraphics[width=0.6\linewidth]{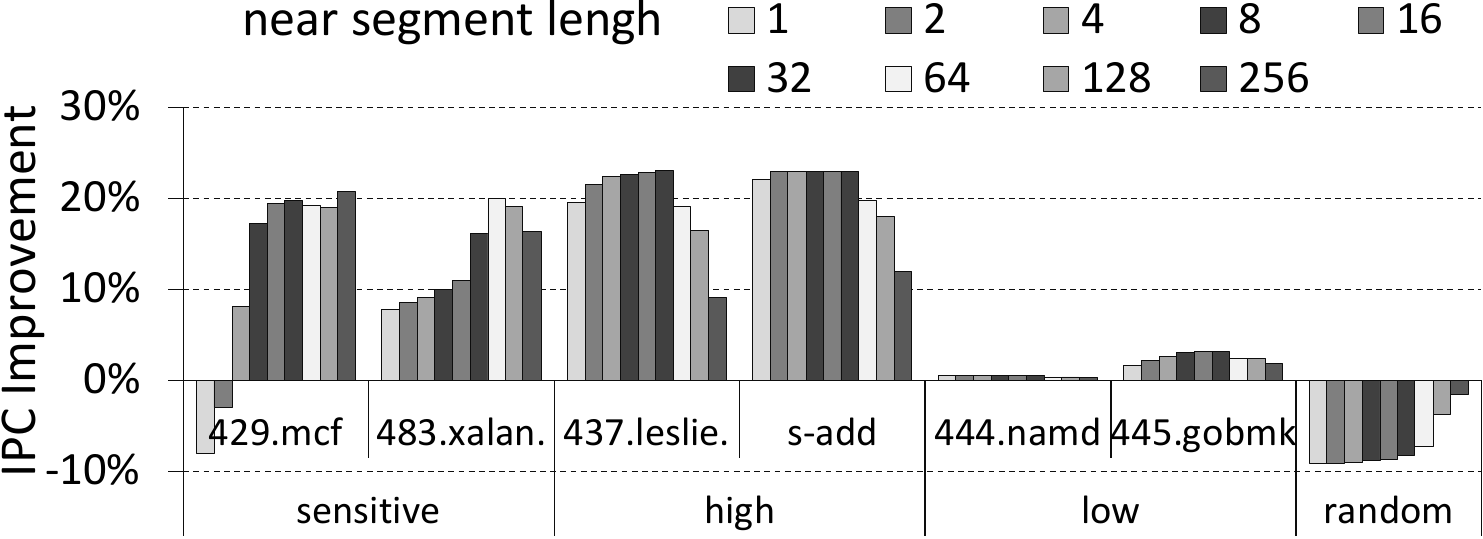}
	\caption{Four Application Groups Based on Performance Characteristics}
	\label{fig:tldram_near_length}
\end{figure}

In Figure~\ref{fig:tldram_near_length}, we categorize benchmarks in four groups
based on the performance characteristics with varying the near segment length.
We first divide into two groups based on the sensitivity of the performance to
the near segment length. Some benchmarks increase performance (IPC) with
increasing near segment length (especially, for relatively short near segments,
e.g., 1 -- 64 rows in the near segment out of 512 rows in a subarray), which we
call {\tt sensitive} group (e.g., {\em mcf} and {\em xalancbmk}). This is
because these benchmarks have relatively large working set (the number of DRAM
rows which need to be accessed in a short period of the execution time)
compared to the near segment capacity. Therefore, these benchmarks show less
performance at short near segment length (e.g., 1 or 2 rows for each near
segment over 512 rows in a subarray) because the working set of each benchmark
does not fit into the near segment capacity, requiring frequent data migration
between the near and far segments. Increasing the near segment length leads to
preserve more rows in the near segment, resulting in better performance.

For the other benchmarks, the performance does not much changes with increasing
the near segment length. There are two cases. One case is that the working set
of benchmarks is much larger than the near segment capacity. We only have one
application in this case, {\em random} benchmark ({\tt random} group). The
other case is that the working set of benchmarks is much smaller than the near
segment capacity (e.g, 1 or 2 rows in the near segment) and thus fits into the
near segment capacity. We subdivide into these benchmarks (having small working
set) into two groups based on the performance improvement with TL-DRAM -- {\tt
high} and {\tt low}. We observe that benchmarks in the {\tt high} group are
memory-intensive (high MPKI), while benchmarks in the {\tt low} group are less
memory-intensive (low MPKI).

For most benchmarks (except for {\em random} and {\em mcf}), increasing the
number of rows per near segment beyond 64 reduces the performance benefits due
to the increased near segment access latency, which is similar to the trend of
Figure~\ref{fig:result_single_sensitive}.

Based on these observations, we conclude that benchmarks have different
performance characteristics with different latency and capacity of the near
segment. We extend this application's characteristic to multi-program workloads
in Section~\ref{sec:result_dual}.

\subsection{Effect of Far Segment Latency: Inclusive Cache}
\label{sec:result_random_anal}

\begin{figure}[h]
	\centering
  \includegraphics[width=0.6\linewidth]{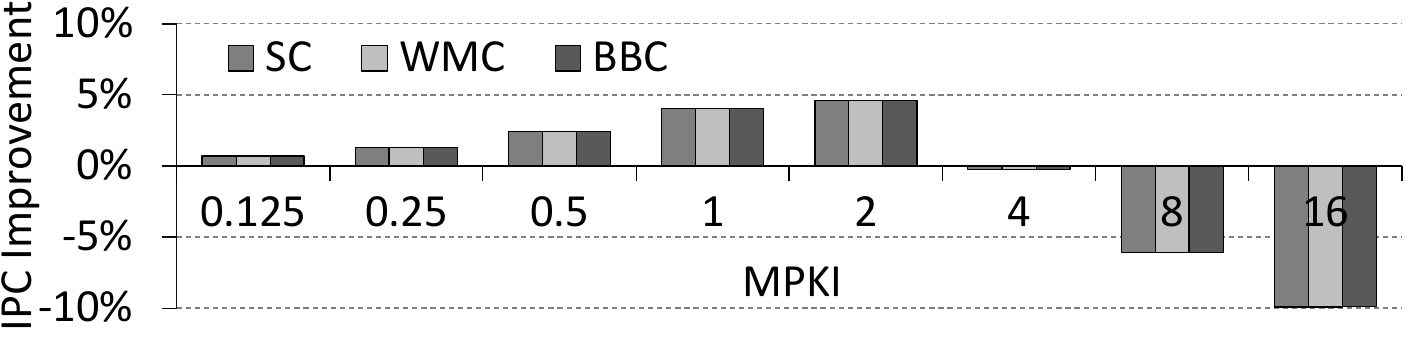}
	\caption{TL-DRAM Performance on Benchmark {\em random}}
	\label{fig:result_random}
\end{figure}

As we describe in Section~\ref{sec:latency_analysis_qualitative}, the \trcd of
the far segment is {\em lower} than the \trcd of conventional DRAM. Therefore,
even if most of the accesses are to the far segment, if the accesses are
sufficiently far apart such that \trc is {\em not} a bottleneck, TL-DRAM can
still improve performance. We study this effect using our {\em random}
microbenchmark \mycolor{(similar to GUPS~\cite{gups})}, which has very little
data reuse and hence usually accesses the far segment.
Figure~\ref{fig:result_random} shows the performance improvement of our
proposed mechanisms compared to the baseline with varying memory intensity
(MPKI) of {\em random}. As the figure shows, when the memory intensity is low
($<2$), the reduced \trcd of the far segment dominates, and our mechanisms
improve performance by up to 5\%. However, further increasing the intensity of
{\em random} makes \trc the main bottleneck as evidenced by the degraded
performance due to our proposed mechanisms.

\subsection{Sensitivity to Channel Count} \label{sec:result_sensitivity}

For 1-core systems with 1, 2 and 4 memory channels, TL-DRAM provides 12.8\%,
13.8\% and 14.2\% performance improvement compared to the baseline. The
performance improvement of TL-DRAM increases with increasing number of
channels. This is because, with more channels, the negative impact of channel
conflicts reduces and bank access latency becomes the primary bottleneck.
Therefore, TL-DRAM, which reduces the average bank access latency, provides
better performance with more channels.

\subsection{\mycolor{Sensitivity to CPU Frequency and DRAM Data Rate}}
\label{sec:sensitivity_frequency}

\begin{figure}[h]
	\centering
	\subcaptionbox{Performance vs. CPU Frequency\label{fig:tldram_cpu_freq}}[1.0\linewidth] {
		\centering
		\includegraphics[height=1.0in]{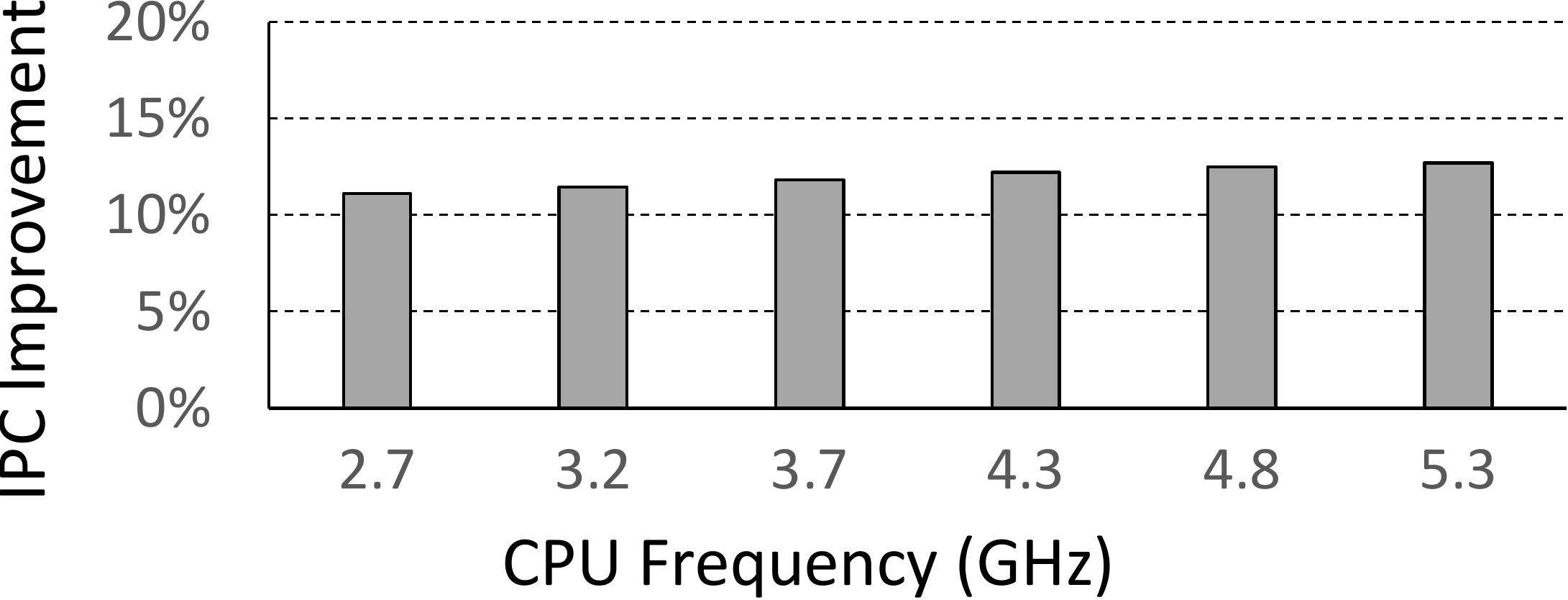}
	}

	\subcaptionbox{Performance vs. DRAM Data Rate\label{fig:tldram_data_rate}}[1.0\linewidth] {
		\centering
		\includegraphics[height=1.0in]{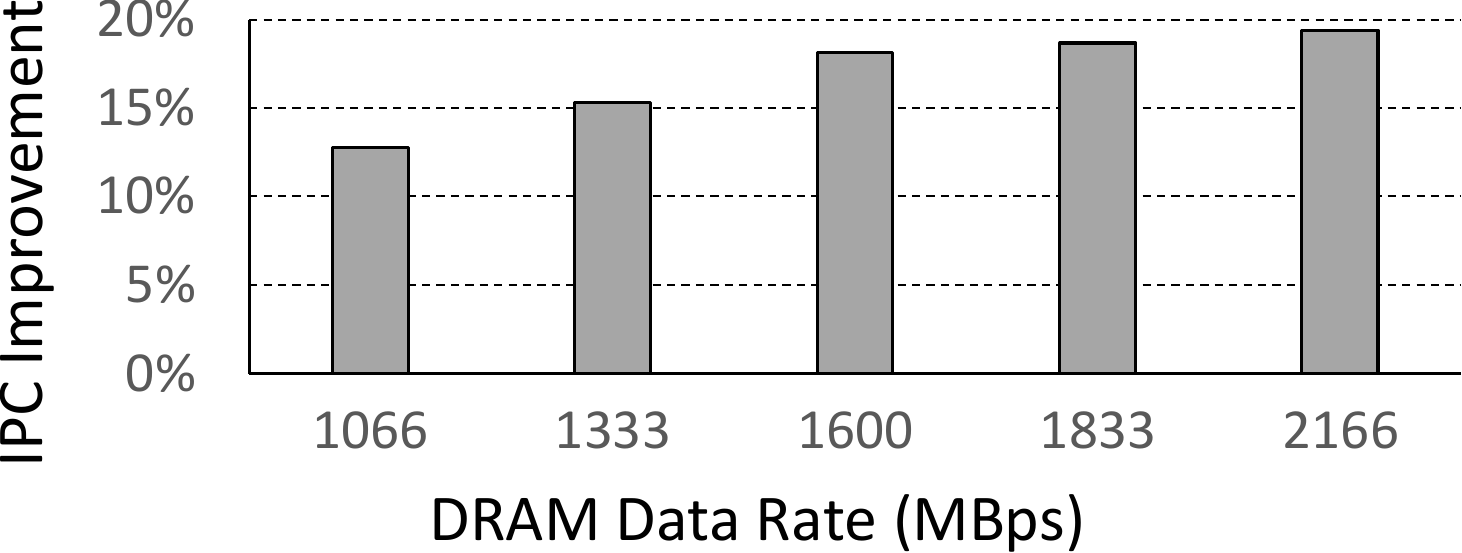}
	}

	\caption{Sensitivity to CPU Frequency and DRAM Data Rate}
	\label{fig:sensitivity_frequency}
\end{figure}

\mycolor{We analyze the sensitivity to CPU frequency and DRAM data rate.
Figure~\ref{fig:tldram_cpu_freq} shows the average system performance
improvement of the single core workloads with different CPU frequencies from
2.7GHz to 5.3GHz. At the higher CPU frequencies, CPU generates memory requests
more frequently, leading to providing more benefits with our mechanisms.
Compared to 11.1\% performance improvement at 2.7GHz CPU frequency, our
mechanism provides 12.7\% performance improvement.}

\mycolor{Figure~\ref{fig:tldram_data_rate} shows the average system performance
improvement of the single core workloads with different DRAM data rate. At
higher DRAM data rate, the DRAM channel takes less time to transfer a
cacheline, leading to reducing the memory channel conflicts. As expected,
compared to 12.7\% performance improvement on DDR3-1066 (533MHz DRAM clock
frequency and 1066Mbps per-pin data rate), DDR3-2166 (1066MHz DRAM clock
frequency and 2166Mbps per-pin data rate) provides much more performance
improvement (19.4\% the average performance improvement).}

\mycolor{Based on these sensitivity analyses, we conclude that {\em i)} our
mechanisms provides significant performance improvement across different CPU
frequencies and different DRAM data rates, and {\em ii)} our mechanisms
provides more performance improvement at higher performance systems (e.g.,
systems that have higher CPU frequencies and memory systems that have higher
data rate).}

\subsection{Multi-Core Results: Inclusive Cache} \label{sec:result_dual}

\begin{figure}[h]
	\centering

	\subcaptionbox{Workloads without Random Benchmark\label{fig:inclusive_normal}}[0.98\linewidth] {
		\includegraphics[width=0.8\linewidth]{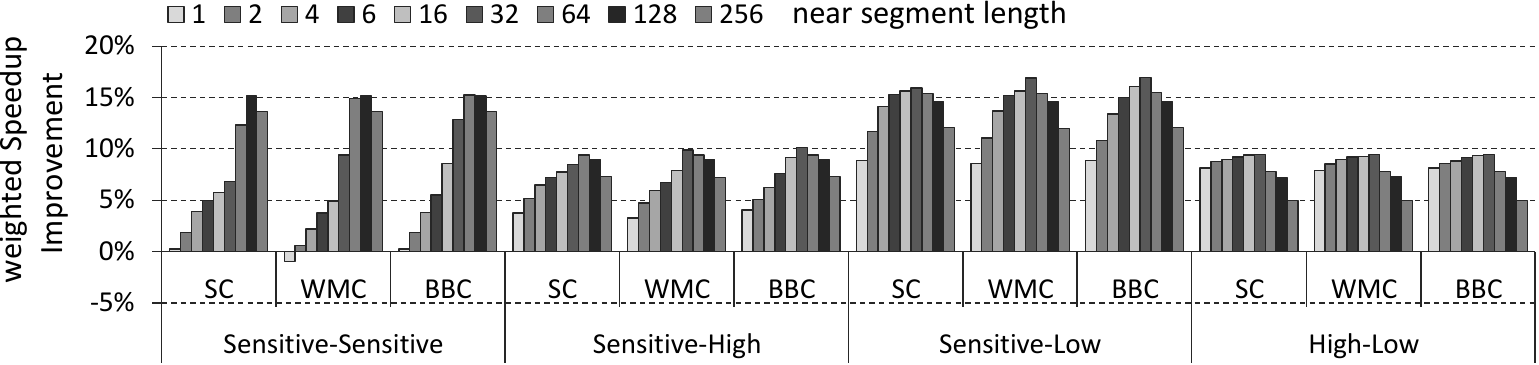}
	}

	\subcaptionbox{Workloads with Random Benchmark\label{fig:inclusive_rand}}[0.98\linewidth] {
		\includegraphics[width=0.8\linewidth]{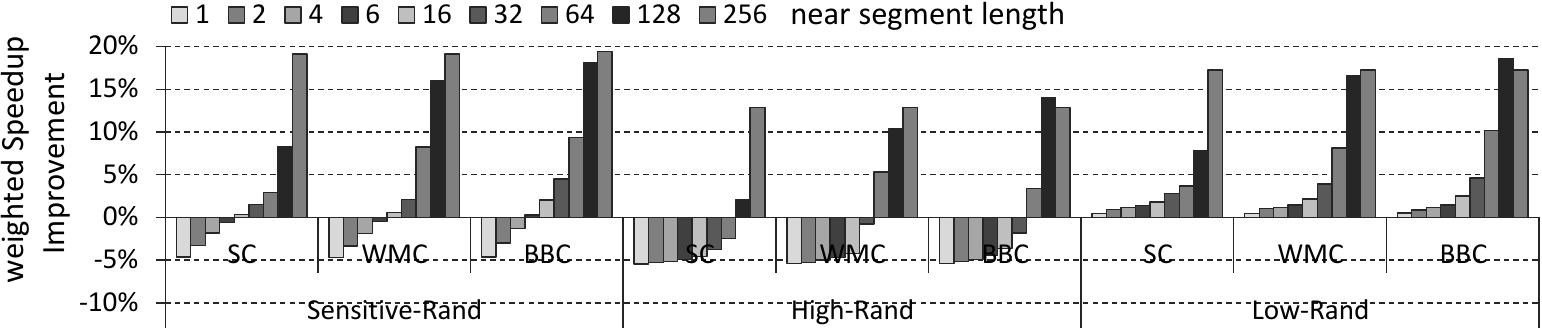}
	}

	\caption{System Performance: 2-Core, Inclusive Cache}
	\label{fig:inclusive_cache}
\end{figure}

Figure~\ref{fig:inclusive_cache} shows the system performance improvement with
our proposed mechanisms compared to the baseline. We build dual-core workloads
by combining two benchmarks from the four benchmark categories -- {\tt
sensitive}, {\tt high}, {\tt low}, and {\tt random} -- as we defined in
Section~\ref{sec:result_near_length}. We plot results in two separate figures,
{\em i)} workloads without including {\tt random} benchmark in
Figure~\ref{fig:inclusive_normal}, and {\em ii)} workloads with including {\tt
random} benchmark in Figure~\ref{fig:inclusive_rand}.

We make three major observations from Figure~\ref{fig:inclusive_normal}. First,
when a workload includes at least one benchmark that is sensitive to the near
segment capacity (left three cases in Figure~\ref{fig:inclusive_normal}), the
improvement in weighted speedup increases with increasing near segment
capacity. Second, when neither benchmark is sensitive to the near segment
capacity, the weighted speedup improvement is less sensitive to the near
segment capacity ({\tt High}-{\tt Low} group in
Figure~\ref{fig:inclusive_normal}). Third, for most workloads, the weighted
speedup improvement decreases at very large near segment capacity (e.g, 128 or
256 rows per near segment) due to the increased near segment access latency.

Figure~\ref{fig:inclusive_rand} shows the weighted speedup for workloads, which
consists of a {\tt random} benchmark and a benchmark from other categories. We
make two major observations. First, the improvement in weighted speedup
increases with increasing near segment capacity, which is similar to our
observation from the {\tt sensitive-sensitive} workloads in
Figure~\ref{fig:inclusive_normal}. Second, almost all workload categories and
near segment capacities, \mbbc performs comparably to or significantly better
than \msc, emphasizing the advantages of our benefit-based near segment
management policy.

\begin{figure}[h]
	\centering
	\includegraphics[height=1.2in]{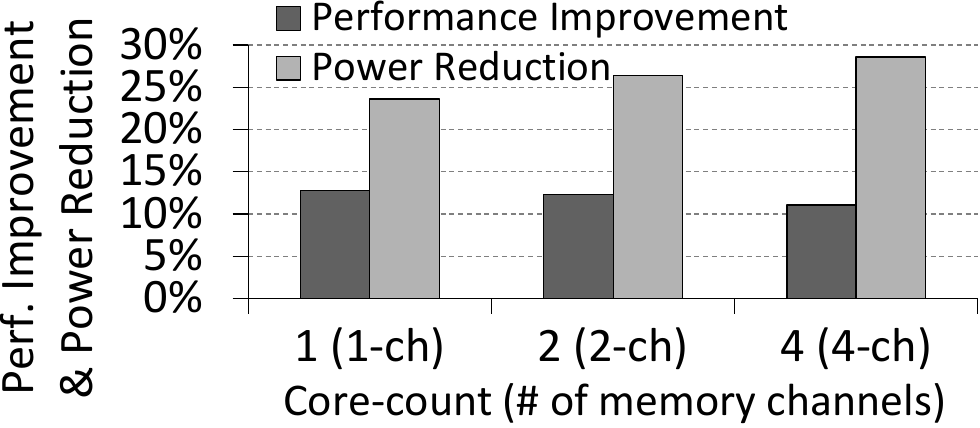} \\
	\captionof{figure}{Inclusive Cache Analysis (BBC)}
	\label{fig:result_inclusive}
\end{figure}

As shown in Figure~\ref{fig:result_inclusive}, on average, \mbbc improves
weighted speedup by 12.3\% and reduces power consumption by 26.4\% compared to
the baseline. We observe similar trends on 4-core systems, where BBC improves
weighted speedup by 11.0\% and reduces power consumption by 28.6\% on average.

\subsection{Exclusive Cache} \label{sec:result_exclusive}

\begin{figure}[h]
	\centering
	\includegraphics[height=1.2in]{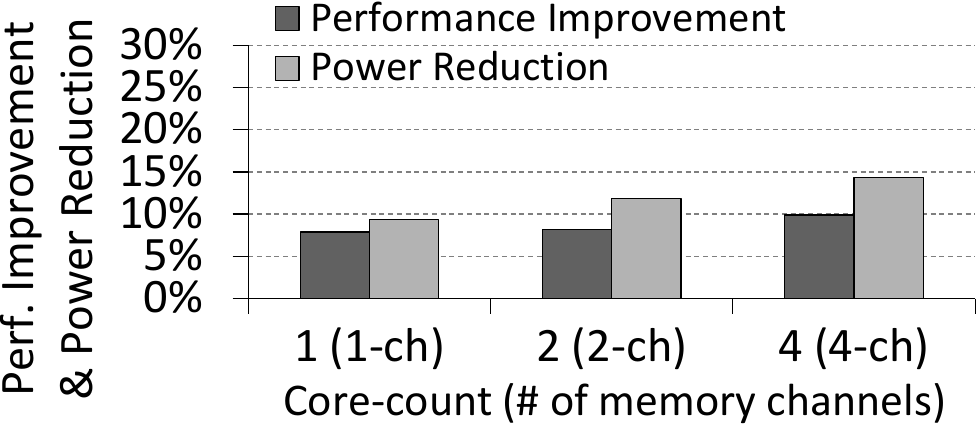} \\
	\captionof{figure}{Exclusive Cache Analysis (WMC)}
	\label{fig:result_exclusive}
\end{figure}

Figure~\ref{fig:result_exclusive} shows the performance improvement of the TL-DRAM
with the exclusive caching mechanism (Section~\ref{sec:alternative}) and 32 rows
in the near segment over the baseline. For 1-/2-/4-core systems,
TL-DRAM with exclusive caching improves performance by 7.9\%/8.2\%/9.9\% and
reduces power consumption by 9.4\%/11.8\%/14.3\%. The performance improvement
due to exclusive caching is lower compared to that of inclusive caching due to
the increased caching latency, as explained in
Section~\ref{sec:alternative}. Unlike inclusive caching, where BBC outperforms
SC and WMC, WMC performs the best for exclusive caching. This is because unlike
BBC and SC that cache any row that is accessed in the far segment, WMC caches a
row only if it is wait-inducing. As a result, WMC reduces bank unavailability
due to the increased caching latency.

\subsection{Profile-Based Page Mapping} \label{sec:result_profile}

\begin{figure}[h]
	\centering
	\includegraphics[height=1.2in]{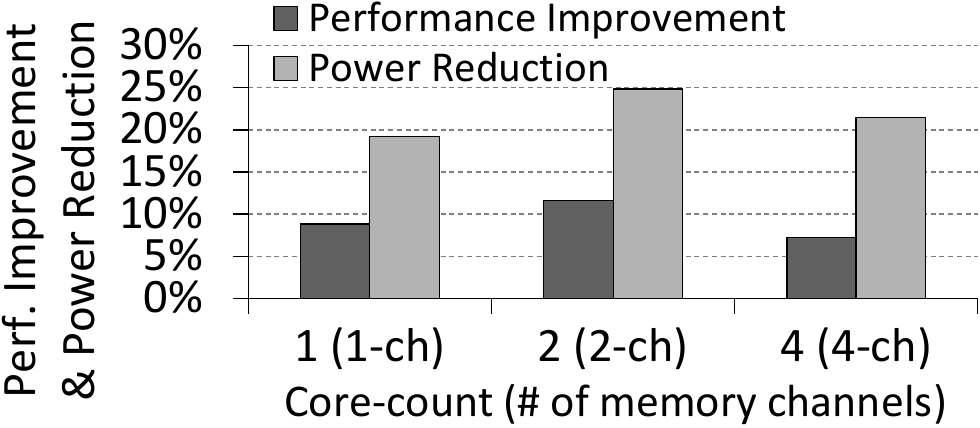} \\
	\captionof{figure}{Profile-Based Page Mapping}
	\label{fig:result_profile}
\end{figure}

Figure~\ref{fig:result_profile} shows the performance improvement of TL-DRAM
with profile-based page mapping over the baseline \mycolor{which uses normal
DRAM}. The evaluated memory subsystem has 64 rows in the near segment.
Therefore, the top 64 most frequently accessed rows in each subarray, as
determined by a profiling run, are allocated in the near segment. For
1-/2-/4-core systems, TL-DRAM with profile-based page mapping improves
performance by 8.9\%/11.6\%/7.2\% and reduces power consumption by
19.2\%/24.8\%/21.4\%. These results indicate a significant potential for such a
profiling based mapping mechanism. We leave a more rigorous evaluation of such
a profiling based mapping mechanism to future work.

	\section{Summary}

Existing DRAM architectures present a trade-off between cost-per-bit and access
latency. One can either achieve low cost-per-bit using long bitlines or low
access latency using short bitlines, but not both. In this chapter, we
introduce Tiered-Latency DRAM (TL-DRAM), a DRAM architecture that provides both
low latency and low cost-per-bit. The key idea behind TL-DRAM is to segment a
long bitline using an isolation transistor, creating a segment of rows with low
access latency while keeping cost-per-bit on par with commodity DRAM.

We present mechanisms that take advantage of our TL-DRAM substrate by using its
low-latency segment as a hardware-managed cache. Our most sophisticated cache
management algorithm, Benefit-Based Caching (\mbbc), selects rows to cache that
maximize access latency savings. Our evaluation results show that our proposed
techniques significantly improve both system performance and energy efficiency
across a variety of systems and workloads.

We conclude that TL-DRAM provides a promising low-latency and low-cost
substrate for building main memories, on top of which existing and new caching
and page allocation mechanisms can be implemented to provide even higher
performance and higher energy efficiency.

  \chapter{{Adaptive-Latency DRAM:\\
Optimizing DRAM Latency to the Common Operating Conditions}}
\label{ch:aldram}

\let\thefootnote\relax\footnotetext{This work has been published in HPCA
2015~\cite{lee-hpca2015}. This dissertation includes more discussions and
evaluation results in Sections~\ref{sec:manydimm} beyond the HPCA 2016 paper.
We provide detailed characterization of each DRAM module online at the SAFARI
Research Group website~\cite{safari-aldram}.}
\let\thefootnote\svthefootnote

When a DRAM chip is accessed, it requires a certain amount of time before
enough charge can move into the cell (or the bitline) for the data to be
reliably stored (or retrieved). To guarantee this behavior, DRAM manufacturers
impose a set of minimum latency restrictions on DRAM accesses, called {\em
timing parameters}, as we demonstrated in Chapter~\ref{ch:bak}. Ideally, timing
parameters should provide just enough time for a DRAM chip to operate
correctly. In practice, however, DRAM manufacturers {\em pessimistically
incorporate a very large margin} into their timing parameters to ensure correct
operation under {\em worst-case} conditions. This is because of two major
concerns. First, due to {\em process variation}, some outlier cells suffer from
a larger RC-delay than other cells, and require more time to be charged. For
example, an outlier cell could have a very narrow connection (i.e., contact) to
the bitline, which constricts the flow of charge and increases the
RC-delay~\cite{lee-iedm1996}. Second, due to {\em temperature dependence}, all
cells suffer from a weaker charge-drive at high temperatures, and require more
time to charge the bitline. DRAM cells are intrinsically leaky, and lose some
of their charge even when they are not being accessed. At high temperatures,
this leakage is accelerated exponentially~\cite{restle-iedm1992,
yaney-iedm1987, mori-iedm2005, khan-sigmetrics2014, liu-isca2013}, leaving a
cell with less charge to drive the bitline when the cell is accessed ---
increasing the time it takes for the bitline to be charged.

Consequently, timing parameters prescribed by the DRAM manufacturers are
dictated by the {\em worst-case cells} (the slowest cells) operating under the
{\em worst-case conditions} (the highest temperature of
85\celsius~\cite{jedec-ddr3}). Such pessimism on the part of the DRAM
manufacturers is motivated by their desire to {\em i)} increase chip yield and
{\em ii)} reduce chip testing time. The manufacturers, in turn, are driven by
the extremely cost-sensitive nature of the DRAM market, which encourages them
to adopt pessimistic timing parameters rather than to {\em i)} discard chips
with the slowest cells or {\em ii)} test chips at lower temperatures.
Ultimately, the burden of pessimism is passed on to the end-users, who are
forced to endure much greater latencies than what is actually needed for
reliable operation under common-case conditions.

We first characterize \DIMMs DRAM modules from three manufacturers to expose
the excessive margin that is built into their timing parameters. Using an
FPGA-based testing platform~\cite{liu-isca2013, khan-sigmetrics2014,
kim-isca2014}, we then demonstrate that DRAM timing parameters can be shortened
to reduce DRAM latency without sacrificing any observed degree of DRAM
reliability. We are able to reduce latency by taking advantage of the two large
gaps between the worst-case and the ``common-case.'' First, most DRAM chips are
{\em not} exposed to the worst-case temperature of 85\celsius: according to
previous studies~\cite{liu-hpca2011, elsayed-sigmetrics2012,
elsayed-techreport2012} and our own measurements
(Section~\ref{sec:factors_temp}), the ambient temperature around a DRAM chip is
typically less than 55\celsius. Second, most DRAM chips do {\em not} contain
the worst-case cell with the largest latency: the slowest cell for a typical
chip is still faster than that of the worst-case chip
(Section~\ref{sec:profiling}).

Based on our characterization, we propose Adaptive-Latency DRAM (\ALD), a
mechanism that dynamically optimizes the timing parameters for different
modules at different temperatures. \ALD exploits the {\em additional charge
slack} present in the common-case compared to the worst-case, thereby
preserving the level of reliability (at least as high as the worst-case)
provided by DRAM manufacturers. We evaluate \ALD on a real
system~\cite{amd-4386, amd-bkdg} that allows us to dynamically reconfigure the
timing parameters at runtime. We show that \ALD improves the performance of a
wide variety of memory-intensive workloads by 14.0\% (on average) without
introducing errors. Therefore, we conclude that \ALD improves system
performance while maintaining memory correctness and without requiring changes
to DRAM chips or the DRAM interface.

	\section{Charge \& Latency Interdependence} \label{sec:leakage}

\begin{figure}[h]
	\centering
	\includegraphics[width=0.6\linewidth]{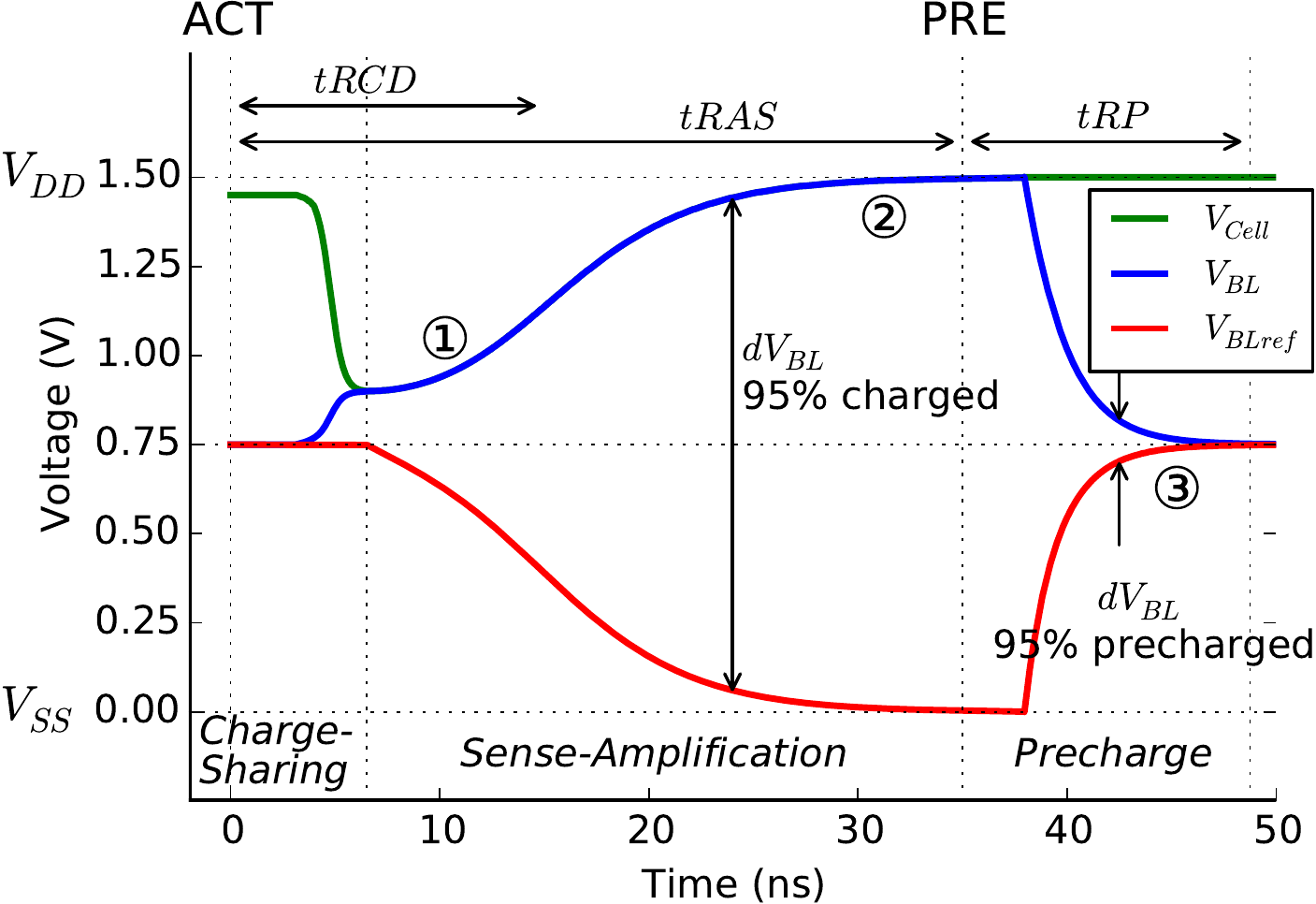}
	\caption{Phases of DRAM Voltage Levels}
	\label{fig:dram_signals}
\end{figure}

As we explained, the operation of a DRAM cell is governed by two important
concepts: {\em i)}~the quantity of charge and {\em ii)}~the latency it takes
to move charge. These two concepts are closely related to each other --- one
cannot be adjusted without affecting the other. To establish a more
quantitative relationship between charge and latency,
Figure~\ref{fig:dram_signals} presents the voltage of a cell and its bitline as
they cycle through the precharged state, charge-sharing state,
sense-amplification state, restored state, and back to the precharged state
(Section~\ref{sec:background}).\footnote{Using 55nm DRAM
parameters~\cite{vogelsang-micro2010, rambus-power}, we simulate the voltage
and current of the DRAM cells, sense-amplifiers, and bitline equalizers (for
precharging the bitline). To be technology-independent, we model the DRAM
circuitry using NMOS and PMOS transistors that obey the well-known MOSFET
equation for current-voltage (SPICE)~\cite{razavi}. We do not model secondary
effects.} This curve is typical in DRAM operation, as also shown in prior
works~\cite{son-isca2013, lee-hpca2013,
gillingham-jssc1991,
keeth00}. The timeline starts with an
\cmdact at 0 ns and ends with the completion of \cmdpre at 48.75 ns. From the
figure, we identify three specific periods in time when the voltage changes
slowly: {\em i)}~start of sense-amplification (part \ding{192}), {\em
ii)}~end of sense-amplification (part \ding{193}), and {\em iii)}~end of
precharging (part \ding{194}). Since charge is correlated with voltage, these
three periods are when the charge also moves slowly. In the following, we
provide three observations explaining why these three periods can be shortened
for typical cells at typical temperatures --- offering the best opportunity for
shortening the timing parameters.

{\em Observation 1. At the start of the sense-amplification phase, the higher
the bitline voltage, the quicker the sense-amplifier is jump-started.} Just as
the amplification phase starts, the sense-amplifier detects the bitline voltage
that was increased in the previous charge-sharing phase (by the cell donating
its charge to the bitline). The sense-amplifier then begins to inject more
charge into the bitline to increase the voltage even further --- triggering a
positive-feedback loop where the bitline voltage increases more quickly as the
bitline voltage becomes higher. This is shown in Figure~\ref{fig:dram_signals}
where the bitline voltage ramps up faster and faster during the initial part of
the amplification phase. Importantly, if the bitline has a higher voltage to
begin with (at the start of sense-amplification), then the positive-feedback is
able to set in more quickly. Such a high initial voltage is comfortably
achieved by typical cells at typical temperatures because they donate a large
amount of charge to the bitline during the charge-sharing phase (as they have a
large amount of charge). As a result, they are able to reach states \ding{204}
and \ding{205} (Figure~\ref{fig:operation} in Chapter~\ref{ch:bak}) more
quickly, creating the opportunity to shorten \trcd and \tras.

{\em Observation 2. At the end of the sense-amplification phase, nearly half
the time (42\%) is spent on injecting the last 5\% of the charge into the
cell}. Thanks to the positive-feedback, the middle part of the amplification
phase (part between \ding{192} and \ding{193} in Figure~\ref{fig:dram_signals})
is able to increase the bitline voltage quickly. However, during the later part
of amplification (part \ding{193} in Figure~\ref{fig:dram_signals}), the
RC-delay becomes much more dominant, which prevents the bitline voltage from
increasing as quickly. In fact, it takes a significant amount of extra delay
for the bitline voltage to reach \vdd (Figure~\ref{fig:dram_signals}) that is
required to {\em fully} charge the cell. However, for typical cells at typical
temperatures, such an extra delay may not be needed --- the cells could already
be injected with {\em enough} charge for them to comfortably share with the
bitline when they are next accessed. This allows us to shorten the later part
of the amplification phase, creating the opportunity to shorten \tras and \twr.

{\em Observation 3. At the end of the precharging phase, nearly half the time
(45\%) is spent on extracting the last 5\% of the charge from the bitline}.
Similar to the amplification phase, the later part of the precharging phase is
also dominated by the RC-delay, which causes the bitline voltage to decrease
slowly to \hvdd (part \ding{194} in Figure~\ref{fig:dram_signals}). If we
decide to incur less than the full delay required for the bitline voltage to
reach exactly \hvdd, it could lead to two different outcomes depending on which
cell we access next. First, if we access the {\em same} cell again, then the
higher voltage left on the bitline works in our favor. This is because the cell
--- which is filled with charge --- would have increased the bitline voltage
anyway during the charge-sharing phase. Second, if we access a {\em different}
cell connected to the same bitline, then the higher voltage left on the bitline
may work as a handicap. Specifically, this happens only when the cell is devoid
of any charge (e.g., storing a data of `0'). For such a cell, its
charge-sharing phase operates in the opposite direction, where the cell steals
some charge away from the bitline to decrease the bitline voltage.
Subsequently, the voltage is ``amplified'' to 0 instead of \vdd. Nevertheless,
typical cells at typical temperatures are capable of comfortably overcoming the
handicap --- thanks to their large capacitance, the cells are able to steal a
large amount of charge from the bitline. As a result, this creates the
opportunity to shorten \trp.

	\section{Charge Gap: Common-Case vs.~Worst-Case}
\label{sec:factors}

Based on the three observations, we understand that {\em timing parameters can
be shortened if the cells have enough charge.} Importantly, we showed that such
a criterion is easily satisfied for typical cells at typical temperatures. In
this section, we explain what it means for a cell to be ``typical'' and why it
has more charge at ``typical'' temperatures. Specifically, we examine two
physical phenomena that critically impact a DRAM cell's ability to receive and
retain charge: {\em i)}~process variation and {\em ii)}~temperature
dependence.

\subsection{Process Variation: Cells Are Not Created Equal}
\label{sec:factors_process}

\begin{figure}[h]
	\center
	\subcaptionbox{Existing DRAM\label{fig:dram}}[0.47\linewidth] {
		\includegraphics[width=.45\linewidth]{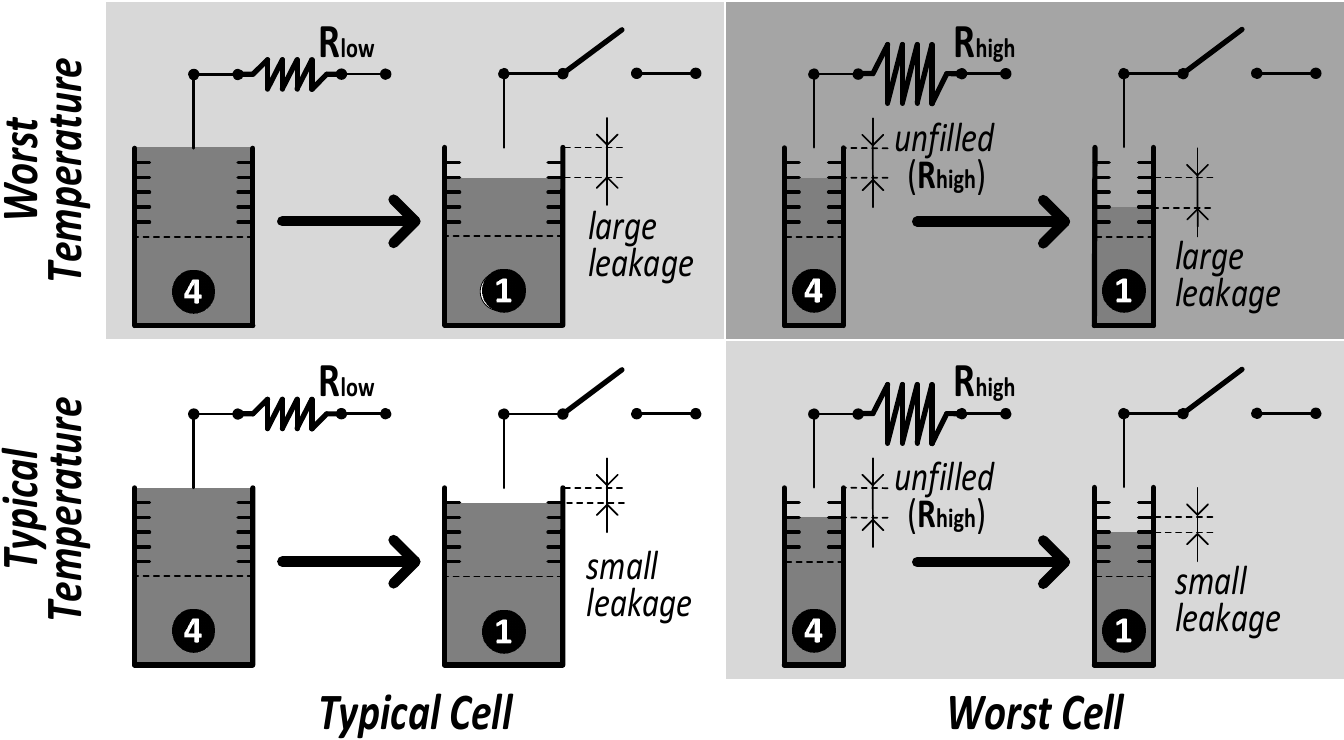}
	}
	\subcaptionbox{Our Proposal (Adaptive-Latency DRAM)\label{fig:aldram}}[0.47\linewidth] {
		\includegraphics[width=.45\linewidth]{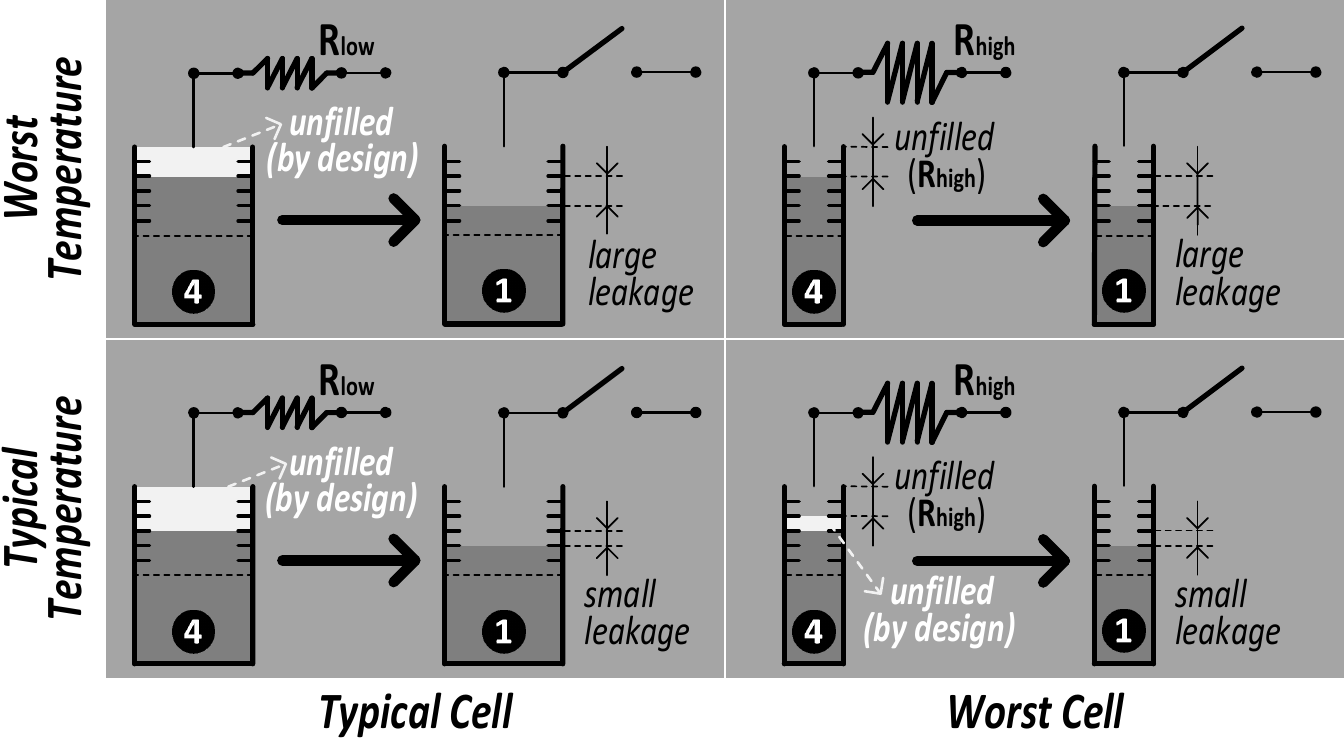}
	}
	\caption{Effect of Reduced Latency: Typical~vs.~Worst \\(Darker Background
	means Less Reliable)}
	\label{fig:common-worst}
\end{figure}

{\em Process variation} is a well-known phenomenon that introduces deviations
between a chip's intended design and its actual
implementation~\cite{friedberg-isqed2005, lee-iedm1996, smruti-tsm2008}. DRAM
cells are affected by process variation in two major aspects: {\em i)}~cell
capacitance and {\em ii)}~cell resistance. Although every cell is designed to
have a large capacitance (to hold more charge) and a small resistance (to
facilitate the flow of charge), some deviant cells may not be manufactured in
such a manner~\cite{liu-isca2012, hamamoto-ted1998, kim-edl2009, li-tcsi2011,
khan-sigmetrics2014, liu-isca2013, kang-memforum2014}. In
Figure~\ref{fig:dram}, we illustrate the impact of process variation using two
different cells: one is a typical cell conforming to design (left column) and
the other is the worst-case cell deviating the most from design (right column).

As we see from Figure~\ref{fig:dram}, the worst-case cell contains less charge
than the typical cell in state \ding{205} (Restored state, as was shown in
Figure~\ref{fig:operation} in Chapter~\ref{ch:bak}). This is because of two
reasons. First, due to its {\em large resistance}, the worst-case cell cannot
allow charge to flow inside quickly. Second, due to its {\em small
capacitance}, the worst-case cell cannot store much charge even when it is
full. To accommodate such a worst-case cell, existing timing parameters are set
to a large value. However, {\em worst-case cells are relatively rare.} When we
analyzed \DIMMs DRAM modules, the overwhelming majority of them had
significantly more charge than what is necessary for correct operation
(Section~\ref{sec:profiling} will provide more details).

\subsection{Temperature Dependence: Hot Cells Are Leakier}
\label{sec:factors_temp}

{\em Temperature dependence} is a well-known phenomenon in which cells leak
charge at almost double the rate for every 10\celsius\xspace increase in
temperature~\cite{restle-iedm1992, yaney-iedm1987, mori-iedm2005, khan-sigmetrics2014,
liu-isca2013}. In Figure~\ref{fig:dram}, we illustrate the impact of temperature
dependence using two cells at two different temperatures: {\em i)} typical
temperature (55\celsius, bottom row), and {\em ii)} the worst-case temperature
(85\celsius, top row) supported by DRAM standards.

\begin{figure}[h]
	\centering
	\includegraphics[height=1.4in]{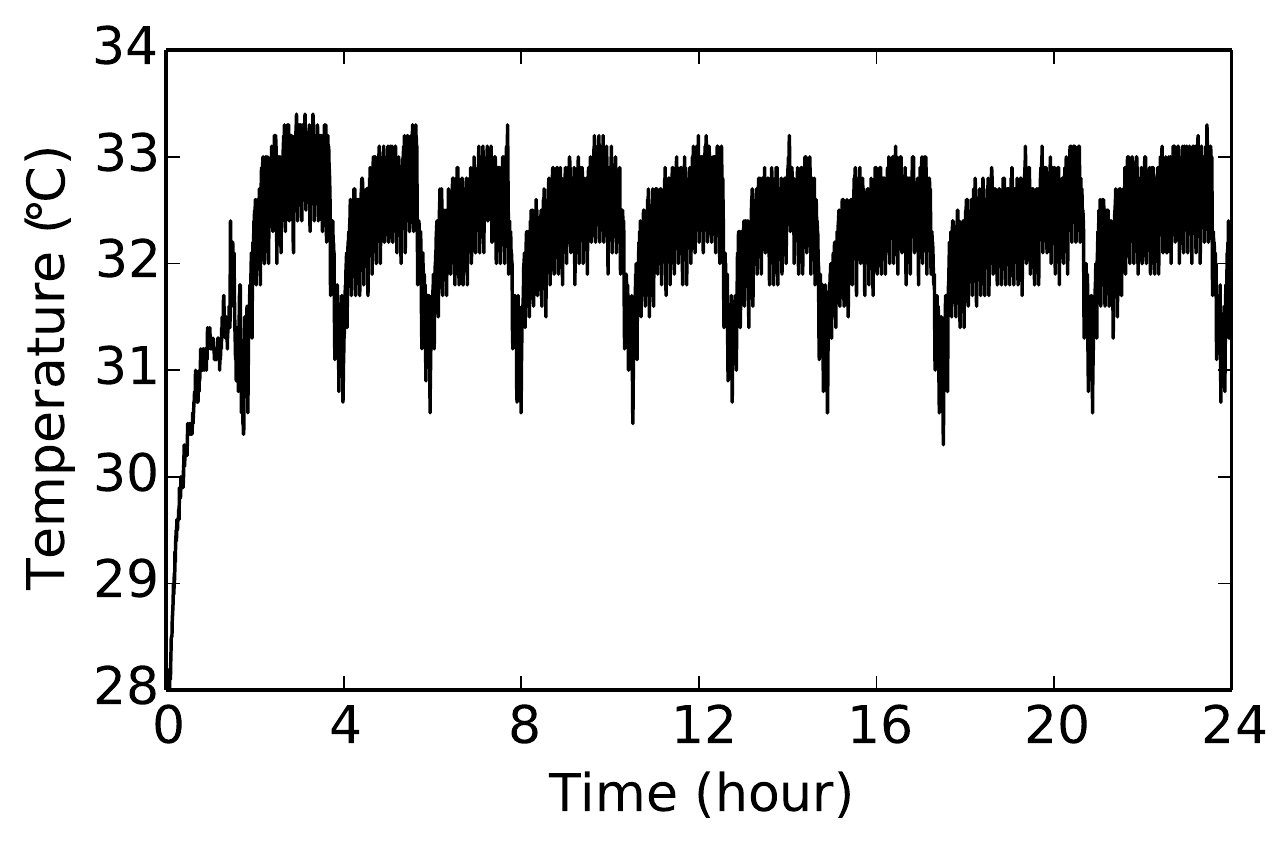}
	\includegraphics[height=1.4in]{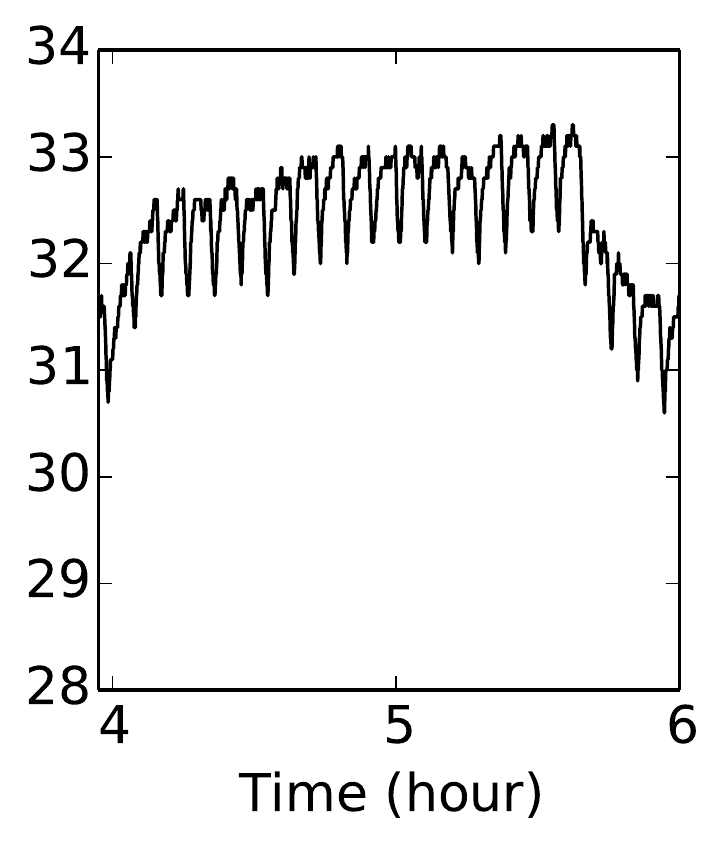}
  \caption{DRAM Temperature in a Server Cluster}
	\label{fig:motivation_temperature}
\end{figure}

\begin{figure}[h]
	\centering
	\includegraphics[height=1.45in]{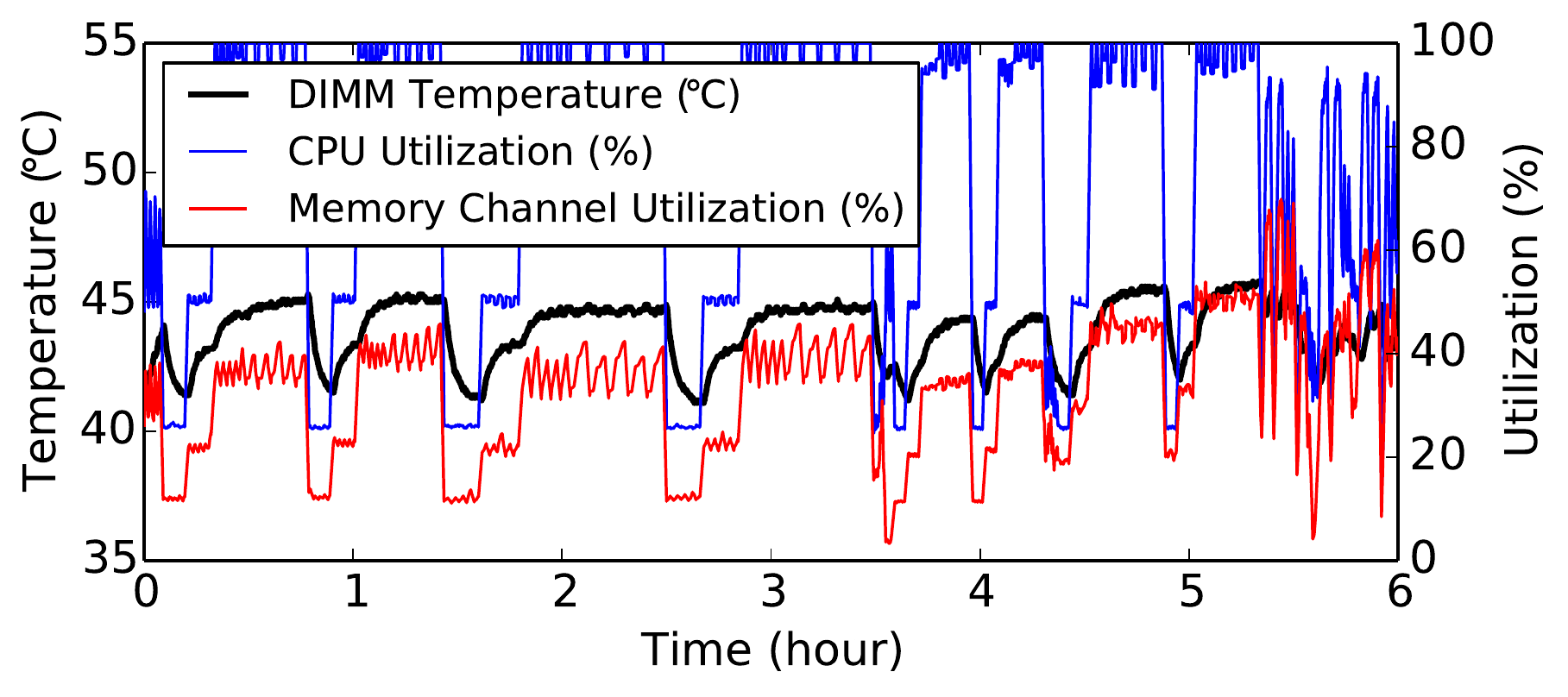}
  \caption{DRAM Temperature in a Desktop System}
	\label{fig:motivation_desktop}
\end{figure}

As we see from the figure, both typical and worst-case cells leak charge at a
faster rate at the worst-case temperature. Therefore, not only does the
worst-case cell have less charge to begin with, but it is left with {\em even
less} charge because it leaks charge at a faster rate (top-right in
Figure~\ref{fig:dram}). To accommodate the combined effect of process variation
{\em and} temperature dependence, existing timing parameters are set to a very
large value. However, {\em most systems do not operate at 85\celsius
}~\cite{liu-hpca2011, elsayed-sigmetrics2012,
elsayed-techreport2012}.\footnote{\label{footnote:temperature} Figure 22
in~\cite{elsayed-techreport2012} and Figure 3 in~\cite{liu-hpca2011} show that
the maximum temperature of DRAM chips at the highest CPU utilization is
60--65\celsius. While some prior works claim a maximum DRAM temperature over
80\celsius~\cite{zhu-itherm2008}, each DRAM module in their system dissipates
15W of power. This is very aggressive nowadays --- modern DRAM modules
typically dissipate around 2--6W (see Figure 8 of~\cite{hp-server}, 2-rank
configuration same as the DRAM module configuration of~\cite{zhu-itherm2008}).
We believe that continued voltage scaling and increased energy efficiency of
DRAM have helped reduce the power consumption of the DRAM module. While old
DDR1/DDR2 use 1.8--3.0V power supplies, newer DDR3/DDR4 use only 1.2--1.5V. In
addition, newer DRAMs adopt more power saving techniques (i.e., temperature
compensated self refresh, power down modes~\cite{micron_lowpower,
jedec_lowpower}) that were previously used only by Low-Power DRAM (LPDDR).
Furthermore, many previous works~\cite{liu-hpca2011, zhu-itherm2008,
lin-sigmetrics2008, lin-sigmetrics2008, david-icac2011} propose
hardware/software mechanisms to maintain a low DRAM temperature and energy.} We
measured the DRAM ambient temperature in a server cluster running a
memory-intensive benchmark, and found that the temperature {\em never} exceeds
34\celsius\xspace\xspace --- as well as never changing by more than
0.1\celsius\xspace per second. We show this in
Figure~\ref{fig:motivation_temperature}, which plots the temperature for a
24-hour period (left) and also zooms in on a 2-hour period (right). In
addition, we repeated the measurement on a desktop system that is not as well
cooled as the server cluster. As Figure~\ref{fig:motivation_desktop} shows,
even when the CPU was utilized at 100\% and the DRAM bandwidth was utilized at
40\%, the DRAM ambient temperature never exceeded 50\celsius. Other
works~\cite{liu-hpca2011, elsayed-sigmetrics2012, elsayed-techreport2012}
report similar results, as explained in detail in
Footnote~\ref{footnote:temperature}. From this, we conclude that the majority
of DRAM modules are likely to operate at temperatures that are much lower than
85\celsius, which slows down the charge leakage by an order of magnitude or
more than at the worst-case temperature.

\subsection{Reliable Operation with Shortened Timing}
\label{sec:factors_reliable}

As explained in Section~\ref{sec:leakage}, the amount of charge in
state~\ding{202} (i.e., the precharged state in Figure~\ref{fig:operation} in
Chapter~\ref{ch:bak}) plays a critical role in whether the correct data is
retrieved from a cell.  That is why the worst-case condition for correctness is
the top-right of Figure~\ref{fig:dram}, which shows the least amount of charge
stored in the worst-case cell at the worst-case temperature in
state~\ding{202}. However, DRAM manufacturers provide reliability guarantees
even for such worst-case conditions. In other words, the amount of charge at
the worst-case condition is still greater than what is required for
correctness.

If we were to shorten the timing parameters, we would also be reducing
the charge stored in the cells. It is important to note, however, that
we are proposing to exploit {\em only} the {\em additional slack} (in
terms of charge) compared to the worst-case. This allows us to provide
as strong of a reliability guarantee as the worst-case.

In Figure~\ref{fig:aldram}, we illustrate the impact of shortening the
timing parameters in three of the four different cases: two different
cells at two different temperatures. The lightened portions inside the
cells represent the amount of charge that we are giving up by using the
shortened timing parameters. Note that we are not giving up any charge
for the worst-case cell at the worst-case temperature. Although the
other three cells are not fully charged in state~\ding{205}, when they
eventually reach state~\ding{202}, they are left with a similar amount
of charge as the worst-case (top-right). This is because these cells are
capable of either holding more charge (typical cell, left column) or
holding their charge longer (typical temperature, bottom row).
Therefore, optimizing the timing parameters (based on the amount of
existing slack) provides the opportunity to reduce overall DRAM latency
while still maintaining the reliability guarantees provided by the DRAM
manufacturers.

In Section~\ref{sec:profiling}, we present the results from our
characterization study where we quantify the slack in \DIMMs DRAM modules.
Before we do so, we first propose our mechanism for identifying and enforcing
the shortened timing parameters.

	\section{Adaptive-Latency DRAM} \label{sec:mechanism}

Our mechanism, Adaptive-Latency DRAM (\ALD), allows the memory controller to
exploit the latency variation across DRAM modules at different operating
temperatures by using customized (aggressive) timing parameters for each DRAM
module/temperature combination. Our mechanism consists of two steps: {\em i)}
{\em identification} of the best timing parameters for each DRAM
module/temperature, and {\em ii)} {\em enforcement}, wherein the memory
controller dynamically extracts each DRAM module's operating temperature and
uses the best timing parameters for each DRAM module/temperature combination.

\subsection{Identifying the Best Timing Parameters}
\label{sec:mechanism_identify}

Identifying the best timing parameters for each DRAM module at different
temperatures is the more challenging of the two steps. We propose that DRAM
manufacturers identify the best timing parameters at different temperatures for
each DRAM chip during the testing phase and provide that information along with
the DRAM module in the form of a simple table. Since our proposal only involves
changing four timing parameters (\trcd, \tras, \twr, and \trp), the size of the
table for, e.g., five, different temperature points is small and such a table
can potentially be embedded in the Serial Presence Detect circuitry (a ROM
present in each DRAM module~\cite{jedec-spd}). We expect this approach to have
low cost as DRAM manufacturers already have an elaborate testing mechanism to
identify faulty DRAM modules. An alternative approach to perform this profiling
step is to do it at the end-user system using \mycolor{an} online testing while
the system is running. We leave the exploitation of such an online testing
mechanism to future work.

\subsection{Enforcing Dynamic Timing Parameters} \label{sec:mechanism_enforce}

Dynamically enforcing the best timing parameters at the memory controller is
fairly straightforward. The memory controller populates a hardware table with
the best timing parameters for different temperatures for all the DRAM modules
connected to the controller. The memory controller divides time into regular
intervals (e.g., 256 ms). At the beginning of each interval, it extracts the
temperature of each DRAM module. It then employs the best timing parameters
corresponding to the temperature of each DRAM module for all accesses during
the remainder of the interval.

This approach should work well in practice as temperature does not change very
frequently in real systems --- our measurements on real server and desktop
systems indicate that temperature changes at the rate of at most
0.1\celsius\xspace per second. In addition, existing DRAM designs such as
LPDDR3~\cite{jedec-lpddr3}, and the recently announced DDR4~\cite{jedec-ddr4}
already have in-DRAM temperature sensors to minimize self-refresh energy. By
accessing the temperature value of in-DRAM temperature sensors during
auto-refresh (usually performed every 7.8us), our mechanism monitors DRAM
temperature without any performance penalty and frequently enough to detect
even drastic temperature changes.

In Section~\ref{sec:evaluation}, we evaluate the benefits of \ALD on a real
system equipped with a memory controller whose timing parameters can be
dynamically changed. Our evaluation shows that \ALD can significantly improve
system performance for a wide variety of applications.

	\section{DRAM Latency Profiling Methodology} \label{sec:profile}

In this section, we describe our FPGA-based DRAM profiling (i.e., testing)
infrastructure and the methodology we use to characterize DRAM modules for
variation in access latency.

\subsection{Profiling Infrastructure} \label{sec:infrastructure}

\begin{figure}[h]
	\centering
	\subcaptionbox{Full System\label{fig:infra_system_aldram}}[0.4\linewidth] {
		\includegraphics[width=2.5in]{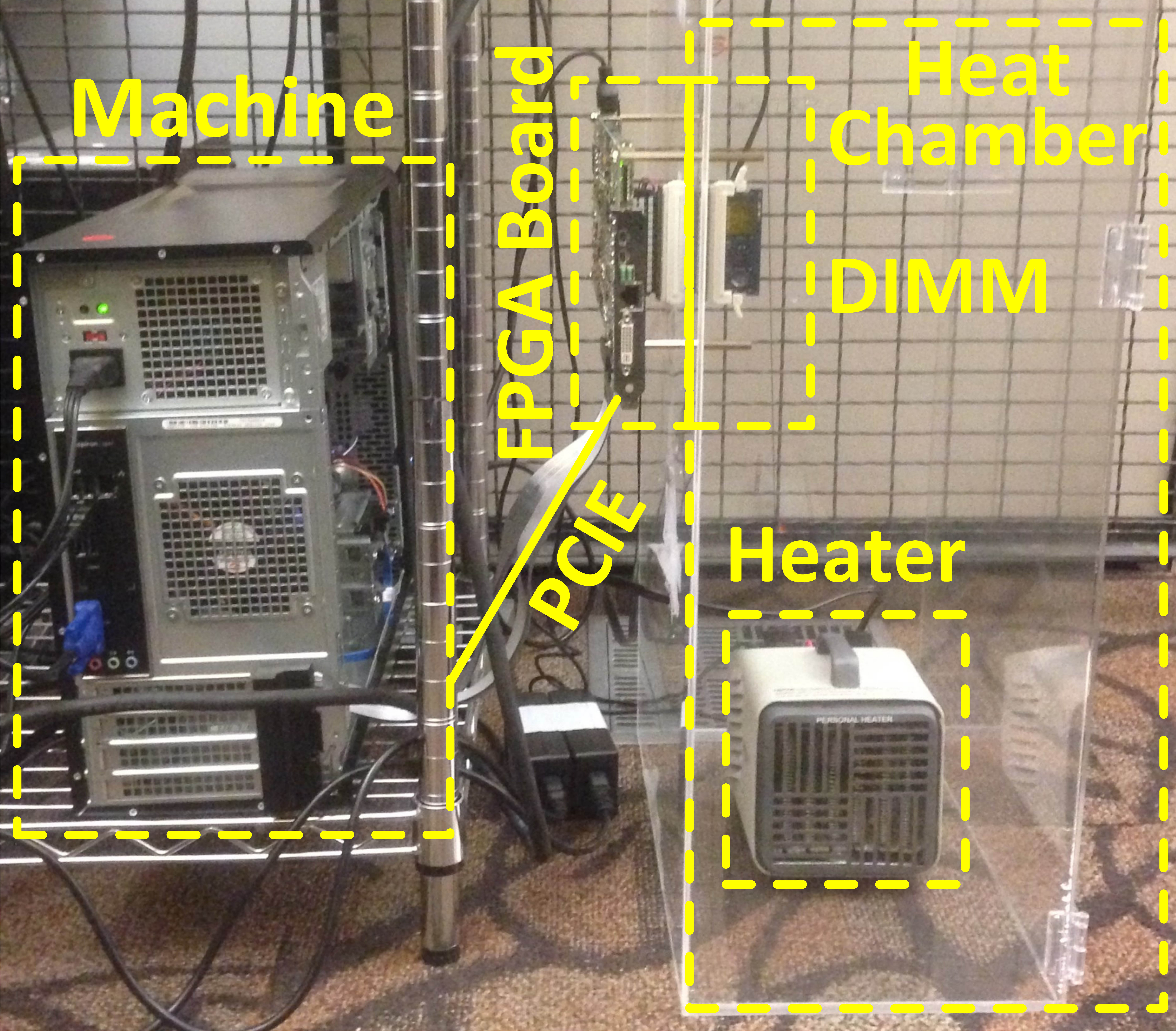}
	}
	\hspace{0.3in}
	\subcaptionbox{FPGA Board\label{fig:infra_fpga_aldram}}[0.4\linewidth] {
		\includegraphics[width=2.5in]{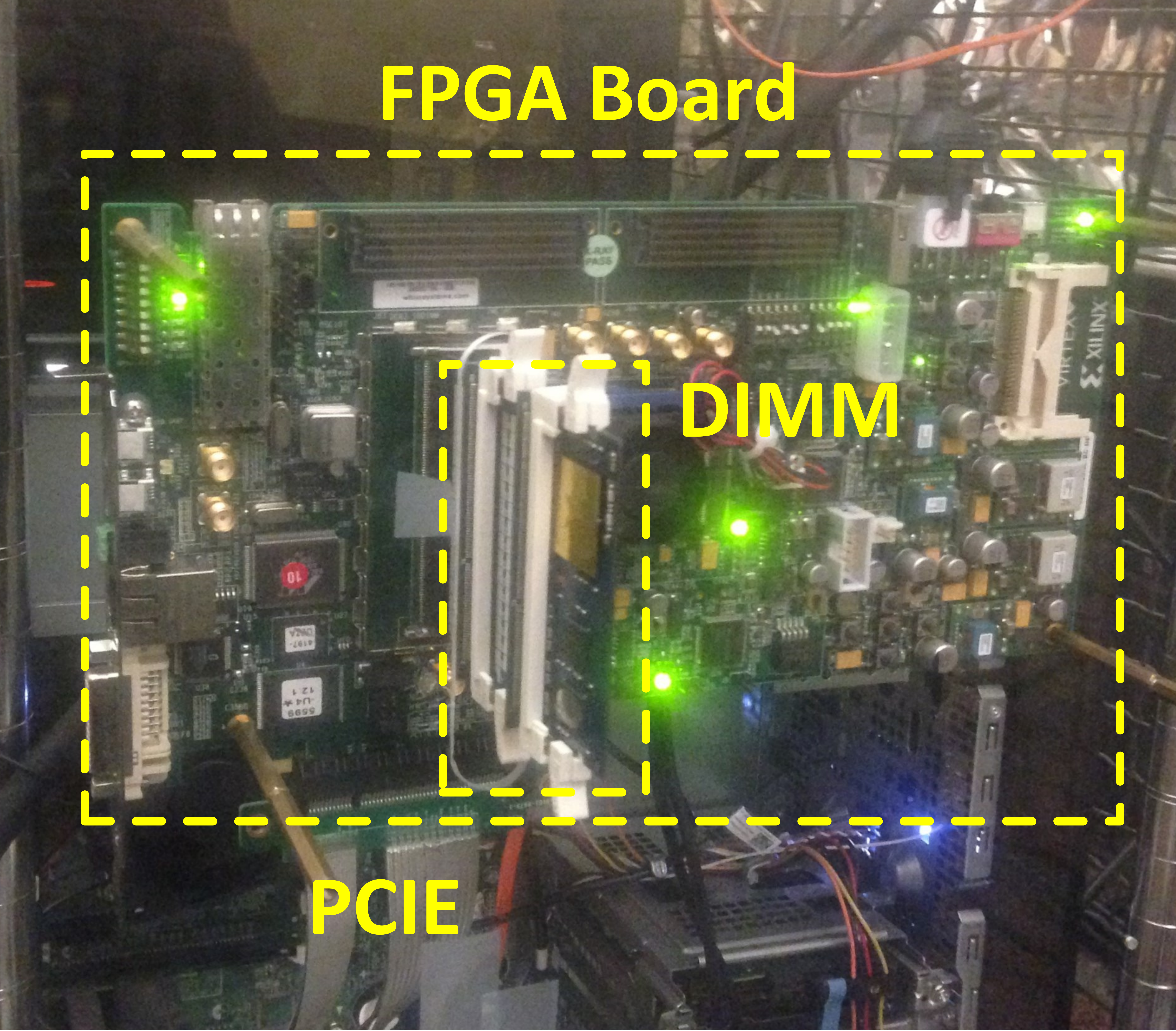}
	}
	\caption{FPGA-Based DRAM Test Infrastructure} \label{fig:infra_aldram}
\end{figure}

To analyze the characteristics of latency variation in DRAM, we have built an
FPGA-based testing platform~\cite{chandrasekar-date2014, liu-isca2013,
khan-sigmetrics2014, kim-isca2014, qureshi-dsn2015, chang-sigmetrics2016,
khan-dsn2016} using Xilinx ML605 boards~\cite{ml605} connected to a host
machine over a PCI-e bus~\cite{pcie}, as shown in
Figure~\ref{fig:infra_aldram}. The FPGA has a DDR3 DRAM memory
controller~\cite{mig} (Figure~\ref{fig:infra_fpga_aldram}). We customize the
memory controller so that we can change DRAM timing parameters from the host
machine at runtime. To test DRAM modules at different operating temperatures,
we built a heat chamber consisting of a heat encloser, a heater, and a
temperature controller with a thermo-couple sensor
(Figure~\ref{fig:infra_system_aldram}). During the test, the temperature within
the heat chamber is maintained within $\pm$0.5{\celsius} of the target
temperature. In all, we present results for \DIMMs DRAM modules that are
produced by three different DRAM manufacturers during the last 5 years (from
2010 to 2014). The majority of these are DDR3-1600 SO-DIMMs with standard
timing parameters: \trcd~$=$~13.75 ns, \tras~$=$~35 ns, \twr~$=$~15 ns, and
\trp~$=$~13.75 ns.

\subsection{Profiling Mechanism} \label{sec:pattern}

The goal of our profiling experiments is to determine the amount by which
different timing parameters can be reduced without inducing any errors. For
this purpose, we devise two types of tests: one for testing read operations
and another for write operations. The read test aims to analyze reductions in
\trcd, \tras, and \trp, whereas the write test aims to analyze reductions in
\twr. Both types of tests are carefully designed to induce a reasonable amount
of coupling-noise between circuit components (i.e., cells, bitlines, and
wordlines), as well as to generate power-noise. We now describe the steps
involved in each of these tests in detail. Each test involves a test address,
a test data pattern, and a target set of timing parameters.

{\bf Read Test.} For a read test, we first write the test data pattern to the
target address with the standard DRAM timing parameters. We follow it with a
read to the target address with the {\em target}, smaller, timing parameters.
This step could potentially corrupt the data if the cells involved in the
operation are not restored with enough charge or if the corresponding bitlines
are not precharged sufficiently. To test for data corruption, we wait for the
refresh interval such that DRAM cells have the smallest charge due to charge
leakage over time. Then, we perform another read operation to the target
address. If the latter read data matches the originally-written test data, the
test is considered successful --- i.e., for the corresponding address, the
timing parameters can be reduced to the target timing parameters. Otherwise,
the test is considered a failure and the system logs the list of errors.

{\bf Write Test.} For a write test, we first write the inverted test data
pattern to the target address with the standard DRAM timing parameters. We then
write the test data pattern with the target timing parameters. The original
write (inverted test pattern) is to ensure that the test write actually flips
the values on the bitlines, thereby making the \twr timing parameter relevant.
After the test write, we wait for the refresh interval and perform a
verification read. We check if the verification succeeded and log the list of
errors, if any.

{\bf Coupling and Power-Noise Effects.} We carefully design the tests to be
close to the worst-case in coupling and power-noise (to stress the reliability
of our proposal). From the standpoint of a single row, each test {\it i)}
writes/reads data to/from the row, and then {\it ii)} reads data from the row
for verification (access -- wait -- verify). However, from the standpoint of
the entire DRAM chip, we test different rows in an overlapped manner, staggered
by \trc (access, access, $\cdots$, access, verify, verify, $\cdots$, verify).
We perform this test sequence twice with two different orders of row addresses
(increasing/decreasing). Due to this overlapped manner of testing multiple
rows, the tests exhibit the following effects.

\squishlist

	\item{{\em Bitline-to-Cell Coupling:} Non-activated cells and bitlines are
	coupled with each other in as many as half of the cells per bitline (i.e.,
	256 times a 512-cell bitline).}

	\item{{\em Wordline-to-Wordline Coupling:} Each wordline is coupled with
	adjacent wordlines in as many columns as in a row (i.e., 256 times for
	a 4k-cell row and 64-bit per access).}

	\item{{\em Power-Noise:} Power noise is close to the worst-case due to the
	maximum number of allowed row activations per time interval (i.e., activating
	a row every \trc)}.

\squishend

{\bf Data Patterns.} We repeat both tests for all DRAM addresses with eight
different checkered data patterns~\cite{vandegoor-delta2002} ({\tt 0000}, {\tt
0011}, {\tt 0101}, {\tt 1001}, {\tt 0110}, {\tt 1010}, {\tt 1100}, {\tt 1111}).
To test the bitline precharge effect, we use exclusive-OR data patterns (i.e.,
{\tt 0101} vs.~{\tt 1010}) for adjacent rows. Therefore, bitlines are amplified
in the opposite direction from the previous row activation.

{\bf Iterating Tests.} We repeat tests with a large number of reduced target
timing parameters. We accumulate the erroneous cells from all the test
patterns. Since cells of the same DRAM row are activated and precharged
together, we run through all the DRAM addresses in the column-major order to
reduce interference between tests of the cells in the same row. We perform 10
iterations of each test before deciding whether it is free from errors.

\mycolor{as we described, we carefully modeled the worst-case test scenarios by
{\em i)} using minimum value of timing parameters for both read and write test,
{\em ii)} generating couplings and power-noise as much as possible, and {\em
iii)} using multiple data patterns. These test scenarios are similar to the
most error prone test patterns related to DRAM timing parameters in
Memtest~\cite{memtest}, while our test scenarios could not cover all test
patterns that can be used by memtest and DRAM companies. Therefore, we expect
that using our test scenarios can uncover most of the DRAM errors which are
related to reducing DRAM timing parameters.}

	\section{DRAM Latency Profiling Results and Analysis} \label{sec:profiling}

In this section, we present and analyze the results of our profiling
experiments. We first present the effect of reducing individual timing
parameters on a single representative module
(Section~\ref{sec:profiling_individual}). We then present the results of
reducing multiple timing parameters simultaneously
(Section~\ref{sec:timing_all}), analyze the timing slack present
(Section~\ref{sec:temp_margin}), and quantify timing parameter reductions
(Section~\ref{sec:detailprofile}) using this representative module. Finally, we
summarize our profiling results for all \DIMMs modules
(Section~\ref{sec:manydimm}).

\subsection{Effect of Reducing Individual Timing Parameters}
\label{sec:profiling_individual}

As discussed in Section~\ref{sec:factors}, DRAM access latency is tightly
coupled with the amount of charge in the accessed cell. To verify this on real
DRAM chips, we need to adjust the amount of charge intentionally.
Unfortunately, there is no way to quantitatively change the charge stored in
DRAM cells. Instead, we indirectly adjust the charge amount by enabling the
leakage of DRAM cells. Note that DRAM cells have less charge in two cases: {\em
i)}~at longer refresh intervals and {\em ii)}~at higher temperatures. Using
these characteristics, we design aggressive test environments for analyzing the
effect of reducing each timing parameter by: {\em i)}~sweeping the refresh
interval from the DRAM standard (64 ms) to very long refresh intervals (up to
512 ms) at the highest temperature in the DRAM standard (85\celsius), and {\em
ii)}~sweeping the temperature at a very long refresh interval (512 ms).

\begin{figure}[h]
	\center
	\subcaptionbox{Sweeping the Refresh Interval (Temperature: $85$\celsius)\label{fig:ret_timing}}[1\linewidth] {
		\includegraphics[height=1.52in]{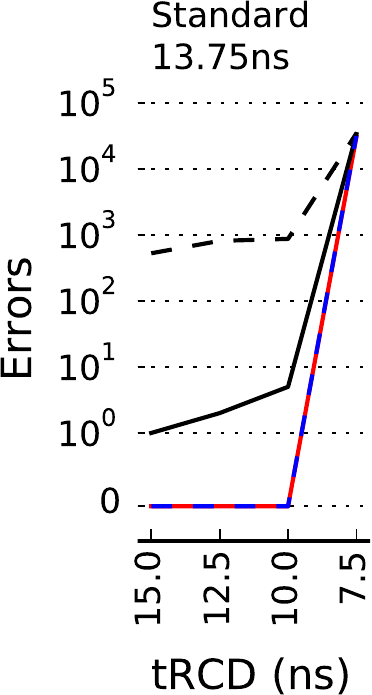}\hspace{0.05in}
		\includegraphics[height=1.52in]{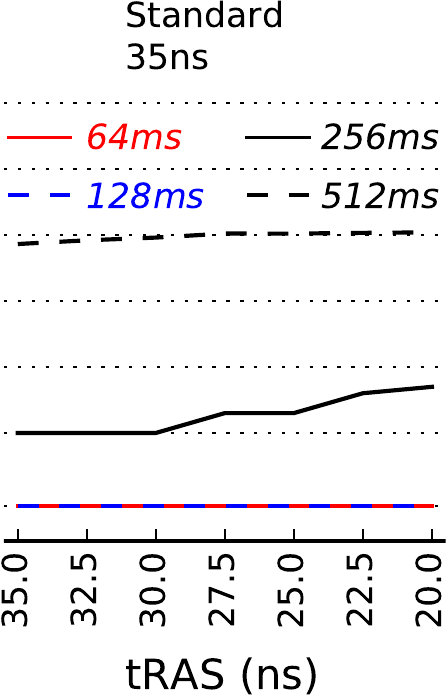}\hspace{0.05in}
		\includegraphics[height=1.52in]{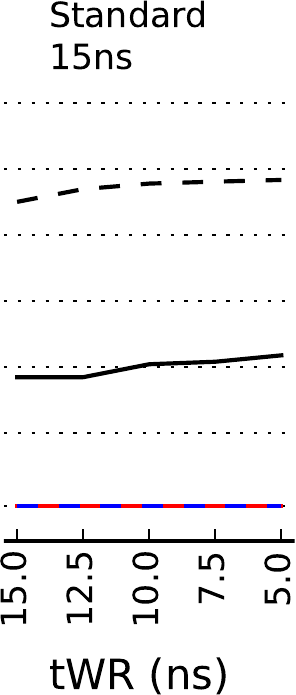}\hspace{0.05in}
		\includegraphics[height=1.52in]{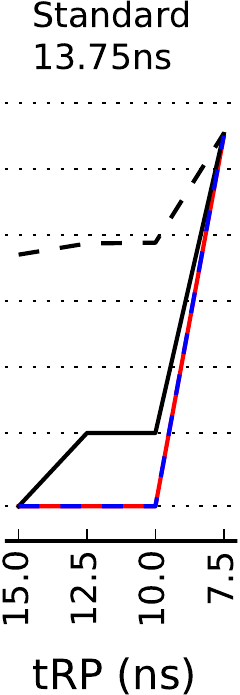}
	}
	\subcaptionbox{Sweeping the Temperature (Refresh Interval: 512ms)\label{fig:temp_timing}}[1\linewidth] {
		\includegraphics[height=1.52in]{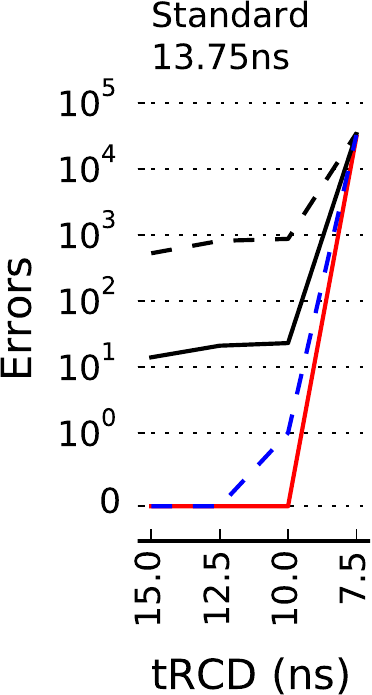}\hspace{0.05in}
		\includegraphics[height=1.52in]{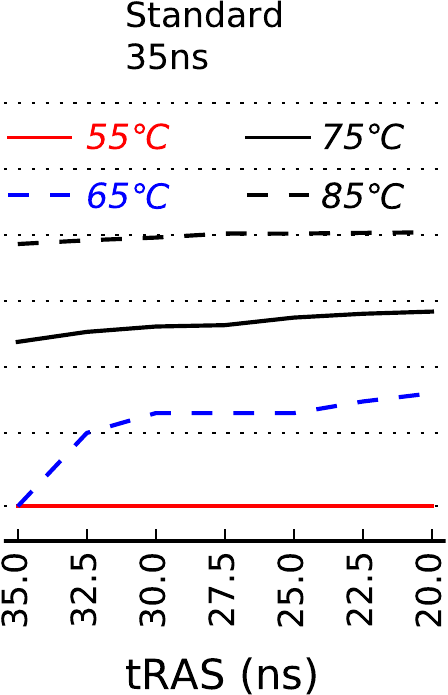}\hspace{0.05in}
		\includegraphics[height=1.52in]{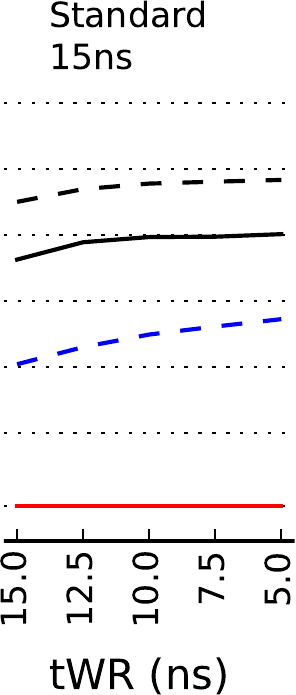}\hspace{0.05in}
		\includegraphics[height=1.52in]{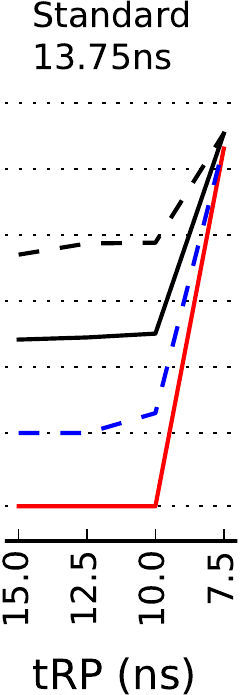}
	}
	\caption{Effect of Varying Each Timing Parameter on Error Count}
	\label{fig:variable-timing}
\end{figure}

{\bf Sweeping the Refresh Interval:} Figure~\ref{fig:ret_timing} shows the
number of errors when we reduce each of the four timing parameters (\trcd,
\tras, \twr, and \trp) at different refresh intervals while the temperature is
kept at 85\celsius. \mycolor{X-axis in the figure represents the value of the
tested timing parameter in linear scale, and Y-axis represents the error count
in logarithm scale (except for the zero label point which represents {\em no}
error. We use this hybrid logarithm scale in Y-axis for all later figures as
well.} We make three key observations from the figure. First, it is possible to
reduce the timing parameters significantly without incurring errors. In
particular, at the standard refresh interval of 64 ms, the four timing
parameters can be reduced by (3.75 ns, 15 ns, 10 ns, 3.75 ns). Second, as the
refresh interval increases, the number of errors starts to increase and the
range of timing parameters for error-free operation decreases.  In fact,
increasing the refresh interval beyond 256 ms results in errors even with the
standard timing parameters. Third, reducing \trcd or \trp below a certain point
results in a drastic increase in the number of errors. This is because doing so
causes a functional failure of DRAM --- e.g., reducing \trcd below 10 ns does
not give DRAM enough time to even activate the wordline completely, resulting
in a malfunction of the entire row.

{\bf Sweeping the Temperature:} Figure~\ref{fig:temp_timing} shows the number
of errors with reduced timing parameters at different temperatures
(55\celsius\xspace to 85\celsius) while the refresh interval is kept at 512 ms.
Note that the refresh interval is 8 times as long as its standard value of 64
ms, meaning that the number of errors shown in the figure are much larger than
what they should be for commodity DRAM. From the figure, we find that the
number of errors decreases at lower temperatures, similar to how it decreases
at lower refresh intervals in the previous experiment. Therefore, we conclude
that there is a significant opportunity to reduce timing parameters at lower
temperatures.

\subsection{Effect of Reducing Multiple Timing Parameters} \label{sec:timing_all}

The results of the previous experiment showed that there is a significant
opportunity for reducing each timing parameter individually. However, we
hypothesize that reducing one timing parameter may also decrease the
opportunity to reduce another timing parameter simultaneously. For example, if
we reduce the value of \tras, the charge on a cell may not be fully restored by
the end of the sense-amplification process. This may decrease the opportunity
to reduce \trp, as the bitlines may have to be fully precharged to ensure that
it can be reliably perturbed by the partially charged cell.

To study the potential for reducing multiple timing parameters simultaneously,
we run our tests with all possible combinations of timing parameters in the
following range: \trcd (12.5--10 ns), \tras (35--20 ns), \twr (15--5 ns), \trp
(12.5--10 ns). For each combination, we also vary the temperature from
55\celsius\xspace to 85\celsius\xspace and the refresh interval from 64 ms to
960 ms.

\begin{figure}[h]
		\centering
		\subcaptionbox{Read\label{fig:read_combined}}[0.45\linewidth] {
			\includegraphics[height=1.4in]{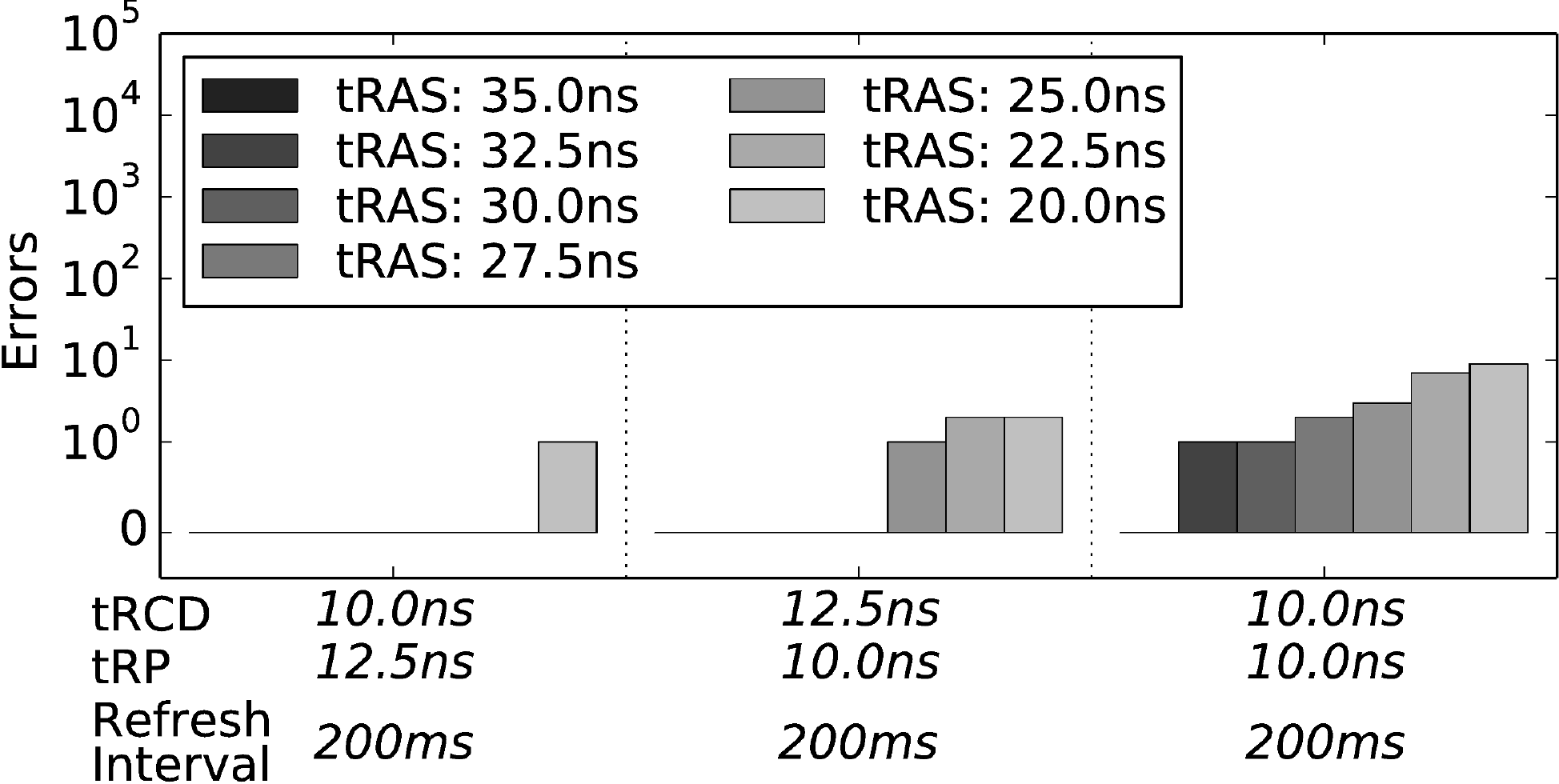}
		}
		\subcaptionbox{Write\label{fig:write_combined}}[0.45\linewidth] {
			\includegraphics[height=1.4in]{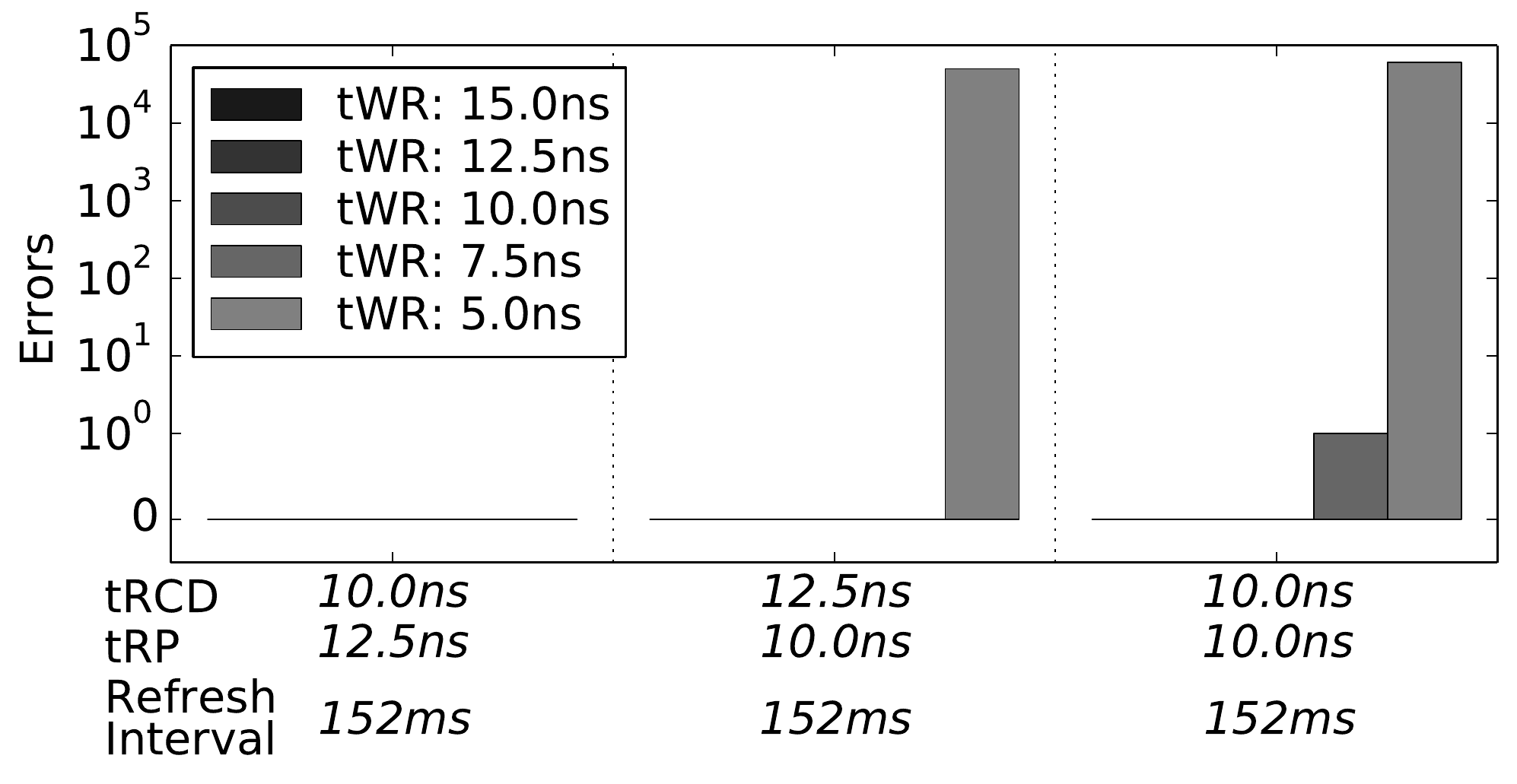}
		}
		\caption{Error Counts When Varying Multiple Timing Parameters Simultaneously} \label{fig:combined}
\end{figure}

Figure~\ref{fig:combined} presents the number of errors for the read and write
tests for multiple such combinations at 85\celsius. Our results validate our
hypothesis. For example, although it was possible to {\em individually} reduce
\tras to 20 ns at 85\celsius\xspace and 200 ms refresh interval, \tras can only
be reduced to 32.5 ns if both \trcd and \trp are {\em also} reduced to 10 ns.
In this section, we only present the results for specific combinations to show
the effect of reducing multiple timing parameters clearly. In
Section~\ref{sec:detailprofile}, we will present the test results with all
timing parameter combinations and resulting potential reductions.

\subsection{Effect of Temperature on Timing Slack} \label{sec:temp_margin}

As discussed in Section~\ref{sec:factors}, the length of the timing parameters
is dictated by the weakest cell with the shortest retention time at
$85$\celsius. However, ensuring enough charge is stored in such a cell may not
be enough to guarantee reliable operation --- this is because the cell could be
affected by other failure mechanisms that are difficult to foresee. Just to
name one example, some DRAM cells are known to suffer from a phenomenon known
as {\em variable retention time} (VRT)~\cite{liu-isca2013,
khan-sigmetrics2014}, in which their retention time could change unpredictably
between short and long values. As a counter-measure against such failures, DRAM
manufacturers provide a built-in {\em safety-margin} in retention time, also
referred to as {\em a guard-band}~\cite{wang-ats2001, ahn-asscc2006,
khan-sigmetrics2014}. This way, DRAM manufacturers are able to guarantee that
even the weakest cell is insured against various other modes of failure.

\begin{figure}[h]
		\centering
		\subcaptionbox{Read Test (Varying \tras)\label{fig:full_read}}[0.4\linewidth] {
			\includegraphics[height=3.0in]{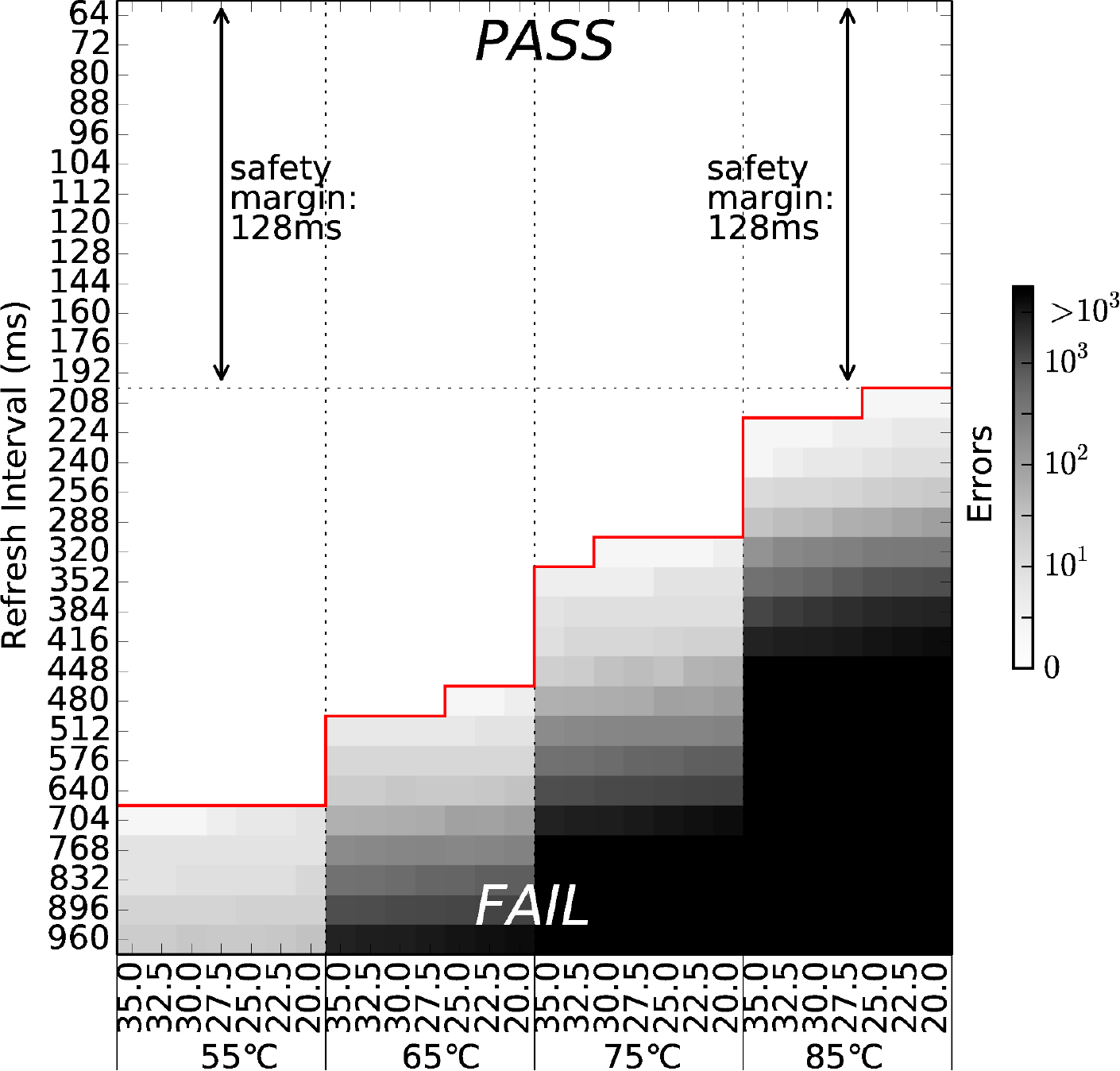}
		}
		\subcaptionbox{Write Test (Varying \twr)\label{fig:full_write}}[0.4\linewidth] {
			\includegraphics[height=3.0in]{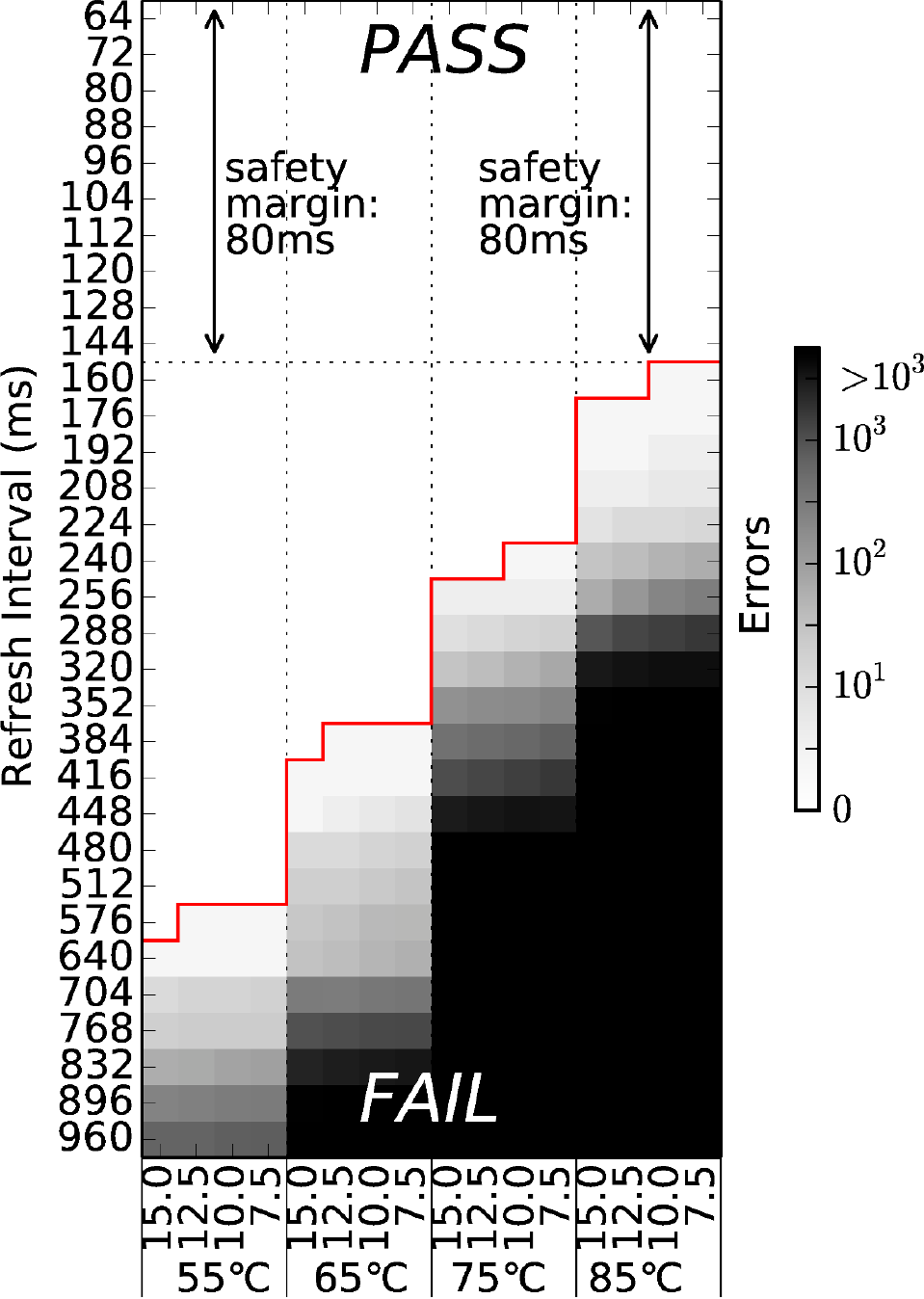}
		}
		\caption{Error Counts When Varying Temperature, Refresh Interval, and
		\tras/\twr (\trcd/\trp: 12.5 ns)}
		\label{fig:profile_full}
\end{figure}

In order to quantify the safety-margin, we sweep the refresh interval from 64
ms to 960 ms. The safety-margin incorporated by the DRAM manufacturers is the
difference between the highest refresh interval that exhibits no errors at
85\celsius\xspace and the standard refresh interval (64 ms).
Figure~\ref{fig:profile_full} plots the number of errors (in gray scale) for
varying refresh intervals for different temperatures (55\celsius\xspace to
85\celsius). For each temperature, we also vary \tras
(Figure~\ref{fig:full_read}) and \twr (Figure~\ref{fig:full_write}). A box at
($x$, $y$) represents the number of errors at a refresh interval $y$ and
\tras/\twr of $x$. A white box indicates no errors, and a darker box indicates
a large number of errors. The red line separates the error-free region from the
region with at least one error.

We make several observations from the figure. First, as expected, for each
temperature, increasing the refresh interval (i.e., going down on the $y$ axis)
leads to more errors. Second, for a given refresh interval and temperature,
reducing the \tras/\twr timing parameters (i.e., going right on the $x$ axis
for each temperature) also increases the number of errors. Third, at
85\celsius, the highest refresh interval for error-free operation is 192 ms for
the read test and 144 ms for the write test --- this implies a safety-margin of
128 ms for reads and 80 ms for writes. Fourth, the slack on the retention time
increases with decreasing temperature, because retention time increases with
decreasing temperature (Section~\ref{sec:factors}). As shown in the figure, at
55\celsius, the margin on the retention time is 576 ms for reads and 448 ms for
writes --- these values are at least 4$\times$ higher than their safety-margins
(at 85\celsius). In summary, there is significant room for reducing DRAM
latency at lower temperatures while still ensuring reliable DRAM operation.

\subsection{Potential Timing Parameter Reductions While Maintaining the
Safety-Margin} \label{sec:detailprofile}

So far, we have discussed the effect of reducing timing parameters both
individually and simultaneously. We also have studied the safety-margin and the
effect of the operating temperature on the slack in timing parameters. In this
section, we study the possible timing parameter reductions of a DRAM module
while maintaining the safety-margin.

\begin{figure}[h]
	\centering
	\subcaptionbox{Maximum Error-Free Refresh Interval at 85\celsius\xspace (Bank/Chip/Module)\label{fig:safety_detail}}[0.9\linewidth] {
		\includegraphics[width=0.6\linewidth]{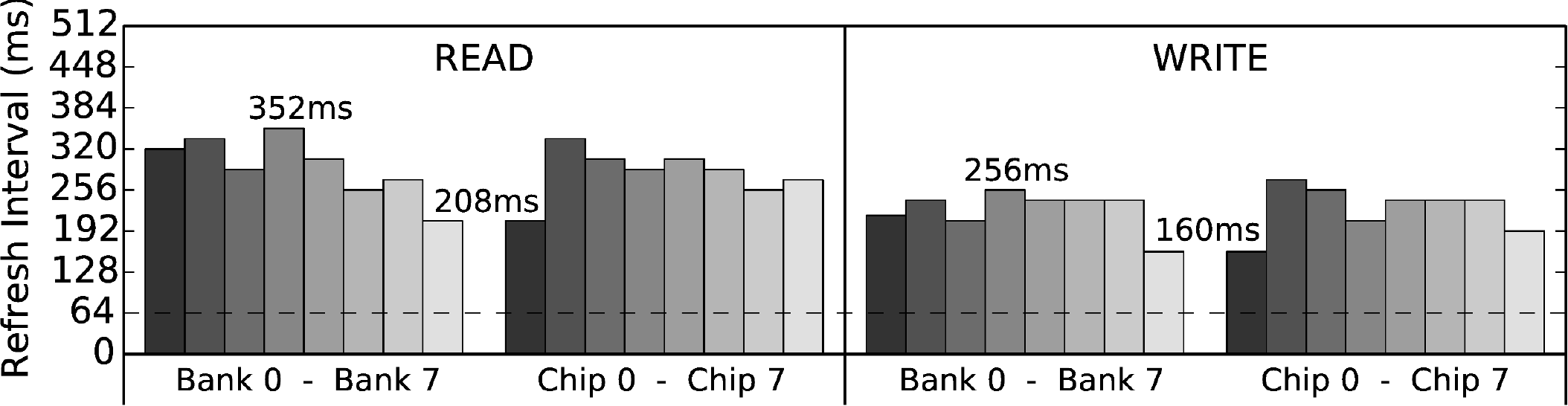}
	}

	\subcaptionbox{Read Latency (Refresh Interval: 200 ms)\label{fig:latency_detail_read}}[0.9\linewidth] {
		\includegraphics[width=0.6\linewidth]{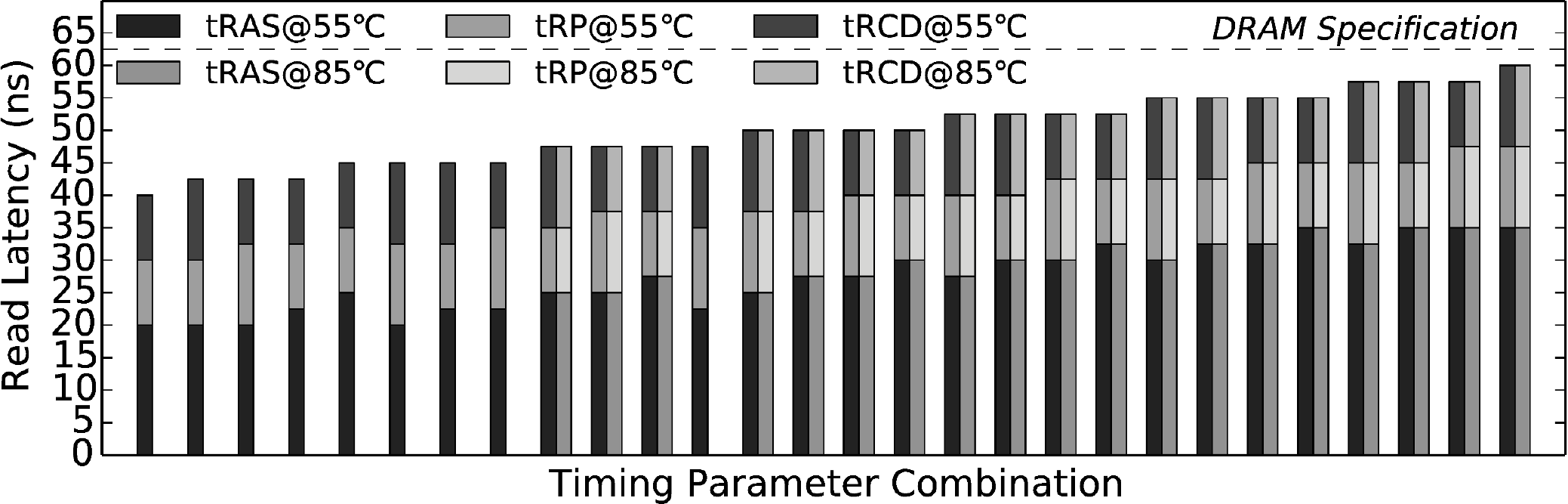}
	}

	\subcaptionbox{Write Latency (Refresh Interval: 152 ms)\label{fig:latency_detail_write}}[0.9\linewidth] {
		\includegraphics[width=0.6\linewidth]{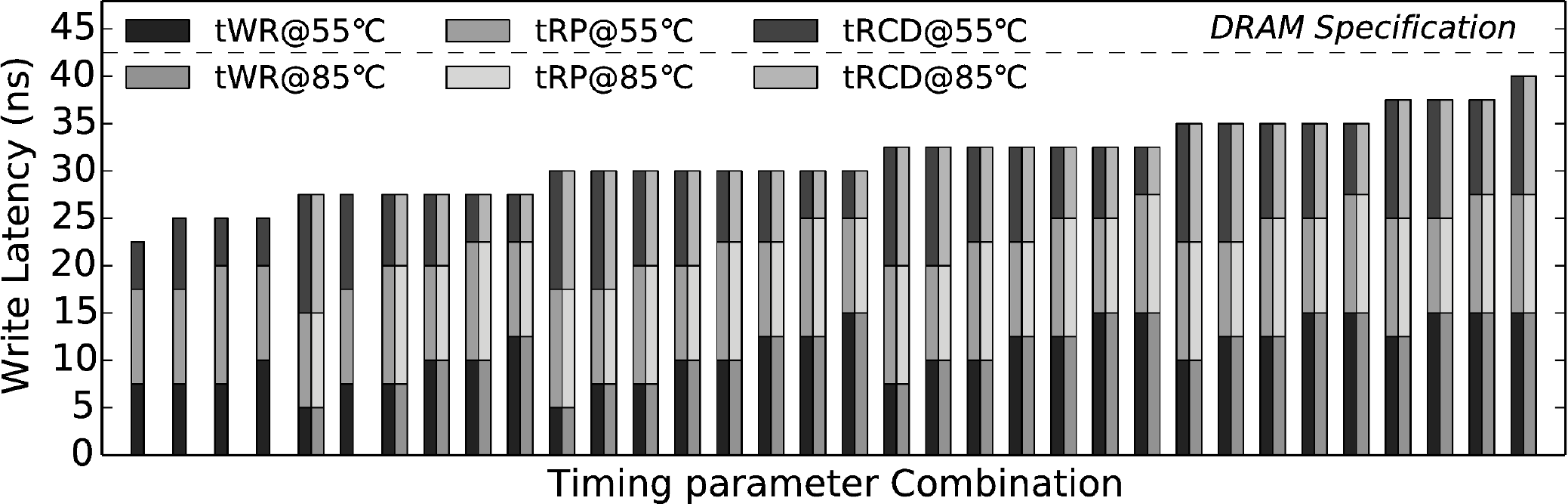}
	}
	\caption{Latency Reductions While Maintaining the Safety-Margin} \label{fig:latency_detail}
\end{figure}

We first measure the safety-margin of a DRAM module by sweeping the refresh
interval at the worst operating temperature (85\celsius), using the standard
timing parameters. Figure~\ref{fig:safety_detail} plots the maximum refresh
intervals of each bank and each chip in a DRAM module for both read and write
operations. We make several observations. First, the maximum refresh intervals
of both read and write operations are much larger than the DRAM standard (208
ms for the read and 160 ms for the write operations vs.~the 64 ms standard).
Second, for the smaller architectural units (banks and chips in the DRAM
module), some of them operate without incurring errors even at much higher
refresh intervals than others (as high as 352 ms for the read and 256 ms for
the write test). This is because the error-free retention time is determined by
the worst single cell in each architectural component (i.e., bank/chip/module).

Based on this experiment, we define the {\em safe refresh interval} for a DRAM
module as the maximum refresh interval that leads to no errors, minus an
additional margin of 8 ms, which is the increment at which we sweep the refresh
interval. The safe refresh interval for the read and write tests are 200 ms and
152 ms, respectively. We then use the safe refresh intervals to run the tests
with all possible combinations of timing parameters. For each combination, we
run our tests at two temperatures: 85\celsius\xspace and 55\celsius.

Figure~\ref{fig:latency_detail_read} plots the error-free timing parameter
combinations (\trcd, \tras, and \trp) in the read test. For each combination,
there are two stacked bars --- the left bar for the test at 55\celsius\xspace
and the right bar for the test at 85\celsius\xspace. Missing bars indicate that
the test (with that timing parameter combination at that temperature) causes
errors. Figure~\ref{fig:latency_detail_write} plots same data for the write
test (\trcd, \twr, and \trp).

We make three observations. First, even at the highest temperature of
85\celsius, the DRAM module reliably operates with reduced timing parameters
(24\% for read, and 35\% for write operations). Second, at the lower
temperature of 55\celsius, the potential latency reduction is even higher (36\%
for read, and 47\% for write operations). These latency reductions are possible
{\em while} maintaining the safety-margin of the DRAM module. From these two
observations, we conclude that there is significant opportunity to reduce DRAM
timing parameters {\em without compromising reliability}. Third, multiple
different combinations of the timing parameters can form the same overall value
of the timing parameters. For example, three different combinations of (\tras,
\trp, and \trcd) show the same overall value of 42.5 ns. This might enable
further optimization for the most critical timing parameter at runtime. We
leave the exploitation of such a fine-grained optimization to future work.

\subsection{Effect of Process Variation on Timing Slack} \label{sec:manydimm}

So far, we have discussed the effect of temperature and the potential to reduce
various timing parameters at different temperatures for a single DRAM module.
The same trends and observations also hold true for all of the other modules.
In this section, we analyze the effect of process variation by studying the
results of our profiling experiments on \DIMMs DRAM modules. We also present
results for intra-chip/inter-chip process variations by studying the process
variation across different banks/chips within each DRAM module.

\begin{figure}[h]
	\centering
	\subcaptionbox{Read Retention Time (DIMM)\label{fig:read_safety}}[0.48\linewidth] {
		\includegraphics[width=2.8in]{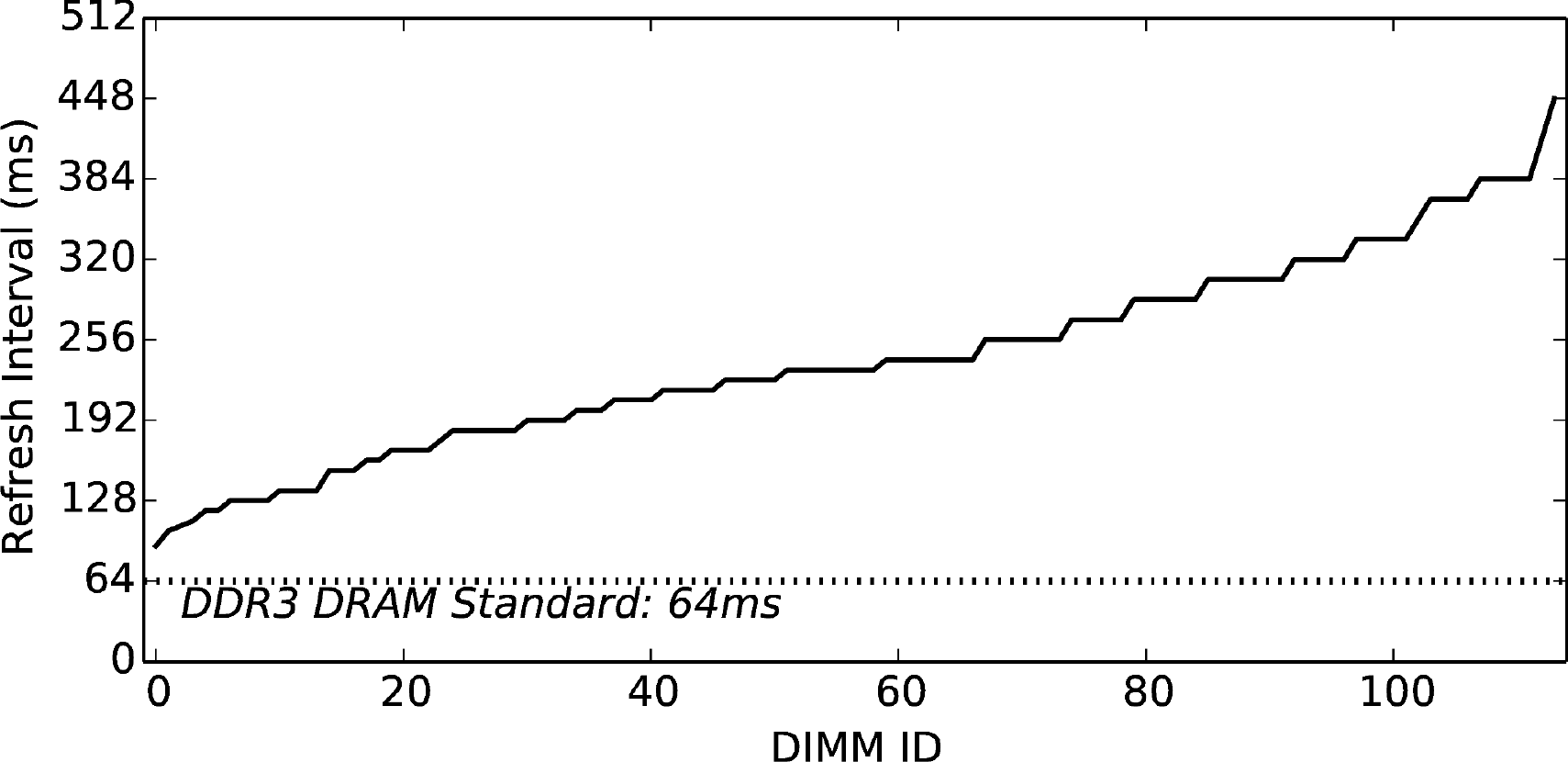}
	}
	\subcaptionbox{Write Retention Time (DIMM)\label{fig:write_safety}}[0.48\linewidth] {
		\includegraphics[width=2.8in]{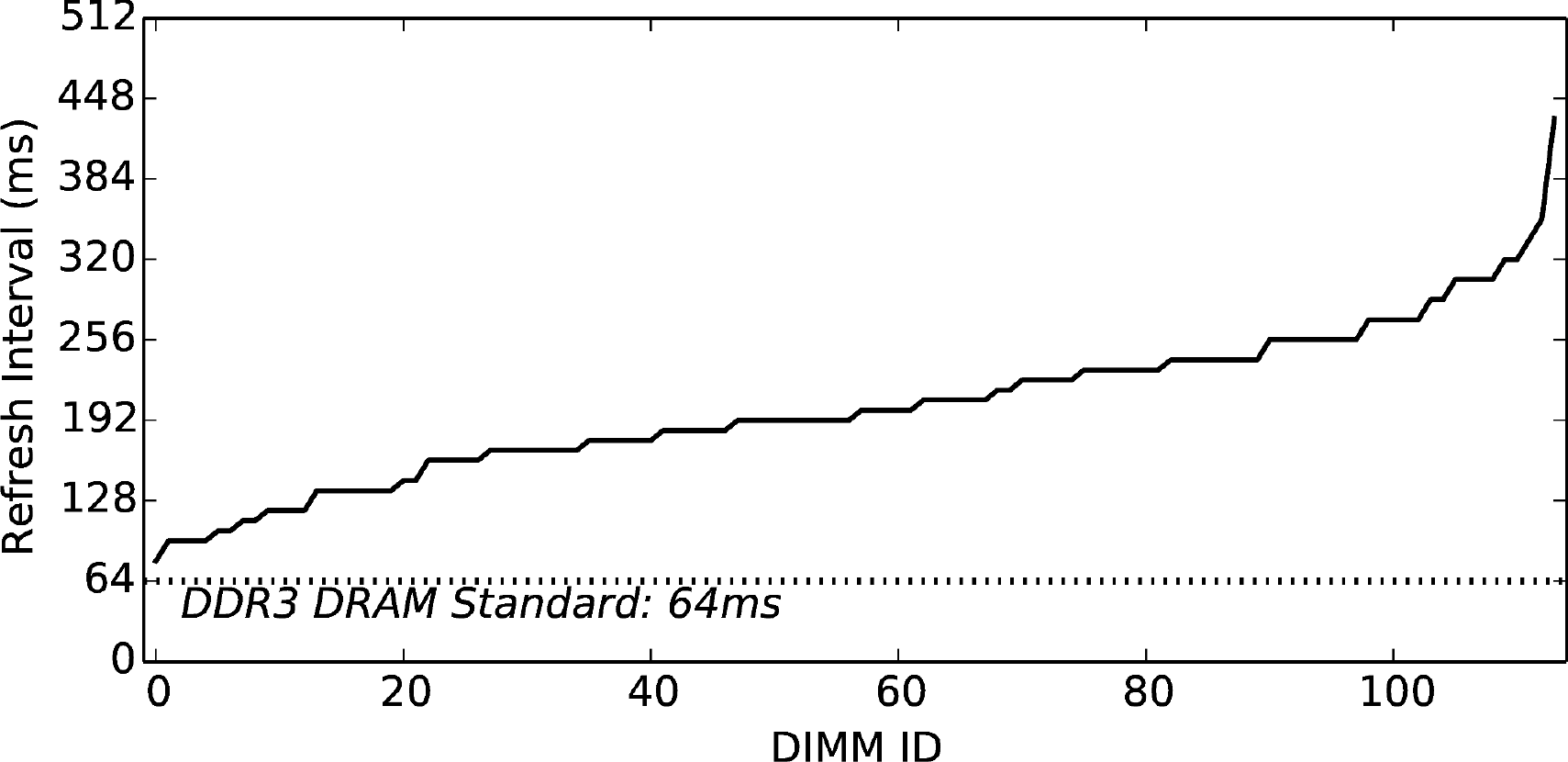}
	}

	\subcaptionbox{Read Retention Time (Bank)\label{fig:read_safety_bank}}[0.48\linewidth] {
		\includegraphics[width=2.8in]{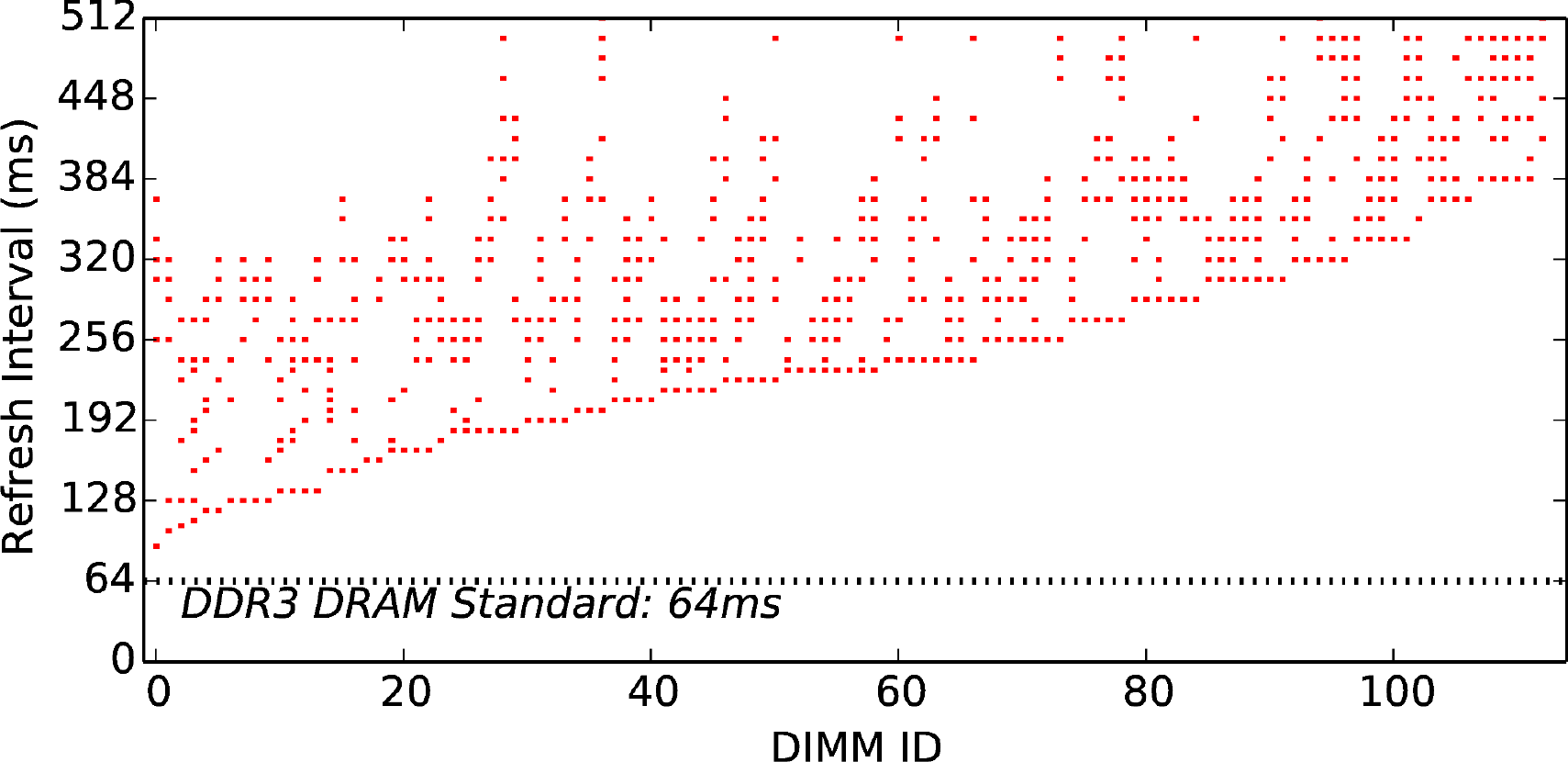}
	}
	\subcaptionbox{Write Retention Time (Bank)\label{fig:write_safety_bank}}[0.48\linewidth] {
		\includegraphics[width=2.8in]{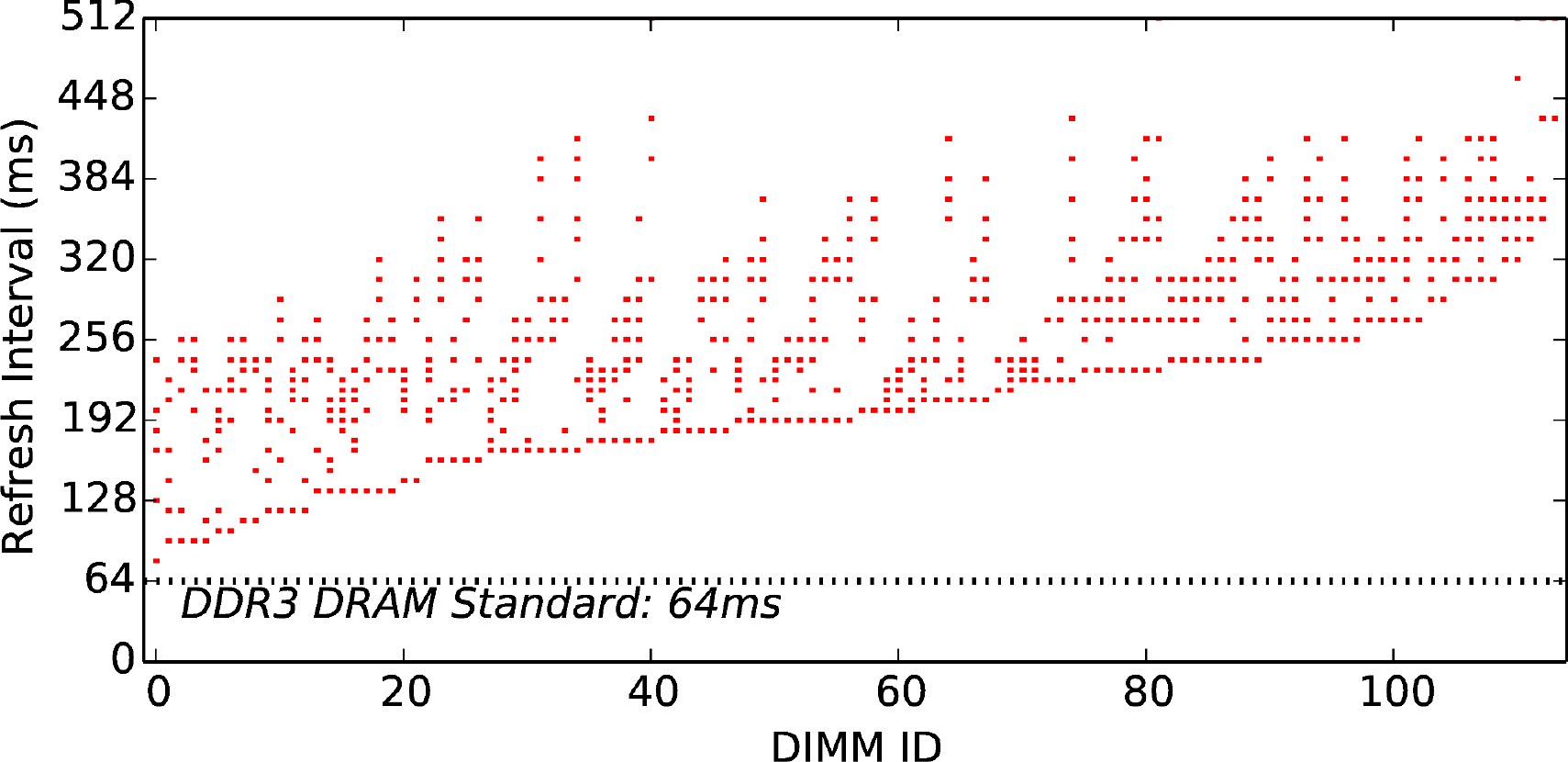}
	}

	\subcaptionbox{Read Retention Time (Chip)\label{fig:read_safety_chip}}[0.48\linewidth] {
		\includegraphics[width=2.8in]{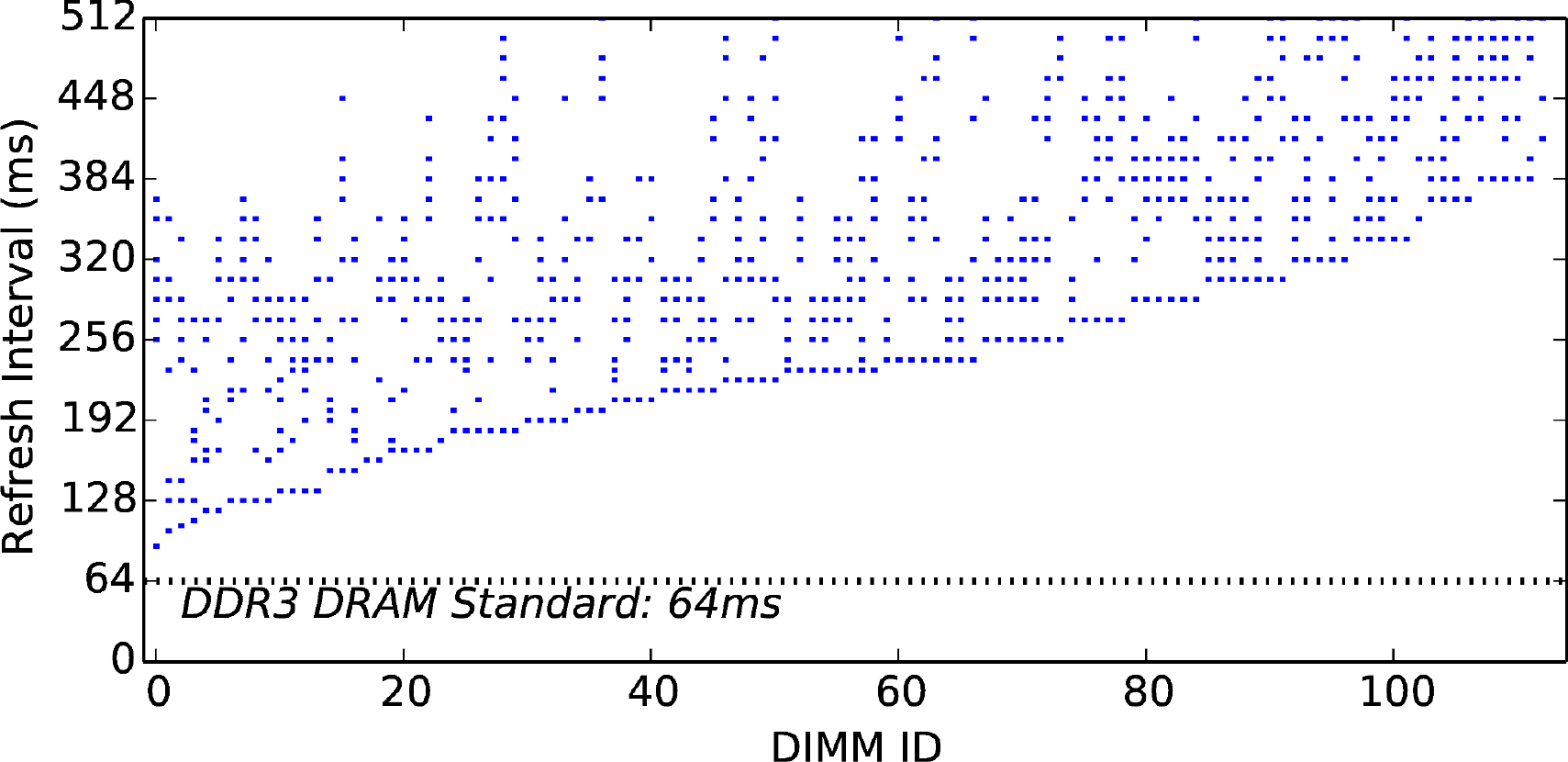}
	}
	\subcaptionbox{Write Retention Time (Chip)\label{fig:write_safety_chip}}[0.48\linewidth] {
		\includegraphics[width=2.8in]{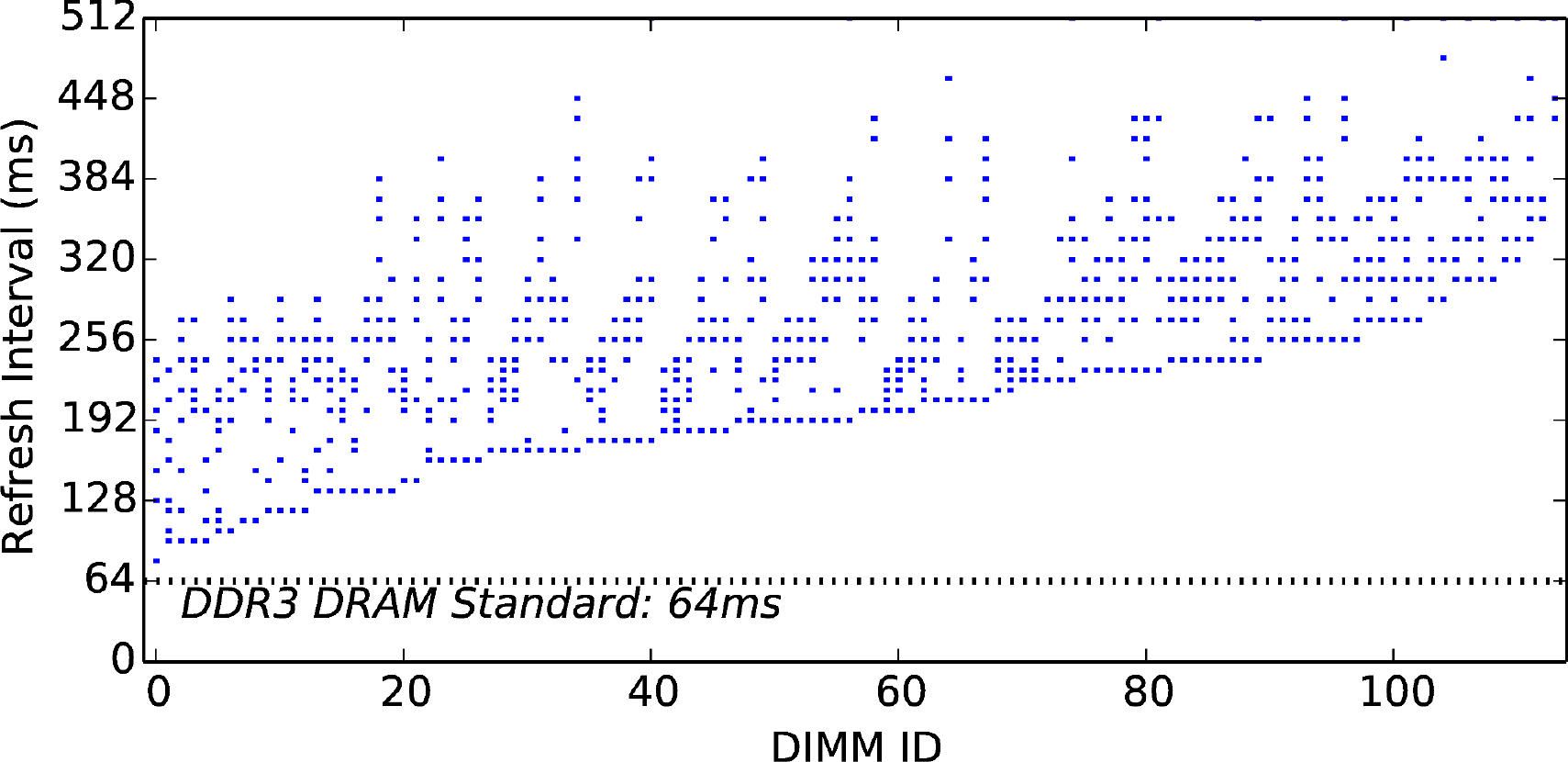}
	}

	\caption{Retention Time of Multiple DIMMs} \label{fig:multiple_dimms_retention}
\end{figure}

Figure~\ref{fig:read_safety} plots the highest refresh interval that leads to
correct operation across all cells at 85\celsius\xspace in {\em each DRAM
module} for the read test (Figure~\ref{fig:write_safety} for the
write test). Our key observation is that although there exist a few modules
which {\em just} meet the timing parameters (with a low safety-margin), a vast
majority of the modules very comfortably meet the standard timing parameters
(with a high safety-margin). This indicates that a majority of the DRAM modules
have significantly higher safety-margins than the worst-case module {\em even
at the highest-acceptable operating temperature of 85\celsius}.

We do the same analysis of the cell retention time in banks and chips. The red
dots in Figure~\ref{fig:read_safety_bank} show the highest refresh interval
that leads to correct operation across all cells within {\em each bank} for all
eight banks. The blue dots in Figure~\ref{fig:read_safety_chip} show the
highest refresh interval that leads to correct operation across all cells
within {\em each chip} for all DRAM chips in a DRAM module (usually eight chips
per each DRAM module). Figures~\ref{fig:write_safety_bank}
and~\ref{fig:write_safety_chip} plot the same data for the write test.

We make two key observations. First, the effect of process variation is even
higher for banks and chips within the same DRAM module, explained by the large
spread of the red and blue dots in each DRAM module. We provide the average
value of the retention time difference (highest retention time $-$ lowest
retention time of banks/chips in a DRAM module) across 115 DRAM modules. For
read test, the average value is 134.8ms/160.7ms for banks/chips in a DRAM
module, respectively. For write test, the average value is 111.2ms/138.64ms for
banks/chips in a DRAM module, respectively. Considering that the average values
are close to twice of the standard refresh interval (64ms), the effect of
process variation across banks/chips are significant. This indicates that a
majority of the DRAM banks/chips have significantly higher safety-margins than
the worst-case bank/chip.

Since banks within a DRAM module can be accessed independently with different
timing parameters, one can potentially imagine a mechanism that more
aggressively reduces timing parameters at a bank granularity and not just the
DRAM module granularity. In some memory architectures~\cite{zheng-micro2008}
that provide an individual control channel to each chip in a DRAM module, one
can imagine a mechanism that each DRAM chip in a DRAM module has different
timing parameters, which reduces DRAM latency more aggressively. We leave these
for future work.

To study the potential of reducing timing parameters for each DRAM module, we
sweep all possible combinations of timing parameters (\trcd/\tras/\twr/\trp)
for all the DRAM modules at both the highest acceptable operating temperature
(85\celsius) and a more typical operating temperature (55\celsius). We then
determine the acceptable DRAM timing parameters for each DRAM module for both
temperatures while maintaining its safety-margin of each DRAM module.

\begin{figure}[h]
	\centering
	\subcaptionbox{Read Latency\label{fig:read_latency}}[0.4\linewidth] {
		\includegraphics[width=2.0in]{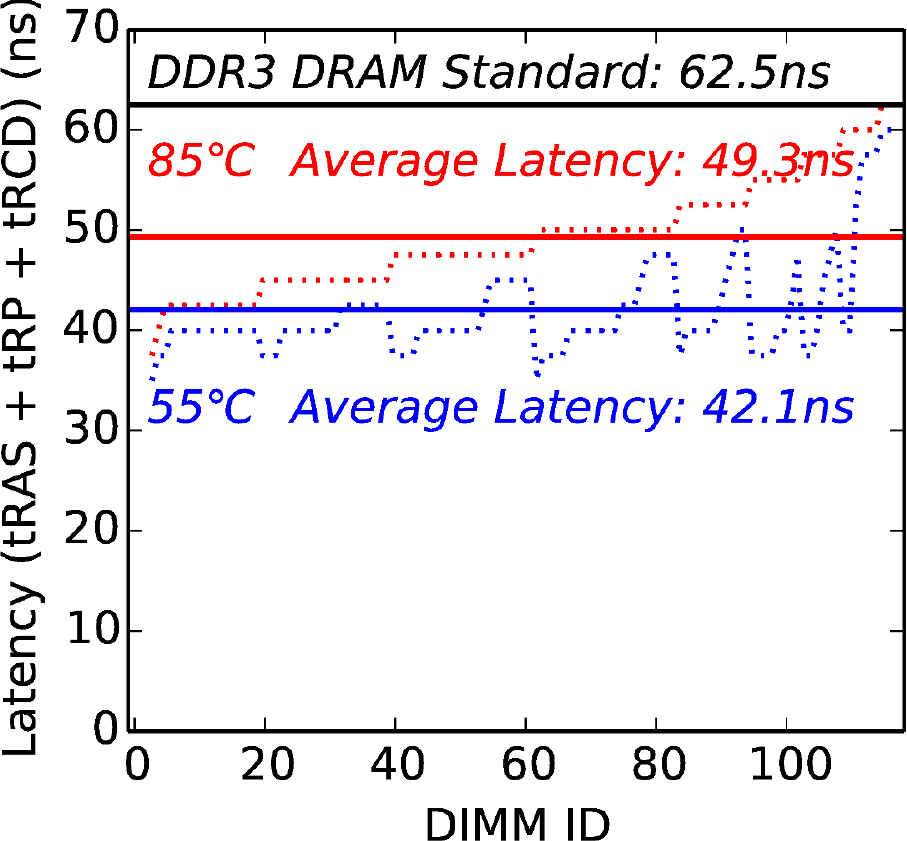}
	}
	\subcaptionbox{Write Latency\label{fig:write_latency}}[0.4\linewidth] {
		\includegraphics[width=2.0in]{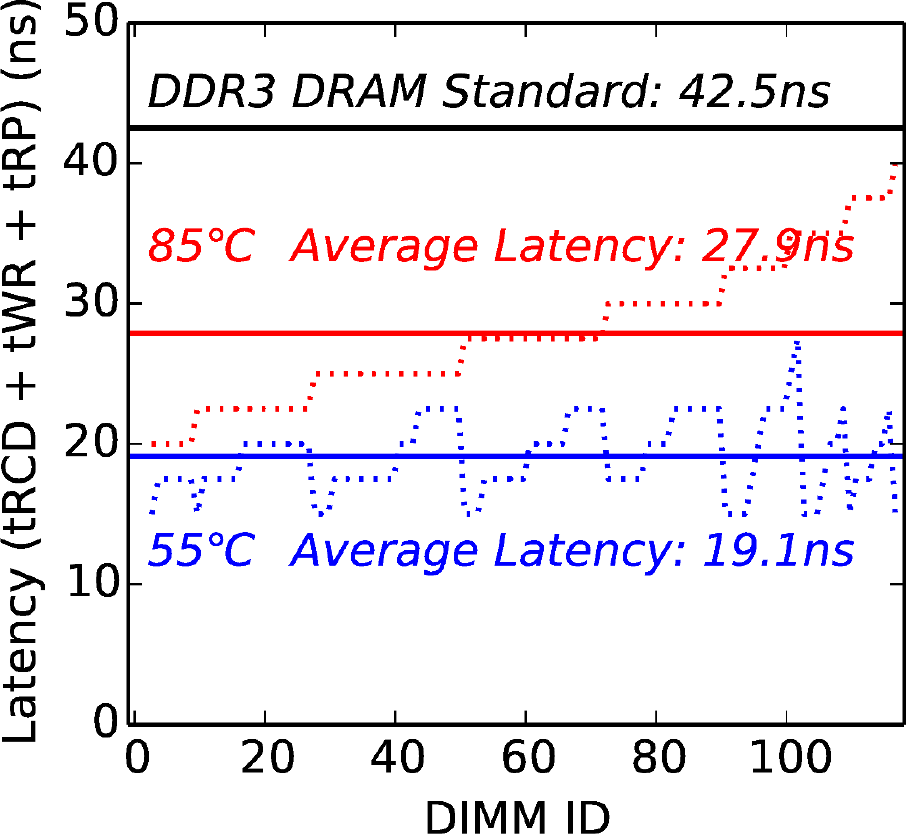}
	}
	\caption{Latency Analysis of Multiple DRAM modules} \label{fig:multiple_dimms}
\end{figure}

Figures~\ref{fig:read_latency} and~\ref{fig:write_latency} show the results of
this experiment for the read test and write test respectively. The y-axis plots
the sum of the relevant timing parameters (\trcd, \tras, and \trp for the read
test and \trcd, \twr, and \trp for the write test). The solid black line shows
the latency sum of the standard timing parameters (DDR3 DRAM specification).
The dotted red line and the dotted blue line show the most acceptable latency
parameters for each DRAM module at 85\celsius\xspace and 55\celsius,
respectively. The solid red line and blue line show the average acceptable
latency across all DRAM modules.

We make two observations. First, even at the highest temperature of 85\celsius,
DRAM modules have a high potential of reducing their access latency: \ReadHot\%
on average for read, and \WriteHot\% on average for write operations. This is a
direct result of the possible reductions in timing parameters
\trcd/\tras/\twr/\trp --- \trcdHot\%/\trasHot\%/\twrHot\%/\trpHot\% on average
across all the DRAM modules. In this dissertation, we present only the {\em
average} potential reduction for each timing parameter. We provide detailed
characterization of each DRAM module online at the SAFARI Research Group
website~\cite{safari-aldram}.

As a result, we conclude that process variation and lower temperatures enable a
significant potential to reduce DRAM access latencies. Second, we observe that
at lower temperatures (e.g., 55\celsius) the potential for latency reduction is
even greater (\ReadCold\% on average for read, and \WriteCold\% on average for
write operations), where the corresponding reduction in timing parameters
\trcd/\tras/\twr/\trp are \trcdCold\%/\trasCold\%/\twrCold\%/\trpCold\% on
average across all the DRAM modules.

Given that DRAM latency is a significant bottleneck for system performance,
this reduction in timing parameters will directly translate to improvement in
overall system performance (as we will show in Section~\ref{sec:performance}).

\subsection{Analysis of the Repeatability of Cell Failures}

In order to adopt a mechanism that exploits the DRAM latency variation in real
systems, it is important to understand whether the latency-margin related
errors remain consistent over time. For this purpose, we specifically measured
how many cells consistently experience errors under different evaluation
conditions. We perform this consistency test for five different scenarios
listed in Table~\ref{tbl:profile_correlation}. For each scenario, we choose a
refresh interval for which there are at least 1000 errors. We then repeat each
test for 10 iterations and report the percentage of cells that appear in all 10
iterations.

As summarized in Table~\ref{tbl:profile_correlation}, the first four scenarios
show that a very high fraction (more than 95\%) of the erroneous cells
consistently experience an error over multiple iterations of the same test.
Even though the overlap ratio is high, it is not 100\%, which means that the
characteristic of some cells may be changing over time. We believe that these
effects could be related to the VRT phenomenon (Section~\ref{sec:temp_margin}).
For the fifth scenario (where parameters are set separately for reads and
writes), the repeatability of errors is the lowest at 71\%. We hypothesize that
this is the result of different power-noise conditions (between activation and
write) for these two different operations. This is why the read and write
operations need to be profiled separately, since they are likely to sensitize
errors in different sets of cells.

\begin{table}[h]
\centering
\small{
	\begin{tabular}{lc}
		\toprule
	Scenario & Overlap (\%) \\
 		\midrule
	\multirow{1}{*}{10 iterations of the same test}																	& 96.94 \\
	\multirow{1}{*}{10 iterations of eight different data patterns}									& 96.01 \\
	\multirow{1}{*}{10 iterations of eight timing-parameter combinations}						& 98.99 \\
	\multirow{1}{*}{10 iterations at two different temperatures~(85~vs.~65\celsius)}& 96.18 \\
	\multirow{1}{*}{10 iterations of two different test types (read~vs.~write)} 		& 71.59 \\
 		\bottomrule
\end{tabular}
}\\

\caption{Repeatability and Consistency of Erroneous Cells} \label{tbl:profile_correlation}
\vspace{0.1in}
\end{table}

\section{Real-System Evaluation} \label{sec:evaluation}

We evaluate \ALD on a real system, whose detailed configuration is listed in
Table~\ref{tbl:result_system}. We chose this system for its AMD processor,
which offers dynamic software-based control over DRAM timing parameters at
runtime~\cite{amd-4386,amd-bkdg}. We paired the processor with one or more DRAM
modules that have ECC (error-correction code) support. For the purpose of
minimizing performance variation, we disabled several optimization features of
the system (e.g., dynamic core frequency scaling, adaptive DRAM row policy, and
prefetching).

\begin{table}[h]
\centering
\small{
	\begin{tabular}{lll}
	\toprule
	\multirow{1}{*}{System}& \multicolumn{2}{l}{Dell PowerEdge R415~\cite{dell-r415}}\\
 	\midrule
	\multirow{1}{*}{Processor}& \multicolumn{2}{l}{AMD Opteron 4386 (8 cores, 3.1GHz, 8MB LLC)~\cite{amd-4386,amd-bkdg}}\\
 	\midrule
							& \multicolumn{2}{l}{6 $\times$ 4GB ECC UDIMM (Single-/Dual-Rank)}\\
	 Main				& \multicolumn{2}{l}{DDR3-1600 (800MHz clock rate, 1.25ns cycle time)}\\
	 Memory 		& \multicolumn{2}{l}{ Default (\trcd/\tras/\twr/\trp):
	 						13.75/35.0/15.0/13.75ns}\\
							& \multicolumn{2}{l}{ Reduced (\trcd/\tras/\twr/\trp):
							{\bf 10.0/23.75/10.0/11.25ns}}\\
 	\midrule
		Storage		& \multicolumn{2}{l}{128GB SSD (random read/write speed: 97K/90K IOPS)}\\
 	\midrule
		Operating System				& \multicolumn{2}{l}{Linux 3.11.0-19-generic}\\
 	\bottomrule
\end{tabular}
}\\
\caption{Evaluated System Configuration} \label{tbl:result_system}
\end{table}

\subsection{Tuning the Timing Parameters} \label{sec:parameter}

First, we evaluate the possible latency reduction in DRAM modules without
losing any data integrity. We stress the system with memory intensive workloads
(99.1\% CPU utilization with STREAM~\cite{stream, moscibroda-usenix2007}
running in all eight cores) while reducing the timing parameters. The minimum
values of the timing parameters that do not introduce any errors at
55\celsius~or less define the maximum acceptable reduction in latency in the
system. Table~\ref{tbl:result_system} shows that the potential reduction is
27\%/32\%/33\%/18\% for \trcd/\tras/\twr/\trp, respectively. During the
evaluation, the observed DRAM temperature range is
30\celsius\xspace--39\celsius\xspace (always less than 55\celsius, even at a
very high CPU and memory utilization). Therefore, we need only one set of DRAM
timing parameters for our real system evaluation.

\subsection{Performance Improvement} \label{sec:performance}

\begin{figure}[h]
	\centering
	\includegraphics[width=1\linewidth]{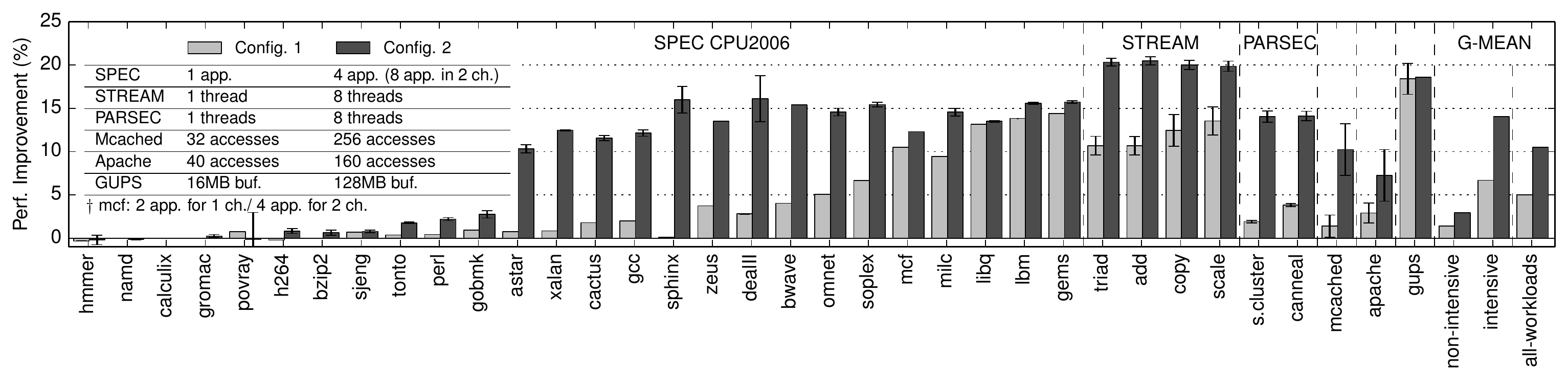}
	\captionof{figure}{Real-System Performance Improvement with AL-DRAM
	(Each Error Bar Shows the Standard Deviation across Multiple Runs)}
	\label{fig:result_1r1c}
\end{figure}

Figure~\ref{fig:result_1r1c} shows the performance improvement of reducing the
timing parameters in the memory system with one rank and one memory channel. We
run a variety of different applications (SPEC, STREAM~\cite{stream,
moscibroda-usenix2007}, PARSEC~\cite{parsec}, Memcached~\cite{memcached},
Apache~\cite{apache}, and GUPS~\cite{gups}) in two different configurations.
The first one (Config.~1) runs only one thread and, the second one (Config.~2)
runs multiple applications/threads (as described in the figure). We run each
configuration 30 times (only SPEC is executed 3 times due to the large
execution time), and present the average performance improvement across all the
runs and their standard deviation as an error bar. Based on the last-level
cache misses per kilo instructions (MPKI), we categorize our applications into
memory-intensive or non-intensive groups, and report the geometric mean
performance improvement across all applications from each group. We draw three
key conclusions from Figure~\ref{fig:result_1r1c}. First, \ALD provides
significant performance improvement over the baseline (as high as 20.5\% for
the very memory-bandwidth-intensive STREAM applications~\cite{stream,
moscibroda-usenix2007}). Second, when the memory system is under higher
pressure with multi-core/multi-threaded applications, we observe significantly
higher performance (than in the single-core case) across all applications from
our workload pool. Third, as expected, memory-intensive applications benefit
more in performance than non-memory-intensive workloads (14.0\% vs.~2.9\% on
average). We conclude that by reducing the DRAM timing parameters using our
approach, we can speed up a system by 10.5\% (on average across all 35
workloads on the multi-core/multi-thread configuration).

\subsection{Reliability of Reduced Timing Parameters}

By reducing the timing parameters, we are also stripping away the excessive
margin in DRAM's electrical charge. The remaining margin should be enough for
DRAM to achieve correctness by overcoming process variation and temperature
dependence (as we discussed in Section~\ref{sec:factors_reliable}). To verify
the correctness of our experiments, we ran our workloads for 33 days non-stop, and
examined their and the system's correctness with reduced timing parameters.
Using the reduced timing parameters over the course of 33 days, our real system
was able to execute 35 different workloads in both single-core and multi-core
configurations (see Figure~\ref{fig:result_1r1c}) while preserving correctness
and being error-free.

Note that these results do {\em not} absolutely guarantee that no errors can be
introduced by reducing the timing parameters. Our real-system experiments are
limited in their statistical significance, since they involve a small sample
population (six ECC DRAM modules) over a relatively short test duration (33
days). However, DRAM manufacturers {\em already have the necessary testing
methodology to guarantee reliable operation with reduced timing parameters that
are appropriately chosen.} Existing industrial-grade methodology for measuring
and ensuring reliability (at different timing parameters) is typically based on
millions of hours of aggregate test time, which is clearly beyond the scope of
this work but is also clearly doable by DRAM manufacturers. Thus, we believe
that we have demonstrated a proof-of-concept which shows that DRAM latency can
be reduced at little-to-no apparent impact on DRAM reliability.

\subsection{Sensitivity Analysis}

\begin{figure}[h]
	\centering
	\includegraphics[width=0.7\linewidth]{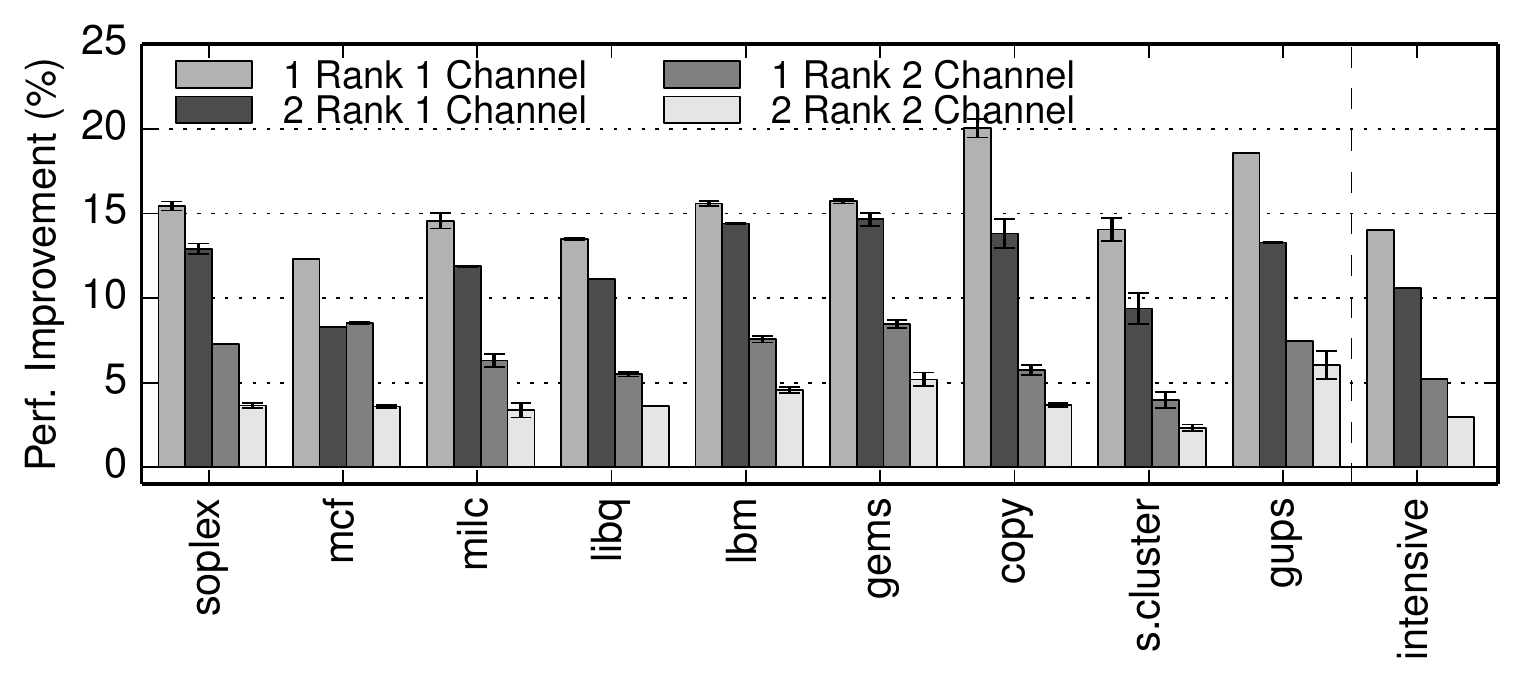}
	\captionof{figure}{AL-DRAM Performance Improvement on a Real System
	with Different Rank and Channel Configurations}
	\label{fig:result_config}
\end{figure}

{\bf Number of Channels, and Ranks:} We analyze the impact of increasing the
number of ranks and channels on the performance achieved with \ALD in
Figure~\ref{fig:result_config}. Note that adding ranks enables more memory
parallelism while keeping the total memory bandwidth constant, but adding
channels also increases the total memory bandwidth. The 2-channel systems we
evaluate are overprovisioned for the workloads we evaluate as our workloads
exert little pressure on memory in such systems. We make two major
observations. First, \ALD significantly improves system performance even on
highly-provisioned systems with large numbers of channels, and ranks:
10.6\%/5.2\%/2.9\% on average across our memory-intensive/multi-core workloads
in a 2-rank 1-channel/1-rank 2-channel/2-rank 2-channel system, respectively.
Second, the benefits \ALD are higher when the system is more memory
parallelism- and bandwidth-constrained, as expected, because memory latency
becomes a bigger bottleneck when limited parallelism/bandwidth causes
contention for memory. As on-chip computational power continues to increase at
a much faster rate than the amount of off-chip memory
bandwidth~\cite{ipek-isca2008, solihin-isca2009} due to the limited pin count, and
as future applications become increasingly data-intensive, it is very likely
that future systems will be increasingly memory bandwidth
constrained~\cite{mutlu-memcon2013}. We conclude that \ALD will likely become more
effective on such future systems.

{\bf Heterogeneous Workloads:} We evaluate the performance impact of \ALD for
heterogeneous workloads generated by combining two different applications from
those listed in Figure~\ref{fig:result_config}. We observe that \ALD provides
significant weighted speedup improvement over the baseline system: 7.1\% on
average in a 1-rank 2-channel system.

{\bf Row Policy:} Performance improvement can depend on the row buffer
management policy used in the memory system. We simulate our mechanism,
AL-DRAM, to analyze the sensitivity of our results to two policies (as our
tested system does not have any flexibility to change row policy): open-row \&
closed-row~\cite{kim-hpca2010}. We use a modified version of
Ramulator~\cite{kim-cal2015}, a fast cycle-accurate DRAM simulator that is
available publicly~\cite{ramulator, safari-tools}, and Ramulator releases an
open-source implementation of AL-DRAM. We use Ramulator combined with a
cycle-level x86 multi-core simulator. We use Ramulator as part of a cycle-level
in-house x86 multi-core simulator, whose front-end is based on
Pin~\cite{luk-pldi2005}.

We evaluate 51 multi-core workloads (randomly-selected from SPEC CPU2006,
STREAM, TPC, and GUPS) in a 1-rank 2-channel system. Our evaluation shows that
\ALD provides similar performance improvements for both policies (11.4\%/11.0\%
improvement for open/closed row policies over the baseline system).

{\bf Energy Efficiency:} By reducing the timing parameters and overall
execution time, \ALD improves energy efficiency. \mycolor{Unfortunately, we do
not have an infrastructure to measure the DRAM energy consumption from the
tested real system.} Instead, we estimate the DRAM energy consumption of \ALD
using a DRAM energy calculator~\cite{micron-power}. When using 4GByte modules
with DDR3-1600, \ALD reduces DRAM power consumption by 5.8\% (in a current
specification, $I_{\mathit{DD1}}$~\cite{samsung-spec}). The major energy
reduction is attributed to the reduction in row activation time. \mycolor{We
leave the DRAM energy consumption measurement in real systems to future work.}

	\section{Summary}

The standard DRAM timing constraints are grossly overprovisioned to ensure
correct operation for the cell with the lowest retention time at the highest
acceptable operating temperature. We make the observation that {\em i)} a
significant majority of DRAM modules do {\em not} exhibit the worst case
behavior and that {\em ii)} most systems operate at a temperature much lower
than the highest acceptable operating temperature, enabling the opportunity to
significantly reduce the timing constraints.

Based on these observation, in this chapter, we introduce Adaptive-Latency DRAM
(\ALD), a simple and effective mechanism for dynamically optimizing the DRAM
timing parameters for the current operating condition without introducing any
errors. \ALD dynamically measures the operating temperature of each DRAM module
and employs timing constraints optimized for {\em that DRAM module at that
temperature}. Results of our latency profiling experiments on \DIMMs modern
DRAM modules show that our approach can significantly reduce four major DRAM
timing constraints by \trcdCold\%/\trasCold\%/\twrCold\%/\trpCold\xspace
averaged across all \DIMMs DRAM modules tested. This reduction in latency
translates to an average 14\% improvement in overall system performance across
a wide variety of memory-intensive applications run on a real multi-core
system.

We conclude that \ALD is a simple and effective mechanism, which exploits the
large margin present in the standard DRAM timing constraints, to reduce DRAM
latency.

	\chapter{{AVA-DRAM:\\
Lowering DRAM Latency by Exploiting Architecture Variation}}
\label{ch:avadram}

\let\thefootnote\relax\footnotetext{We provide detailed characterization of
each DRAM module online at the SAFARI Research Group
website~\cite{safari-avadram}.}
\let\thefootnote\svthefootnote

In this chapter, we introduce new techniques to improve DRAM latency by taking
advantage of the variation in cell latencies. We observe that there is
variability in DRAM cells based on their location, which has not been exposed
or leveraged by any previous works. Some DRAM cells can be accessed faster than
others depending on their physical location. We refer to this variability in
cells' access times, caused by the physical organization of DRAM, as {\em
architectural variation}.

\begin{figure}[h]
	\centering
	\includegraphics[height=1.60in]{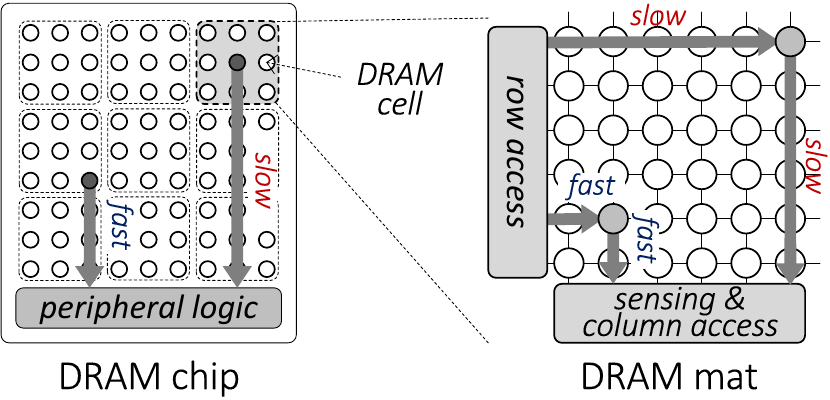}
	\caption{Architectural Variation in a DRAM Chip}
	\label{fig:intro}
\end{figure}

Architectural variation arises from the difference in the distance between the
cells and the peripheral logic that is used to access these cells
(Figure~\ref{fig:intro}). The wires connecting the cells to peripheral logic
exhibit large resistance and large capacitance~\cite{lee-hpca2013,
lee-hpca2015}. Consequently, cells experience different RC delays based on
their distances from the peripheral logic. Cells logic closer to peripheral
experience smaller delay and can be accessed faster than the cells located
farther from the peripheral logic.

Architectural variation in latency is present in both vertical and horizontal
directions in a 2D DRAM cell array (called a mat): {\em i)} each vertical {\em
column of cells} is connected to a component called a {\em sense amplifier} and
{\em ii)} each horizontal {\em row of cells} of a mat is connected to a {\em
wordline driver}. Variations in the vertical and horizontal dimensions,
together, divide the cell array into heterogeneous latency regions, where cells
in some regions require larger access latencies for reliable operation. This
variation in latency has direct impact on the reliability of the cells.
Reducing the latency {\em uniformly across all regions} in DRAM would improve
performance, but can introduce failures in the {\em inherently slower} regions
that have to be accessed longer for correct DRAM operation. We refer to these
inherently slower regions of DRAM as {\em architecturally vulnerable regions}.

{\em The goal} of this work is twofold. First, to experimentally demonstrate
the existence of architectural variation in modern DRAM chips and identify the
architecturally vulnerable regions. Second, to develop new mechanisms that
leverage this variation to reduce DRAM latency while providing reliability at
low cost.

\noindent {\bf Identifying architecturally vulnerable regions.} Towards
achieving our goals, we first identify the architecturally vulnerable regions
of DRAM. Doing so is not a trivial task due to two major challenges. First,
{\em identifying architecturally vulnerable regions requires a detailed
knowledge of DRAM internals.} Modern DRAM cells are organized in a hierarchical
manner, where cells are subdivided into multiple mats and these mats are
organized as a matrix (Figure~\ref{fig:intro}). Due to this hierarchical
organization, we show that the vulnerability of cells {\em does} not
necessarily increase linearly with increasing row and column addresses, but
depends on {\em i)} the location of the cell in that mat and {\em ii)} the
location of the mat in the chip.

Second, {\em identifying architecturally vulnerable regions is difficult due to
the current DRAM interface that does not expose how data corresponding to an
address is mapped inside of DRAM}. Even though certain regions in DRAM might be
architecturally vulnerable, internal scrambling and remapping of rows and
columns scatters and distributes that region all over the address
space~\cite{khan-dsn2016}. In this work, we provide a detailed analysis on how
to identify the vulnerable regions despite the limitations posed by DRAM
interface.

To understand the architectural variation of latency in modern DRAM chips, we
use an FPGA-based DRAM testing infrastructure to characterize \dimms~DRAM
modules. Our experimental study shows that {\em i)} Modern DRAM chips exhibit
architectural latency variation in both row and column directions, {\em ii)}
Architectural vulnerability gradually increases in the row direction within a
mat and repeats the variability pattern in every mat. {\em iii)} Some columns
are more vulnerable than others based on the internal hierarchical design of
the specific DRAM chip.

\noindent {\bf Reducing DRAM latency by exploiting the knowledge of
architecturally vulnerable regions.} We develop two new mechanisms that exploit
the architecturally vulnerable regions to enable low DRAM latency with high
reliability and at low cost. First, we propose to reduce the DRAM latency at
runtime, by identifying the lowest possible latency that still ensures reliable
operation. To this end, we develop an online DRAM testing mechanism, called
{\em AVA Profiling}. The key idea is to periodically test {\em only} the
architecturally vulnerable regions to find the minimum possible DRAM latency
(for reliable operation), as these regions would exhibit failures earlier than
others when reducing the latency and therefore, would indicate the latency
boundary where further reduction in latency would hurt reliability.

Second, to reduce the DRAM latency even further beyond the point of reliable
operation, we propose a mechanism to reduce multi-bit failures while operating
at a lower latency, called {\em AVA Shuffling}. The key idea is to leverage the
error characteristics in architecturally vulnerable regions to remap or shuffle
data such that the failing bits get spread over multiple ECC code words and
become correctable by ECC.

	\section{Architectural Variation} \label{sec:arch}

In this work, we show that DRAM access latency varies based on the location of
the cells in the DRAM hierarchy. Intuitively, transferring data from the cells
near the IO interfaces incurs less time than transferring data from the cells
farther away from the IO interfaces. We refer to this variability in the cell
latency caused by the physical organization of DRAM as {\em architectural
variation}.

\noindent {\bf Properties of Architectural Variation.} Architectural variation
has specific characteristics that clearly distinguish it from other known types
of variation observed in DRAM cells (e.g., process variation and temperature
dependency~\cite{lee-hpca2015, chandrasekar-date2014}). Architectural variation
in DRAM has the following four characteristics.

\squishlist

	\item {\bf Predetermined at design time.} Architectural variation depends on
	the internal DRAM design. As a result, it is predetermined at {\em design
	time}. This is unlike other types of variation, (e.g., process variation and
	temperature induced variation~\cite{lee-hpca2015, chandrasekar-date2014}),
	which depend on the manufacturing process after design.

	\item {\bf Static distribution.} The distribution of architectural variation
	is static. For example, a cell closer to the sense amplifier is {\em always}
	faster than a cell that is farther away from the sense amplifier, assuming
	there are no other sources of variation (e.g., process variation). On the
	other hand, prior works show that variability due to process variation
	follows a random distribution~\cite{lee-hpca2015, chandrasekar-date2014}.

	\item {\bf Constant.} Architectural variation depends on the physical
	organization, which remains constant over time. Therefore, it is different
	from other types of variation that change over time (e.g., variable retention
	time~\cite{liu-isca2013, khan-sigmetrics2014, kim-edl2009}, wear-out due to
	aging~\cite{sridharan-sc2012, sridharan-sc2013, meza-dsn2015,
	schroeder-tdsc2010, li-atc2007, hwang-asplos2012, wee-jssc2000, min-vlsi2001,
	tanabe-jssc1992}).

	\item {\bf Similarity in DRAMs with the same design.} DRAMs that share the
	same internal design and organization exhibit similar architectural variation,
	unlike process variation that manifests significantly differently in different
	DRAM chips with the same design.\footnote{\mycolor{To increase yield, modern
	DRAM integrates redundant rows and columns in its banks (usually 1 to 3\% of
	total cells). When uncovering faulty cells during manufacturing time, those
	faulty rows and columns are remapped to the redundant rows and columns
	(row/column repair). Since remapped rows and columns are different in
	different DRAM chips, we expect to observe small variations in architectural
	variation of different chips which share the same design.}}

\squishend

\noindent {\bf Architectural Variation in the DRAM Hierarchy.} As we mentioned
earlier, architectural variation arises from the difference in distance between
the DRAM cells and the peripheral logic that is used to access the cells. As
DRAM is internally organized as a multi-level hierarchy (in the form of chips,
banks and ranks), architectural variation exists at multiple levels.

In this work, we focus on the architectural variation within and across mats,
as they are the smallest units in DRAM and it is inherently difficult for the
manufacturers to minimize this variation by dividing the mats into even smaller
units. Furthermore, such architectural variation in mats is becoming only worse
with technology scaling. That is mainly because integrating more cells within a
DRAM chip requires increasing the length of the wordline and bitline to
amortize the area of wordline drivers and sense amplifiers across more cells.
In a majority of today's DRAMs, a mat consists of 512$\times$512 cells, while
more recently, DRAM with mats consisting of 1024$\times$1024 cells has been
introduced~\cite{lim-isscc2012}.

{\bf The goal} of this work is to {\em i)} experimentally demonstrate,
characterize, and understand the architectural variation in modern DRAM chips
and {\em ii)} leverage this variation and our understanding of it to reduce DRAM
latency at low cost in a reliable way. Unfortunately, detecting the
architecturally vulnerable regions is not trivial and depends on two factors
{\em i)} how bitline and wordline drivers are organized internally and {\em ii)}
how data from a cell is accessed through the DRAM interface. In order to define
and understand the architectural variation in modern DRAM chips, we investigate
three major research questions related to the impact of DRAM organization,
interface, and operating conditions on architectural variation and provide
answers to these questions in the following sections
(Section~\ref{sec:overview_existence}--\ref{sec:overview_sensitivity}).

\subsection{Impact of DRAM Organization} \label{sec:overview_existence}

The first question we answer is: {\em how does the DRAM organization affect the
architecturally vulnerable regions?} Based on a detailed understanding of DRAM
internal organization, we provide hypotheses on the characteristics of
architectural variation and systematic methodologies to identify these
characteristics in modern DRAM chips. Using these methodologies, we
experimentally validate our hypotheses with the results we provide in
Section~\ref{sec:profile}.

\begin{figure}[h]
	\centering
	\subcaptionbox{Conceptual Bitline\label{fig:arch_row_concept}}[0.4\linewidth] {
		\includegraphics[height=1.40in]{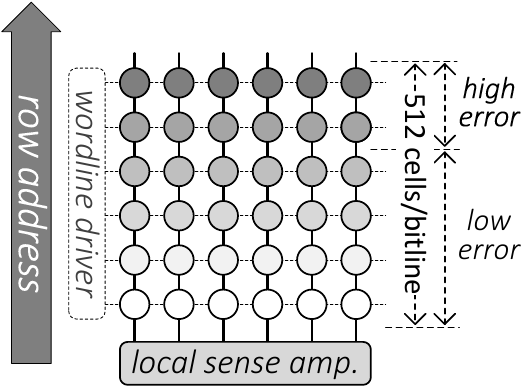}
	}
	\subcaptionbox{Open Bitline Scheme\label{fig:arch_row_real}}[0.4\linewidth] {
		\includegraphics[height=1.40in]{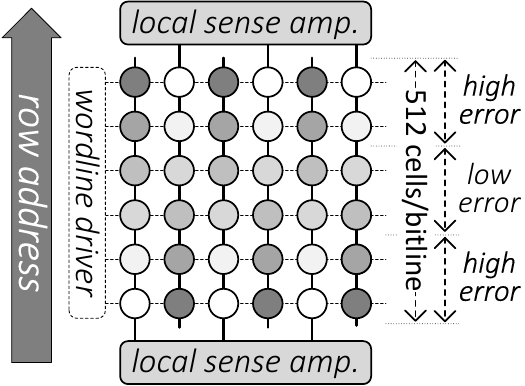}
	}
	\caption{Architectural Variation due to Row Organization}
	\label{fig:arch_row}
\end{figure}

{\bf Effect of Row Organization on Architectural Variation.} As discussed in
Chapter~\ref{ch:bak}, a mat consists of a 2D array of DRAM cells along with
peripheral logic needed to access this data. In the vertical direction, DRAM
cells, connected through a bitline, share a local sense amplifier (typically,
512 cells~\cite{vogelsang-micro2010}). As a result, variation in access latency
gradually increases as the distance of a row from the local sense amplifier
increases (due to the longer latency of propagation delay through the bitline).
This variation can be exposed by overclocking the DRAM by using smaller values
for DRAM timing parameters. Cells in the rows closer to the local sense
amplifiers can be accessed faster, so they exhibit no failures due to
overclocking. On the contrary, cells located farther away from the sense
amplifier need longer time to access, and might start failing when smaller
values are used for the timing parameters. As a result, accessing rows in
ascending order starting from the row closest to the sense amplifiers should
exhibit gradual increase in failures due to architectural variation, as shown
in Figure~\ref{fig:arch_row_concept}. In this figure, \mycolor{darker colors
indicate} slower cells, which are more vulnerable to failures, if we reduce
access latency.

In the open-bitline scheme~\cite{inoue-jssc1988}, alternate bitlines within a
mat are connected to two different rows of sense amplifiers (the top and bottom
of the mat), as shown in Figure~\ref{fig:arch_row_real}. In this scheme, even
cells and odd cells in a row located at the edge of the mat exhibit very
different distances from their corresponding sense amplifiers, leading to
different access latencies. On the other hand, cells in the middle of a mat
have similar distance from both the top and bottom sense amplifiers, exhibiting
similar latencies. Due to this organization, we observe that there are more
failures in rows located on both ends of a mat, but there is a gradual decrease
in failures for rows in the middle of the mat.

Based on these, we define two characteristics of vulnerable regions across the
rows when we reduce DRAM latency uniformly. First, {\bf the number of failures
would gradually increase with increased distance from the sense amplifiers}.
Second, {\bf this gradual increase in failures would periodically repeat in
every mat (every 512 rows)}. We experimentally demonstrate these
characteristics in Section~\ref{sec:profile_bitline}.

\begin{figure}[h]
	\centering
	\includegraphics[height=1.40in]{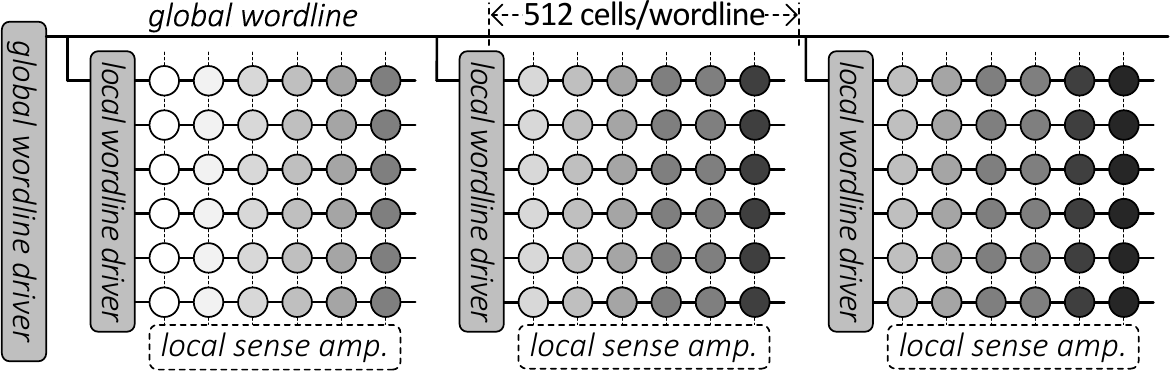}
	\caption{Architectural Variation in Column Organization}
	\label{fig:arch_col}
\end{figure}

{\bf Effect of Column Organization on Architectural Variation.} As we discussed
in Section~\ref{sec:background}, the wordline drivers in DRAM are organized in a
hierarchical manner, where a global wordline is connected to {\em all} mats
within a row and then a local wordline driver activates a {\em single} row
within a mat. This {\em hierarchical wordline organization} leads to latency
variation at two levels. First, a wordline in a mat located closer to the global
wordline driver starts activating a row earlier than a mat located farther away
from the global wordline driver ({\em architectural variation due to the global
wordline}). Second, within a mat, a cell closer to the local wordline driver
gets activated faster than a cell farther away from the local wordline driver
({\em architectural variation in local wordline}). Therefore, columns that have
the same distance from the local wordline driver, but located in two different
mats will have different latency characteristics (Figure~\ref{fig:arch_col},
where darker color indicates slower cells, which are more vulnerable to
failures, if we reduce access latency).

Based on these, we define two characteristics of vulnerable regions across
columns when we reduce DRAM latency uniformly. First, {\bf although some
columns are more vulnerable than others, the number of failures would {\em not}
gradually increase with ascending column numbers}. Second, {\bf the failure
characteristics observed with ascending column numbers would be similar for all
rows}. We experimentally demonstrate these characteristics in
Section~\ref{sec:profile_wordline}.

\subsection{Impact of the Row/Column Interface} \label{sec:overview_interface}

Our second question is: {\em how does the row/column interface affect the
ability to identify the architecturally vulnerable regions in DRAM?}
Unfortunately, identifying architecturally vulnerable regions becomes
challenging due to a limited understanding of how data corresponding to an
address is mapped inside DRAM. While it is possible to identify vulnerable
regions based on location, exposing and exploiting such information through the
row/column DRAM addressing interface is challenging due to two reasons.

{\bf Row Interface (Row Address Mapping).} DRAM manufacturers internally
scramble the row addresses in DRAM making the address known to the system
different from the actual physical address~\cite{vandegoor-delta2002}. As a
result, consecutive row addresses issued by the memory controller, can be mapped
to entirely different regions of DRAM. Unfortunately, the mapping of the row
addresses is not {\em exposed} to the system and varies across products from
different generations and manufacturers. In the previous section we showed that
if latency is reduced, accessing rows in mats in ascending row numbers would
exhibit gradual increase in failures. Unfortunately, due to row remapping,
accessing rows in ascending order of addresses known to memory controller will
exhibit irregular and scattered failure characteristics.

\begin{figure}[h]
	\centering
	\subcaptionbox{Data Mapping\label{fig:arch_col_data_mapping}}[0.35\linewidth] {
		\includegraphics[height=1.6in]{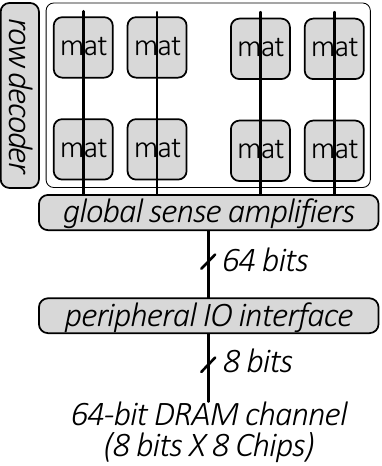}
	}
	\subcaptionbox{Data Burst (Data Out)\label{fig:arch_col_burst}}[0.55\linewidth] {
		\includegraphics[height=1.6in]{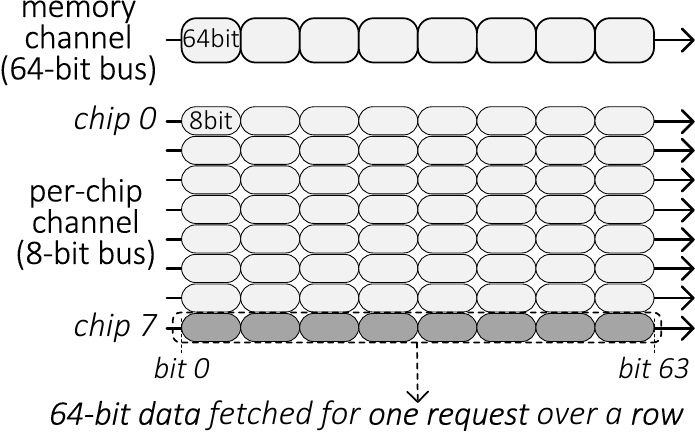}
	}
	\caption{Accessing Multiple Mats in a Data Burst}
	\label{fig:arch_col_interface}
\end{figure}

{\bf Column Interface (Column Address Mapping).} In the current column
interface, the column addresses issued by the memory controller do not access
consecutive columns in a mat, making it challenging to identify the vulnerable
regions in a wordline. When a column address is issued, 64-bit of data from a
row is transferred with global bitlines (typically, 64-bit
width~\cite{vogelsang-micro2010}). Then, this data is transferred in eight 8-bit
bursts over the IO channel as shown in Figure~\ref{fig:arch_col_interface}.
However, data transferred with each column address comes from different mats,
making it impossible to always access {\em consecutive} physical columns in a
mat by simply {\em increasing} the column address.

In this work, we provide alternate ways to identify architecturally vulnerable
DRAM regions using the current row/column interface in DRAM. We describe the
key ideas of our methods below. Section~\ref{sec:profile_rowint}
and~\ref{sec:profile_colint} provide the detailed analysis with experimental
validation of our methods in real DRAM chips.

\squishlist

	\item {\em Inferring vulnerable rows from per-row failure count.} In order to
	identify the gradual increase in architecture variability with increasing row
	addresses in mats (in terms of internal DRAM physical address), we try to
	reverse engineer the row mapping in DRAM. We hypothesize the mapping for one
	mat and then verify that mapping in other DRAM mats in different chips that
	share the same design. The key idea is to correlate the number of failures to
	the physical location of the row. For example, the most vulnerable	row would
	be the one with the most failures and hence should be located at the edge of
	the mat. Section~\ref{sec:profile_rowint} provides experimental validation of
	our method.

	\item {\em Inferring vulnerable columns from per-bit failure count in the IO
	channel.} As we explained earlier, a column access transfers 64 bits of data
	from a chip over the IO channel. These 64 bits come from 64 bitlines that are
	distributed over different mats across the entire row. Our key idea to
	identify the vulnerable bitlines in the column direction is to examine each
	bit in the 64-bit data burst. We expect that due to architectural variation,
	some bits in a 64-bit burst that are mapped to slower bitlines than others,
	are more vulnerable than other bits. In Section~\ref{sec:profile_colint}, we
	experimentally identify the location of bits in bursts that consistently
	exhibit more failures, validating the existence of architectural variation in
	columns. 

\squishend

\subsection{Impact of Operating Conditions} \label{sec:overview_sensitivity}

The third question we seek to answer is: {\em Does architectural variation in
latency show similar characteristics at different operating conditions?} DRAM
cells get affected by temperature and refresh interval. Increasing the
temperature or refresh interval increases the leakage in cells, making them more
vulnerable to failure. However, as all the cells in DRAM get similarly affected
by changes in operating condition, we observe that the trends due to
architectural variation remain similar at different temperatures and refresh
intervals, even though the absolute number of failures may change. We provide
the detailed experimental analysis of architectural variation at different
operating conditions in Section~\ref{sec:profile_sensitivity}.

	\section{DRAM Testing Methodology}\label{sec:pmethod}

In this section, we describe our FPGA-based DRAM testing infrastructure and the
testing methodology we use for our experimental studies in
Section~\ref{sec:profile}.

\begin{figure}[h]
	\centering
	\subcaptionbox{Full System\label{fig:infra_system_avadram}}[0.4\linewidth] {
		\includegraphics[width=2.5in]{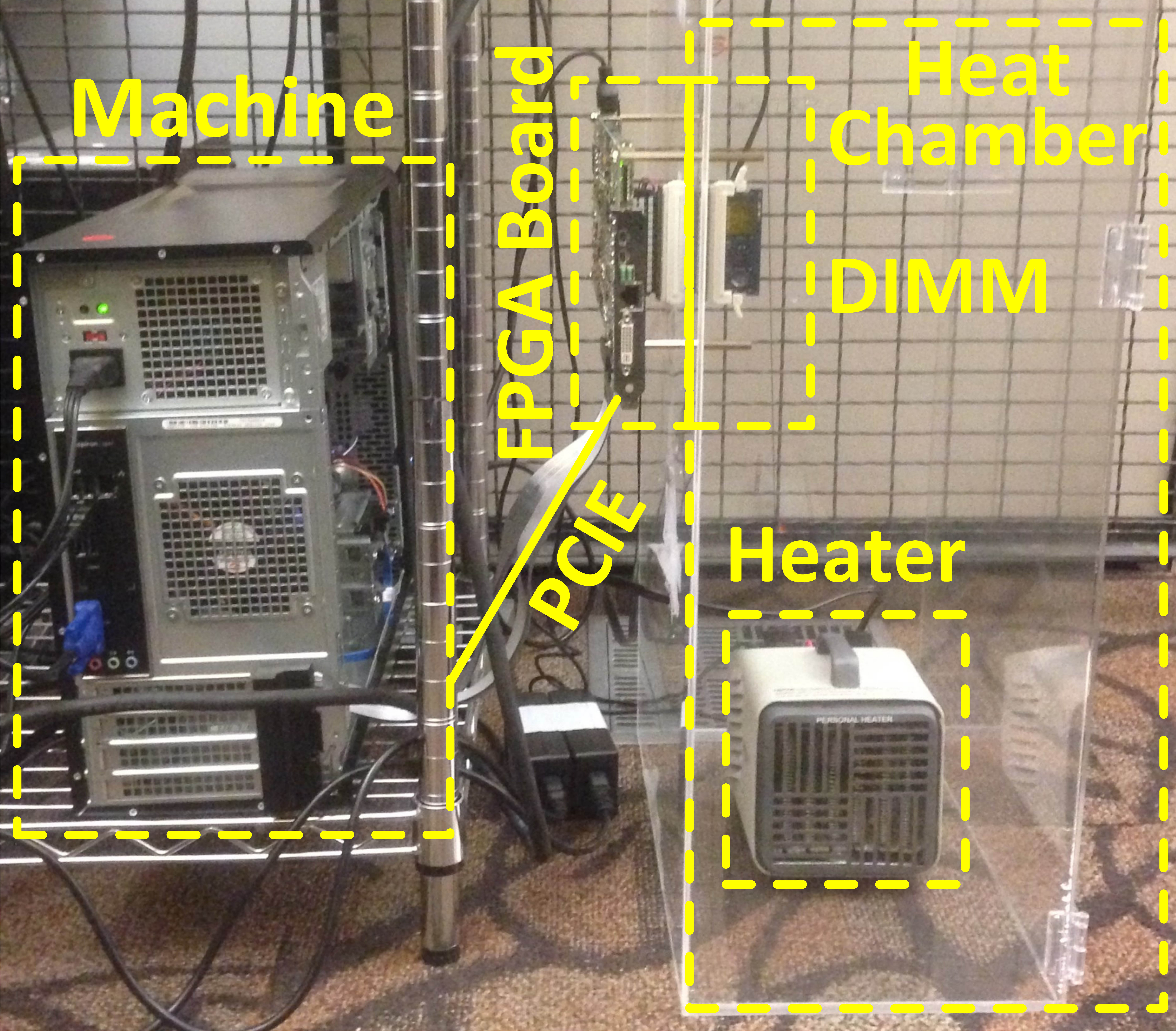}
	}
	\hspace{0.3in}
	\subcaptionbox{FPGA Board\label{fig:infra_fpga_avadram}}[0.4\linewidth] {
		\includegraphics[width=2.5in]{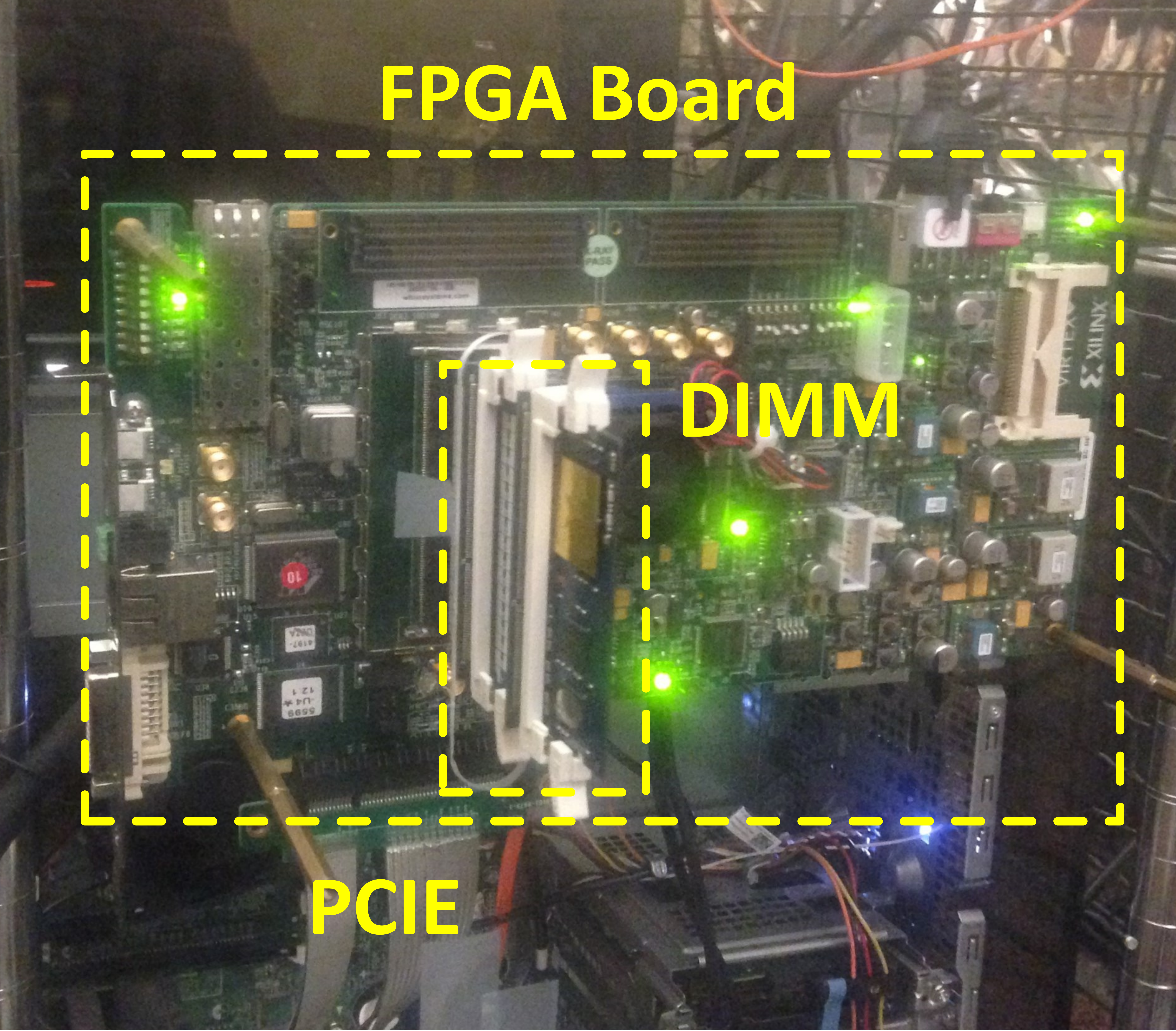}
	}
	\caption{FPGA-Based DRAM Test Infrastructure} \label{fig:infra_avadram}
\end{figure}

{\bf FPGA-Based DRAM Testing Infrastructure.} We build infrastructure similar
to that used for AL-DRAM in Chapter~\ref{ch:aldram} and previous
work~\cite{chandrasekar-date2014, liu-isca2013, khan-sigmetrics2014,
kim-isca2014, qureshi-dsn2015, khan-dsn2016, chang-sigmetrics2016}. Our
infrastructure provides the ability to: {\em i)} generate test patterns with
flexible DRAM timing parameters, {\em ii)} provide an interface from a host
machine to the FPGA test infrastructure, and {\em iii)} maintain a stable DRAM
operating temperature during experiments. We use a Xilinx ML605
board~\cite{ml605} that includes an FPGA-based memory controller which is
connected to a DDR3 SODIMM socket (Figure~\ref{fig:infra_fpga_avadram}). We
designed the memory controller~\cite{mig} with the flexibility to change DRAM
parameters. We connect this FPGA board to the host machine through a PCI-e
interface~\cite{pcie}, as shown in Figure~\ref{fig:infra_system_avadram}. We
manage the FPGA board from the host machine and preserve the test results in
the host machine's storage. In order to maintain a stable operating temperature
for the DRAM modules, during our experiments, we place the FPGA board in a heat
chamber that consists of a temperature controller, a temperature sensor, and a
heater which enables us to test at different temperatures
(Figure~\ref{fig:infra_system_avadram}).

{\bf Profiling Methodology.} The major purpose of our experiments is to
characterize architectural variation in latency. We would like to {\em i)}
determine the characteristics of failures when we reduce timing parameters
beyond the error-free operation regions, and {\em ii)} observe any correlation
between the error characteristics and the internal architecture/organization of
the DRAM modules. To this end, we analyze the error characteristics of DRAM by
lowering DRAM timing parameters below the values specified for error-free
operation.

An experiment consists of three major steps: {\em i) writing background data},
{\em ii) changing timing parameters}, and {\em iii) verifying cell content}. In
Step 1, we write a certain data pattern to the entire DRAM module with standard
DRAM timing parameters, ensuring that correct (the intended) data is written
into all cells. In Step 2, we access DRAM with the changed timing parameters,
\mycolor{and wait for the {\em refresh interval} such that DRAM cells have the
smallest charge due to charge leakage over time.} In Step 3, we verify the
content of the DRAM cells after the timing parameters are changed. To pass
verification, a DRAM cell must maintain its data value until the next refresh
operation. If the data read in Step 3 does not match the data written in Step
1, we log the addresses corresponding to the failure and the order of bits in
the failed address.

{\bf Data Patterns.} In order to exercise worst-case latency behavior, we use a
row stripe pattern, wherein a test pattern is written in odd rows and an
inverted test pattern is written in even rows~\cite{vandegoor-delta2002}. This
pattern drives the bitlines in opposite directions when accessing adjacent
rows. The patterns we have used in our tests are {\tt 0000}, {\tt 0101}, {\tt
0011}, and {\tt 1001}. We perform the test twice per pattern, once with the
test data pattern and once with the inverted version of the test data pattern,
in order to test every cell in charged (e.g., data 1) and non-charged states
(e.g., data 0). We report the sum of failures from these two cases for each
test. We perform 10 iterations of the same test for each DRAM module to make
sure the errors are consistent.

We evaluate three DRAM timing parameters, \trcd, \tras, and \trp. For each
timing parameter, our evaluations start from the standard DRAM timing
parameters (13.75/35.0/13.75ns for \trcd/\tras/\trp,
respectively)~\cite{micron-dram} and reduce the timing parameters to the lowest
values that our DRAM infrastructure allows (5ns for \trcd and \tras, and \trcd
+ 10ns for \tras). We use \dimms~DRAM modules (DDR3-1600~\cite{micron-dram}),
comprising 768 DRAM chips, from three DRAM manufacturers for our experiments.

	\section{DRAM Test Results and Analysis} \label{sec:profile}

In this section, we present the results of our profiling studies that
demonstrate the presence of architectural variation in both the vertical
(bitline) and horizontal (wordline) directions. We {\em i)} show the existence
of architectural variation in Sections~\ref{sec:profile_bitline}
and~\ref{sec:profile_wordline}, {\em ii)} analyze the impact of the row and
column interface in Sections~\ref{sec:profile_rowint}
and~\ref{sec:profile_colint}, and {\em iii)} characterize the impact of
operating conditions on architectural variation in
Section~\ref{sec:profile_sensitivity}. We then provide a summary of our
analysis on architectural latency variation across 96 DRAM modules
(Section~\ref{sec:profile_dimms}).

\subsection{Architectural Variation in Bitlines} \label{sec:profile_bitline}

As we explain in Section~\ref{sec:overview_existence}, we expect different
error characteristics for cells connected to a bitline, depending on the
distance from the local sense amplifiers. To demonstrate the existence of
architectural variation in a bitline, we design a test pattern that sweeps the
row address.

\begin{figure}[h]
	\centering
	\subcaptionbox{\trp 12.5ns\label{fig:profile_row_125ns}}[0.45\linewidth] {
		\includegraphics[height=1.5in]{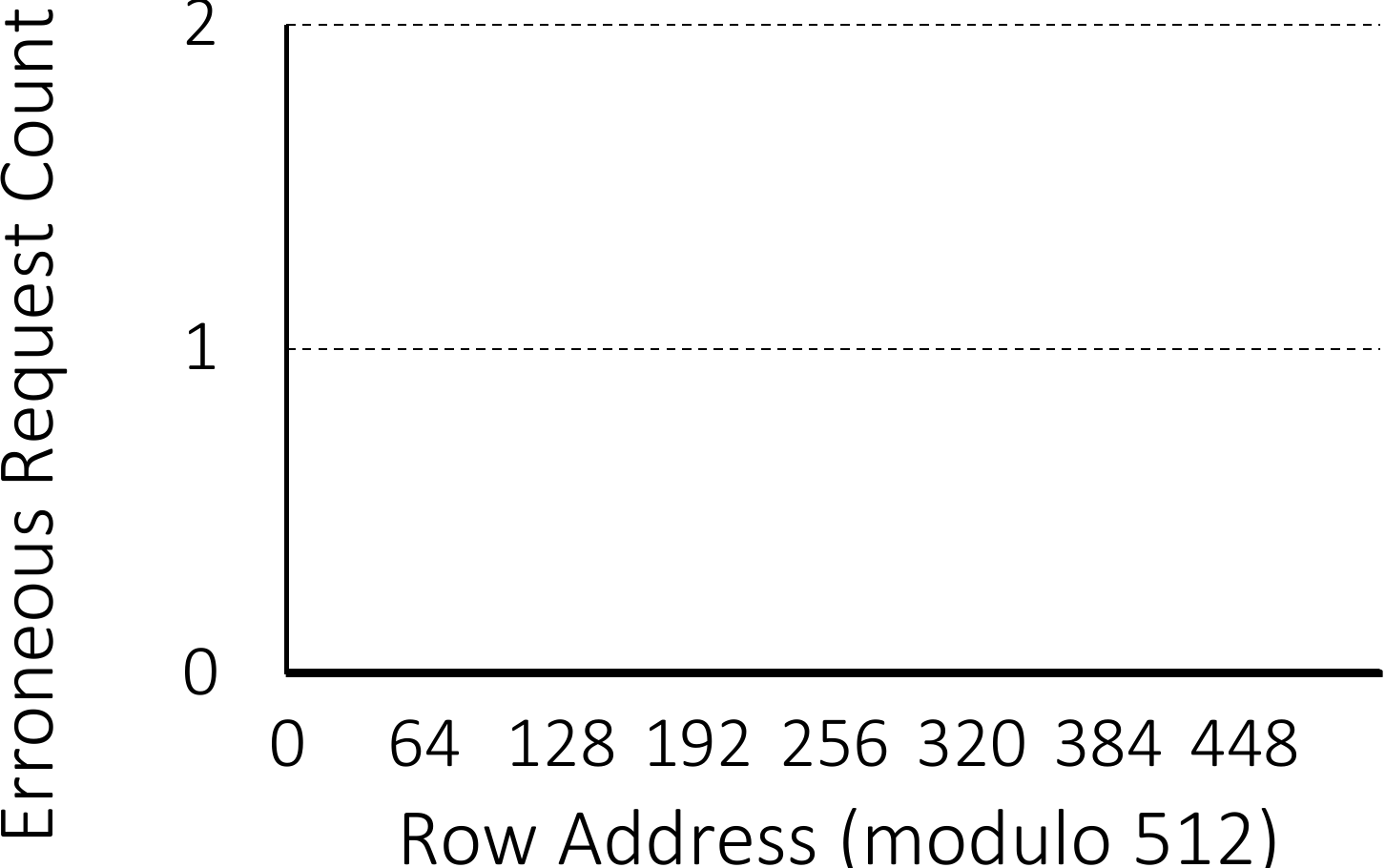}
	}
	\subcaptionbox{\trp 10.0ns\label{fig:profile_row_10ns}}[0.45\linewidth] {
		\includegraphics[height=1.5in]{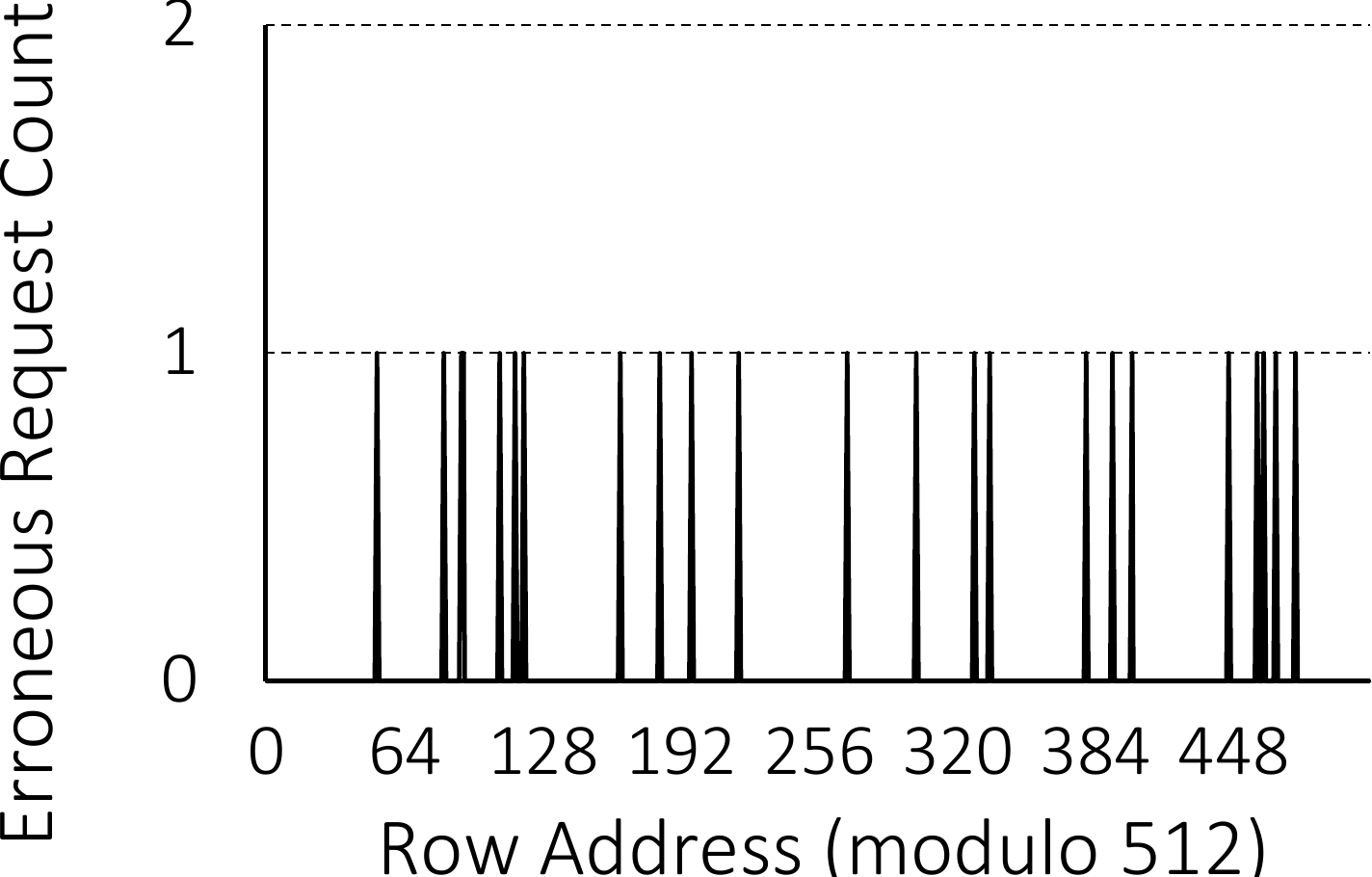}
	}

	\subcaptionbox{\trp 7.5ns\label{fig:profile_row_75ns}}[0.45\linewidth] {
		\includegraphics[height=1.5in]{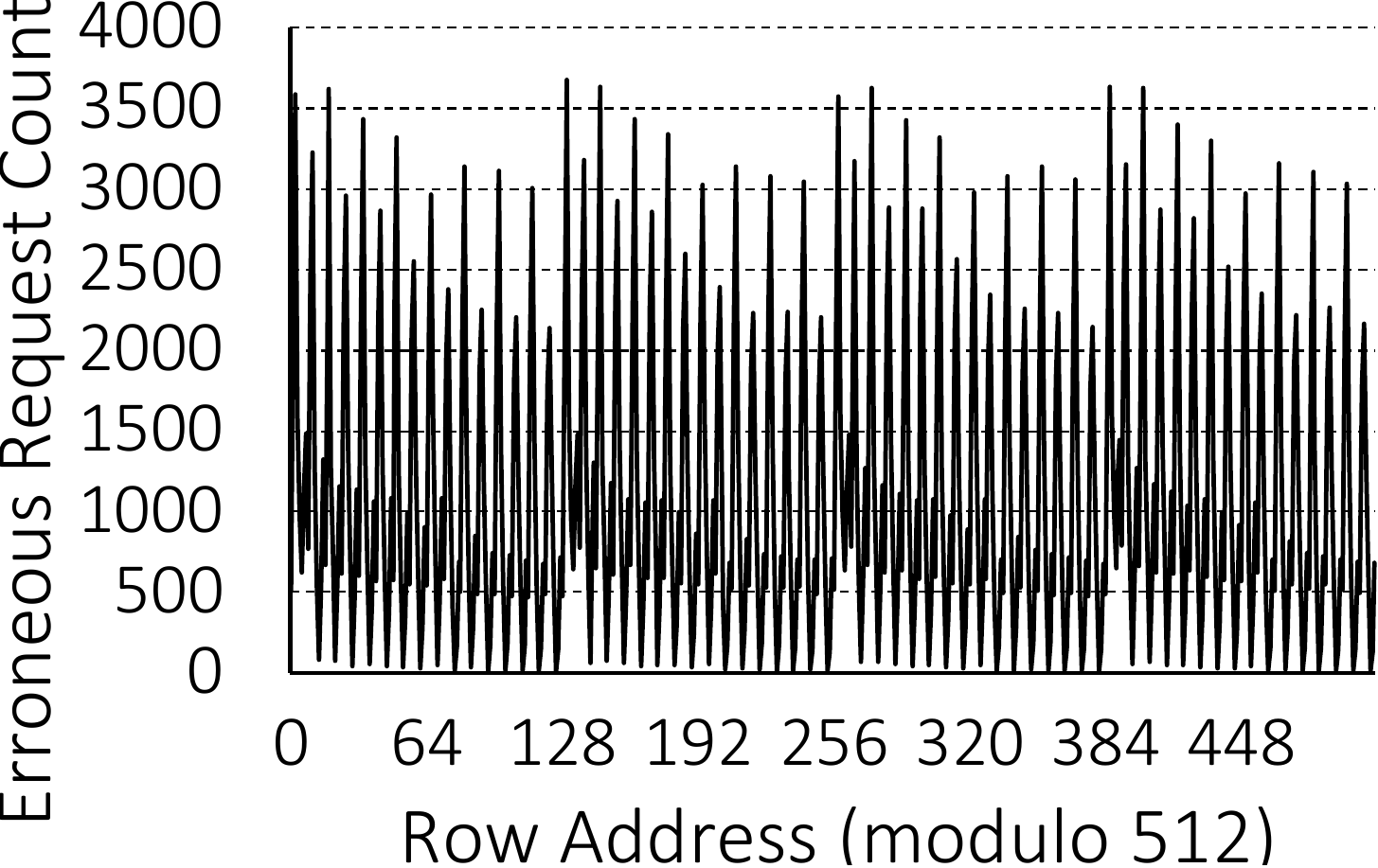}
	}
	\subcaptionbox{\trp 5.0ns\label{fig:profile_row_5ns}}[0.45\linewidth] {
		\includegraphics[height=1.5in]{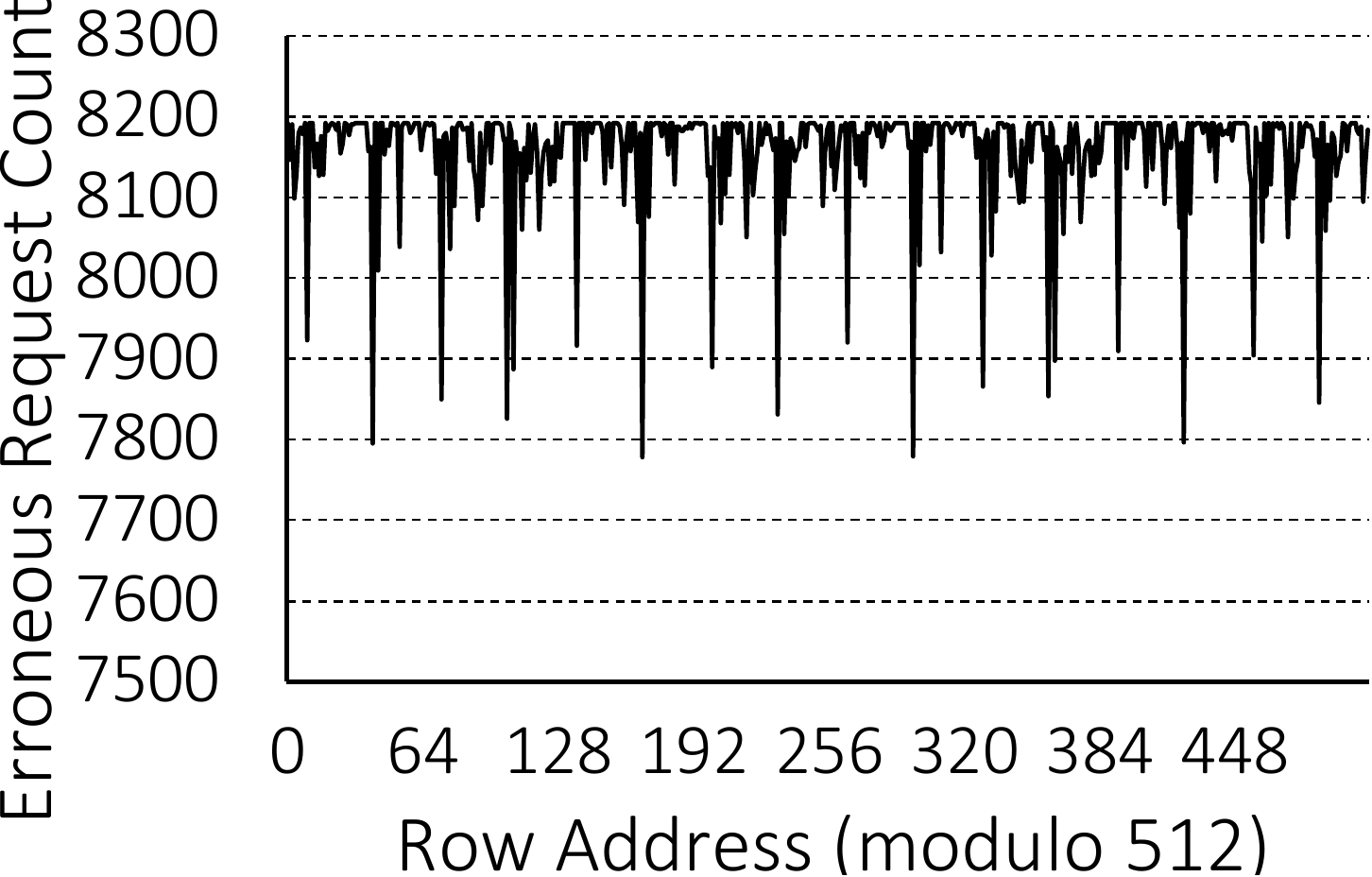}
	}
	\caption{Erroneous Request Count when Sweeping Row Addresses with Reduced
	\trp Timing Parameter}
	\label{fig:profile_row_trp}
\end{figure}

\begin{figure}[h]
	\centering
	\subcaptionbox{Sorted \& Aggregated\label{fig:profile_row_sort}}[0.3\linewidth] {
		\includegraphics[height=1.6in]{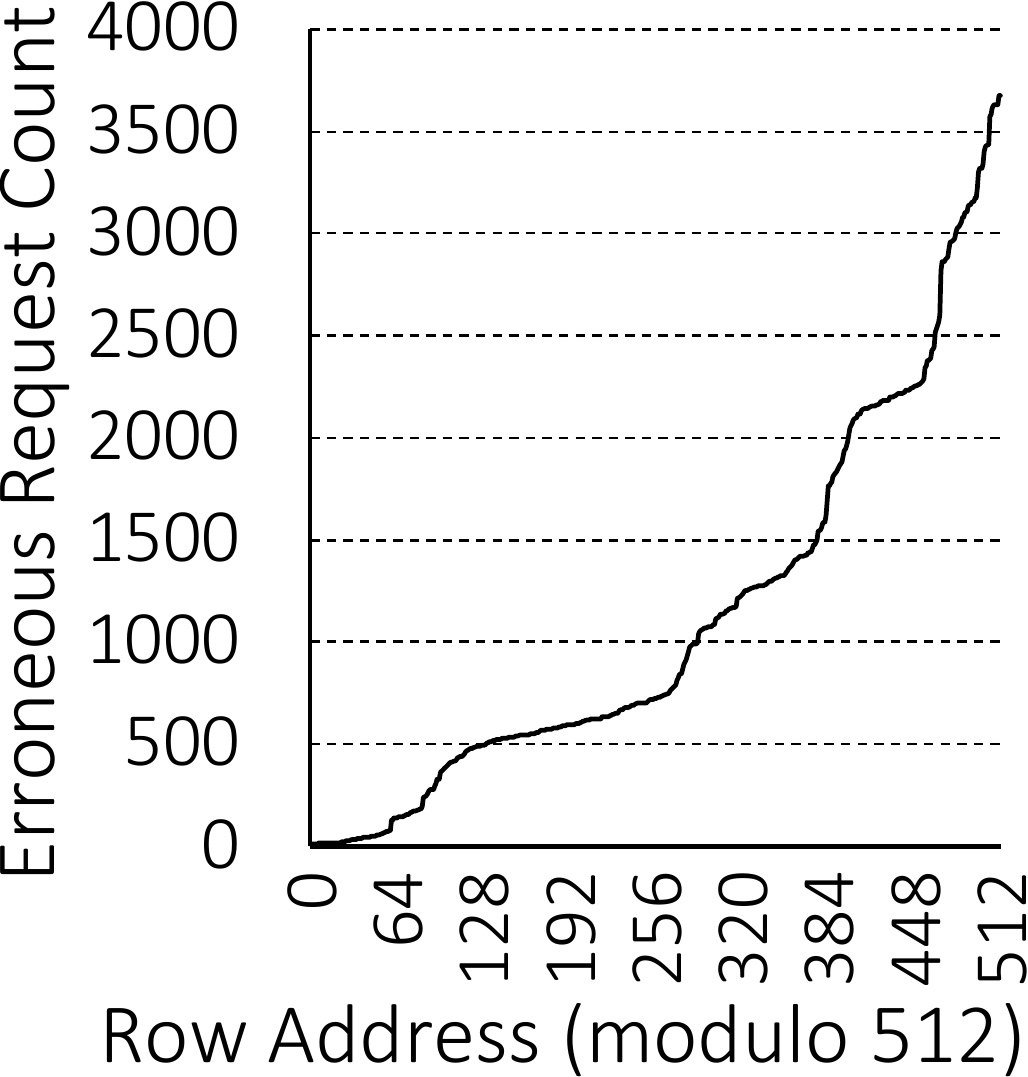}
	}
	\subcaptionbox{Erroneous Request Counts with Rows Sorted within 512-Row Groups\label{fig:profile_row_order}}[0.68\linewidth] {
		\includegraphics[height=1.6in]{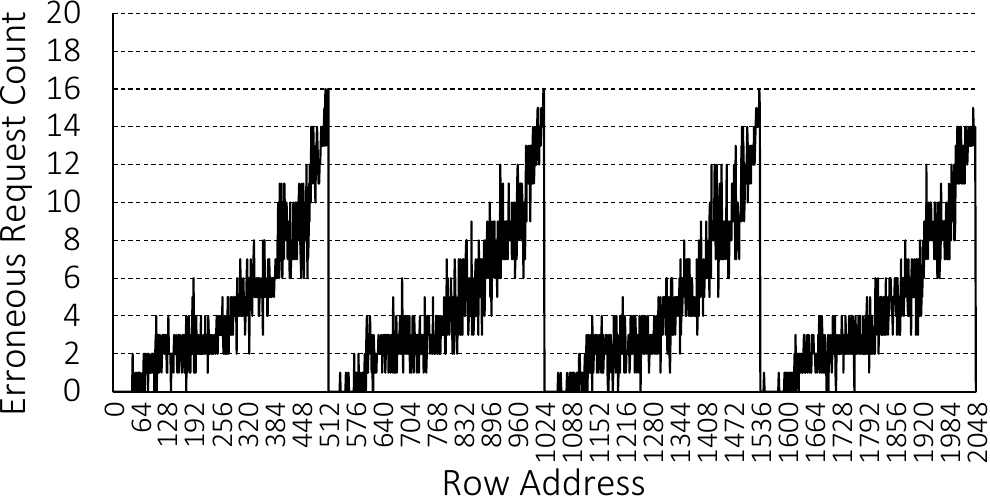}
	}
	\caption{Periodicity in Error Request Count (\trp 7.5ns)}
	\label{fig:dram_org}
\end{figure}

{\bf Per-Row Error Count with Row Address Sweeping.}
Figure~\ref{fig:profile_row_trp} plots the error count for three values of a
DRAM timing parameter, \trp (whose standard value is 13.75ns), with a refresh
interval of 256 ms (greater than the normal 64 ms refresh interval to
\mycolor{emphasize} the effects of access latency) and an ambient temperature
of 85\celsius. We tested all rows (and 16 columns) in a DRAM module and plot
the number of erroneous accesses \mycolor{for every modulo 512
rows}.\footnote{\mycolor{ Even though there are redundant cells (rows), DRAM
does not allow direct access to redundant cells. Therefore, we can only access
a 512$\times$512 cell mat ($2^n$ data chunk). Figure~\ref{fig:profile_row_trp}
plots the number of erroneous requests in every 512 cell chunk.}} We
accumulate errors every modulo 512 rows because, {\em i)} each bitline is
connected to 512 cells, and {\em ii)} our expectation that the architectural
variation pattern will repeat every 512 cells (we provide each row's error
count in Figure~\ref{fig:profile_row_order}). We draw two key observations.
First, reducing a timing parameter below its standard value induces errors,
and reducing it further induces more errors. At a \trp of 10.0ns (3.75ns
reduction from the standard value), the number of errors is small, while at a
\trp of 5.0ns, we observe a large number of errors. Second, we observe error
count variation across rows only at 7.5ns (from 0 to more than 3500 in
Figure~\ref{fig:profile_row_75ns}), while most errors are {\em randomly}
distributed at 10.0ns (Figure~\ref{fig:profile_row_10ns}) and most rows show
very high error counts at 5.0ns (Figure~\ref{fig:profile_row_5ns}).

{\bf Periodicity in Per-Row Error Count.} To understand these trends better, we
sort the error counts in every row address modulo 512, with a \trp of 7.5ns
(data in Figure~\ref{fig:profile_row_75ns}), as shown in
Figure~\ref{fig:profile_row_sort}. We then apply this sorting to every group of
512 rows (e.g., first row to 512th row as the first group, 513th row to 1024th
row as the second group, and so on). We plot the error counts of individual
rows in Figure~\ref{fig:profile_row_order}. The reason why we do this sorting
is because we expect periodicity in error counts, which is not shown in
Figure~\ref{fig:profile_row_75ns}. Figures~\ref{fig:profile_row_sort} and
\ref{fig:profile_row_order} show similar characteristics: error increases
periodically across 512 rows. Therefore, we conclude that {\em error count
shows periodicity with row address}, confirming our expectation that there is
predictable architectural variation in the latency of cells across a bitline.

\subsection{Architectural Variation in Wordlines} \label{sec:profile_wordline}

As we explained in Section~\ref{sec:overview_existence}, we expect
architectural variation across cells in a wordline, depending on the distance
from the wordline driver. To confirm the existence of architectural variation
across a wordline, we use a similar evaluation methodology as the one used in
Section~\ref{sec:profile_bitline}, except that {\em i)} we sweep the column
address instead of the row address, {\em ii)} merge errors in the same column
across multiple rows (128 columns in a row). In order to minimize the impact of
variation across a bitline and focus on variation across a wordline, as shown
in Section~\ref{sec:profile_bitline}, we test columns in only 16 rows.

\begin{figure}[h]
	\centering
	\subcaptionbox{\trp 10ns \& Aggregated\label{fig:profile_col_addr_std}}[0.45\linewidth] {
		\includegraphics[height=1.5in]{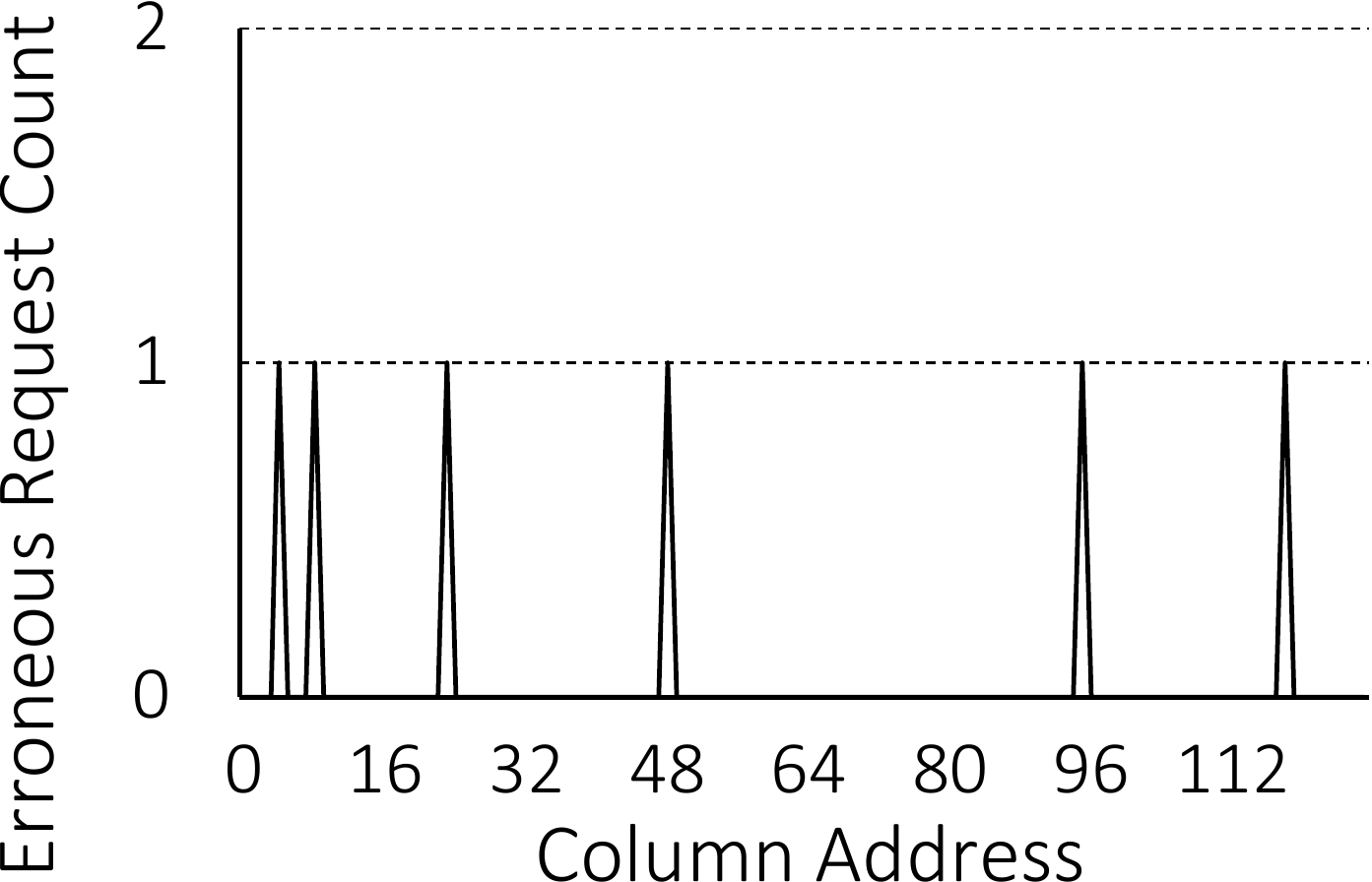}
	}
	\subcaptionbox{\trp 7.5ns \& Aggregated\label{fig:profile_col_addr_low}}[0.45\linewidth] {
		\includegraphics[height=1.5in]{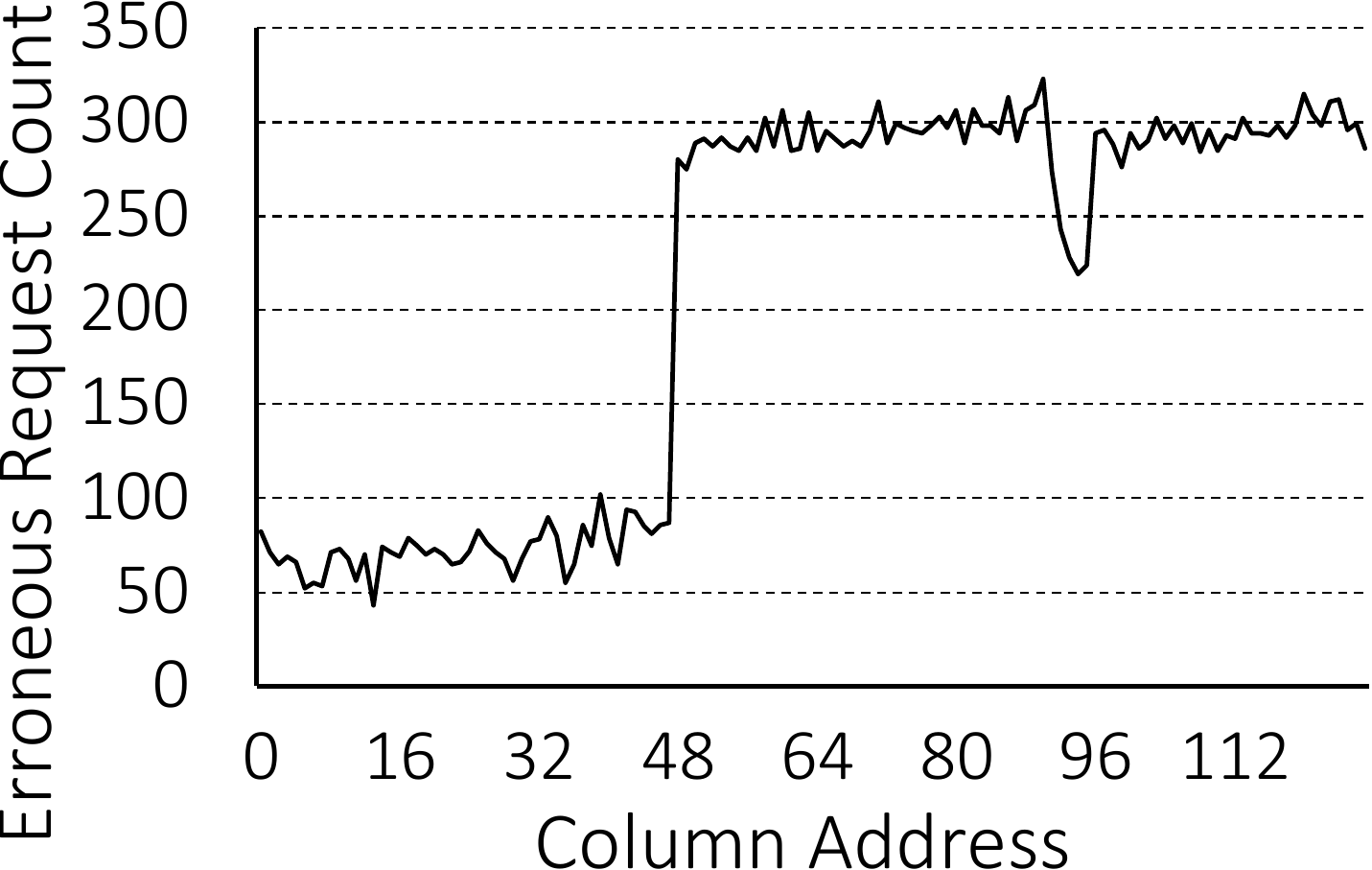}
	}

	\subcaptionbox{\trp 7.5ns \& Case 1\label{fig:profile_col_case1}}[0.45\linewidth] {
		\includegraphics[height=1.5in]{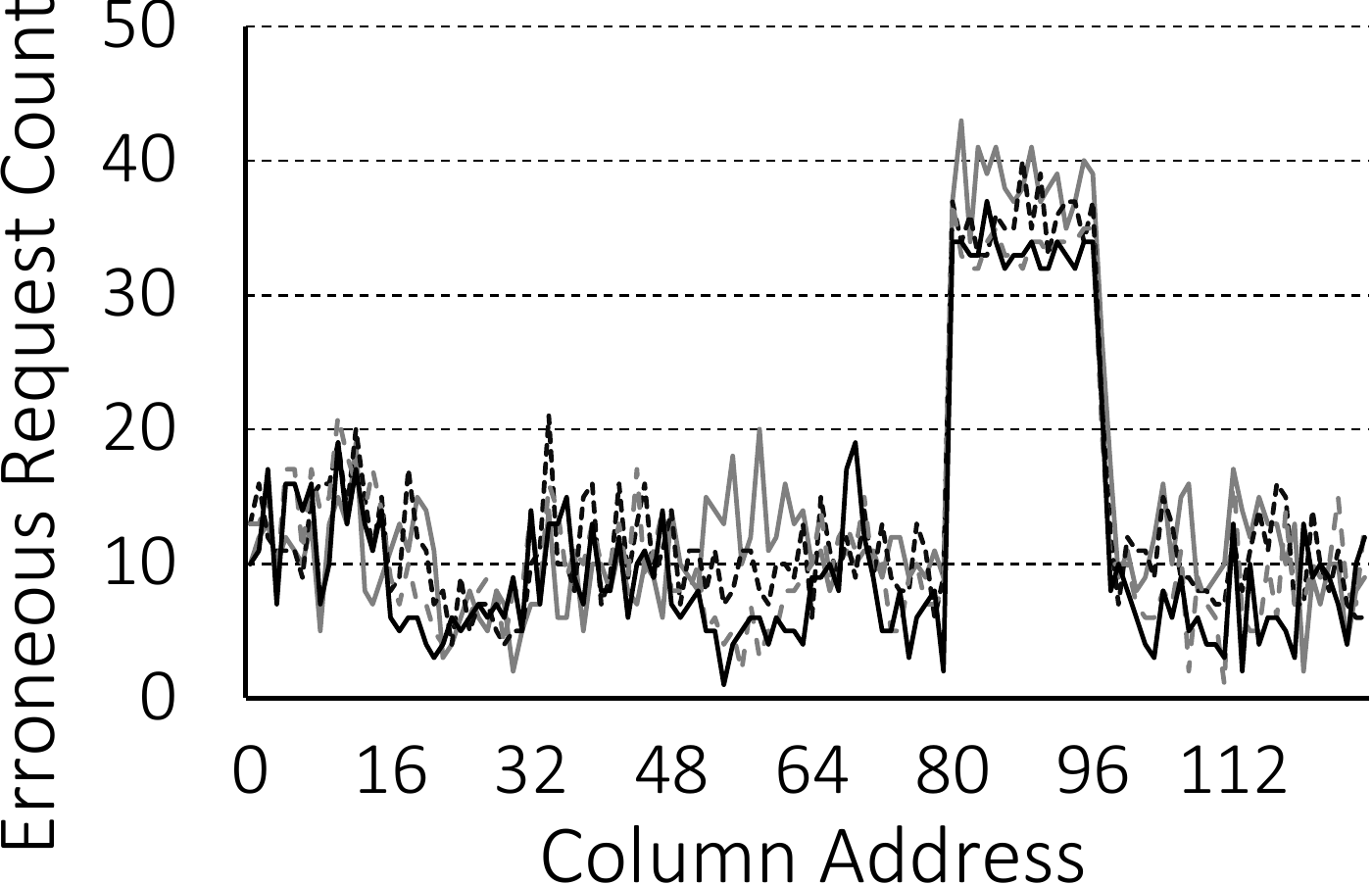}
	}
	\subcaptionbox{\trp 7.5ns \& Case 2\label{fig:profile_col_case2}}[0.45\linewidth] {
		\includegraphics[height=1.5in]{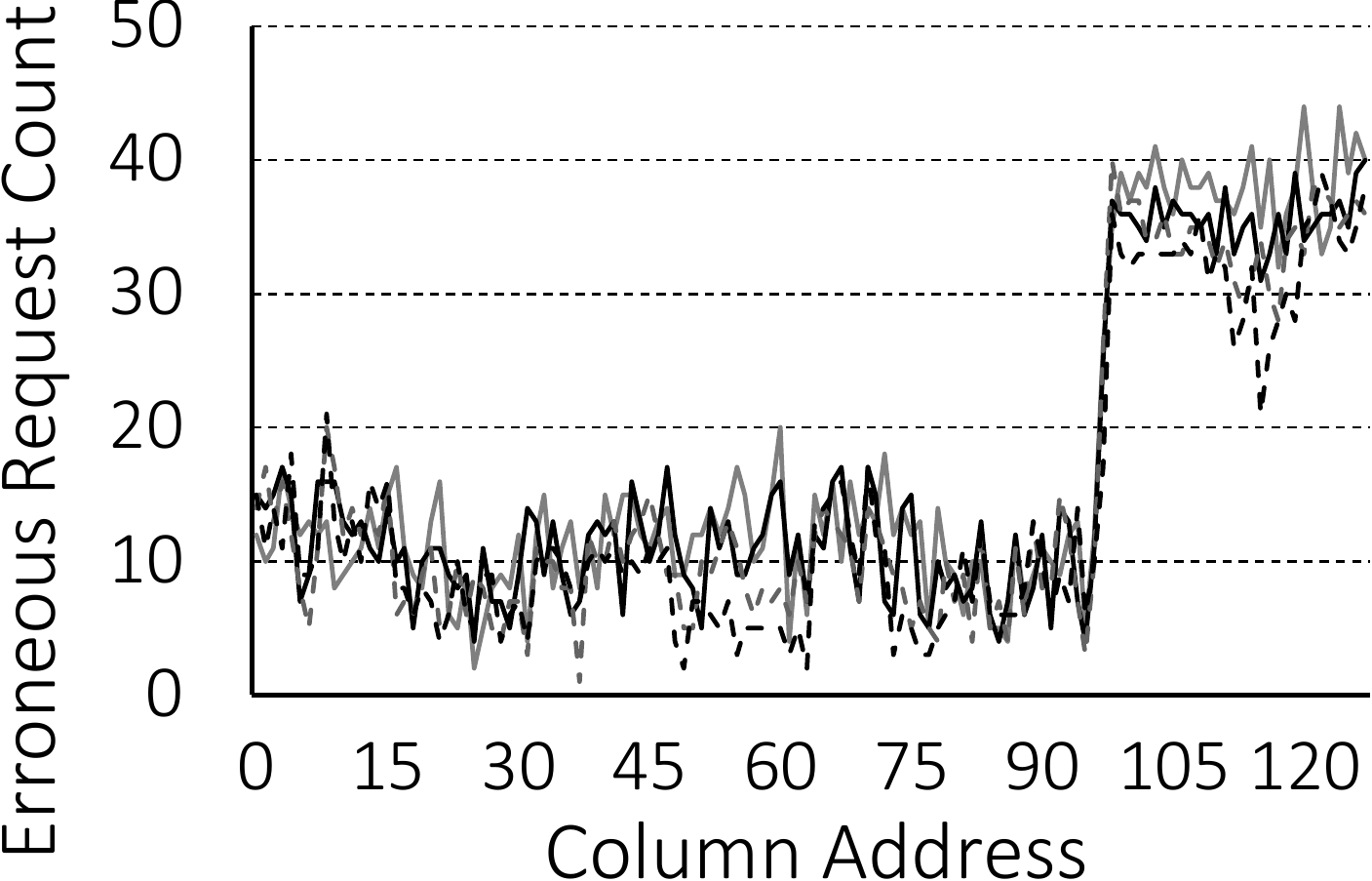}
	}
	\caption{Erroneous Request Count when Sweeping Column Addresses with Reduced
	\trp Timing Parameter}
	\label{fig:profile_col_addr}
\end{figure}

{\bf Per-Column Error Count with Column Address Sweeping.}
Figure~\ref{fig:profile_col_addr} provides results with two \trp values (10ns
and 7.5ns). Similar to the evaluation with sweeping row addresses, we see that
the number of errors is small and the distribution is random when \trp is
reduced by a small amount, as shown in Figure~\ref{fig:profile_col_addr_std}.
However, the number of errors is large when \trp is reduced significantly, as
shown in Figure~\ref{fig:profile_col_addr_low}. We observe variations in error
counts across different column addresses at a \trp of 7.5ns. Besides other
variations, there is a large jump near the 48th column and a dip in error count
near the 96th column, as shown in Figure~\ref{fig:profile_col_addr_low}.

To understand these, we separately plot each row's error count, which displays
different patterns. We provide two such types of patterns (from multiple rows)
in Figures~\ref{fig:profile_col_case1}~and~\ref{fig:profile_col_case2}. In one
such type, shown in Figure~\ref{fig:profile_col_case1}, the error count
drastically increases at around the 80th column and drops at around the 96th
column (There are other types of patterns with similar shapes but with the
jumps/drops happening at different locations). In the type of pattern shown in
Figure~\ref{fig:profile_col_case2}, the error count drastically increases at
96th column and stays high. We attempt to correlate such behavior with the
internal organization of DRAM.

\begin{figure}[h]
	\centering
	\includegraphics[width=0.8\linewidth]{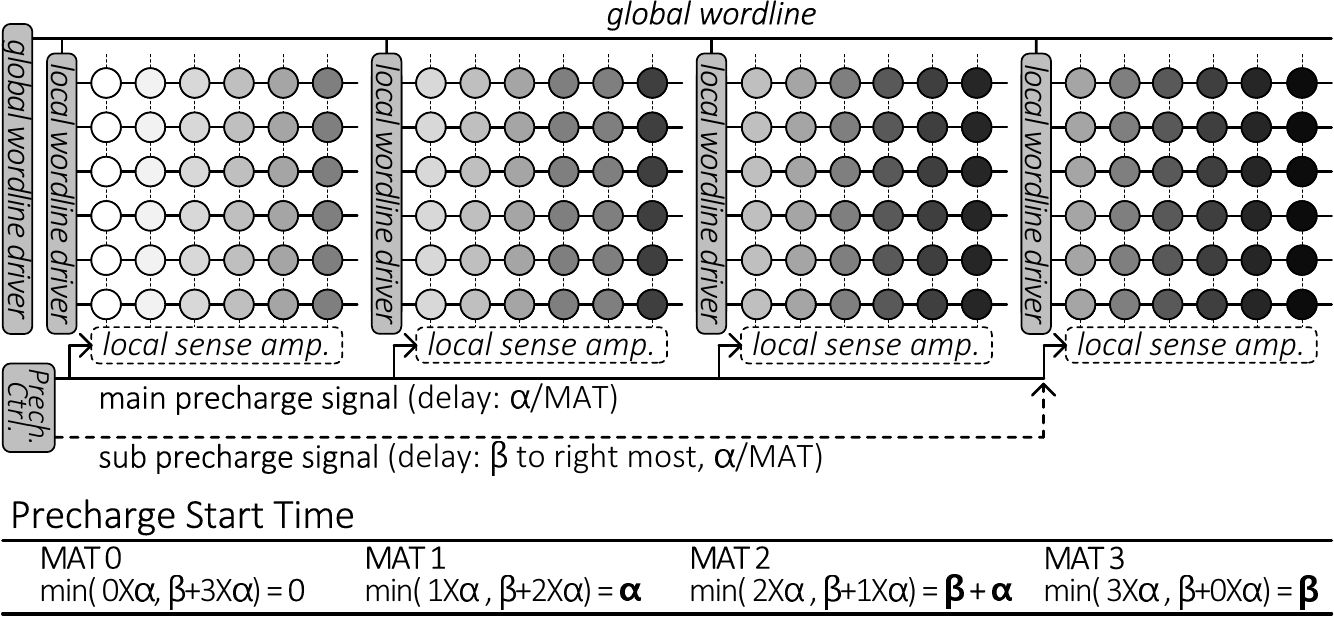}
	\caption{Architectural Variation due to Precharge Control}
	\label{fig:prech_signal}
\end{figure}

Figure~\ref{fig:prech_signal} shows an illustration of how the precharge
control signal flows across mats. The timing parameter \trp dictates how long
the memory controller should wait after it issues a precharge command before it
issues the next command. When a precharge command is issued, the precharge
signal propagates to the local sense amplifiers in each mat, leading to
propagation delay (higher for sense amplifiers that are farther away). To
mitigate this variation in the delay of the precharge control signal, DRAM uses
two signals, {\em i)} main precharge signal -- propagating from left to right,
and {\em ii)} sub precharge signal -- that directly reaches the right and
propagates from right to left.

The main and sub precharge signals arrive at different times at the different
mats due to parasitic capacitance on the propagation path. The main precharge
signal is delayed by $\alpha$ per mat going from left to right, while the sub
precharge signal is delayed by $\beta$ per mat $\alpha > \beta$, since the sub
precharge signal does not have any load going from left to right. However,
after that, the sub precharge signal exhibits a delay of $\alpha$ per mat when
propagating through mats from right to left. The sense amplifiers in a mat
respond to the faster one of the two precharge signals. For instance, in the
illustration in Figure~\ref{fig:prech_signal}, mat 3 receives the precharge
signal the last. Hence, accesses to it would exhibit more errors than accesses
to other mats if \trp is reduced. Such control signal delays result in the kind
of jumps in errors at particular column addresses we see in real DRAM chips
(e.g., Figures~\ref{fig:profile_col_addr_low}, \ref{fig:profile_col_case1},
\ref{fig:profile_col_case2}). We conclude that error count varies across
columns, based on the column's distance from the wordline and control signal
drivers. While such control signal delays explain why such jumps occur,
knowledge of the exact location of mats and how they are connected to the
control signals is necessary to tie back the control signal propagation to the
specific column addresses at which the jumps occur.

\subsection{Effect of the Row Interface} \label{sec:profile_rowint}

As shown in Figure~\ref{fig:profile_row_75ns}, the error count across a bitline
does not linearly increase with increasing {\em DRAM-external row address} --
the address issued by the memory controller over the memory channel, while we
observe periodicity when rows are sorted by error count, in
Section~\ref{sec:profile_bitline}. This is mainly because the DRAM-external row
address is {\em not} directly mapped to the internal row address in a DRAM
mat~\cite{liu-isca2013, khan-dsn2016}. Without information on this mapping, it
is difficult to tie the error count periodicity to specific external row
addresses. In this subsection, we estimate the most-likely mapping between the
DRAM-external row address and the DRAM-internal row address ({\em estimated row
mapping}) based on the observed error count. We then analyze the similarity of
the estimated row address mapping across multiple DRAM modules manufactured by
the same DRAM company (in the same time frame).

\begin{figure}[h]
	\centering
	\subcaptionbox{Internal Address\label{fig:addr_physical_row}}[0.4\linewidth] {
		\includegraphics[height=1.5in]{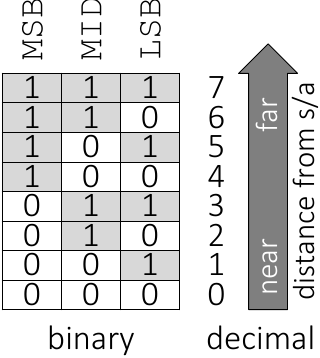}
	}
	\subcaptionbox{External Address\label{fig:addr_logical_row}}[0.5\linewidth] {
		\includegraphics[height=1.5in]{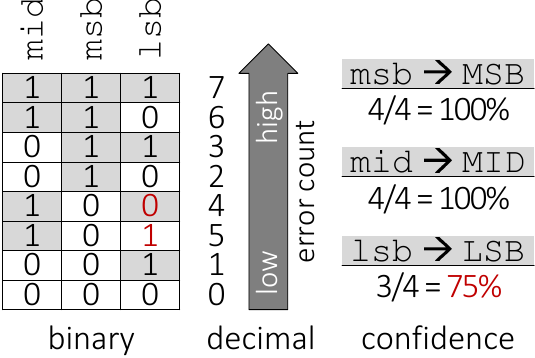}
	}
	\caption{DRAM-External Address vs. DRAM-Internal Address}
	\label{fig:arch_tran_row}
\end{figure}

{\bf Methodology for Estimating Row Address Mapping.} We explain our estimation
methodology using a simple example shown in Figure~\ref{fig:arch_tran_row},
which has a 3-bit row address (eight rows per mat).
Figure~\ref{fig:addr_physical_row} shows the DRAM-internal row address in both
decimal and binary, increasing in the order of distance between the row and the
local sense amplifier.

Figure~\ref{fig:addr_logical_row} shows DRAM-external row addresses which are
{\em ranked based on the error counts}. As observed, the order is not the same
as the DRAM-internal address order in Figure~\ref{fig:addr_physical_row}. To
determine the estimated external to internal row mapping, we try to find the
bit values across bit positions that display the largest similarity. For
instance, {\tt MSB} in the internal address has four consecutive ``1''s and
``0''s (in order from largest to smallest rows). When comparing this with the
external address, we see that the middle bit of the external address matches
this order exactly (100\% confidence). We compare the external and internal
address bits and identify which bit positions in the external address map to
which bit positions in the internal address. The estimated mapping (in the
logical address) is indicated by dark boxes when the expected bit is ``1'' and
light boxes when the expected bit is ``0''. There are cases when this mapping
does not match with the actual external address (indicated in red).

\begin{figure}[h]
	\centering
	\includegraphics[width=0.8\linewidth]{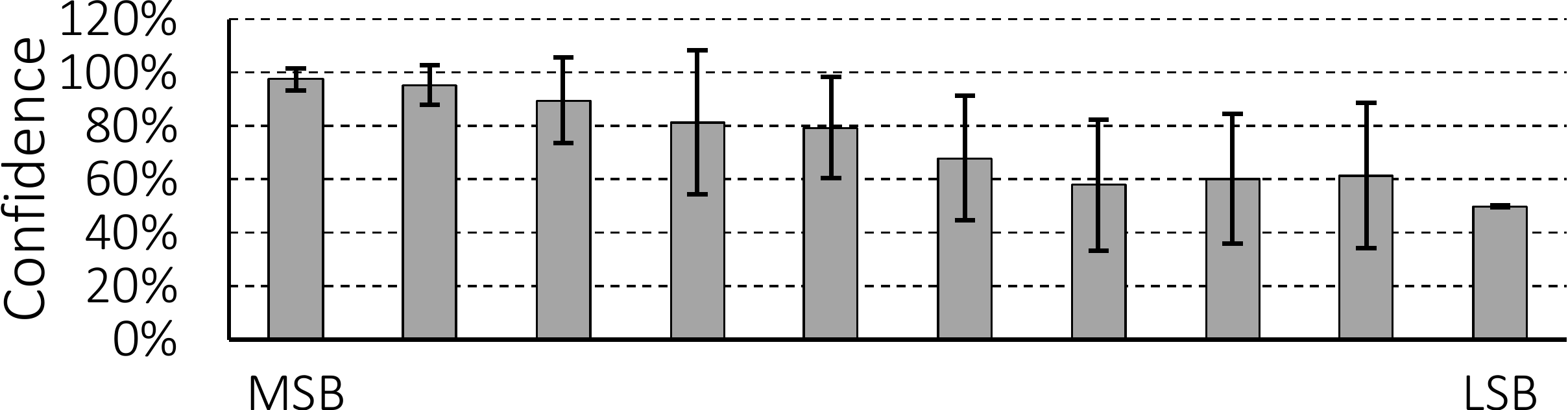}
	\caption{Confidence in Estimated Row Mapping for Each Bit}
	\label{fig:profile_addr_tran}
\end{figure}

{\bf Estimated Row Address Mapping in Real DRAM modules.} We perform such an
external to internal address mapping comparison and mapping exercise on eight
DRAM modules manufactured by the same company in a similar time frame.
Figure~\ref{fig:profile_addr_tran} shows the average confidence level of the
estimated row mapping along with the standard deviation over the eight chips.
We make three observations. First, all DRAM modules show the same estimated row
mapping (with fairly high confidence) for at least the five most significant
bits. This result shows that DRAM modules manufactured by the same company at
the same time have similar architectural variation. Second, the confidence
level is almost always less than 100\%. This is because process variation
introduces perturbations besides architectural variation, which can change the
ranking of rows (determined based on error counts). Third, the confidence level
drops gradually from {\tt MSB} to {\tt LSB}. This is also due to the impact of
process variation. The noise from process variation and \mycolor{row repair}
can change row ranking and grouping by error count. Address bits closer to {\tt
MSB} tend to divide rows into groups at a larger granularity than address bits
closer to {\tt LSB}. Therefore, the higher order bits show higher confidence.
Based on these observations, we conclude that DRAMs that have the same design
display similar error characteristics due to architectural latency variation.

\mycolor{DRAM modules from two companies show this similarity of the estimated
row address mapping. However, extracting the estimated row address mapping by
this mechanism does not work for DRAM modules from a company. We expect that
DRAMs from the company may have a symmetric organization in terms of external
addresses. For example, {\tt MSB} = 0 rows are distributed evenly over a mat.
We leave the correlation between external addresses and internal organization
for the DRAM modules from the company to future work.}

\subsection{Effect of the Column Interface} \label{sec:profile_colint}

\begin{figure}[h]
	\centering
	\includegraphics[width=0.8\linewidth]{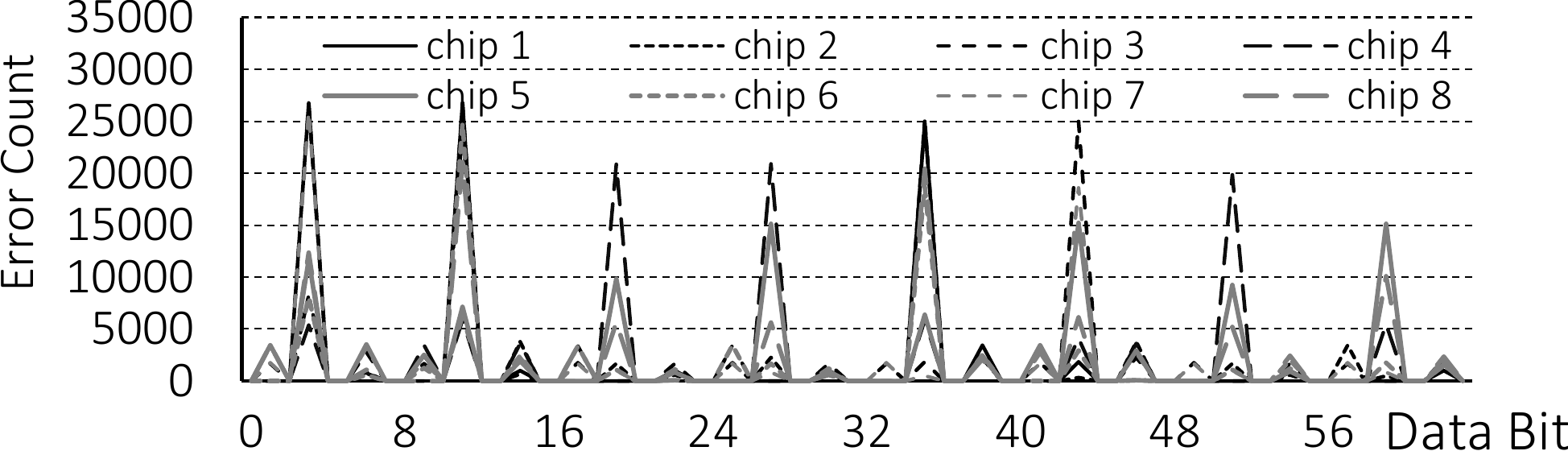}
	\caption{Error Count in Bit Positions}
	\label{fig:profile_col_io}
\end{figure}

Another way to observe the error characteristics in the wordline organization
is using the {\em mapping between the global sense amplifier and the IO
channel}. As we explained, global sense amplifiers in a DRAM chip concurrently
read 64-bit data from different locations of a row, leading to variation in
errors. Figure~\ref{fig:profile_col_io} plots errors in 64-bit data-out (as
shown in Figure~\ref{fig:arch_col_interface}) in the IO channel (For example,
first eight bits (bit 0 -- 7) are the first burst of data transfer). We provide
two conclusions. First, there is large variation in the amount of errors in the
IO channel. For example, more than 26K errors happen in the third bit while no
errors in the first bit of the IO channel. Second, the error characteristics of
eight DRAM chips show similar trends. Section~\ref{sec:mech_shuffling} uses
these observations to develop a new error correction mechanism.

\subsection{Effect of Operating Conditions} \label{sec:profile_sensitivity}

\begin{figure}[h]
	\centering
	\subcaptionbox{Varying Retention Time\label{fig:profile_ret}}[0.45\linewidth] {
		\includegraphics[height=1.5in]{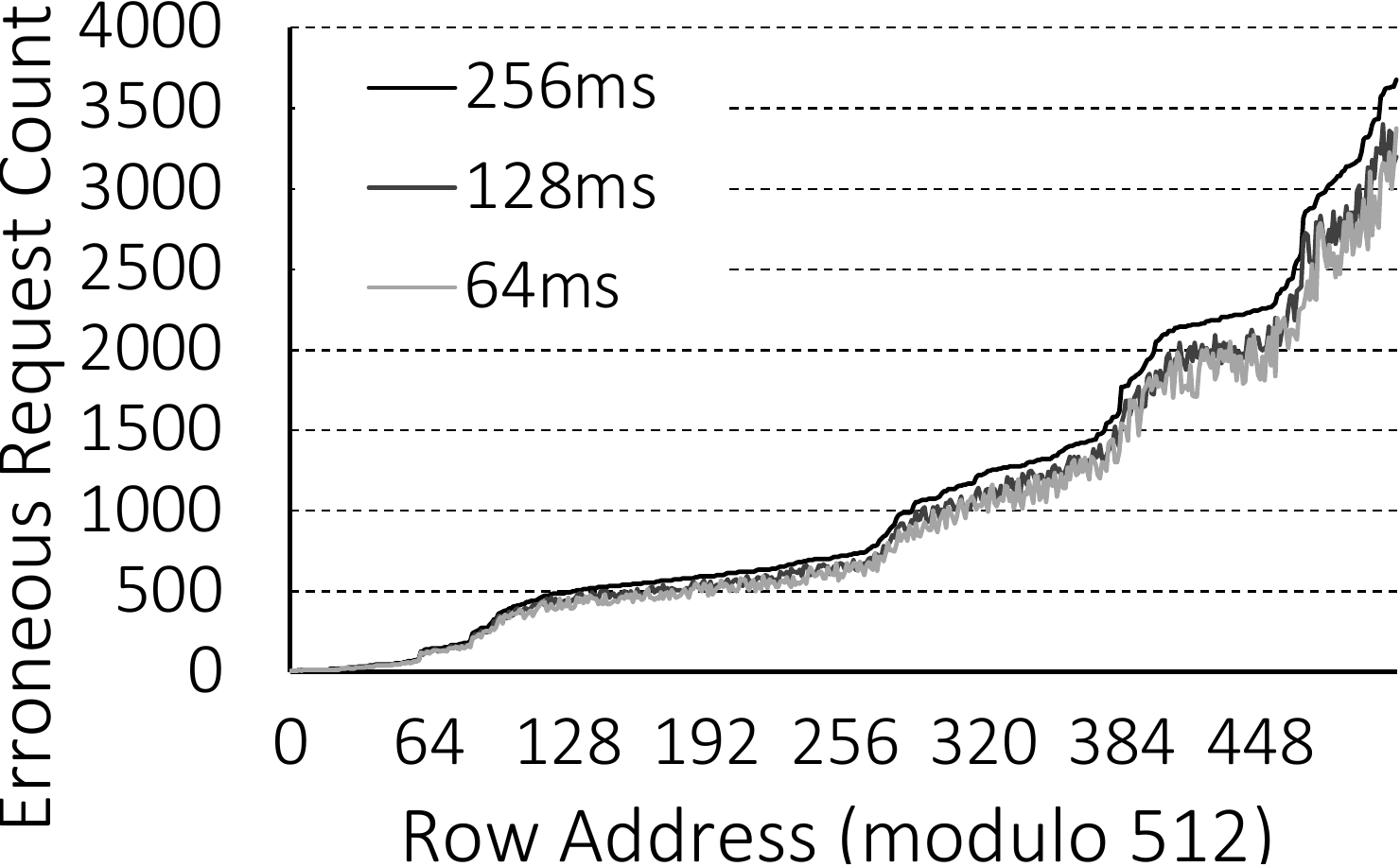}
	}
	\subcaptionbox{Varying Temperature\label{fig:profile_temp}}[0.45\linewidth] {
		\includegraphics[height=1.5in]{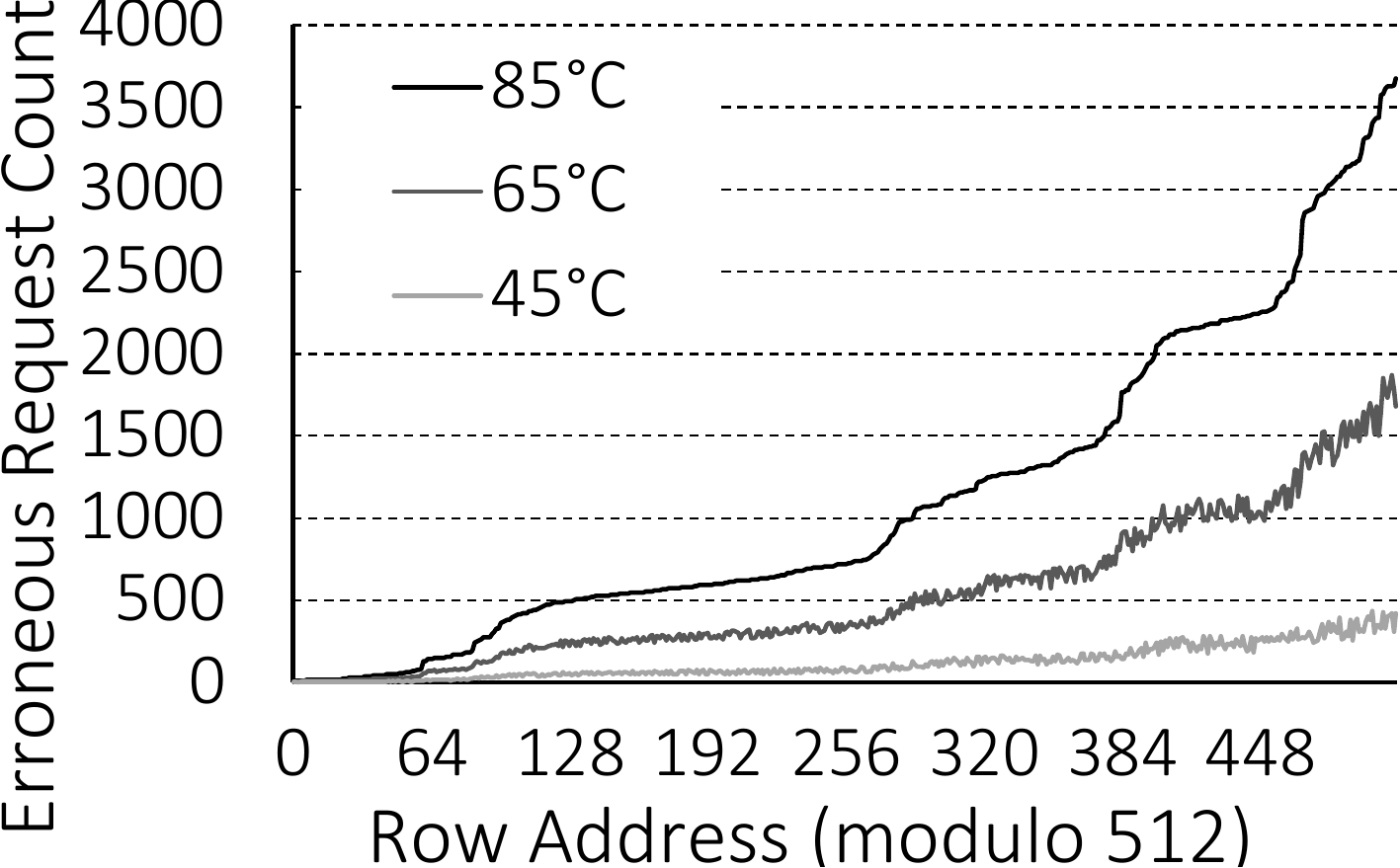}
	}
	\caption{Architectural Variation vs. Operating Conditions}
	\label{fig:profile_sensitivity}
\end{figure}

Figure~\ref{fig:profile_sensitivity} shows the error count sensitivity to the
refresh interval and the operating temperature by using the same method as row
sweeping (accumulating errors in modulo 512 rows as done in
Section~\ref{sec:profile_bitline}). We make three observations. First, neither
the refresh interval nor temperature changes the overall trends of
architectural variation (e.g., variability characteristics in different row
addresses remain the same, though the absolute number of errors changes).
Second, reducing the refresh interval or the ambient temperate leads to fewer
errors.\footnote{\mycolor{We observe this trend over most of timing
parameters.  One possible exception against our observations is the tWR timing
parameter which is worse at low temperatures due to increased cell contact
resistance.  However, this effect is known to happen below 25\celsius. In the
temperature range we evaluate (45\celsius~to 85\celsius), the effect of cell
leakage (more pronounced at high temperatures) is dominant, leading to more
errors at higher temperatures. We will incorporate this discussion in the
paper.}} Third, the variability in cells is much more sensitive to the ambient
temperature than the refresh interval. When changing the refresh interval, the
total error count does not change drastically (exhibits only 15\% decrease in
error count with 4X reduction in refresh interval). On the other hand,
changing the ambient temperature has a large impact on the total error count
(90\% decrease in the total error count with 45\celsius~change in
temperature). This is due the fact that frequent refreshes impact only the
cells and make them faster, whereas reducing temperature makes not only the
cells but also the peripheral circuits faster. Based on these observations, we
conclude that temperature or refresh does not change the trends in variability
of the cells, however they impact the total number of failures in vulnerable
regions at different rates.

\subsection{Summary Results of \dimms~DRAM modules} \label{sec:profile_dimms}

In this work, we profile \dimms~DRAM modules with 768 chips from three
different manufacturers to characterize the architectural variation in modern
DRAM chips. We observe similar trends and characteristics in modules from the
same generation, though the absolute number of failures are different. \mycolor{In
Table~\ref{tbl:summary_ava}, we summarize how many DRAM modules we observe
architectural variation out of the test DRAM modules from three major DRAM
vendors.}

\begin{table}[h]
	\centering
	\small{\setlength{\tabcolsep}{12pt}
	\begin{tabular}{ccc} \toprule
							& \trcd (Observed/Tested) & \trp (Observed/Tested)	\\ \midrule
		Vendor A	& 27/30 (90.0\%) 					& 26/30 (86.7\%)					\\
		Vendor B	& 16/30 (53.3\%)					& 4/30 (13.3\%)						\\
		Vendor C	& 14/36 (38.9\%)					& 27/36 (75.0\%)					\\ \midrule
		Total			& \multicolumn{2}{c}{72/96 (75.0\%)}								\\ \bottomrule
	\end{tabular}}
	\caption{DRAM Modules, Observed Architectural Variation}
	\label{tbl:summary_ava}
\end{table}

We make two observations from this table. \mycolor{First, most of the tested
DRAM modules exhibit architectural variation (72 out of 96 modules, 75\%).}
Second, we did not observe architectural variation in \navdimms~DRAM modules.
However, we believe that this is due to the limitation in our infrastructure
that can only reduce timing parameters at a coarser granularity (step of 2.5
ns). As a result, sometimes it is possible to miss the timing where
architectural variation is clearly visible and enter a region where latency is
low enough to make all cells fail. We believe that in real machines where
state-of-the-art DRAM uses much lower clock period (e.g., DDR3-2133: 0.94ns),
architectural variation will be prevalent.

Thus, we have experimentally demonstrated that architectural variation is
prevalent across a large number of DRAM modules and our observations hold true
in most of the modules. We conclude that most modern DRAM chips are amenable to
reducing latency by exploiting architectural variation.

	\section{Mechanisms to Reduce Latency} \label{sec:case}

So far, we have described the phenomenon of architectural variation, studied the
reasons behind it, and provided the experimental results demonstrating the
presence and characteristics of architectural variation. In this section, we
focus on leveraging architectural variation to achieve low DRAM latency while
maintaining reliability. Specifically, we propose two techniques, {\em i)}
architectural-variation-aware online DRAM profiling (AVA Profiling) to determine
how much DRAM latency can be safely reduced while still achieving failure-free
operation and {\em ii)} architectural-variation-aware data shuffling (AVA
Shuffling) to avoid uncorrectable failures (due to lower latency) in systems
with ECC.

\subsection{Architectural Variation Aware Online Latency Profiling}
\label{sec:mech_lowlatency}

Previous works (including our second mechanism, AL-DRAM, in
Chapter~\ref{ch:aldram}) observe that the standard DRAM timing parameter values
are determined based on the worst-case impact of process variation and leverage
this observation to reduce overall DRAM latency during normal operating
conditions~\cite{lee-hpca2015, chandrasekar-date2014}. For instance, AL-DRAM (in
Chapter~\ref{ch:aldram}) observed large process and temperature dependency among
different DRAM cells' access latencies. As a result, DRAM timing parameters can
be lowered using the slowest cells in each DRAM module at current operating
temperatures instead of using standard DRAM timing parameters. These works have
two shortcomings. They {\em i)} assume DRAM manufacturers would determine
reliable timing parameters for multiple operating conditions, {\em ii)} do not
take into account DRAM latency changes over time due to aging and wear out.

One solution to both shortcomings is to profile DRAM characteristics online.
However, this solution has large performance overhead~\cite{singh-mtdt2005,
elm-mtdt1994, rahman-prdc2011}. Our work develops a {\em dynamic} and {\em low
cost} DRAM profiling technique that leverages architectural variation in DRAM.
We call this technique {\em Architectural Variation Aware Online DRAM Profiling}
(or simply {\em AVA Profiling}). The key idea is to {\em i)} separate errors
into two categories, those caused by architectural variation and those caused by
process variation, and then {\em ii)} employ different error mitigation
techniques for these two categories.

\begin{figure}[h]
	\centering
	\subcaptionbox{Process Variation\label{fig:failure_rand}}[0.3\linewidth] {
		\includegraphics[height=1.6in]{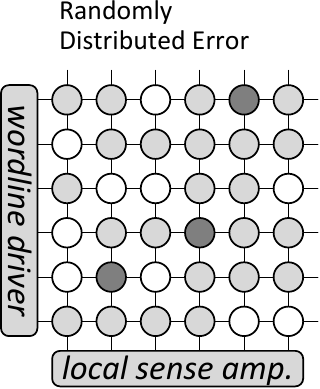}
	}
	\subcaptionbox{Architectural Variation\label{fig:failure_arch}}[0.3\linewidth] {
		\includegraphics[height=1.6in]{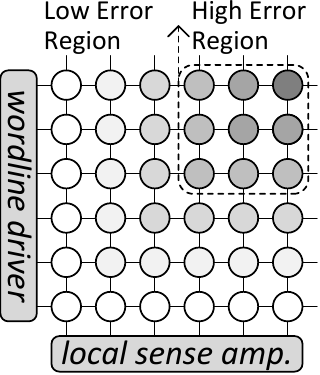}
	}
	\subcaptionbox{Process$+$Architectural Variation\label{fig:failure_both}}[0.35\linewidth] {
		\includegraphics[height=1.6in]{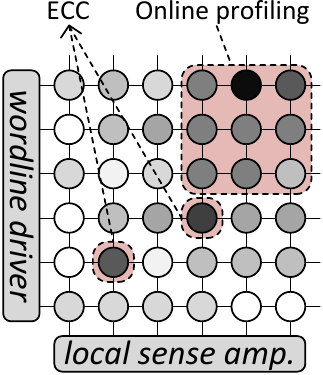}
	}
	\caption{Latency Variation in a Mat (Darker: higher latency)}
	\label{fig:failure}
\end{figure}

{\bf Architectural Variation vs. Process Variation.} The error characteristics
from {\em i)} process variation and {\em ii)} architectural variation are very
different. First, the errors caused by process variation are usually randomly
distributed over the entire DRAM chip~\cite{lee-hpca2015,
chandrasekar-date2014}. Figure~\ref{fig:failure_rand} illustrates random errors
caused by process variation. As darker cells hold less charge, leading to higher
access latency, these darker cells start failing earlier than the lighter cells.
Because these errors are random, they often affect only individual DRAM cells
around them. In this case, existing ECC mechanisms such as SECDED, can detect
and recover these random errors, leading to better DRAM reliability.

The errors caused by {\em architectural variation} are concentrated in specific
regions. When considering only the impact of architectural variation, as shown
in Figure~\ref{fig:failure_arch}, cells that are farther away from the row
driver and local sense amplifier have less charge and hence higher access
latency. Therefore, a mat can be divided into a high error region (the rightmost
and topmost region in a mat) and a low error region (the other regions besides
the high error region). When timing parameters are aggressively reduced, the
first set of errors tend to occur in the high error region.  Furthermore, these
errors tend to be multi-bit errors because high error cells are clustered in a
small region of a mat. These multi-bit errors exceed the error correction
capabilities of a simple ECC (e.g., SECDED) and require a complicated ECC which
usually incurs high area and latency overhead. To avoid these undesirable
multi-bit errors, we periodically profile the high error regions alone, which
incurs much less overhead than profiling the entire DRAM, and tune timing
parameters appropriately based on this profile.

\begin{figure}[h]
	\centering
	\includegraphics[width=0.7\linewidth]{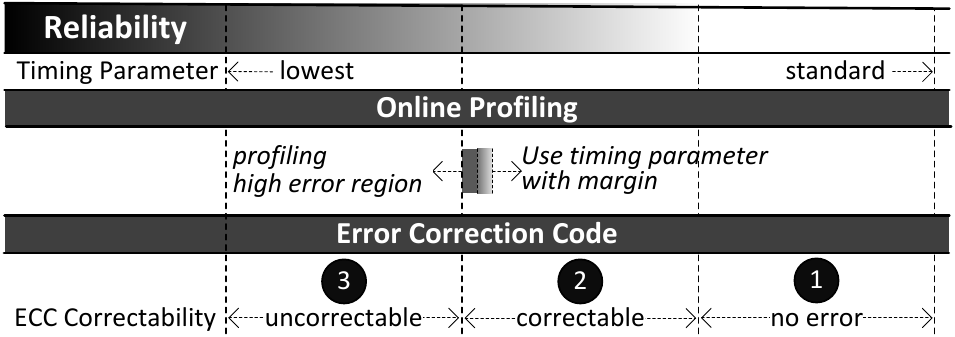}
	\caption{Architectural Variation Aware Online Profiling}
	\label{fig:mech_lowlatency}
\end{figure}

{\bf AVA Profiling Mechanism.} Our AVA Profiling mechanism combines ECC with
online profiling in a synergistic manner, with the goal of reducing DRAM latency
while maintaining high reliability. Figure~\ref{fig:mech_lowlatency} illustrates
reliability reduction due to lowering DRAM timing parameters from the standard
value (right most) to the lowest value (left most). Darker regions represent
more failures (less reliability). Conventional ECC techniques divide the
reliability spectrum into three regions: {\em i)} error free region
(\ding{202}), {\em ii)} correctable error region (\ding{203}), and {\em iii)}
uncorrectable error region (\ding{204}). ECC seeks to maintain a system in the
error free or correctable error regions. However, determining whether a system
is in the correctable region or not is not straightforward. We propose to
achieve this by using online profiling that is aware of architectural variation.

Due to architectural variation, there is a specific region that requires the
highest access latency in DRAM. We rely on DRAM manufacturers to provide
information on the location of this slowest region. The AVA-profiling-based
memory system, then, uses this slowest region to perform online latency
profiling (we call it the {\em latency test region}). Note that the actual data
is not stored in this region. The AVA profiler {\em periodically} accesses this
latency test region and determines the smallest value of DRAM timing parameters
required for reliable operation in this latency test region (e.g., not causing
multi-bit errors). The system then adds a small margin to the timing parameters
obtained from this profiling (e.g., one clock cycle increase) to determine the
timing parameters for the other regions ({\em data region}). Our mechanism
profiles only the slowest region that is most impacted by architectural
variation, thereby incurring low profiling overhead, while achieving low DRAM
latency {\em and} high reliability.

To implement AVA Profiling, the latency test regions (e.g., a row per mat) need
to be identified and avoided when repairing/remapping. Simply avoiding
repair/remapping for the latency test regions (the most vulnerable regions) can
avoid complexity associated with repair/remapping. Since these latency test
regions are reserved only for test and not for data storage, the overhead in
doing so is low. While we show the correlation between error count and
row/column addresses to demonstrate the existence of architectural variation in
Section~\ref{sec:profile}, such correlation is not always necessary to implement
the proposed mechanisms. Our mechanism only needs to know the addresses of the
most vulnerable regions. DRAM companies are very likely to know the most
vulnerable regions (based on their knowledge of the design) of their DRAM
products. Such vulnerable regions can be assigned as test regions, which are
reserved only for test and are not used for general data storage. During the
repair process, such test regions are avoided, preventing repair/remapping
issues.

In order to profile the test region, many possible mechanisms can be employed.
For example, a simple and low overhead mechanism is to integrate the online test
into the DRAM refresh operation. Such a mechanism can keep the implementation
simple, while having minimal effect on memory system performance.

{\bf AVA Profiling with Other Latency Variations in DRAM.} So far, we explained
our mechanism to cover two types of latency variation in DRAM chips, process
variation and architectural variation. However, there can be other latency
variations in DRAM, e.g., PVT variation and VRT (variable retention time). We
have designed our mechanisms with careful consideration of these variations as
well by adopting two error mitigation techniques (ECC and online profiling)
together.

As explained, we divide the DRAM failures into two categories: {\em i)}
localized failures (caused by architectural variation and chip-to-chip process
variation) and {\em ii)} random failures (caused by cell-to-cell process
variation and variable retention time (VRT)), and exploit two different error
mitigation techniques to tackle these two categories of failures, i.e., online
profiling for localized failures and ECC for random failures.

Since the physical dimension of a mat is very small (e.g., 1415.6 $um^2$ for a
$6F^2$ 30nm technology 512 cell $\times$ 512 cell mat), we assume that the
effect of voltage and temperature variation might be similar across a mat.
Chip-to-chip process variation should be the same across each mat. The
cell-to-cell process variation and VRT effects can be covered by ECC.

{\bf Optimizing Performance and Reliability.} Compared to a server system that
uses ECC only for uncertain errors (e.g., alpha particle errors), systems that
employ our mechanism for better performance might have higher failure rate. In
such systems that need high reliability, AVA Profiling can be adapted to suit
the system's higher reliability needs. For example, when determining timing
parameters, AVA Profiling can consider both the profiled lower latency bound in
test regions and the frequency of ECC correctable errors, keeping ECC
correctable errors as low as the required threshold.

\subsection{Architectural Variation Aware Shuffling} \label{sec:mech_shuffling}

\begin{figure}[h]
	\centering
	\subcaptionbox{Conventional Mapping\label{fig:mech_map_ori}}[0.45\linewidth] {
		\includegraphics[height=1.9in]{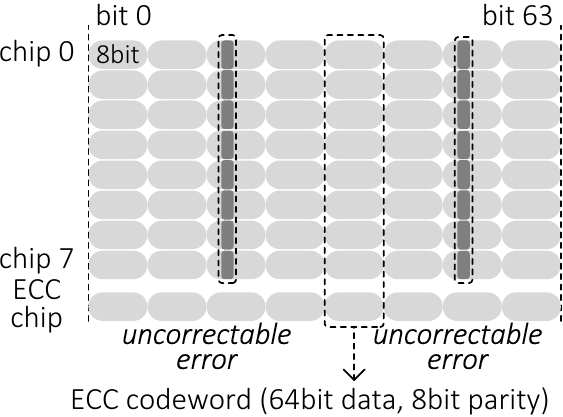}
	}
	\subcaptionbox{Proposed Mapping\label{fig:mech_map_new}}[0.45\linewidth] {
		\includegraphics[height=1.9in]{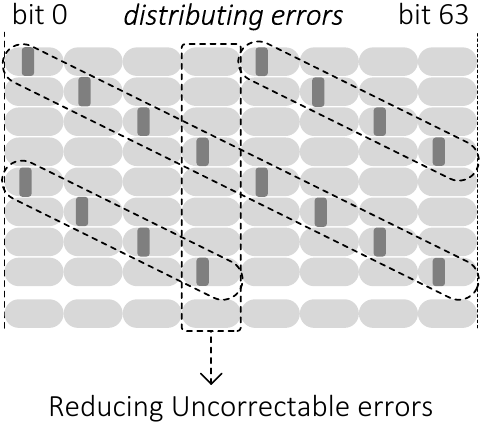}
	}
	\caption{Architectural Variation Aware Data Shuffling}
	\label{fig:mech_map}
\end{figure}

Our second approach focuses on leveraging architectural variation to mitigate
uncorrectable errors in memory systems with ECC. As we observe in
Section~\ref{sec:profile_colint}, when data is read out of a memory channel,
data in specific locations tends to fail more frequently. This happens because
data is delivered from locations which are distributed across a wordline. Due to
the architectural variation in wordline and control signals, it takes longer to
access cells in specific locations compared to cells in other locations, leading
to multi-bit errors in memory systems with ECC.  Figure~\ref{fig:mech_map_ori}
shows the effect of architectural variation in systems with ECC. Data in the
darker grey regions ({\em high-error bit}) tends to be more error-prone than
data in the lighter grey regions. Unfortunately, these high-error bits are
concentrated in a similar location across different chips and hence, they are
part of the same data-transfer burst.

We tackle this problem and mitigate potential uncorrectable errors by leveraging
awareness of architectural variation. Our key idea is {\em to distribute the
high-error bits across different ECC code words}. We call this mechanism {\em
architectural-variation-aware data shuffling (AVA Shuffling).} There are three
potential ways in which such a shuffling mechanism can be implemented. One way
to achieve this is by using eight DRAM chips that have different data-out
mappings. This would require changing the DRAM chip internally, when it is being
manufactured. In this case, since the data mapping is changed internally in the
DRAM chips to shuffle the high-error bits across different ECC code words, the
address decoding mechanism for reads and writes can remain identical across DRAM
chips. The second way is to shuffle the address mapping of DRAM chips within a
DRAM module. We achieve this by connecting the address bus bits in a different
order for different DRAM chips in a DRAM module, leading to different column
addresses being provided by different DRAM chips. Using these two mechanisms, we
can achieve data shuffling in the data output from DRAM, as shown in
Figure~\ref{fig:mech_map_new}. The third possible mechanism is to shuffle the
data from a cache line access in the memory controller such that the high-error
bits are distributed across different code words. The advantage of implementing
the shuffling in the memory controller is that the shuffling mechanism can be
changed. However, this would work only if the memory controller performs error
correction after receiving all the data for a cache line request, as performance
could be degraded from waiting for all data to reach the memory controller.

Interleaving data (or bit) over multiple ECC codeword is not new, but, the
awareness of the most vulnerable regions enables more robustness over a bit
interleaving scheme that is not aware of the most vulnerable regions. AVA
Shuffling, for instance, interleaves data to distribute the most vulnerable
regions over different ECC codewords, leveraging knowledge of architectural
variation. Awareness of architectural variation can enable other efficient
shuffling mechanisms too, for example, using different data (burst) shuffling
orders in different chips in a DRAM module.

\begin{figure}[h]
	\centering
	\includegraphics[width=0.9\linewidth]{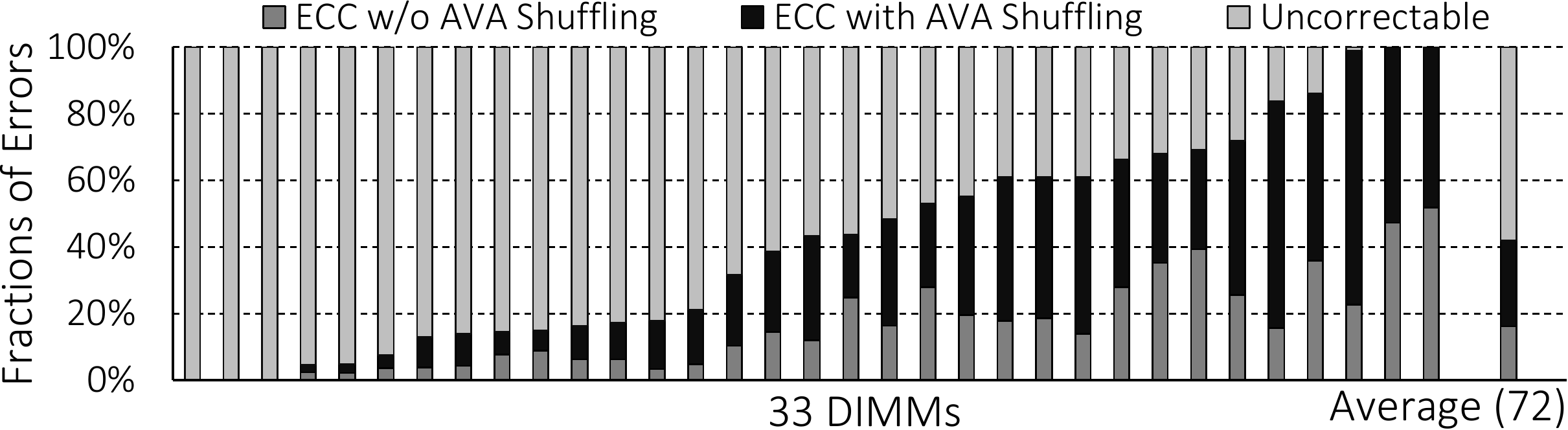}
	\caption{ECC with/without AVA Shuffling}
	\label{fig:mech_map_correction}
\end{figure}

Figure~\ref{fig:mech_map_correction} shows the fraction of correctable errors
using SECDED with/without AVA Shuffling. The Y-axis represents the total
percentage of errors with lower DRAM timing parameters, and the X-axis
represents 33 (randomly selected) DRAM modules.\footnote{The operating
conditions where selected to make sure that there are actually errors, so that
ECC is useful.} We make two observations. First, our mechanism corrects 26\% of
the errors which are not correctable by using {\em only} conventional ECC,
leading to further latency reduction for 24 DRAM modules out of 96 DRAM modules.
Second, both error correction mechanisms (ECC with/without AVA Shuffling) show
correlation in the error correction rate. For example, when ECC without AVA
Shuffling shows high correction rate, ECC with AVA Shuffling also shows high
error correction rate.

\subsection{AVA Profiling/Shuffling vs. Having Stronger ECC}
\label{sec:ava_vs_ecc}

The goal of our proposal is {\em not} replacing ECC mechanisms, but, in
contrast, complementing existing ECC mechanisms to achieve better performance
and reliability.

First, having ECC alone (regardless of ECC strength) is not enough to guarantee
correct operation with maximum latency reduction, since it is not possible to
determine the smallest value for each timing parameter without profiling. AVA
Profiling can do so, enabling maximum latency reduction while leveraging ECC
support to correct random failures. Second, AVA Shuffling enables greater
reliability with any ECC mechanism by distributing possible errors over
different ECC codewords. Third, our work opens up new research opportunities to
exploit architectural variation in combination with different ECC schemes. For
example, variable-strength ECC can exploit awareness of architectural variation
by adjusting ECC strength based on error probability indications/predictions
from architectural variation.

\subsection{DRAM Latency \& Performance Analysis} \label{sec:mech_result}

{\bf DRAM Latency Profiling.} We profile \dimms~DRAM modules, comprising 768
DRAM chips, for potential latency reduction. We use the same test methodology,
described in Section~\ref{sec:pmethod}, which is also similar to the profiling
methodology of Chapter~\ref{ch:aldram} and a previous
work~\cite{chandrasekar-date2014}. We measure the latency reduction of four
timing parameters (\trcd, \tras, \trp, and \twr).

\begin{figure}[h]
	\centering
	\subcaptionbox{READ (\tras$-$\trp$-$\trcd)\label{fig:profile_timing_rd}}[0.45\linewidth] {
		\includegraphics[height=1.5in]{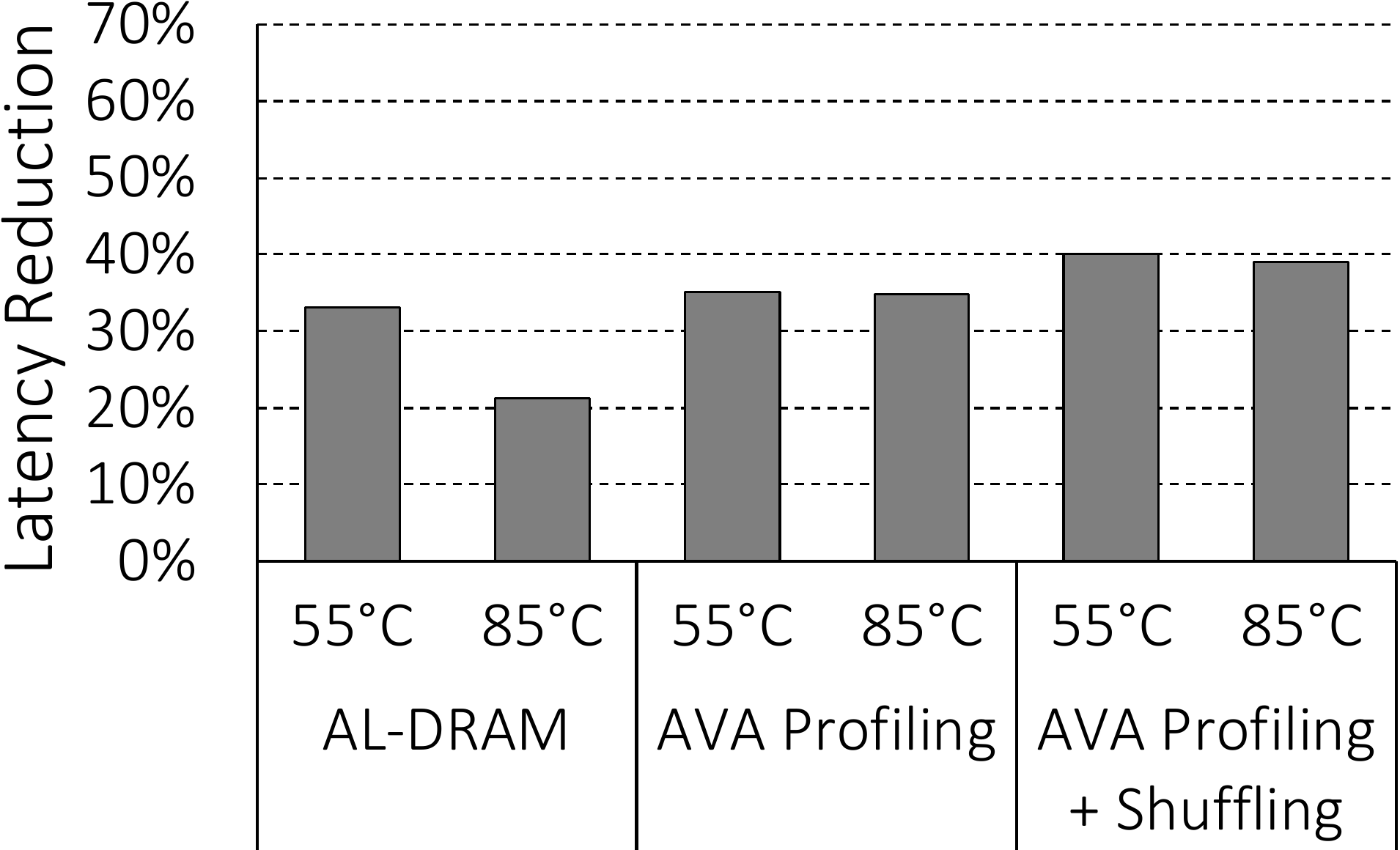}
	}
	\subcaptionbox{WRITE (\twr$-$\trp$-$\trcd)\label{fig:profile_timing_wr}}[0.45\linewidth] {
		\includegraphics[height=1.5in]{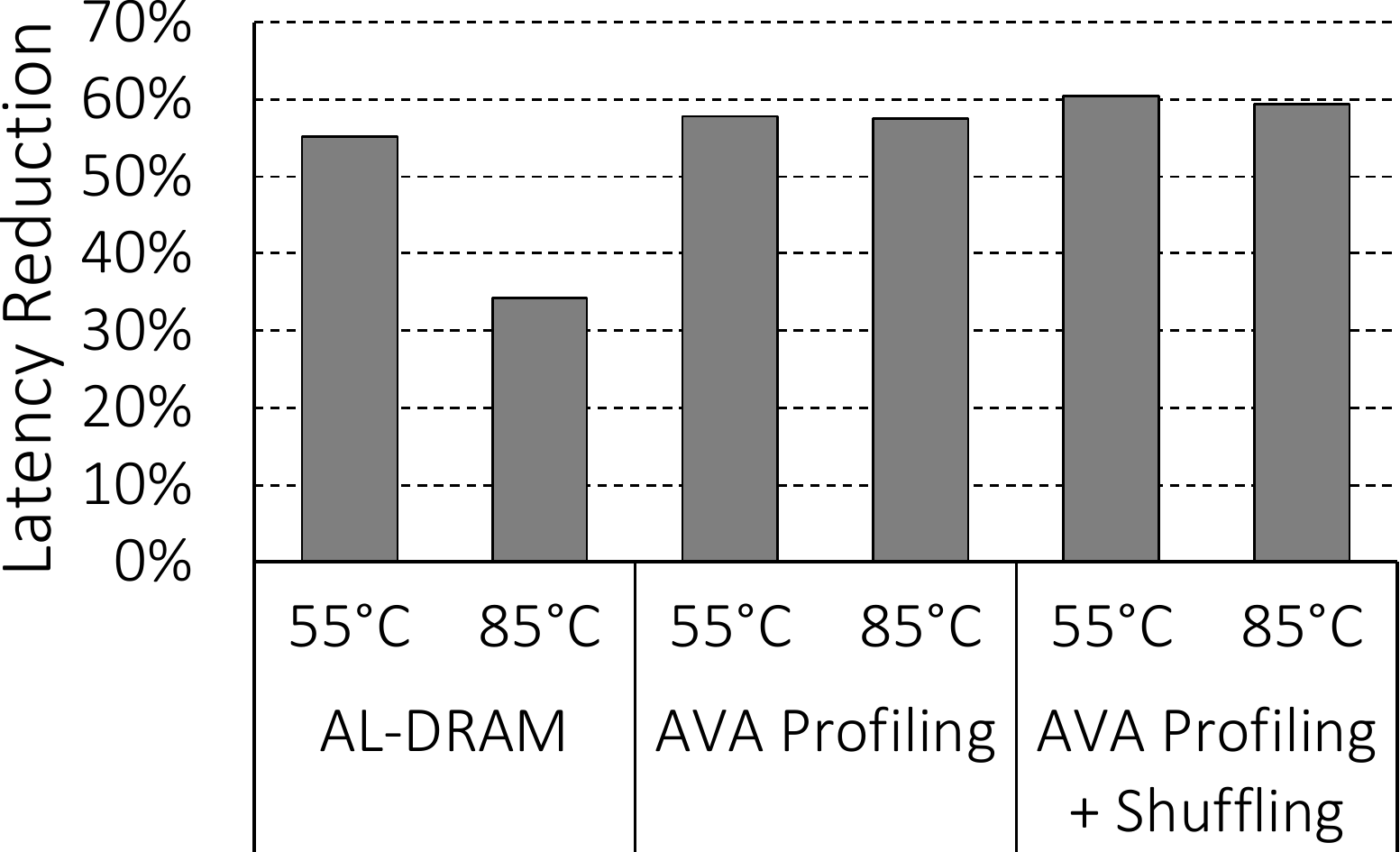}
	}
	\caption{Read and Write Latency Reduction}
	\label{fig:profile_rd_wr}
\end{figure}

Figure~\ref{fig:profile_rd_wr} shows the average latency reduction for DRAM read
and write operations with three mechanisms --- AL-DRAM (our second mechanism in
Chapter~\ref{ch:aldram}), AVA Profiling, and AVA Profiling with Shuffling --- as
the sum of the corresponding timing parameters. We compare these mechanisms at
two operating temperatures, 55\celsius~and 85\celsius.  AL-DRAM mechanism can
reduce the latency for read/write operations by 33.0\% and 55.2\% at 55\celsius,
and 21.3\% and 34.3\% at 85\celsius, respectively.  To integrating ECC (SECDED),
our architectural variation aware online profiling mechanism further reduces the
corresponding latencies by 35.1\% and 57.8\% at 55\celsius, and 34.8\% and
57.5\% at 85\celsius, respectively. Using AVA Shuffling on top of AVA Profiling
enables more latency reduction (by 1.8\% on average). We conclude that our
proposed mechanisms are able to achieve better latency reduction compared to
AL-DRAM at different temperatures. This is mainly because ECC (and also ECC with
AVA Shuffling) can correct many single-bit errors in a ECC codeword.

\begin{figure}[h]
	\centering
	\subcaptionbox{\trcd\label{fig:profile_trcd}}[0.45\linewidth] {
		\includegraphics[height=1.5in]{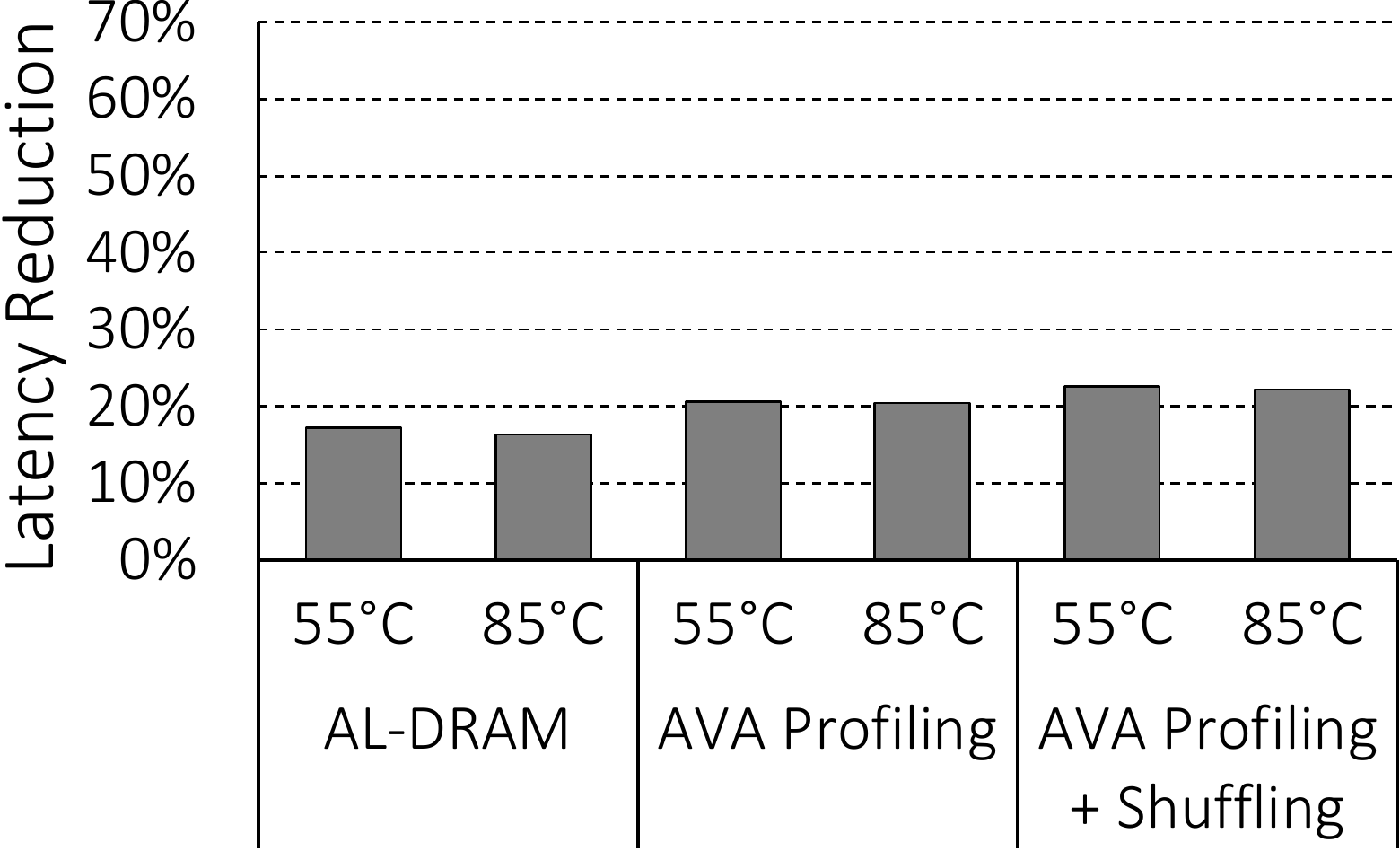}
	}
	\subcaptionbox{\tras\label{fig:profile_tras}}[0.45\linewidth] {
		\includegraphics[height=1.5in]{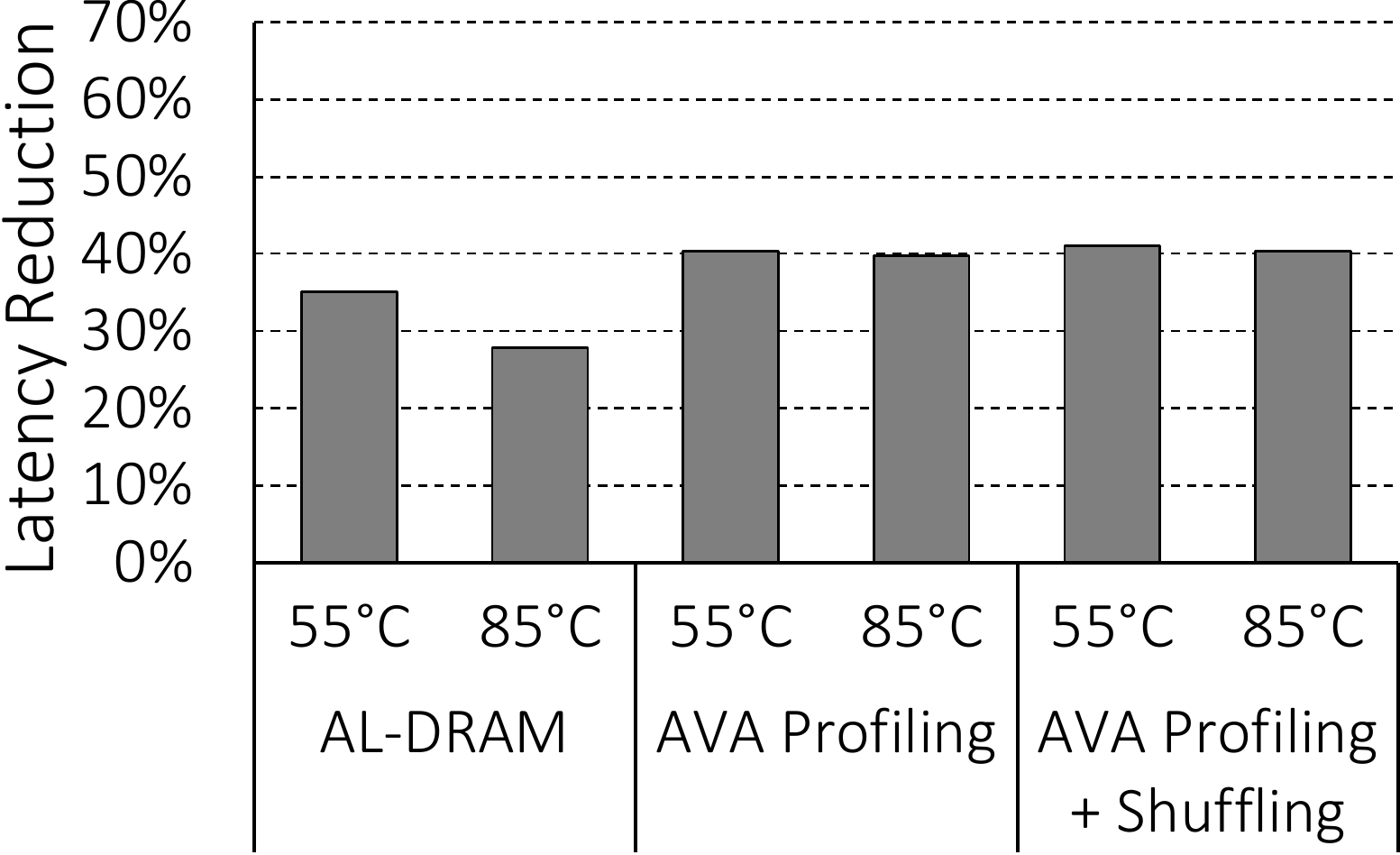}
	}

	\subcaptionbox{\trp\label{fig:profile_trp}}[0.45\linewidth] {
		\includegraphics[height=1.5in]{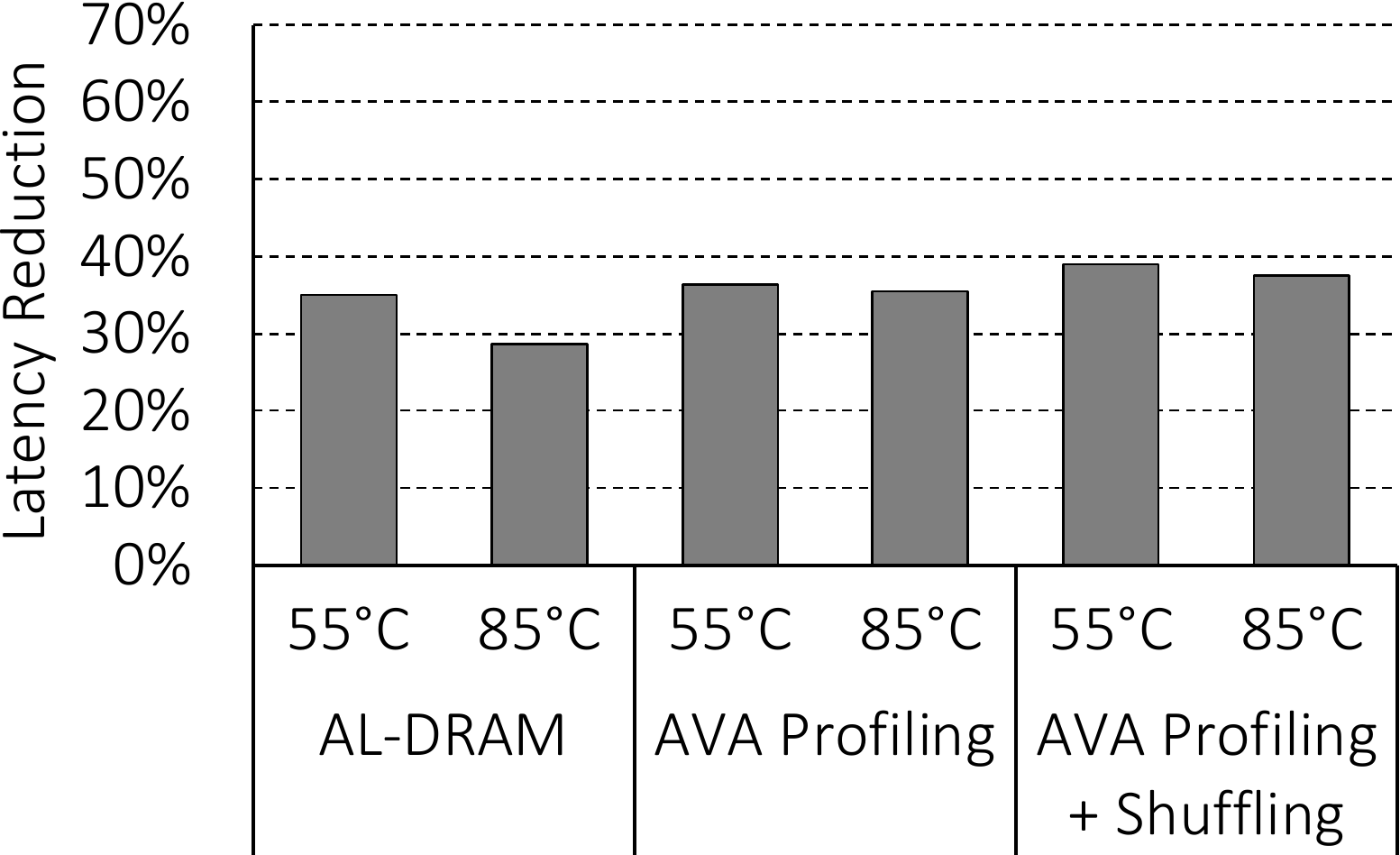}
	}
	\subcaptionbox{\twr\label{fig:profile_twr}}[0.45\linewidth] {
		\includegraphics[height=1.5in]{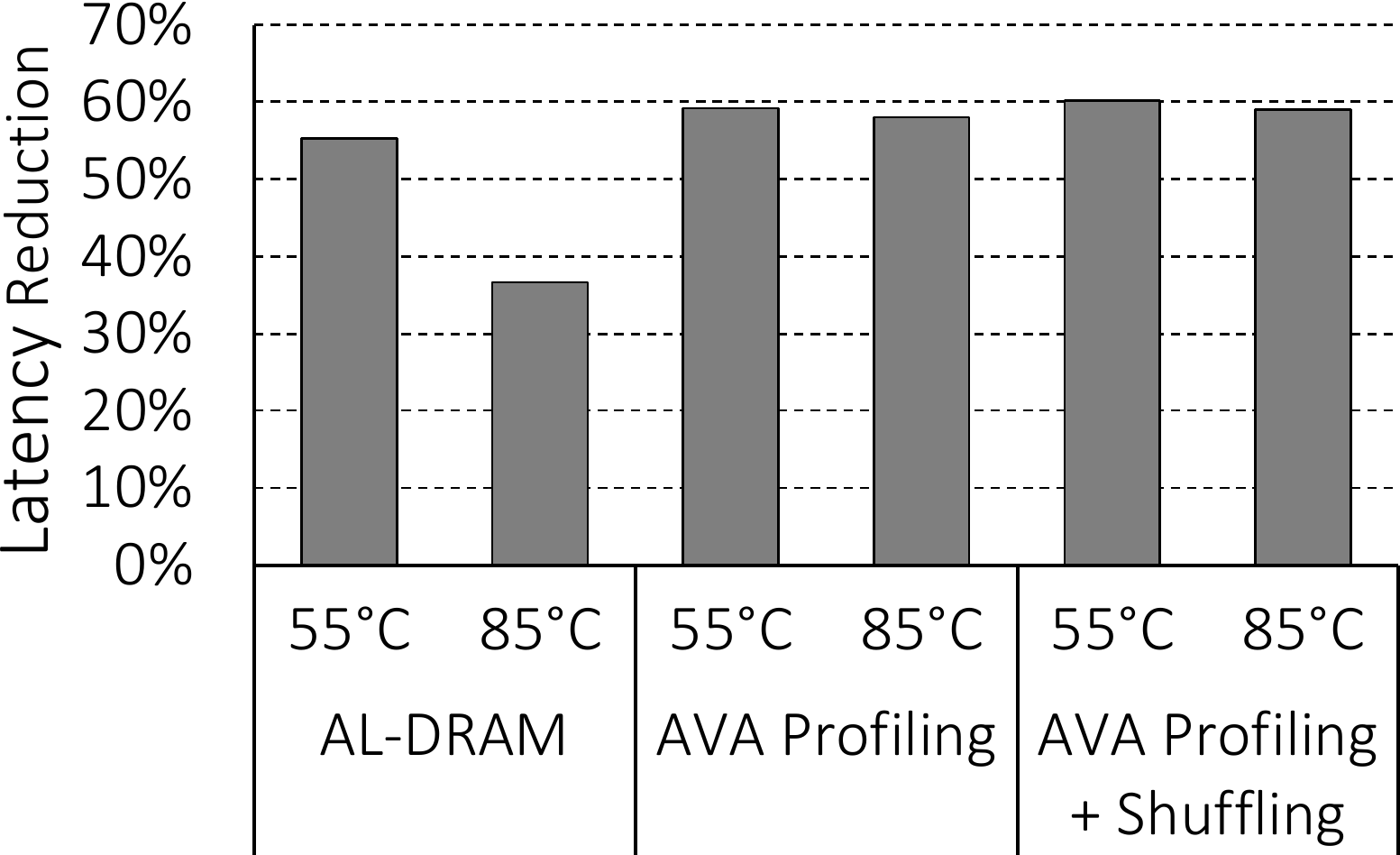}
	}
	\caption{Latency Reduction for each Timing Parameter}
	\label{fig:profile_timings}
\end{figure}

Figure~\ref{fig:profile_timings} shows average latency reduction in each timing
parameter, \trcd, \tras, \trp, and \twr. We compare AVA-DRAM to AL-DRAM at two
operating temperatures, 55\celsius~and 85\celsius. We observe similar trends
with these timing parameters as with the read/write latency results. In this
dissertation, we present only the {\em average} potential reduction for each
timing parameter. We provide detailed characterization of each DRAM module
online at the SAFARI Research Group website~\cite{safari-avadram}.

While our sample size (96 DIMMs and 768 DRAM chips) may not be large enough to
identify the exact amount of architectural variation and possible latency/error
reduction across all DRAM designs, we strongly believe that our results are
consistent enough to support our hypothesis of the existence of architectural
variation. As shown in Section~\ref{sec:profile_dimms}, 75\% of modules show
significant amount of architectural variation which we leverage to enable
latency and error reduction.

{\bf Performance Evaluation.} We evaluate the performance improvement of using
our AVA Profiling mechanism. We use a modified version of
Ramulator~\cite{kim-cal2015}, a fast, cycle-accurate DRAM simulator that is
publicly available~\cite{ramulator}. We use Ramulator combined with a
cycle-level x86 multi-core simulator. Table~\ref{tbl:system} shows the system
configuration we model. We built a cycle-accurate memory model with detailed
timing parameters. We accurately modeled the latency reduction in our
evaluations. Our baseline system has original standard timings (e.g., tRCD
13.75ns for DDR3-1600) and we use reduced timing parameters (e.g., tRCD 11.25ns)
to evaluate our mechanisms.

For workloads, we use Pinpoints tool~\cite{luk-pldi2005, patil-micro2004} to
collect traces. We use 32 benchmarks from SPEC CPU2006~\cite{spec},
stream~\cite{stream}, TPC~\cite{tpc} and GUPS~\cite{gups}, each of which is used
for a single-core workload. We construct 32 two-, four-, and eight-core
workloads -- a total of 96 multi-core workload (randomly selected from the 32
benchmarks). We measure single-core performance using instructions per cycle
(IPC) and multi-core performance using weighted
speedup~\cite{snavely-asplos2000, eyerman-ieeemicro2008} metric. We simulate 100
million instructions after caches are warmed up.

\begin{table}[ht]
	\centering
 	\begin{tabular}{ll}
		\toprule
		Component & Parameters \\
		\midrule
		\multirow{2}{*}{Processor}				& 8 cores, 3.2GHz, 3-wide issue,\\
																			& 8 MSHRs/core, 128-entry inst. window\\
		\multirow{2}{*}{Last-level cache} & 64B cache-line, 16-way associative,\\
   															      & 512KB private cache-slice per core\\
		\multirow{1}{*}{Memory} 					& 64/64-entry read/write queues/controller,\\
									{controller}				& FR-FCFS scheduler\\
		\multirow{1}{*}{Memory system} 		& DDR3-1600~\cite{ddr3}, 2 channels, 2 ranks-per-channel\\
		\bottomrule
	\end{tabular}
\caption{Configuration of Simulated Systems} \label{tbl:system}
\end{table}

\begin{figure}[h]
	\centering
	\includegraphics[width=0.70\linewidth]{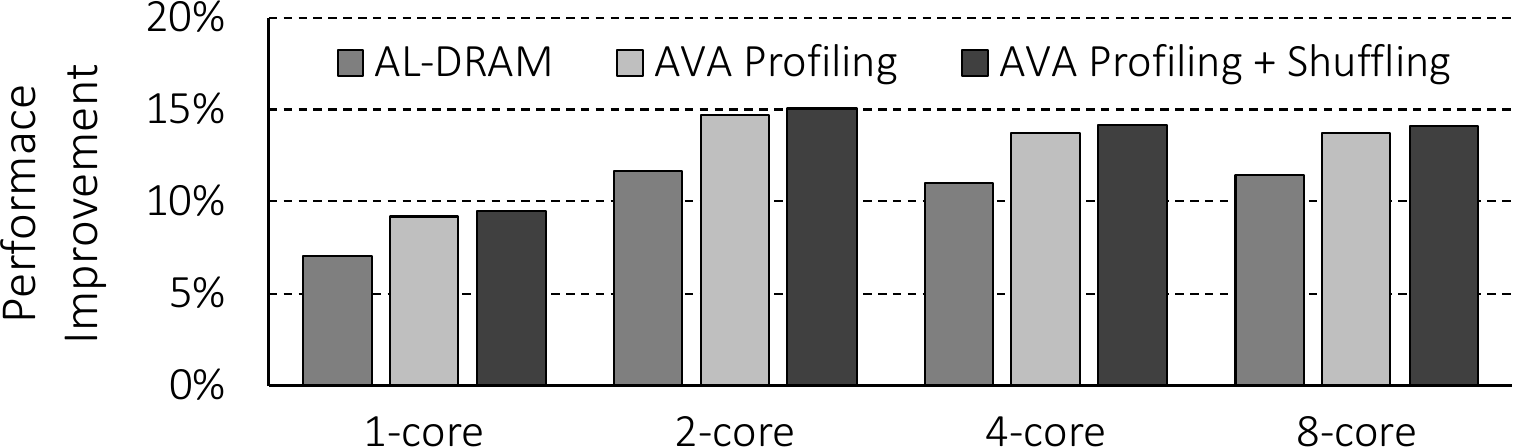}
	\caption{Performance Improvement with AVA-DRAM}
	\label{fig:mech_performance}
\end{figure}

Figure~\ref{fig:mech_performance} shows the performance improvement with AVA
Profiling and AVA Shuffling. We draw two major conclusions. First, AVA Profiling
provides significant performance improvement over the baseline DRAM
(9.2\%/14.7\%/13.7\%/13.8\% performance improvement in
single-/two-/four-/eight-core systems, respectively). This improvement is mainly
due to the reduction in DRAM latency. Second, using AVA Profiling and AVA
Shuffling together provides even better performance improvements (by 0.5\% on
average) due to additional latency reductions with AVA Shuffling. We achieve
this performance while maintaining the DRAM reliability by dynamically
monitoring and optimizing DRAM latency (AVA Profiling).

We also observe that AL-DRAM provides good performance improvement (as high as
11.5\% in two-core system, as also shown in Chapter~\ref{ch:aldram}). As we
discussed in Section~\ref{sec:mech_lowlatency}, AL-DRAM has two major
shortcomings, which our mechanism overcomes by performing low cost online
profiling by exploiting architectural variation. Considering that there are
significant parts of DRAM suffer from aging or post-packaging
failures~\cite{sridharan-sc2012, sridharan-sc2013, meza-dsn2015,
schroeder-tdsc2010, li-atc2007, hwang-asplos2012}, which AL-DRAM cannot handle,
we conclude that our mechanisms provide both better performance and reliability
at low cost over AL-DRAM.

	\section{Summary} \label{ch6:summary}

In this chapter, we provide the first experimental demonstration and
characterization of the variation in memory access latency due the internal
organization of DRAM, which we call {\em architectural variation}. We find that
there is significant variation in the access latency of DRAM cells required for
reliable operation, depending on where cells reside within the DRAM chip. We
show that cell latency is especially affected by how close the cell is to the
peripheral structures that are used to access the cell. We show the existence
and characteristics of architectural variation by the profiling of 96 modern
DRAM memory modules from three major manufacturers.

Based on the extensive understanding of architectural variation, we develop
AVA-DRAM, which consists of two new methods to reduce DRAM latency reliably at
low cost. AVA Profiling is the first technique that can dynamically find and
use the lowest latency at which DRAM can operate reliably. AVA Shuffling
further reduces latency by lowering it to a point that causes multi-bit errors,
while ensuring reliable operation by shuffling data such that these errors
become correctable by ECC. We demonstrate that AVA-DRAM can greatly reduce DRAM
read/write latency, leading to a significant system performance improvement on
a variety of workloads and system configurations.

We conclude that exploiting architectural latency variation inherent in DRAM
using our new techniques provides a promising, reliable, and low-cost way of
significantly reducing DRAM latency.

	\chapter{System Design Guidelines\\ for Heterogeneous Memory Systems}
\label{ch:system_guide}

To integrate low-latency DRAM architectures proposed in this dissertation in
future systems, one required modification is to expose DRAM latency
configurations and operating conditions (e.g., DRAM operating temperature, and
latency variation in different DRAM chips and different areas in a DRAM chip)
to the system (e.g., processor). We call this information {\em configuration
information}. To this end, providing an efficient interface for transferring
such configuration information from DRAM to the system is one of the key
features that is required. In this chapter, we provide guidelines to
seamlessly integrate such an interface within future systems. In \mycolor{the}
following sections, we first explain the required information for enabling
various heterogeneous-latency DRAMs proposed in this dissertation, and then
describe various options to construct an efficient interface between the DRAM
and the system. Note that other works also pointed at the necessity of
constructing such interfaces~\cite{mutlu-imw2013}.

\section{Required Information for Enabling Heterogeneous-Latency DRAM}

In Chapters~\ref{ch:tldram},~\ref{ch:aldram}, and~\ref{ch:avadram}, we
proposed three memory systems that enable heterogeneous latency in DRAM. Each
of these proposals has different configuration information required for
enabling low-latency DRAM architectures. In this Section, we describe examples
for the configuration information.

\subsection{Information for Tiered-Latency DRAM}

Tiered-Latency DRAM divides a subarray into two different latency segments,
the near segment (fast segment) and the far segment (slow segment). To exploit
the Tiered-Latency DRAM based memory system, the processor needs to know
following configuration information.

\squishlist

	\item {\bf Subarray Organization.} As we explained in
	Section~\ref{sec:background}, a DRAM bank is divided into multiple
	subarrays. \mycolor{The processor should be made aware of the subarrays.}
	For example, 4GByte DDR3 DRAM~\cite{micron} consists of 8 banks, each of
	which has 64K rows (each of which is mapped to a value of 16-bit row
	addresses). Assuming that each subarray consists of 512 rows, each bank has
	128 subarrays. Therefore, the 16-bit row address can be divided into two
	portions, {\em i)} address bits for selecting subarrays (7 bits {\em
	subarray address}) and {\em ii)} address bits for selecting rows in a
	subarray (9 bits {\em row address in a subarray}). This information should
	be exposed to the processor.

	The easiest way is to integrate the subarray address and the row address (in
	a subarray) separately into the DRAM specification. For example, similar to
	the conventional DRAM specifications that have bank addresses and row
	addresses separately, the Tiered-Latency DRAM specification might have bank
	addresses, {\em subarray} address, and row address. Alternatively,
	Tiered-Latency DRAM might provide the subarray configuration directly to the
	processor. \mycolor{A simple example data format that contains the subarray
	configuration is a bit vector, each bit of which represents whether the
	corresponding row address bit is {\em i)} subarray address or {\em ii)} row
	address in a subarray. For example, a bank consists of 32 rows in total (16
	bits) and is subdivided into 64 subarrays (7 bits), each of which consists
	of 512 rows (9 bits). Then, the 16-bit row address can be divided into 7-bit
	subarray address and 9-bit row address in each subarray. TL-DRAM can provide
	this information to the processor as a 16-bit vector of {\tt
	1111111000000000b} where ``1'' represents the corresponding row address bit
	is a part of {\em subarray address} and ``0'' represents the corresponding
	row address bit is a part of {\em row address in a subarray.}}

	\item {\bf Near Segment and Far Segment Organization.} Tiered-Latency DRAM
	then subdivides a subarray into the near segment and the far segment. For
	example, 512 rows in a subarray can be divided into 32 rows in the near
	segment and 480 rows in the far segment. In this organization, 512 rows in a
	subarray consist of 32 32-row groups, and one of the 32-row groups can be
	the near segment.

	There can be many data formats to transfer this configuration information from
	TL-DRAM to the processor (specifically, the memory controller in the
	processor). One simple example is \mycolor{to provide} the address range of
	the near segment in a 512-row subarray from TL-DRAM to the memory controller.
	\mycolor{For the case of 32-row near segment out of 512-row subarray (total
	9-bit row address),} the near segment can be assigned the range of row
	addresses from {\tt 000000000b} to {\tt 000001111b} (initial 32 rows). The
	amount of data to be transferred to the memory controller in this case is only
	18 bits, which can be transferred over a specialized interface.

	\item {\bf Operating Mode of Tiered-Latency DRAM.} Tiered-Latency DRAM is a
	substrate that can be leveraged in many ways. Since Tiered-Latency DRAM can
	support multiple modes \mycolor{(e.g., using the near segment as
	hardware-managed exclusive cache to the far segment, using the near segment
	as hardware-managed inclusive cache to the far segment, profile-based page
	mapping to the near segment, and so on)}, the system is required to select a
	current operation mode. This can be integrated to {\em Mode Register Set},
	which already exists in the conventional DRAM~\cite{micron}.

\squishend

\subsection{Information for Adaptive-Latency DRAM}

As we explained in Chapter~\ref{ch:aldram}, Adaptive-Latency DRAM leverages
both DRAM operating conditions (e.g., DRAM operating temperature) and process
variation in DRAM to optimize DRAM timing parameters. To this end, the
following information needs to be exposed to the processor.

\squishlist

	\item {\bf DRAM Operating Conditions.} We show that DRAM can be accessed
	with different timing parameters in different operating conditions. There
	can be many operating conditions, for example, operating temperature (which
	is leveraged by AL-DRAM), power consumption, voltage level, and so on.
	These are pieces of information that are different in different DRAM chips.
	To leverage the information for better performance, the information needs to
	be exposed to the processor. For example, \mycolor{in} an AL-DRAM-based
	heterogeneous memory system, the processor needs to know the {\em operating
	temperature} of each DRAM module. The amount of information to be
	transferred depends on the resolution of temperature and frequency of
	transfer. For example, 10-bit temperature information can provide 0.1 degree
	resolution in the range of temperature, 0 -- 100\celsius.

	\item {\bf DRAM Timing Parameters for each Operating Condition.} To leverage
	process variation, each DRAM module has a set of timing parameters for each
	operating condition (e.g., operating temperature) and requires transfer of
	the timing parameters to the processor. Each timing parameter can be
	represented \mycolor{using} several bits (usually \mycolor{fewer} than 10
	bits) since \mycolor{most timing parameter values} are less than 1024. A
	request of 64 bytes can transfer 512 bits in total, which are enough to
	transfer 51 timing parameters per request.

\squishend

\subsection{Information for AVA-DRAM}

As we explained in Chapter~\ref{ch:avadram}, AVA-DRAM leverages the subarray
organization to reduce DRAM latency \mycolor{while} maintaining DRAM
reliability. The key observation in AVA-DRAM is that the distance between
cells and peripheral logic leads to variation in latency. The required
information to leverage AVA-DRAM is as follows.

\squishlist

	\item {\bf Inherently Slower Cells in a DRAM Mat.} AVA-DRAM leverages the
	existence of the inherently slower cells in a DRAM mat. A naive way to
	expose this information is \mycolor{by} making the order of addresses in
	line with the distance from peripheral logic. This works for row addresses,
	each of which maps to a row in a bank. However, the hierarchical
	organization of the column access path might not be easy to represent the
	full organization with simple information. Instead, we propose that DRAM
	provides \mycolor{to the processor} the {\em exact address} \mycolor{range}
	of the inherently slower cell region. In this approach, DRAM provides
	multiple sets of row and column addresses which show the worst latency in a
	subarray. For this, the required amount of data for a set is a 9-bit row
	address and a 10-bit column address.

\squishend

\section{Interface to Heterogeneous-Latency DRAM}

To transfer the required {\em configuration information} from DRAM to the
memory controller, we consider multiple constraints: {\em i)} additional
overhead for integrating \mycolor{this information into the} interface, {\em
ii)} performance impact for transferring data, and {\em iii)} security issues
related to exposing internal DRAM organization \mycolor{to the processor}. We
briefly describe each as follows.

{\bf Low Area Overhead.} A naive way to transfer the data is \mycolor{to
integrate a new set of I/O interfaces} (e.g., specific wire connections and
corresponding peripheral logic) for the configuration information. However,
this approach might significantly increase the DRAM chip area \mycolor{and
energy consumption by increasing} the number of wire connections between DRAM
modules and the system. An alternative way is \mycolor{to use the} existing
data bus in the memory channel. To this end, \mycolor{we might need to add}
{\em i)} specialized commands for accessing the configuration information and
{\em ii)} peripheral logic to support the new commands. This approach does not
require any additional physical wires in the memory channel. However, since it
is not possible to access data for executing instructions during the transfer
of the configuration information of DRAM operating conditions and
organization, this approach might lead to performance degradation. Next, we
discuss the performance impact \mycolor{of} transferring the configuration
information from the DRAM to the system \mycolor{using} the existing memory
channel.

{\bf Low Performance Impact.} As we described, using the existing data bus for
transferring the configuration information consumes memory channel bandwidth
periodically, leading to performance degradation. There are two factors to
determine how much the performance impact is. The first factor is how much
data is needed to be transferred. Considering that the amount of data for
configuration is very modest, we expect that the performance degradation might
not be significant.

The second factor is how frequently the configuration information needs to be
transferred. There are two categories of the configuration information based
on the frequency of transferring data, {\em i)} static configuration
information and {\em ii)} dynamic operating conditions. The organization of
DRAM (e.g., subarray organization, the near and far segment organization, and
so forth) does not change over time. Therefore, the static configuration
information needs to be transferred {\em only once} \mycolor{at boot} time.
However, since the dynamic operating conditions change over time, DRAM needs
to update the processor with its new operating conditions. The frequency of
the updates depends on how frequently the operating conditions change. As we
described in Section~\ref{sec:factors_temp}, since the operating temperatures
in real systems do {\em not} change drastically (\mycolor{e.g.,} less than
0.1\celsius\xspace per second even in the worst case), we expect that the
performance degradation is even less \mycolor{for this type of information
transfer}. For AVA-DRAM, \mycolor{we have not answered the question of} how
frequently the latency parameters should be updated to avoid potential errors.
However, it might be in the order of few hours, few days and few months,
\mycolor{which leads to small} performance impact. We leave this question
\mycolor{of} ``how frequently should the latency parameters be updated'' to
future work.

{\bf Security Issues.} While exposing the internal organization of DRAM is the
easiest way for leveraging heterogeneous-latency DRAM architectures, it can
potentially lead to security issues. Kim et al.~\cite{kim-isca2014} showed
that accessing a row frequently enough before refresh \mycolor{can lead} to
errors in its adjacent rows. Exposing the internal DRAM organization and
address mapping might lead to even more security holes. To address this issue,
a possible alternative design is that {\em i)} the processor accesses DRAM
with a standard and simplified address interface, and {\em ii)} DRAM maps the
issued addresses to the physical rows and columns with address {\em
scrambling}. In this design, the processor can access different latency
regions without exposing DRAM organization to \mycolor{outside} of DRAM,
maintaining the \mycolor{existing security of the memory system.}

\section{Summary}

In this chapter, we describe the required configuration information that needs
to be communicated between the DRAM and the system to leverage
heterogeneous-latency DRAM \mycolor{architectures proposed in this
dissertation}. We provide the \mycolor{various} considerations \mycolor{that
need to be taken into account} to implement the interfaces to communicate the
configuration information. We hope that these discussions help \mycolor{in the
design} of \mycolor{the system-DRAM} interface to take advantage of
heterogeneous-latency DRAM.

	\chapter{Conclusions and Future Research Directions}
\label{ch:conclusion}

In this dissertation, we present three techniques to lower DRAM latency at low
cost. These techniques enable or exploit heterogeneity in DRAM. They are {\em
i)} Tiered-Latency DRAM, which enables heterogeneous bitlines at low cost by
dividing the long bitline into two different latency segments, {\em ii)}
Adaptive-Latency DRAM, which optimizes DRAM latency for the common operating
conditions, and {\em iii)} AVA-DRAM, which lowers DRAM latency by exploiting
architectural variation in the internals of DRAM organization.

We first present a new DRAM architecture, Tiered-Latency DRAM (TL-DRAM), that
provides both low latency and low cost-per-bit in Chapter~\ref{ch:tldram}. Our
key observation is that existing DRAM architectures present a trade-off between
cost-per-bit and access latency. One can either achieve low cost-per-bit using
long bitlines or low access latency using short bitlines, but not both. Our key
idea to leverage this trade-off for lowering DRAM latency is to segment a long
bitline using an isolation transistor, creating a segment of rows with low
access latency while keeping cost-per-bit on par with commodity DRAM. We
present mechanisms that take advantage of our TL-DRAM substrate by using its
low-latency segment as a hardware-managed cache. Our most sophisticated cache
management algorithm, Benefit-Based Caching (\mbbc), selects rows to cache that
maximize access latency savings. We show that our proposed techniques
significantly improve system performance by 12.8\% and reduce energy
consumption by 23.6\% across a variety of systems and workloads.

While TL-DRAM reduces DRAM latency significantly, it requires changing existing
DRAM architecture, which might limit the applicability of the proposed
techniques (even though the changes are low cost, i.e, only 3\% additional DRAM
area).

Towards achieving the goal of lowering DRAM latency without changing the DRAM
architecture, we approach to leverage {\em latency slack} that exists in modern
DRAM. To this end, we build an FPGA-based DRAM test infrastructure that can
profile DRAM characteristics, which {\em i)} can flexibly change DRAM timing
parameters, {\em ii)} apply specific data and access patterns, and {\em iii)}
maintain the operating conditions (e.g., ambient temperature). Based on
characterizations of 115 DRAM modules, we introduce two new mechanisms to
reduce DRAM latency without changing DRAM architecture.

We first propose Adaptive-Latency DRAM (\ALD), a simple and effective mechanism
for dynamically tailoring the DRAM timing parameters for the current operating
condition without introducing any errors in Chapter~\ref{ch:aldram}. The
standard DRAM timing constraints are grossly overprovisioned to ensure correct
operation for the cell with the lowest retention time at the highest acceptable
operating temperature. We make the observation that a significant majority of
DRAM modules do {\em not} exhibit the worst case behavior and that most systems
operate at a temperature much lower than the highest acceptable operating
temperature, enabling the opportunity to significantly reduce the timing
constraints. Based on these observations, \ALD dynamically measures the
operating temperature of each DRAM module and employs timing constraints
optimized for {\em that DRAM module at that temperature}. Results of our
latency profiling experiments on \DIMMs modern DRAM modules show that our
approach can significantly reduce four major DRAM timing constraints by
\trcdCold\%/\trasCold\%/\twrCold\%/\trpCold\xspace averaged across all \DIMMs
DRAM modules tested. This reduction in latency translates to an average 14\%
improvement in overall system performance across a wide variety of
memory-intensive applications run on a real multi-core system.

While \ALD is a simple and effective mechanism to reduce DRAM latency, it has
two overheads that may limit its use. First, it requires DRAM manufacturers
would determine reliable timing parameters for multiple operating conditions,
increasing test cost during the manufacturing time. Second, it do not take into
account DRAM latency changes over time due to aging and wear out.

To reduce DRAM latency on a more dynamic and achievable manner, we leverage our
experimental demonstration and characterization of the variation in memory
access latency due the internal organization of DRAM, which we call {\em
architectural variation}, in Chapter~\ref{ch:avadram}. We find that there is
significant variation in the access latency of DRAM cells required for reliable
operation, depending on {\em where} cells reside within a DRAM chip. We show
that cell latency is especially affected by how close the cell is to the
peripheral structures that are used to access the cell.

Building upon an extensive understanding of architectural variation developed
by the characterization of 96 modern DRAM modules from three major
manufacturers, we develop AVA-DRAM, which leverages architectural variation,
which consists of two new mechanisms to reliably reduce DRAM latency at low
cost. AVA Profiling is the first technique that can dynamically find and use
the lowest latency at which DRAM can operate reliably. AVA Shuffling further
reduces latency by lowering it to a point that causes multi-bit errors, while
ensuring reliable operation by shuffling data such that these errors become
correctable by ECC. We demonstrate that AVA-DRAM can greatly reduce DRAM
read/write latency (by 40.0\%/60.5\%, respectively), leading to a significant
system performance improvement (by 14.7\%/13.7\%/13.8\% on 2-/4-/8-core system,
respectively) on a variety of workloads.

We conclude that our three mechanisms provide promising low-latency low-cost
DRAM designs. We have shown that each of them significantly improves overall
system performance. We believe that both our new characterization of DRAM
latency (across a large number of real DRAM chips) and the mechanisms we have
developed on the basis of our analysis of DRAM architecture and our
experimental DRAM characterization enable the development of other mechanisms
for improving DRAM latency and perhaps reliability.

\section{Future Research Directions} \label{sec:future}

We describe potential research directions to further reduce DRAM latency and
mitigate the negative impact of high DRAM latency in the next sections.

\subsection{Optimizing Timing Parameters in 3D-Stacked DRAM}

3D-stacked DRAMs~\cite{jedec-hbm, hmc, hmc10, hmc11, lee-taco2016,
loh-isca2008} are likely to provide high-bandwidth and energy-efficient memory
systems in the future. Similar to commodity DRAM, it is possible to optimize
latency in 3D-stacked DRAM for the common case. To this end, the first step is
investigating the operating conditions to determine the worst-case and
common-case conditions in 3D-stacked DRAM. Then, the next step is developing
mechanisms to reduce access latency for the common-case.

In addition to the possible latency reduction for the common-case, we expect
that 3D-stacked DRAM might exhibit much higher reduction across the stack due
to two major reasons. First, integrating DRAM chips on top of each other
dissipates heat to the adjacent layer, which creates a temperature gradient
across the layers. As a result, 3D-stacked DRAMs might exhibit much higher
variations in operating temperature within DRAM. Second, the power delivery
network of 3D-stacked DRAM drives power from bottom to top, creating a
heterogeneous voltage supply across the layers. As variation in voltage and
temperature directly impact the latency of DRAM access, these two variations
might enable more opportunities to reduce access latency in 3D-stacked DRAM.

Considering the variations present across different layers of 3D-stacked DRAM,
we believe that investigating these variations at different operating
conditions leads to devising new 3D-stacked DRAM architectures that enable
higher latency reduction, and developing mechanisms that adaptively optimize
access latency for each layer.

\subsection{Optimizing Refresh Operations for the Common-Case}

Retention time of a DRAM cell depends on the data stored in the neighboring
cells, referred to as a phenomenon called {\em data pattern dependence}.
Depending on the data content stored in DRAM, some rows can exhibit much higher
retention time than others~\cite{liu-isca2013, khan-sigmetrics2014}. In today's
systems, this heterogeneity in retention time is not exploited to reduce
refresh operations. All DRAM rows are refreshed at the same rate determined by
the worst-case data patterns tested during the manufacturing time. Typical
program contents from different applications may not exhibit these worst-case
patterns frequently and may not require to be refreshed at the nominal rate. We
can leverage this longer retention time of the common-case data patterns in
programs to reduce the refresh overhead.

We believe that investigating the retention time in real DRAM chips with
different program content using our FPGA-based testing infrastructure and
analyzing the possible reduction in refresh rate for the common case would
enable us to develop dynamic refresh mechanisms that minimize refresh
operations in different regions of memory depending on the current program
content.

\subsection{System Design for Heterogeneous-Latency DRAM}

We have shown in this dissertation that future low-latency DRAM architectures
enable faster access by either {\em i)} creating a smaller faster region within
DRAM or by {\em ii)} exploiting the variations present in different regions of
memory at different operating conditions. Both of these approaches enable a
heterogeneous-latency DRAM design, where some regions provide faster access to
data. In order to maximize the benefits of such a heterogeneous latency DRAM,
it is necessary to design appropriate interfaces and software-hardware
collaborative mechanisms across the system stack.

We believe that investigating the following two approaches leads to designing
an end-to-end system leveraging these heterogeneous DRAMs. First approach is
designing a system that takes advantage of the faster segments to maximize the
latency benefits. We can maximize the use of faster segment by allocating
frequently used or more critical pages to this region. Exposing the latency
variations to the memory controller and system software and allocating critical
pages appropriately can provide better system performance. These mechanisms
introduce heterogeneous-latency aware mapping and partitioning mechanisms that
effectively leverage the additional memory latency tiers.

Second approach is investigating system-level mechanisms to leverage the
operating conditions that result in faster DRAM access. For example, hot spots
in DRAM can increase the access time of cells belonging to that region. A
hardware-software collaborative technique can map pages in a way that accesses
are spread out to different regions. This mechanism can ensure that DRAM can be
operated at the lowest possible latency by avoiding hot spots. Similarly, a
system can be designed to avoid the worst-case pattern by using intelligent
coding to map that data to the common-case pattern, thereby avoiding the
worst-case refresh rates.

We conclude and hope that this dissertation, with the analysis \&
characterization of many DRAM chips it provides and the new low-latency
low-cost DRAM architectures it introduces, enables many new ideas in DRAM
design for low latency and high reliability.

\subsection{New Interfaces for Heterogeneous Main Memory}

As we discussed in Chapter~\ref{ch:system_guide}, heterogeneous main memory
should have a specialized interface to expose its organization (e.g., subarray
organization, fast/slow regions, and so forth) and operating conditions (e.g.,
temperature) to the processor. Fundamentally, the heterogeneous main memory
system has four pieces of information that needs to be transferred through
memory channel: {\em i)} command and address, {\em ii)} data for write, {\em
iii)} data for read, and {\em iv)} the configuration information. We believe
that this dissertation opens up a new research direction to investigate the new
DRAM interfaces for future heterogeneous main memory systems.

The key question is how the wire connections in the memory channel should be
organized. There may be three major approaches at a high level, One extreme
approach is providing a separate and dedicated wire connection for each of the
four information. The other extreme approach is transferring all information
through one unified wire connection. For example, the conventional DDR
interface transfers data for read and write through the same wire connection.
Alternatively, the two extreme approaches can be integrated into a flexible
memory channel, which can change the role of wire connections based on the
operating conditions (e.g., operating temperature, strength of voltage supply,
and so on) and the memory characteristics of the current workloads. We leave
this research question, what is the most efficient interface for future
heterogeneous main memory, to future work.

\subsection{Reducing Latency of Emerging Memory Technologies}

Due to difficulties in DRAM scaling, several new technologies are being heavily
investigated as potential alternatives to DRAM that can replace or augment DRAM
as main memory~\cite{lee-cacm2010, mutlu-superfri2015, meza-weed2013,
ren-micro2015, pelley-isca2014, kang-msst2015, lu-iccd2014}. These technologies
include Phase Change Memory (PCM)~\cite{raoux-ibm2008, lee-isca2009,
qureshi-isca2009, qureshi-micro2009, dhiman-dac2009, lee-ieeemicro2010,
lee-cacm2010, meza-iccd2012, yoon-taco2014, wong-ieee2010}, Spin-Transfer
Torque Magnetic Memory (STT-MRAM)~\cite{kultursay-ispass2013, meza-iccd2012,
li-arxiv2015}, Resistive RAM~\cite{wong-ieee2012} or
memristors~\cite{chiu-jssc2012, niu-aspdac2012}, and Conductive Bridging Memory
(CB-RAM)~\cite{kund-iedm2005}. Most of these technologies are at least as slow
as DRAM, and in many cases slower, and therefore they likely exacerbate the
main memory latency bottleneck tackled in this dissertation. We believe the
techniques developed in this dissertation can inspire similar or related
approaches to reduce latency in such emerging main memory technologies. We also
believe the techniques developed in this dissertation can be adapted to NAND
Flash memory technology~\cite{luo-msst2015, cai-date2013, cai-itj2013,
cai-iccd2012, cai-sigmetrics2014, cai-iccd2013, cai-dsn2015, cai-hpca2015,
meza-sigmetrics2015, lu-tc2015}, to reduce the latency of flash memory chips.
In particular, exploiting heterogeneity in both emerging and existing
technologies seems promising beyond DRAM. We leave the exploration of this
exciting and promising direction to future works and dissertations.

\section{Final Summary}\label{sec:final_summary}

We conclude and hope that this dissertation, with the analysis \&
characterization of many DRAM chips it provides and the new low-latency
low-cost DRAM architectures it introduces, enables many new ideas in DRAM
design for low latency and high reliability.

\small
\singlespacing
\bibliographystyle{abbrv}
\bibliography{ref}

\end{document}